\documentclass[prd,aps,showpacs,onecolumn,preprintnumbers,nofootinbib,amssymb]{revtex4}
\UseRawInputEncoding
\usepackage{graphicx}
\usepackage{epsf}
\usepackage{amsmath}
\usepackage{epstopdf}
\usepackage{bm}
\usepackage{color}
\usepackage{tabularx}
\usepackage{enumitem}
\usepackage{float}
\usepackage{array,booktabs}
\usepackage{footnote}
\usepackage{threeparttable}
\usepackage{amssymb,epsf}
\usepackage{latexsym}
\usepackage{epsfig}
\usepackage{multirow}
\usepackage{bm}
\usepackage{caption}
\usepackage{xcolor}
\usepackage[colorlinks=true,linkcolor=blue]{hyperref}
\def\e3p{$\eta \rightarrow 3 \pi$}

\def\d_2{\left(u_3 h_0^2\right)}
\def\e_3^a{\left(u_4 h_0\right)}
\def\C4{u_2}
\def\c2{\left(- u_1 h_0^2\right)}

\begin{document}

\title{Spinless glueballs in generalized linear sigma model}

%\hfill{\normalsize\vbox{%
% }}\\
%{}}

\author{Amir H. Fariborz \footnote{Email: fariboa@sunypoly.edu}\\}

\affiliation{Department of Mathematics and Physics, State University of New York, Polytechnic Institute, Utica, NY 13502, USA}

\date{\today}

\begin{abstract}
Within the framework of the generalized linear sigma model, a comprehensive analysis of scalar and pseudoscalar glueballs and their mixing with mesons is presented.    The Lagrangian of the model contains two chiral nonets,    a quark-antiquark type and a two-quark two-antiquark type.   The pseudoscalar and scalar  glueballs are introduced through their connections with the  axial and the trace anomalies of QCD, respectively.   It is found that in order to satisfy the axial anomaly and at the same time accurately generate all seven eta masses, it is necessary to include at least two pseudoscalar glueballs, a physical one and an unphysical one that gets integrated out and yields an effective instanton-type term which is needed in generation of the eta masses.  At the leading order,  which corresponds to keeping effective terms in the Lagrangian that contain no more than eight underlying quarks and antiquarks, the mass spectrum of the model is worked out and shown to be  in complete agreement with experiment. The quark and glue contents of the isosinglet scalars below 2 GeV,  and of the isosinglet pseudoscalars up to around 2.2 GeV,  are analyzed in detail and their correlations with the scalar  glueball condensate are examined.  Decay widths of isosinglet scalars as well as different self consistencies within this framework  are used to probe the glueball condensate and thereby estimate the quark and glue contents of these states.  In the pseudoscalar sector,  the state that is dominantly made of glue is clearly a state with mass above 2  GeV. In the scalar sector,  the identification of glue contents is less certain and  in principle  the three isosinglets in the 1.5-2.0 GeV can contain substantial glue.   These glue contents are determined as functions of the  scalar glueball condensate which is the key quantity in this analysis.

\end{abstract}


\maketitle

%%%%%%%%%%%%%%%%%%%%%%%%%%%%%%%%%%%%%%%%%%%%%%%%%%%%%%%%%%%%%%%%%%%%%%%%%%%%%%%%%%%%%%%%%%%%%%%%%%%%%%%%%%%%%%%%%%%%%%

\section{introduction}

Although the existence of glueballs was theoretically speculated more than fifty years ago \cite{72_Fritzsch_econf,75_Fritzsch_NCA30}, their direct experimental verification  remains an open question to date. The fact that the ground-state quark-antiquark spectroscopy cannot account for all observed scalar and pseudoscalar states in the low-energy region, suggests that some of these states may have perhaps more complicated substructure such as higher excitations of quark-antiquars, tetraquarks, glue-rich contents, or a hybrid of these  possibilities.   This has led to a considerable body of literature on various aspects of these topics (see for example reviews   \cite{24_PDG,24_PDG_Rev,23_Vadacchino_2305,22_Klempt_2211,21_Llanes_Estrada_EPJST230,16_Pelaez_PR685,13_Ochs_JPG40,07_Klempt_PR454}).  
Different studies of glueballs have included lattice QCD 
\cite{93_Hulsebos_PLB309,99_Morningstar_PRD60,06_Chen_PRD73,12_Gregory_JHEP10,22_Chen_CPC47,24_Morningstar_PoS004}, QCD sum-rules \cite{79_Shifman_NPB147,84_Narison_ZPC84,11_Harnett_NPA850}
chiral perturbation theory \cite{22_Alb_PRL101,11_Nebreda_PRD84}, linear sigma model \cite{79_Schechter_PRD21,80_Rosenzweig_PRD21,81_Rosenzweig_PRD24,11_Janowski_PRD84,13_Eshraim_PRD87,18_Fariborz_PRD98,19_Fariborz_PLB790,19_Fariborz_IJMPA34,21_Fariborz_NPA1015,24_Giacosa_PRD109}, nonlinear sigma model \cite{04_Fariborz_IJMPA19,06_Fariborz_PRD74,15_Fariborz_PRD91,15_Fariborz_PRD92A}, 
Bethe-Salpeter equation \cite{13_Meyers_PRD87,15_Sanchis-Alepuz_PRD92,20_Kaptari_Few61} and many other  approaches 
\cite{83_Carlson_PRD27,86_Jaffe_AP168,96_Szczepaniak_PRL76,03_Szczepaniak_PLB577,05_Chanowitz_PRL95,11_Dudal_PRL106,13_Windisch_PRD87,14_Dudal_PLB732,15_Frere_PRD92,15_Cheng_PRD92,20_Huber_EPJC80,20_Souza_EPJA56,22_Rodas_EPJC82}.

Characterization of glueballs gets naturally entangled with properties of mesons with the same quantum numbers.  The surplus of scalar and pseudoscalar mesons in the low-energy region  requires a classification that extends beyond a simple quark-antiquark picture. The five isosinglet scalars $f_0(500)$, $f_0(980)$, $f_0(1370)$, $f_0(1500)$ and $f_0(1710)$ are more than what is needed to fill two scalar nonets below 2 GeV.     The three states above 1 GeV, $f_0(1370)$, $f_0(1500)$ and $f_0(1710)$,   have long been under scrutiny for  high glue contents, with  some variation of conclusion on which one is likely to be the dominant glue holder \cite{96_Amsler_PRD53,97_Anisovich_PLB395,97_Close_PLB397,98_Narison_PLB509,01_Close_EPJC21,02_Amsler_PLB541,05_Close_PRD71,06_Cheng,00_Lee_PRD61,14_Janowski_PRD90,15_Brunner_PRL115,21_Sar_PLB816,22_Rodaz_EPJ82}. Similarly, the isosinglet pseudoscalars $\eta(547)$, $\eta'(985)$, $\eta(1295)$, $\eta(1405)$, $\eta(1475)$ and $\eta(1760)$ are beyond the four needed  to form two  pseudoscalar nonets below 2 GeV.

To investigate scalar glueballs and their mixings with mesons, it seems necessary to include both mixing with quark-antiquarks as well as with tetraquarks for the following reason. While the tetraquark model of the light scalar mesons below 1 GeV \cite{77_Jaffe_PRD15} is broadly accepted and captures the main features of these states, a {\it pure} tetraquark assignment for these states, tohether with  a {\it pure} quark-antiquark assignment for those above 1 GeV, do not capture the more refined characteristics of these states \cite{00_Black_PRD61}.   This implies the need for consideration of mixing between the tetraquark and quark-antiquark components and  leads to a more complete framework for understanding the scalar mesons below 2 GeV.    Such a mixing is not a mere mathematical convenience and can be intuitively understood by noting that some of the light scalars such as $f_0(500)$ and $K_0^*(700)$ are very broad and  naturally overlap with the nearby states.  Therefore, it seems that a comprehensive framework for study of scalar glueballs should include not only their mixings with the quark-antiquarks, but also with the tetraquarks.   This is the basis of our approach in this work within the framework of the generalized linear sigma model. 
   
Since our objective is to investigate both the scalar and pseudoscalar glueballs, linear sigma model which naturally incorporates scalars and pseudoscalars as chiral partners,  becomes a good choice for our purpose.   To fully span all  desired states, the framework should include a quark-antiquark chiral nonet and a four-quark chiral nonet together with a scalar glueball  and a pseudoscalar glueball.
This  approach to studies of glueballs can be particularly useful as this framework is  formulated in terms of hadronic fields and can be used to analyze experimental data directly.

The generalized linear sigma model without glueballs was first proposed in \cite{01_Black_PRD64} and formulated in \cite{05_Fariborz_PRD72}.  It was  further developed in \cite{09_Fariborz_PRD79} and applied to various low-energy scattering and decay processes in \cite{11_Fariborz_PRD84,14_Fariborz_PRD90,15_Fariborz_PRD92,22_Fariborz_EJPC82} in reasonably close agreement with experiment.   The glueballs were added to this framework in \cite{18_Fariborz_PRD98,19_Fariborz_PLB790} in which the most general Lagrangian was developed.      Since the general Lagrangian includes many terms, for practical calculations a general strategy that can be systematically improved needs to be worked out.     The number of underlying quark and antiquak lines in each term was used (in both cases of with and without glueballs) to classify these terms, and  those with fewer lines were considered to be favored.   In the leading order, terms with eight or less quark and antiquark lines were considered and this will be used in the present work as well.   Even with this approximation, there are about 20 free parameters that have to be determined by inputting various experimental data. 
As the first step in applying this model, the framework was analyzed in the SU(3) flavor limit in \cite{19_Fariborz_IJMPA34,21_Fariborz_NPA1015}
in which the mixing of SU(3) singlets with glueballs were studied in detail.    While the SU(3) limit is more tractable and provides useful insights into the mixing among glueballs and mesons \cite{19_Fariborz_IJMPA34,21_Fariborz_NPA1015},  for a complete description, the SU(3) symmetry breaking has to be taken into account and that is studied in the present work.

In Sec. II, we give a brief review of the generalized linear model without glueballs followed by inclusion of the glueballs in this framework in Sec. III.
In Sec. IV we give the details of the parameter determination and the general predictions.  We analyze the decay widths of the isosinglet scalars in Sec. V,  and then in Sec. VI we use these decay width results,  as well as other constraints, to further filter the general predictions of Sec. IV and present a more refined predictions.   A summary and discussion of the results is given in Sec. VII.   Additional details are given in the appendixes.

%%%%%%%%%%%%%%%%%%%%%%%%%%%%%%%%%%%%%%%%%%%%%%%%%%%%%%%%%%%%%%%%%%%%%%%%%%%%%%%%%%%%%%%%%%%%%%%%%%%%%%%%%%%%%%%%%%%%%%

%%%%%%%%%%%%%%%%%%%%%%%%%%%%%%%%%%%%%%%%%%%%%%%%%%%%%%%%%%%%%%%%%%%%%%%%%%%%%%%%%%%%%%%%%%%%%%%%%%%%%%%%%%%%%%%%%%%%%%

\section{Generalized linear sigma model without glueballs}

%%%%%%%%%%%%%%%%%%%%%%%%%%%%%%%%%%%%%%%%%%%%%%%%%%%%%%%%%%%%%%%%%%%%%%%%%%%%%%%%%%%%%%%%%%%%%%%%%%%%%%%%%%%%%%%%%%%%%%

In this section we  briefly review the generalized linear sigma model without glueballs of Refs. \cite{05_Fariborz_PRD72,09_Fariborz_PRD79} which paves the way for the more complicated case in which the glueballs are included (next section). The generalized linear sigma model Lagrangian is formulated in terms of two chiral nonets:  $M$,  a quark-antiquark type,  and $M'$,  a two quark two antiquark type.   At the quark level,  nonet $M'$ can have three distinct constructions, one of them of molecular type and two others that are of the diquark-antidiquark type with different color and spin constructions \cite{05_Fariborz_PRD72}.   At the schematic quark level, there is a linear relationship among these  three constructions \cite{08_Fariborz_PRD77} and therefore only two of them can be considered independent and thus, at the schematic quark level, $M'$  can in principle be considered as a linear combination of two of these three four-quark composites. 
Of course,  all three constructions transform in the same way under chiral transformation,  and therefore, in this framework which is formulated at the mesonic level, there is no immediate way to identify which specific underlying four-quark construction $M'$ has.   

Under chiral transformation both of these nonets transform in the same way
\begin{eqnarray}
	M & \rightarrow & U_L M U_R^\dagger, \nonumber \\
	M' & \rightarrow & U_L M' U_R^\dagger,
	\label{E_chiral_trans}
\end{eqnarray} 
where the unitary, unimodular transformations $U_L$ and $U_R$ act on the left handed $\left[q_L = \frac{1}{2}\left( 1 + \gamma_5 \right) q\right]$ and right
handed $\left[ q_R = \frac{1}{2}\left( 1 - \gamma_5 \right) q\right]$ quark projections.
Under U(1)$_A$ transformation $q_{aL}
\rightarrow e^{i\nu} q_{aL}$, $q_{aR} \rightarrow e^{-i\nu} q_{aR}$,  and
therefore $M$ and $M'$ transform differently under this transformation
\begin{eqnarray}
	M & \rightarrow & e^{2i\nu} M, \nonumber \\
	M' & \rightarrow & e^{-4i\nu} M'.
	\label{E_U1A_trans}
\end{eqnarray}
The U(1)$_A$ transformation (\ref{E_U1A_trans}) is how we can distinguish the quark-antiquark nonet $M$ versus the four-quark nonet $M'$.

The two chiral  nonets $M$ and $M'$  can be expressed in terms of the scalar nonets $(S,S')$ and the  pseudoscalar nonets $(\phi, \phi')$
consistent with their quark compositions:
\begin{eqnarray}
	M &=& S +i\phi, \nonumber  \\
	M^\prime &=& S^\prime +i\phi^\prime.
	\label{E_M_Mp_def}
\end{eqnarray}
The diagonal elements of $S$ and $S'$ can develop vacuum expectation values
\begin{eqnarray}
	\left< S_a^b \right> & = & \alpha_a \delta_a^b, \nonumber \\
	\quad \quad \left< S_a^{\prime b} \right> & = &
	\beta_a \delta_a^b,
	\label{E_S_Sp_VEVs}
\end{eqnarray}
and induce spontaneous chiral symmetry breaking. In the isospin invariant limit $\alpha_1=\alpha_2$ and $\beta_1=\beta_2$. 

The Lagrangian of the model has the general structure
\begin{equation}
	{\cal L} = - \frac{1}{2} {\rm Tr}
	\left( \partial_\mu M \partial_\mu M^\dagger
	\right) - \frac{1}{2} {\rm Tr}
	\left( \partial_\mu M^\prime \partial_\mu M^{\prime \dagger} \right)
	- V_0 \left( M, M^\prime \right) - V_{\rm SB},
	\label{E_GLSM_Lag}
\end{equation}
where $V_0(M,M')$ contains terms that are invariant under SU(3)$_L \times$ SU(3)$_R$, and $V_{\rm SB}$ is the flavor symmetry breaking term.    In principle, $V_0(M,M')$ can contain many terms (for example, even if $V_0(M,M')$ is limited to renormalizable terms, this potential contains more than 20  terms \cite{05_Fariborz_PRD72}).  Also, some of these terms break U(1)$_{\rm A}$.   At the practical level, an approximation scheme  that determines the leading terms, and that also maps out the subsequent improvements, is needed.   Since effective vertices with too many quark and antiquark lines are expected to be less bound and therefore less favored,   an approximation approach in which terms   in  $V_0(M,M')$ are classified in terms of their total number of quark and antiquark lines was considered in \cite{09_Fariborz_PRD79}.  In addition,  to ensure that $V_0(M,M')$ exactly reflects the underlying  axial anomaly of QCD, following Refs. \cite{80_Rosenzweig_PRD21,81_Rosenzweig_PRD24},   a specific natural log function for generating this anomaly was modeled in \cite{09_Fariborz_PRD79}.   In this approach, the terms that break U(1)$_{\rm A}$ are only included in this natural log function according to the prescription given in  \cite{80_Rosenzweig_PRD21,81_Rosenzweig_PRD24}.
In the generalized linear sigma model of \cite{09_Fariborz_PRD79},  the part that breaks U(1)$_{\rm A}$ anomaly is the extended version of the form given in \cite{80_Rosenzweig_PRD21,81_Rosenzweig_PRD24} in    
which two chiral nonets $M$ and $M'$ are incorporated. As a result, $V_0(M,M')$ takes a form that is consist of a part that is invariant under both  SU(3)$_L \times$ SU(3)$_R$ as well as U(1)$_{\rm A}$ transformations, and a part that is specifically designed to implement the axial anomaly and is given in terms of a specific  natural log functions of terms that break U(1)$_{\rm A}$. 

In the leading order,  $V_0$ contains terms  with eight or fewer quark plus antiquark lines at each effective vertex
and  is given by:
\begin{eqnarray}
	V_0 &=& -c_2 {\rm Tr} \left( M M^\dagger \right)
	+ c_4^a {\rm Tr} \left( M M^\dagger M M^\dagger \right)
	+ d_2 {\rm Tr} \left( M^\prime M^{\prime \dagger} \right)
	+ e_3^a \left( \epsilon_{abc} \epsilon^{def} M_d^a M_e^b {M'}_f^c + h.c. \right)\nonumber \\
	&&	
	+ c_3 
	\left[  
	\gamma_1 
	{\rm ln}\, \left(\frac{{\rm det}\, M}{{\rm det}\, 
		M^{\dagger}}\right)
	+ \left(1 - \gamma_1\right)  
	{\rm ln}\, \left(
	\frac{ {\rm Tr} \, (M {M'}^\dagger) }
	{ {\rm Tr}\,  (M' M^{\dagger})}
	\right)
	\right]^2,
	\label{E_GLSM_V0}
\end{eqnarray}
where all terms  are invariant under chiral transformation (\ref{E_chiral_trans}).  The last term ($c_3$) breaks the U(1)$_{\rm A}$  in a manner that replicates  the known U(1)$_{\rm A}$ anomaly in QCD \cite{80_Rosenzweig_PRD21,81_Rosenzweig_PRD24} and is needed to generate the mass spectrum of the etas.   As we will see in the next section, the $c_3$ term is  derived from coupling of an effective pseudoscalar glueball field to the  terms in the square bracket  together with a  mass term  for the glueball field (with the wrong sign), and then the glueball field is integrated out to obtain the $c_3$ term in (\ref{E_GLSM_V0}).  It is shown in \cite{08_Fariborz_PRD77} that when the $c_3$ term is linearized, it is consistent with the quark-level 't Hooft's  instanton  description of U(1)$_{\rm A}$ anomaly \cite{76_tHooft_PRD76}.

There are also many flavor symmetry breaking terms in  $V_{\rm SB}$, even in the renormalizable case \cite{05_Fariborz_PRD72}.  To limit the number of these terms, we assume that the leading term should mock up both the SU(3)$_L\times$ SU(3)$_R$ and U(1)$_A$
transformation of the quark mass terms in the fundamental
QCD Lagrangian
\begin{equation}
	V_{\rm SB}=-{\rm Tr}\, [A(M+M^{\dagger})]=-2 {\rm Tr}\, (AS),
	\label{E_GLSM_VSB}
\end{equation}
in which the
diagonal matrix $A$ is
\begin{equation}
	A = {\rm diag} (A_1,A_2,A_3).
	\label{E_A_def}
\end{equation}
The diagonal  elements are proportional to the
light quark masses $(m_u, m_d, m_s)$.

Overall, this framework gives a reasonably coherent picture for the properties of the isosinglet and isodoublet pseudoscalars and scalars.   In \cite{09_Fariborz_PRD79}, the mass spectrum was studied  and the substructure of these states were estimated in detail. It was found that the light isotriplet and isodoublet pseudoscalar states [$\pi$ and $K$] are,  expectedly,  of quark-antiquark type while their heavier states [$\pi(1300)$ and $K(1460)$] have significant four-quark admixtures. It was also found that for  the scalar states, there is a significant mixing among quark-antiquark and four-quark components \cite{09_Fariborz_PRD79}, with isotriplet and isodoublet light scalars [$a_0(980)$ and $K_0^*(700)$] being closer to four-quark states,  whereas their heavier states [$a_0(1450)$ and $K_0^*(1430)$]  are closer to  quark-antiquark states. The case of isosinglet scalars and pseudoscalars are more involved due to their mixing with glueballs and the framework of the generalized linear sigma model without glueballs can only provide a rough picture for these states which can be helpful for comparison with the extension of this model when glueballs are included. Moreover, since  states like $f_0^*(500)$ and $K_0^*(700)$ are broad, their final-state interactions with their decay products should be taken into account.  Using K-matrix unitarization method,  these states were probed in appropriate scattering amplitude, such as in $\pi\pi$ scattering in \cite{11_Fariborz_PRD84} in which a mass of $477 \pm 8$ MeV and a decay width of $398 \pm 107$ MeV were found for $f_0(500)$ in agreement with PDG \cite{24_PDG}.   Similarly the generalized linear sigma mode framework was applied to studies of $\pi K$ scattering in \cite{15_Fariborz_PRD92},  and a mass of 670-770 MeV and a decay width of 640-750 MeV were found for $K_0^*(700)$ in agreement with PDG values \cite{24_PDG}.     With the model parameters fixed in $\pi \pi$ and $\pi K$ scatterings \cite{11_Fariborz_PRD84,15_Fariborz_PRD92}, the model was then applied to the analysis of $\pi\eta$ scattering in \cite{22_Fariborz_EJPC82} in which a mass of $984 \pm 6$ MeV and a decay width of  $108 \pm 30$ MeV were found for the $a_0(980)$ in close agreement with PDG \cite{24_PDG}.   

With the success of the generalized linear sigma model without glueballs in describing the properties of  isodoublets and isotriplets, as well as its success in providing an initial understanding  of the isosinglet states,   it is natural to extend the model to include the glueballs in order to refine the predictions of the isosinglet states.    An overview of this extension is given in the next section.

%=====================================================================================================================

%%%%%%%%%%%%%%%%%%%%%%%%%%%%%%%%%%%%%%%%%%%%%%%%%%%%%%%%%%%%%%%%%%%%%%%%%%%%%%%%%%%%%%%%%%%%%%%%%%%%%%%%%%%%%%%%%%%%%%%%

\section{Generalized linear sigma model with glueballs}

%%%%%%%%%%%%%%%%%%%%%%%%%%%%%%%%%%%%%%%%%%%%%%%%%%%%%%%%%%%%%%%%%%%%%%%%%%%%%

\subsection{The Lagrangian}

We extend the Lagrangian described in (\ref{E_GLSM_Lag}) to include both the scalar and the pseudoscalar glueballs.   The general structure of this extended  Lagrangian is studied in \cite{18_Fariborz_PRD98} and will be briefly reviewed in this section. In addition, the part of this extended Lagrangian that models the U(1)$_{\rm A}$ will be further enhanced to a more complete form than what is given in \cite{18_Fariborz_PRD98}.   The Lagrangian contains the following parts:

\begin{eqnarray}
	{\cal L}&=&-\frac{1}{2}{\rm Tr}(\partial^{\mu}M\partial_{\mu}M^{\dagger})-\frac{1}{2}{\rm Tr}(\partial^{\mu}M'\partial_{\mu}{M'}^{\dagger})
	-{1\over 2} (\partial_{\mu} h)(\partial_{\mu} h) - {1\over 2} (\partial_{\mu} g) (\partial_{\mu} g) - V,\nonumber \\
%---------------------------------------------------
    -V &=&  f + f_{\rm A} + f_{\rm S} + f_{\rm SB}.
	\label{E_L_Def_MMp2g}
\end{eqnarray}
where $M$ and $M'$ are defined in (\ref{E_M_Mp_def}), $h$ and $g$ are respectively the scalar and the pseudoscalar glueball fields,   $f(M, M', g, h)$ is the part   that is invariant under chiral, axial and scale transformations.  The terms   $f_{\rm A}$ and $f_{\rm S}$ model the U(1)$_{\rm A}$  axial and scale anomalies, respectively, and  $f_{\rm SB}$ is the explicit symmetry breaker due to quark masses.

In the minimal extension of this model (which, as discussed in previous section,  corresponds to keeping effective terms with no more than eight quark and antiquark lines), function  $f$ contains the following terms        
\begin{eqnarray}
	f(M, M', g, h) &=&
	- \left(
	u_1 h^2 {\rm Tr}[MM^{\dagger}]
	+ u_2{\rm Tr}[MM^{\dagger}MM^{\dagger}]+
	u_3 h^2 {\rm Tr}[M^{\prime}M^{\prime \dagger}]+ u_4 h (\epsilon_{abc}\epsilon^{def}M^a_dM^b_eM^{\prime c}_f+h.c.)+
	\right.
	\nonumber \\
	&&
	\left.\hskip .5cm  u_5 h^4 + u_6  h^2 g^2  + \cdots\right),
	\label{E_f_terms}
\end{eqnarray}
where $u_1\cdots u_6$ are free parameters that will be determined from experimental inputs in the next section.     The scalar glueball field $h$ can develop vacuum expectation value:
\begin{equation}
	\langle h\rangle = h_0
\end{equation}
 and contribute to the spontaneous chiral symmetry breaking, in addition to the spontaneous chiral symmetry breaking generated by the expectation values of $S$ and $S'$ in (\ref{E_S_Sp_VEVs}).

The parts $f_{\rm A}$ and $f_{S}$  model the  well-known axial and trace anomalies of QCD given by:
\begin{eqnarray}
	&&\partial^{\mu}J^5_{\mu}=\frac{{\mathfrak g}^2}{16\pi^2}N_F{\widetilde F}F=G,
	\nonumber\\
	&&\theta^{\mu}_{\mu}=\partial^{\mu}D_{\mu} = -\frac{\beta({\mathfrak g}^2)}{2 {\mathfrak g}}FF=H,
	\label{E_anomalies}
\end{eqnarray}
where $J^5_{\mu}$ is the axial current,  $D_{\mu}$ is the dilation current, ${\mathfrak g}$ is the coupling constant,  $N_F$ is the number of flavors, $F$ and ${\widetilde F}$ are the $SU(3)_C$ field tensor and  is its dual,  and $\beta({\mathfrak g}^2)$ is the beta function.   At the mesonic level, the axial anomaly was first modeled  in \cite{80_Rosenzweig_PRD21,81_Rosenzweig_PRD24} in the context of a single-nonet linear sigma model by introducing a glueball with unphysical mass term and then integrating it out by its equation of motion. This was further  
extended in the generalized linear sigma model with  two chiral nonets in \cite{05_Fariborz_PRD72}, and was subsequently shown in \cite{08_Fariborz_PRD77} that when this axial anomaly term is linearized it agrees with the instanton description of this anomaly by t' Hooft \cite{76_tHooft_PRD76}.   This instanton term is necessary in order to generate the mass spectrum of etas.

Since in this work we intend to study the mixing of glueballs with isosinglet states, we need to keep the pseudoscalar  glueball field in the Lagrangian instead of integrating it  out.  However,  if we only have one pseudoscalar glueball that we do not integrate out, then the instanton term is not generated and consequently we cannot generate the eta masses correctly.  Therefore, we need at least two pseusoscalar glueballs:  One unphysical pseudoscalar glueball  which is integrated out to generate the necessary instanton term, and one physical pseudoscalar glueball that will remain in the Lagrangian and interacts with the meson fields.  Satisfying the axial anomaly determines the  intricate way in which the combined effects of these two glueballs should be taken into account.

The first step in extending the generalized linear sigma model to include glueballs was taken in  \cite{18_Fariborz_PRD98} and further studied in the SU(3) flavor limit in \cite{19_Fariborz_PLB790,19_Fariborz_IJMPA34,21_Fariborz_NPA1015}.     In these works,   for simplicity, the instanton term was neglected.  Since the model was applied in the SU(3) flavor limit,  which provides  only an approximation of the physical picture,  neglecting the instanton term had minimal effect.   In the present study that we consider the SU(3) flavor breaking effects and work in  isospin limit,   the instanton term cannot be neglected.

To formulate the axial anomaly with two pseudoscalar glueballs, let field $G$  in (\ref{E_anomalies}) be a linear combination of a physical pseudoscalar glueball $g$ and another pseudoscalar glueball $g'$ that has an unphysical mass:
\begin{equation}
	G = (1 - \xi) h^3 g  +  \xi h^3 g'
	\label{E_G_def}
\end{equation}
where $0 \le \xi \le 1$ is an a priori unknown parameter that has to be determined from experiment, and gives the percentage of each glueball  field in the total field $G$.  Note also that $G$ in this equation linearly depends on the pseudoscalar glueball fields,  as higher powers of $g$ do not contribute to the two-point mixing terms between  glueballs  and mesonic components made of quarks  in the eta squared mass matrix.   The part of the Lagrangian related to the couplings of $g$ and $g'$ to chiral nonets,  together with the term that supplies  mass to $g'$,   is:
\begin{equation}
{\cal L} = \frac{i}{12} 
\Bigg[(1 - \xi) h^3 g +  \xi h^3 g' \Bigg] 
\left[  
\gamma_1 
{\rm ln}\, \left(\frac{{\rm det}\, M}{{\rm det}\, 
	M^{\dagger}}\right)
+ \left(1 - \gamma_1\right)  
{\rm ln}\, \left(
\frac{ {\rm Tr} \, (M {M'}^\dagger) }
{ {\rm Tr}\,  (M' M^{\dagger})}
\right)
\right] + u'_6 h^2 g'^2
+ \cdots
\label{E_L_G1G2}
\end{equation}	
where the last term is the unphysical mass term for $g'$ field (with $u'_6 > 0$ as will be seen in numerical results in the next section).  Note that the mass term for $g$ is already taken into account in (\ref{E_f_terms}).  Eliminating $g'$ from its equation of motion results in the following effective term for the coupling of $g$ to the chiral nonets and the full term describing the U(1)$_{\rm A}$ anomaly becomes:
\begin{eqnarray}
	f_{\rm A} &=& \frac{i}{12} (1 - \xi) h^3 g
	\left[  
	\gamma_1 
	{\rm ln}\, \left(\frac{{\rm det}\, M}{{\rm det}\, 
		M^{\dagger}}\right)
	+ \left(1 - \gamma_1\right)  
	{\rm ln}\, \left(
	\frac{ {\rm Tr} \, (M {M'}^\dagger) }
	{ {\rm Tr}\,  (M' M^{\dagger})}
	\right)
	\right] \nonumber\\
	&&
	- c_3 \xi^2
	\left[  
	\gamma_1 
	{\rm ln}\, \left(\frac{{\rm det}\, M}{{\rm det}\, 
		M^{\dagger}}\right)
	+ \left(1 - \gamma_1\right)  
	{\rm ln}\, \left(
	\frac{ {\rm Tr} \, (M {M'}^\dagger) }
	{ {\rm Tr}\,  (M' M^{\dagger})}
	\right)
	\right]^2 
\label{E_fA_fianl}
\end{eqnarray}
where $c_3 = -h_0^4/(576 \, u'_6)$.  The two extreme values of $\xi = 0,1$ correspond to the two special cases previously investigated:   $\xi = 0$ corresponds  to the formulation  of axial anomaly in Refs. \cite{18_Fariborz_PRD98,19_Fariborz_PLB790,19_Fariborz_IJMPA34,21_Fariborz_NPA1015} in which the instanton term was not taken into account, and $\xi = 1$ corresponds to the formulation of Ref. \cite{05_Fariborz_PRD72} in which the physical pseudoscalar glueball was not investigated.   In the present  study, equation (\ref{E_fA_fianl}) generalizes these previous studies by modeling the axial anomaly in a manner that contains both a physical pseudoscalar glueball (the first term), as well as  the instanton term (the second term) that is needed to correctly generate the eta masses. 

The function $f_{\rm S}$ that models the trace anomaly is formulated in terms of operators $\det M$ and ${\rm Tr}\left( MM'^\dagger\right)$,  together with   a term that is pure glue \cite{18_Fariborz_PRD98}.  With $H=h^4$ this function is:
\begin{eqnarray}
	f_{\rm S} &=&
	-H \left\{
	\lambda_1 \ln\left(\frac{H}{\Lambda^4}\right)
	+\lambda_2 \left[\ln\left(\frac{\det M}{\Lambda^3}\right)+\ln\left(\frac{\det M^{\dagger}}{\Lambda^3}\right)\right]
	\right.
	\nonumber \\
	&&
	\left.
	\hskip .8cm
	+ \lambda_3\left[ \ln\left(\frac{{\rm Tr} MM^{\prime\dagger}}{\Lambda^2}\right)+\ln\left(\frac{{\rm Tr}M' M^{\dagger}}{\Lambda^2}\right)
	\right]\right\}
	\label{E_fS_final}
\end{eqnarray}
where  $\Lambda$ is a characteristic scale of QCD and has mass dimension one, and   $\lambda_1$, $\lambda_2$ and $\lambda_3$ are  arbitrary parameters with  constraint  $4\lambda_1+6\lambda_2+4\lambda_3=1$ \cite{18_Fariborz_PRD98}.

In the presence of the quark masses,  the trace of improved energy momentum tensor modifies to
\begin{eqnarray}
	\theta^{\mu}_{\mu}=H-\left(1+\gamma_m \right) V_{\rm SB}
	\label{E_trace_VSB}
\end{eqnarray}
where $\gamma_m$ is the anomalous dimension of the   fermion mass operator [in our notation in (\ref{E_L_Def_MMp2g}), $V_{\rm SB}=-f_{\rm SB}$].  Consequently,  
the SU(3) flavor symmetry breaking term should reflect this modification in (\ref{E_trace_VSB}) and is given by \cite{79_Schechter_PRD21}:
\begin{eqnarray}
	f_{\rm SB}={\rm Tr} [MM^{\dagger}]^{1-\frac{\gamma_m}{2}}{\rm Tr}[A(M+M^{\dagger}]
	\label{E_VSB_gen}
\end{eqnarray}
which, under the scale transformation, results in the second term on the right hand side of Eq. (\ref{E_trace_VSB}) . 
Similar to (\ref{E_GLSM_VSB}),  $A={\rm diag}(A_1,A_2,A_3)$ is proportional to the three light quark masses.

\subsection{Minimum equations}

The vacuum structure of the model is defined in terms of the vacuum expectation values (VEVs) of the scalar fields
\begin{eqnarray}
	\alpha_a &=& \langle S_a^a \rangle, \nonumber\\
	\beta_a  &=& \langle {S'}_a^a \rangle, \nonumber \\
	h_0  &=& \langle h \rangle.
	\label{All_condensates}
\end{eqnarray}
The minimum of the potential is determined by differentiating   $V$ with respect to the scalar fields and evaluating them  at the VEVs  of (\ref{All_condensates}) which results in the following equations: 
\begin{eqnarray}
	\left\langle\frac{\partial V}{\partial S^1_1}\right\rangle_0&=&
	4\, u_4 \,h_0\, \left( \alpha_1\,\beta_3+\alpha_3\,\beta_1
	\right) +2\, u_1 \,   h_0^2\alpha_1 + 4\, u_2\,\alpha_1^3
	+4\,{\frac { \left( \beta_1\, \left( \lambda_2+\lambda_3/2 \right)
			\alpha_1+ 1/2\,\alpha_3\,\beta_3\,\lambda_2 \right)  h_0^4}{2\,{
				\alpha_1}^{2}\beta_1+\alpha_1\,\alpha_3\,\beta_3}}
	\nonumber\\
	&&
	+4\, \left( 2\,{\alpha_1}^
	{2}+{\alpha_3}^{2} \right) ^{- \gamma_m/2} \left( -2+\gamma_m \right)  \left( A_1\,\alpha_1 + 1/2\, A_3 \,\alpha_3
	\right) \alpha_1-2\, \left( 2\,{\alpha_1}^{2}+{\alpha_3}^{2} \right) ^{1
		- \gamma_m/2} A_1
	\nonumber\\
	\left\langle\frac{\partial V}{\partial S^3_3}\right\rangle_0&=&
	8\, u_4\, h_0\,\beta_1\,\alpha_1+2\, u_1\, h_0^2
	\alpha_3 + 4\,  u_2 \,\alpha_3^3 + 4\,{\frac { \left( 1/2\,\beta_3\,
			\left( \lambda_2 + \lambda_3 \right) \alpha_3 + \alpha_1\,\beta_1\,\lambda_2
			\right) h_0^4}{2\,\alpha_1\,\alpha_3\,\beta_1+\beta_3\,{\alpha_3}
			^{2}}}
	\nonumber \\
	&&
	+4\, \left( 2\,{\alpha_1}^{2} + {\alpha_3}^{2} \right) ^{-
			\gamma_m/2} \left( -2 +  \gamma_m \right) \alpha_3\, \left( 
		 A_1\,\alpha_1 + 1/2\, A_3 \,\alpha_3 \right) -2\, \left( 2\,{\alpha_1}^{2} + {\alpha_3}^{2} \right) ^{1-  \gamma_m /2} A_3
	\nonumber\\
	\left\langle\frac{\partial V}{\partial {S'}^1_1}\right\rangle_0&=&
	2\,{\frac { \left( 4\,{\alpha_1}^{2}\alpha_3\,\beta_1\, u_4 + \left( 2
			\,{\alpha_3}^{2}\beta_3\, u_4 + 2\,{\beta_1}^{2} h_0\, u_3 + {
					 h_0}^{3}\lambda_3 \right) \alpha_1 + \alpha_3\,\beta_1\,\beta_3\, h_0
			\, u_3 \right)  h_0}{2\,\beta_1\,\alpha_1+\beta_3\,\alpha_3}}
	\nonumber \\
	\left\langle\frac{\partial V}{\partial {S'}^3_3}\right\rangle_0&=&
	2\,{\frac { h_0\, \left( 4\,{\alpha_1}^{3}\beta_1\, u_4 +2\,{
				\alpha_1}^{2}\alpha_3\,\beta_3\, u_4 + 2\,\alpha_1\,\beta_1\,\beta_3\,
				 h_0\, u_3 + \alpha_3\,{\beta_3}^{2} h_0\, u_3 + \alpha_3\,{
					h_0}^{3}\lambda_3 \right) }{2\,\beta_1\,\alpha_1+\beta_3\,\alpha_3}}
	\nonumber \\
	\left\langle\frac{\partial V}{\partial h}\right\rangle_0&=&
	8\,\ln  \left( {\frac {2\,\beta_1\,\alpha_1+\beta_3\,\alpha_3}{{\Lambda}^{
				2}}} \right) { h_0}^{3}\lambda_3+8\,\ln  \left( {\frac {{\alpha_1}^{
				2}\alpha_3}{{\Lambda}^{3}}} \right) { h_0}^{3}\lambda_2+4\,\ln
	\left( {\frac {{ h_0}^{4}}{{\Lambda}^{4}}} \right) { h_0}^{3}
	\lambda_1+ \left( 4\,\lambda_1+4\, u_5 \right) { h_0}^{3}
	\nonumber\\
	&&
	+
	\left( 4\,{\alpha_1}^{2} u_1 + 2\,{\alpha_3}^{2} u_1 +4\,{\beta_1}^
	{2} u_3 + 2\,{\beta_3}^{2}  u_3  \right)  h_0
	+4\, u_4\,\alpha_1\, \left( \alpha_1\,\beta_3+2\,\alpha_3\,\beta_1
	\right).
	\label{E_Min_Eqs_gen}
\end{eqnarray}
The  subscript zero on brackets mean that the derivatives are evaluated  at VEVs (\ref{All_condensates}), and  the five equations respectively represent minimization of the potential with respect to quark-antiquark components (first and second equations), four-quark composites (third and fourth equations), and the scalar glueball field (the last equation).

%%%%%%%%%%%%%%%%%%%%%%%%%%%%%%%%%%%%%%%%%%%%%%%%%%%%%%%%%%%%%%%%%%%%%%%%%%%%%%%%%%%%%%%%%%%%%%%%%%%%%%%%%%%%%%%%%%%%%%

                                   \section{Parameter determination and general predictions}

\subsection{The setup}

Lagrangian (\ref{E_L_Def_MMp2g}) contains 20  a priori unknown parameters ($u_1 \cdots u_6$, $\alpha_1$, $\alpha_3$, $\beta_1$, $\beta_3$, $h_0$, $A_1$, $A_3$, $\gamma_m$,  $\lambda_1$, $\lambda_2$, $\xi$, $\gamma_1$, $c_3$ and $\Lambda$) that should be determined through  a careful comparison with appropriate experimental data.  However, due to the complexities of the mass matrices and the minimum equations that results from the Lagrangian (\ref{E_L_Def_MMp2g}), matching the model predictions to any experimental data leads to solving a complicated system of equations and naturally requires an elaborate numerical analysis.    To keep the computations tractable, it is important to maintain a clear connection to the generalized linear sigma model without glueballs studied  in \cite{09_Fariborz_PRD79}.    To establish this connection,  we note that the trace anomaly term (\ref{E_fS_final}) contributes to all mass matrices because of the  $\lambda_2$ and $\lambda_3$ terms.  Therefore, the minimum extension of the framework of Ref. \cite{09_Fariborz_PRD79} is obtained by suppressing  $\lambda_2$ and $\lambda_3$ terms and saturating the trace anomaly with only the gluons ($\lambda_1$ term). Hence, as our first step in probing these 20 model parameters, in this study we work in the following limit:
\begin{eqnarray}
	\lambda_1 &=& 1\over 4,\nonumber \\
	\lambda_2 &=& \lambda_3 = 0, \nonumber \\
	\gamma_m &=& 2.
\label{E_LTA_def}
\end{eqnarray} 

In this limit, the mass matrices are as follows.  For isotriplet and isodoublet pseudoscalars ($\pi$ and $K$ systems) the mass matrices are:
\begin{eqnarray}
	M_\pi^2  =  \left[\begin{array}{cc}
		4 u_2 \,\alpha_1^2 + 4 u_4 h_0 \beta_3 + 2 u_1 \,  h_0^2 & 4 u_4 h_0 \alpha_3  
		\\
		4 u_4 h_0 \alpha_3  & 2 u_3 \, h_0^2 
	\end{array}\right]
\label{E_Mpi2_LTA}
\end{eqnarray}

\begin{equation}
	M_K^2 = 
	\left[\begin{array}{cc}
		\left(4 \alpha_1^2 - 4 \alpha_1 \alpha_3 + 4 \alpha_3^2\right) u_2 + 2 u_1 h_0^2 + 4 u_4 h_0 \beta_1  & 4 u_4 h_0 \alpha_1  
		\\
		4 u_4 h_0 \alpha_1  & 2 u_3 h_0^2 
	\end{array}\right].
\label{E_MK2_LTA}
\end{equation}

For isotriplet and isodoublet scalars ($a_0$ and $K_0^*$ systems) the mass matrices are:

\begin{equation}
	X_a^2 = 
	\left[\begin{array}{cc}
		12 u_2 \alpha_1^2 - 4 u_4 h_0 \beta_3 + 2 u_1 h_0^2 & - 4 u_4 h_0 \alpha_3  
		\\
		- 4 u_4 h_0 \alpha_3  & 2 u_3 h_0^2 
	\end{array}\right]
\label{E_Xa2_LTA}
\end{equation}

\begin{equation}
	X_\kappa^2 = 
	\left[\begin{array}{cc}
		\left(4 \alpha_1^2 + 4 \alpha_1 \alpha_3 + 4 \alpha_3^2\right) u_2 + 2 u_1 h_0^2 - 4 u_4 h_0 \beta_1  & - 4 u_4 h_0 \alpha_1  
		\\
		- 4 u_4 h_0 \alpha_1  & 2 u_3 h_0^2 
	\end{array}\right].
\label{E_Xkappa2_LTA}
\end{equation}

The mass matrix for the isosinglet scalars ($f_0$ system) the mass matrix is: 

\[
	X_0^2 = 
\left[
	\begin{array}{cccc}
		12 u_2 \alpha_1^2 + 4 u_4 h_0 \beta_3 + 2 u_1 h_0^2 & 4 \sqrt{2} u_4 h_0 \beta_1  & 4 u_4 h_0 \alpha_3  & 4 \sqrt{2} u_4 h_0 \alpha_1 
		\\
		4 \sqrt{2} u_4 h_0 \beta_1  & 12 u_2 \alpha_3^2 + 2 u_1 h_0^2 & 4 \sqrt{2} u_4 h_0 \alpha_1  & 0 
		\\
		4 u_4 h_0 \alpha_3  & 4 \sqrt{2} u_4 h_0 \alpha_1  & 2 u_3 h_0^2 & 0 
		\\
		4 \sqrt{2} u_4 h_0 \alpha_1  & 0 & 0 & 2 u_3 h_0^2 
		\\
		4 \sqrt{2} \left(\alpha_1 \beta_3 u_4 + u_1 h_0 \alpha_1 + \alpha_3 \beta_1 u_4 \right) & 8 u_4 \alpha_1 \beta_1 + 4 u_1 h_0 \alpha_3  & 4 \sqrt{2} \left(\alpha_1 \alpha_3 u_4 + \beta_1 h_0 u_3 \right) & 4 u_4 \alpha_1^2 + 4 u_3 h_0 \beta_3  \\
		 & & & \\
	\end{array}
\right.
	\]
\begin{equation}
\left.
	\begin{array}{ccc}
		 & &  4 \sqrt{2} \left(\alpha_1 \beta_3 u_4 + u_1 h_0 \alpha_1 + \alpha_3 \beta_1 u_4 \right) \\
		 & &   8 u_4 \alpha_1 \beta_1 + 4 u_1 h_0 \alpha_3  \\
		 & &   4 \sqrt{2}\, \left(\alpha_1 \alpha_3 u_4 + \beta_1 h_0 u_3 \right) \\ 
		 & &   4 u_4 \alpha_1^2 + 4 u_3 h_0 \beta_3   \\
		 & &   3 h_0^2 \ln \! \left(\frac{h_0^4}{\Lambda^4}\right)+4 \alpha_1^2 u_1 + 2 \alpha_3^2 u_1 + 4 \beta_1^2 u_3 + 2 \beta_3^2 u_3 + 12 u_5 h_0^2+
		7 h_0^2 
	\end{array}
\right].
\label{E_X02_LTA}
\end{equation}

The mass matrix for the isosinglet pseudoscalars (the $\eta$ system) is considerably more complicated and its matrix elements are: 

\begin{eqnarray}
\left(M_{\eta}^2\right)_{11} & = &
\frac{2}{\alpha_1^2 
	\left(
	      2 \alpha_1 \beta_1  + \alpha_3 \beta_3 
    \right)^2
	    }
\Bigg[
  8 \alpha_1^6 \beta_1^2 u_2 
 + 8 \alpha_1^5 \alpha_3 \beta_1 \beta_3 u_2 
 +  2  \alpha_1^4 \left[
             2 \beta_1^2\left(u_1 h_0^2 - 2 u_4 h_0 \beta_3\right) 
             + \alpha_3^2  \beta_3^2 u_2
      \right]
 \nonumber \\
 &&+4 \alpha_3 h_0 \beta_1  \beta_3  \alpha_1^3
     \left( h_0 u_1 -2 \beta_3 u_4 \right) 
 + \alpha_1^2 \left[\alpha_3^2 h_0 
 \beta_3^2 \left( h_0 u_1 -2 \beta_3  u_4  \right)
 -8 c_3\xi^2 \beta_1^2 \left(1 + \gamma_1 \right)^2 
 \right] 
 \nonumber \\
&&  
 -16 c_3  \alpha_1 \alpha_3  \gamma_1 \xi^2 \beta_1 \beta_3  \left(1 +  \gamma_1\right)  -8 \alpha_3^2 \beta_3^2 c_3\gamma_1^2 \xi^2
\Bigg]
\nonumber \\
% ----------------------------------------------------------------------
\left(M_{\eta}^2\right)_{12} & = & 
- \frac{16 \sqrt{2}}{\alpha_1\alpha_3  \left(2 \alpha_1\beta_1  +\alpha_3 \beta_3  \right)^2 }
\Bigg[
       \alpha_1^3 \alpha_3\beta_1^3 h_0 u_4 
       + \alpha_1^2 \beta_1^2
      \Big(
              h_0 u_4 \beta_3\alpha_3^2
             + c_3\xi^2 \gamma_1  \left(1 + \gamma_1 \right)
       \Big) 
\nonumber\\
&& 
 +   {1\over 4} \alpha_1 \alpha_3 \beta_1 \beta_3  
     \Big(
                     h_0 u_4 \beta_3\alpha_3^2  + 2 c_3 \xi^2 
                       \left(
                              2 \gamma_1^2 + \gamma_1 + 1
                       \right)
     \Big)  
   +  \frac{1}{2}\alpha_3^2 \beta_3^2 c_3 \gamma_1 \xi^2  
\Bigg]
\nonumber\\
% ----------------------------------------------------------------------
\left(M_{\eta}^2\right)_{13} & = & 
-\frac{4}{\left(2 \alpha_1\beta_1 +\alpha_3 \beta_3  \right)^{2}}
\Bigg[
         \alpha_3^3 \beta_3^2 h_0 u_4 
      + 4 \alpha_1 \alpha_3^2 \beta_1  \beta_3  h_0 u_4 
      +  4 h_0 u_4 \beta_1^2 \alpha_1^2 \alpha_3
      +  4 c_3 \xi^2 \beta_3 \alpha_3 \gamma_1 \left(\gamma_1 - 1 \right)
\nonumber \\
&&
      + 4 c_3 \xi^2 \beta_1  \alpha_1  \left(\gamma_1^2 - 1\right)
\Bigg]
\nonumber\\
% ----------------------------------------------------------------------
\left(M_{\eta}^2\right)_{14} & = & 
-
\frac{4 \sqrt{2}}
{\alpha_1\left(2 \alpha_1 \beta_1  +\alpha_3 \beta_3\right)^2 }
\Bigg[
        4 \alpha_1^4 \beta_1^2 h_0 u_4 
      + 4 \alpha_1^3 \alpha_3 \beta_1  \beta_3 h_0 u_4 
      + \alpha_1^2 \alpha_3^2 \beta_3^2 h_0 u_4 
\nonumber \\
&&
      + 2 c_3 \xi^2 \alpha_1 \beta_1 \alpha_3  \left(\gamma_1^2 - 1       \right)   
      + 2 c_3 \xi^2 \beta_3 \alpha_3^2 \gamma_1  \left(\gamma_1 -1  \right)
\Bigg]
\nonumber\\
% ----------------------------------------------------------------------
\left(M_{\eta}^2\right)_{15} & = & 
\frac{
	    \sqrt{2} \left(1 - \xi \right) h_0^3 
	    \Big(  \alpha_1 \beta_1  \left(1 + \gamma_1 \right)  
	           + \alpha_3 \beta_3  \gamma_1 
	    \Big)
	 }
     {12 \beta_1\alpha_1^2 + 6 \alpha_1 \alpha_3\beta_3 
     }
\nonumber\\
% ----------------------------------------------------------------------
\left(M_{\eta}^2\right)_{22} & = & 
2 u_1 h_0^2 + 4 u_2 \alpha_3^2 
-\frac{
	    8 \xi^2 c_3 
	    \left(2  \alpha_1 \beta_1 \gamma_1  
              +\alpha_3 \beta_3
        \right)^2
       }
       {
       	\alpha_3^2 \left(
       	                  2 \alpha_1  \beta_1 + \alpha_3\beta_3
       	          \right)^2
       }
\nonumber\\
% ----------------------------------------------------------------------
\left(M_{\eta}^2\right)_{23} & = & 
-
\frac{4  \sqrt{2} \alpha_1}
     {\alpha_3 \left(2 \alpha_1  \beta_1  +\alpha_3  \beta_3  \right)^2  }
\Bigg[
        \alpha_3^3 \beta_3^2 h_0 u_4 
      + 4 \alpha_1  \alpha_3^2 \beta_1 \beta_3 h_0 u_4 
      + 4 h_0 u_4 \beta_1^2 \alpha_1^2 \alpha_3 
      + 2 c_3 \xi^2 \alpha_3 \beta_3  \left(\gamma_1 - 1 \right)  
\nonumber \\
&&
      + 4 c_3 \xi^2 \beta_1  \alpha_1  \gamma_1 
          \left(\gamma_1 - 1 \right)
\Bigg]
\nonumber\\
% ----------------------------------------------------------------------
\left(M_{\eta}^2\right)_{24} & = & 
\frac{8 c_3 \xi^2 \left(1 - \gamma_1 \right) 
	  \left(2 \alpha_1 \beta_1  \gamma_1  +\alpha_3 \beta_3  \right)}
	 {\left(2 \alpha_1 \beta_1  + \alpha_3 \beta_3 \right)^2}
\nonumber\\
% ----------------------------------------------------------------------
\left(M_{\eta}^2\right)_{25} & = & 
\frac{
	   \left(1 - \xi\right) h_0^3 
	   \left(2 \alpha_1\beta_1\gamma_1  + \alpha_3\beta_3  \right)
	  }
	  {
	   6 \alpha_3 \left(2 \alpha_1\beta_1  + \alpha_3\beta_3 \right)}
\nonumber\\
% ----------------------------------------------------------------------
\left(M_{\eta}^2\right)_{33} & = & 
2 u_3 h_0^2 -\frac{16 c_3 \,\xi^2 \alpha_1^2 \left(1 - \gamma_1 \right)^{2} }{\left(2 \alpha_1 \beta_1 +\alpha_3 \beta_3 \right)^2}
\nonumber \\
% ----------------------------------------------------------------------
\left(M_{\eta}^2\right)_{34} & = & 
-\frac{8 \sqrt{2}  c_3 \,\xi^2 \alpha_1 \alpha_3 \left(1 - \gamma_1 \right)^2  }{\left(2 \alpha_1 \beta_1 + \alpha_3 \beta_3 \right)^2}
\nonumber \\
% ----------------------------------------------------------------------
\left(M_{\eta}^2\right)_{35} & = & 
\frac{\sqrt{2}  h_0^3 \alpha_1 \left(1 - \xi \right)  \left(\gamma_1 -1 \right) }{6 \left(2 \alpha_1 \beta_1 +  \alpha_3 \beta_3\right)}
\nonumber \\
% ----------------------------------------------------------------------
\left(M_{\eta}^2\right)_{44} & = & 
2 u_3 h_0^2 -\frac{8 c_3 \,\xi^2 \alpha_3^2\left(1 - \gamma_1 \right)^2 }{\left(2 \alpha_1 \beta+1 + \alpha_3 \beta_3 \right)^2}
\nonumber \\
% ----------------------------------------------------------------------
\left(M_{\eta}^2\right)_{45} & = & 
\frac{h_0^3 \alpha_3 \left(1 - \xi \right)  \left( \gamma_1 -1 \right) }{6 \left(2 \alpha_1 \beta_1 + 2 \alpha_3 \beta_3\right)}
\nonumber \\
% ----------------------------------------------------------------------
\left(M_{\eta}^2\right)_{55} & = & 2 u_6 h_0^2.
\label{E_Meta2_LTA}
\end{eqnarray}
These mass matrices are subject to the following minimum equations which are obtained from imposing limit (\ref{E_LTA_def}) on minimum equations (\ref{E_Min_Eqs_gen}):
\begin{eqnarray}
\left\langle
{
	{\partial V_0} \over {\partial f_a}
}
\right\rangle_0
& = &
2 \sqrt{2}\, \left(2 u_2 \alpha_1^3 + 2 \alpha_1 \beta_3 h_0 u_4 + u_1 h_0^{2} \alpha_1 + 2 \alpha_3 \beta_1 h_0 u_4 - A_1 \right) = 0 
\nonumber \\
\left\langle
{
	{\partial V_0} \over {\partial f_b}
}
\right\rangle_0
& = &
8 \alpha_1 \beta_1 h_0 u_4 + 4 u_2\alpha_3^3 + 2 u_1 h_0^2 \alpha_3 - 2 A_3 = 0
\nonumber  \\
\left\langle
{
	{\partial V_0} \over {\partial f_c}
}
\right\rangle_0
& = &
2  \sqrt{2} h_0 \left(2 \alpha_1 \alpha_3 u_4 +  \beta_1 h_0 u_3 \right) = 0
\nonumber  \\
\left\langle
{
	{\partial V_0} \over {\partial f_d}
}
\right\rangle_0
& = &
4 \alpha_1^2 h_0 u_4 + 2 u_3 h_0^2 \beta_3 = 0
\nonumber  \\
\left\langle
{
	{\partial V_0} \over {\partial h}
}
\right\rangle_0
& = &
h_0^3 
\ln  
\left( \frac{h_0^4}{\Lambda^4} \right) + 
h_0^3 \left(1 + 4 u_5 \right)   +  
2 h_0 \left(
       2 \alpha_1^2 u_1 +  \alpha_3^2 u_1 +  2\beta_1^2 u_3 +  \beta_3^2 u_3 
      \right)  
\nonumber  \\
&& + 8 \alpha_1 \alpha_3 \beta_1 u_4 + 4 \alpha_1^2 \beta_3 u_4 = 0.
\label{E_MinEq_LTA}
\end{eqnarray}

With the identification
\begin{eqnarray}
	u_1 &=& -{c_2 \over {h_0^2}}, \nonumber\\
	u_2 &=& c_4, \nonumber \\
	u_3 &=& {d_2\over {h_0^2}}, \nonumber\\
	u_4 &=& {e_3^a \over h_0},
	\label{E_u14_de_connect}
\end{eqnarray}
where $c_2$, $c_4$, $d_2$ and $e_3^a$ are the coupling constants in the model  of Ref. \cite{09_Fariborz_PRD79} given in (\ref{E_GLSM_V0}).
The mass matrices (\ref{E_Mpi2_LTA}), (\ref{E_MK2_LTA}), (\ref{E_Xa2_LTA}) and  (\ref{E_Xkappa2_LTA}) reduce to those found in \cite{09_Fariborz_PRD79} in the absence of glueballs. Similarly, the upper $4 \times 4$ part of isosinglet matrices, i.e. $\left(M_\eta^2\right)_{ij}$ and $\left(X_0^2\right)_{ij}$,  $i,j=1\cdots 4$,  as well as the first four  minimum equations (\ref{E_MinEq_LTA}) reduce to those found in \cite{09_Fariborz_PRD79}.  The extension that is being studied in this work is the mixing with the scalar and pseudoscalar glueballs expressed by the fifth elements of the ${\left(X_0^2\right)}_{ij}$ and ${\left(M_\eta^2\right)}_{ij}$   with $i$ or $j = 5$, together with the additional fifth  minimum equation in (\ref{E_MinEq_LTA}), as well as the generalized term for  U(1)$_{\rm A}$ anomaly given in Eq. (\ref{E_fA_fianl}).

\subsection{Numerical analysis}

In the limit of (\ref{E_LTA_def}), as the first step, we compute $\alpha_1, \beta_1, \alpha_3, \beta_3, A_1, A_3, u_2$ as well as the combined parameters $u_1 h_0^2$, $u_3 h_0^2$ and $u_4 h_0$
using  the following experimental inputs:
\begin{eqnarray}
m[a_0(980)] &=& 980 \pm 20\, {\rm MeV},
\nonumber
\\ m[a_0(1450)] &=& 1439 \pm 34\, {\rm MeV},
\nonumber \\
m[\pi(1300)] &=& 1300 \pm 100\, {\rm MeV},
\nonumber \\
m_\pi &=& 137 \, {\rm MeV},
\nonumber \\
F_\pi &=& 131 \, {\rm MeV},
\nonumber \\
{A_3\over A_1} &=& 27 - 30.
\label{E_inputs1}
\end{eqnarray}
In addition, we utilize  the  minimum equations (\ref{E_MinEq_LTA}).   The complete details of the parameter determination are given in Appendix \ref{A_par_determ}.  Clearly, $m[\pi(1300)]$ has a large uncertainty that results in a broad range of experimental target values that need to be numerically scanned and the favored values of $m[\pi(1300)]$ that  give acceptable predictions be determined.    This results in a large numerical set of the following parameters:

\begin{equation}
	S_I = 
	\Biggl\{ \hskip .1cm
	{\mathbf p} \hskip .15cm
	\Bigg|\hskip .15cm {\rm Inputs \hskip 0.1cm of \hskip .1cm  (\ref{E_inputs1}) \hskip .1cm are \hskip .1cm satisfied \hskip .1cm} 
	\Biggr\}
	\label{E_SI}
\end{equation}
where each ${\mathbf p}$ represents an acceptable  group of the following  10 parameters 
\begin{equation}
{\mathbf p} =	\left(\alpha_1, \alpha_3, \beta_1, \beta_3, A_1, A_3, u_2, u_1 h_0^2, u_3 h_0^2, u_4 h_0
\right) 
\label{E_p_def}
\end{equation}  
that have satisfied (\ref{E_inputs1}).   
Since the parameters determined at this step will be inputed into the next step to compute the remaining parameters,  ultimately  a subset of (\ref{E_SI}) and experimental inputs (\ref{E_inputs1}) that give the best overall agreement will be determined.

Next,  as our second step,  we proceed to determine the five parameters:  $h_0$, $c_3$, $\gamma_1$, $u_6$, $\xi$ (note that for a given numerical value of $h_0$ the three parameters $u_1$, $u_3$ and $u_4$ are also simultaneously determined from the values of $u_1 h_0^2$, $u_3 h_0^2$ and $u_4 h_0^2$ already found in set $S_I$).  Parameters $u_5$ and $\Lambda$ get eliminated by the fifth minimum equation in (\ref{E_MinEq_LTA}).   Therefore, all 20 parameters will be determined in this second step of computation.  We refer to the remaining eight parameters as $\mathbf q$:
\begin{equation}
	{\mathbf q} =	\left(h_0, c_3, \gamma_1, u_1, u_3, u_4, u_6, \xi
	\right). 
	\label{E_q_def}
\end{equation}

To determine each acceptable $\mathbf q$, we target   experimental values for the eta masses \cite{24_PDG}:
\begin{eqnarray}
m^{\rm exp.}[\eta (547)] &=& 547.862 \pm
0.017\, {\rm
	MeV},\nonumber \\
m^{\rm exp.}[\eta' (958)] &=& 957.78 \pm 0.06
\, {\rm
	MeV},
\nonumber\\
m^{\rm exp.}[\eta (1295)] &=& 1294 \pm 4\, {\rm
	MeV},\nonumber \\
m^{\rm exp.}[\eta (1405)] &=& 1408.8 \pm 1.8 \,
{\rm
	MeV},
\nonumber \\
m^{\rm exp.}[\eta (1475)] &=& 1476 \pm 4\, {\rm
	MeV},\nonumber \\
m^{\rm exp.}[\eta (1760)] &=& 1751 \pm 15 \,
{\rm
	MeV},
\nonumber \\
m^{\rm exp.}[\eta (2225)] &=& 2221^{+13}_{-10} \,
{\rm 	MeV}.
\label{E_eta_exp}
\end{eqnarray}
Note that the isosinglet pseudoscalar state $X(2370)$ discovered by BESIII 
\cite{11_Ablikim_PRL106} has a mass that differs by only about 3\%  from  the mass of $\eta(2225)$.  Since our analysis of the isosinglet pseudoscalars is based on their mass spectrum,  practically we cannot distinguish $X(2370)$ from $\eta(2225)$ at this stage.   For simplicity, throughout this work, we use $\eta(2225)$ as our heaviest isosinglet pseudoscalar, but this state can be $X(2370)$ or a combination of the two.

With every choice of the 10  parameters from set $S_I$, together with a selection of numerically generated values of $\mathbf q$, the model predicts five eta masses that should be compared with five of the seven experimental masses in (\ref{E_eta_exp}).  Since the two lightest etas,  $\eta(547)$ and $\eta'(958)$,  are well established, we want the two  lightest theoretically predicted eta masses, $m^{\rm theo.}[\eta_i]$, $i = 1,2$,  to match these two experimentally well-known states. The remaining three theoretically predicted masses, $m^{\rm theo.}[\eta_i]$, $i = 3\cdots 5$,  should be compared with three of the  five  experimentally known etas above 1 GeV in the list (\ref{E_eta_exp}).   This means that for every choice of the model parameters (each choice of $\mathbf p$ and $\mathbf q$),  we should examine $5!/\left[3! (5-3)! \right] = 10$ scenarios (defined in Table \ref{T_10_scenarios}) to find the best agreement between theory and experiment.    We perform an extensive numerical analysis (within the 20-dimensional parameter space of the model) that is also guided by both the special limit of this model in the SU(3) flavor limit \cite{18_Fariborz_PRD98,19_Fariborz_PLB790,19_Fariborz_IJMPA34,21_Fariborz_NPA1015} as well as  by  the generalized linear sigma model without glueballs in \cite{09_Fariborz_PRD79}.   We use Monte-Carlo method to generate numerical values of the parameters, compute physical quantities and compare them with their experimental ranges for each of the 10 scenarios.

\begin{table}[!htbp]
	\centering
	\caption{The 10 different scenarios for identifying the five eta states of this model with etas known experimentally.   For all scenarios $\eta_1 = \eta(547)$,  $\eta_2 = \eta'(958)$, and $\eta_3, \eta_4$, and $\eta_5$ are given in this table.}
	\renewcommand{\tabcolsep}{0.4pc} % enlarge column spacing
	\renewcommand{\arraystretch}{1.5} % enlarge line spacing
	\begin{tabular}{c|c}
		\noalign{\hrule height 1pt}
		\noalign{\hrule height 1pt}	
		Scenario & $\eta_3$, $\eta_4$, $\eta_5$  \\
		\noalign{\hrule height 1pt}		
		\noalign{\hrule height 1pt}	
		1        & $\eta(1295)$,  $\eta(1405)$, $\eta(1475)$ \\  
		        \noalign{\hrule height 1pt}		
        2        & $\eta(1295)$,  $\eta(1405)$, $\eta(1760)$ \\  
        		\noalign{\hrule height 1pt}		
        3        & $\eta(1295)$,  $\eta(1405)$, $\eta(2225)$ \\  
        		\noalign{\hrule height 1pt}		
        4        & $\eta(1295)$,  $\eta(1475)$, $\eta(1760)$ \\  
        		\noalign{\hrule height 1pt}		
        5        & $\eta(1295)$,  $\eta(1475)$, $\eta(2225)$ \\  
        		\noalign{\hrule height 1pt}		
        6        & $\eta(1295)$,  $\eta(1760)$, $\eta(2225)$ \\  
        		\noalign{\hrule height 1pt}		
        7        & $\eta(1405)$,  $\eta(1475)$, $\eta(1760)$ \\  
        		\noalign{\hrule height 1pt}		
        8        & $\eta(1405)$,  $\eta(1475)$, $\eta(2225)$ \\  
        		\noalign{\hrule height 1pt}		
        9        & $\eta(1405)$,  $\eta(1760)$, $\eta(2225)$ \\  
        		\noalign{\hrule height 1pt}		
        10        & $\eta(1475)$,  $\eta(1760)$, $\eta(2225)$ \\  
		\noalign{\hrule height 1pt}
	\end{tabular}\\[2pt]
	\label{T_10_scenarios}
\end{table}

To guide the numerical simulations to approach the target values, for each of the 10 scenarios,  we  define  the function
\begin{equation}
\chi_k \left( {\mathbf p}, {\mathbf q} \right) = 
\sum_{i = 1}^5 
\frac{\left| m^{\rm theo.}\left[\eta_i\right] -m^{\rm exp.} 
	\left[ \eta_i\right] \right|}{m^{\rm exp.}[\eta_i]} 
\label{E_chi_def}
\end{equation}
where $\mathbf p$ and $\mathbf q$ are defined in (\ref{E_p_def}) and (\ref{E_q_def}), respectively (with $\mathbf p$ imported from set $S_{I}$ in (\ref{E_SI}) and $\mathbf q$ randomly generated using Monte-Carlo method);  $m^{\rm theo.}[\eta_i]$ are the model predictions for the five eta masses  and $m^{\rm exp.}[\eta_i]$ are five of the experimental values of the eta masses given in (\ref{E_eta_exp}).    The goodness of a set of parameters can be judged by comparing their resulting  $\chi$ value with the quantity
\begin{equation}
\chi_k^{\rm exp.} = \sum_{i = 1}^5  \frac{\Delta m^{\rm exp.}[\eta_i] }{m^{\rm exp.}[\eta_i]} 
\label{E_chi_exp_def}
\end{equation}
where $k=\cdots 10$ is computed for each of the 10 scenarios.   For a set of model parameters ($\mathbf p$ and $\mathbf q$) a scenario is considered acceptable if the resulting chi value  is less than the corresponding experimental chi value for that scenario. However, to generate a larger ensemble of points, we impose a less stringent condition defined by
\begin{equation}
\left(\chi_k\right)_{\rm min} \le \left(\chi_k^{\rm exp.} \right)_{\rm max}
\label{E_chi_goodness}
\end{equation}
where the minimum value of the predicted $\chi_k$ from (\ref{E_chi_def}) among all scenarios ($k=1\cdots 10$) is expected to be lower than the maximum of the experimental $\chi_k^{\rm exp.}$ from (\ref{E_chi_exp_def}).  
This results in a large ensemble of simulations (set $S_{II}$)  containing a subset of points $\mathbf p$ from $S_I$, together with points $\mathbf q$ that collectively satisfy condition  (\ref{E_chi_goodness}):

\begin{equation}
	S_{II} = 
	\hskip .1cm 
	\Biggl\{
	{\mathbf p} \in S_I \bigcup  {\mathbf q} \hskip .15cm
	\Bigg| \hskip .15cm {\rm Condition  \hskip .1cm  (\ref{E_chi_goodness}) \hskip .1cm is \hskip .1cm satisfied} \hskip .1cm
	\Biggr\}
	\label{E_SII}
\end{equation}
where $\mathbf q$ is defined in (\ref{E_q_def}).    This large set of model parameters $\mathbf p$ and $\mathbf q$ can then be used for computation of physical quantities such as the masses, substructures, decay widths, etc, and will be analyzed next.

\subsection{General results and predictions}

Generating  1.2$\times 10^8$ random parameters $\mathbf p$ and $\mathbf q$,  out of which less than 0.04\% satisfy the condition (\ref{E_chi_goodness}),  results in a set $S_{II}$ with  43,486 number of acceptable parameters $\mathbf p$ and $\mathbf q$.  This gives a reasonable data ensemble for  statistical analysis and testing the model predictions.
We then use $S_{II}$ to compute the physical quantities of interest such as the masses and quark and glue substructures. In addition,  since the function $\chi$ in (\ref{E_chi_def}) also  depends on the target experimental ranges for $m[\pi(1300)]$ and $A_3/A_1$,
the simulations probe these experimental data and show that the following ranges are favored in this analysis:
\begin{eqnarray}
m[\pi(1300)] & = & 1.29 - 1.31 {\rm GeV}\nonumber \\
A_3 \over A_1 &=& 28.3 - 30.0. 
\label{E_mPiA31_ranges}
\end{eqnarray}
The predicted masses and substructures are given below.

\subsubsection{Isodoublet and isotriplet states}

Although these states have been studied in the framework of the generalized linear sigma model without glueballs in \cite{09_Fariborz_PRD79},  in this section we revisit their case within the present extended framework that has been augmented with glueballs.  While glueballs do not directly contribute to these states,  as the framework is extended and its parameter space is enlarged, the glueballs do still play an indirect role even on isodoublets and isotriplets.  This indirect role can be easily seen by noting that while in general the characteristics of isodoublet and isotriplet states are described by parameters $\mathbf p$ in set $S_I$, but since the characteristics of isosinglet states are described by both parameters $\mathbf p$ as well as $\mathbf q$, when the condition (\ref{E_chi_goodness}) is imposed (which includes properties of pseudoscalar glueballs) it affects parameters $\mathbf p$ as well, and thereby, indirectly affects the properties of isodoublets and isotriplets. Therefore, while we have studied isodoublets and isotriplets in \cite{09_Fariborz_PRD79}, it is still important to revisit that study with new constraints from  mixing with glueballs, albeit an indirect connection.  An immediate consequence of this indirect connection is the observation that this new study favors a limited ranges of experimental mass for $\pi(1300)$ as well as for the ratio of strange to nonstrange quark masses given in Eq.  (\ref{E_mPiA31_ranges}) compared to their experimental ranges in (\ref{E_inputs1}).

\begin{figure}
	%	\centering
	\includegraphics[height=1.5in]{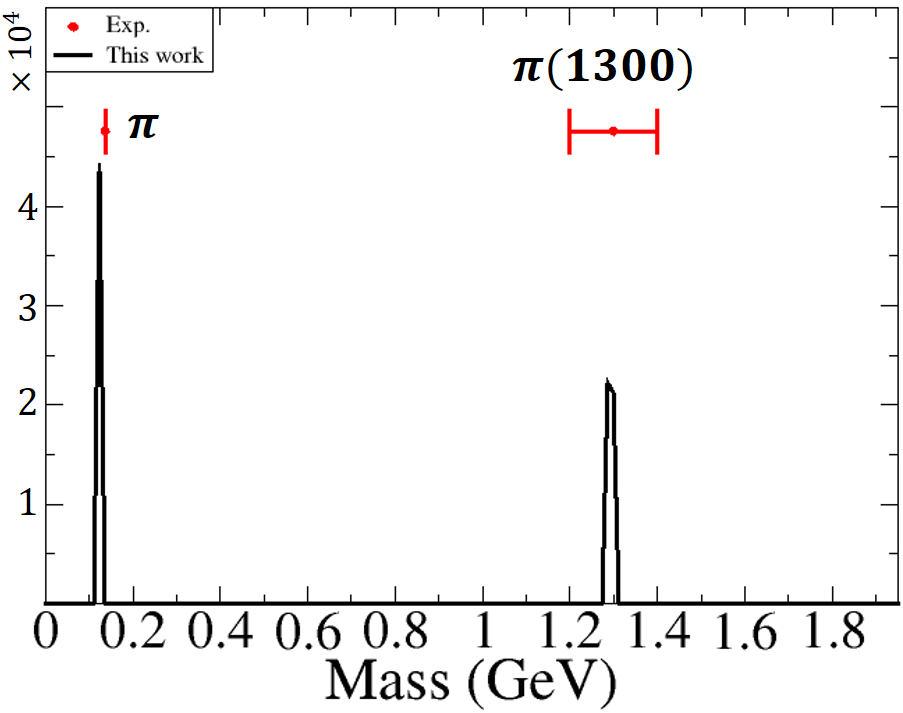}\hspace{0.02\columnwidth}\hskip .3cm
	\includegraphics[height=1.5in]{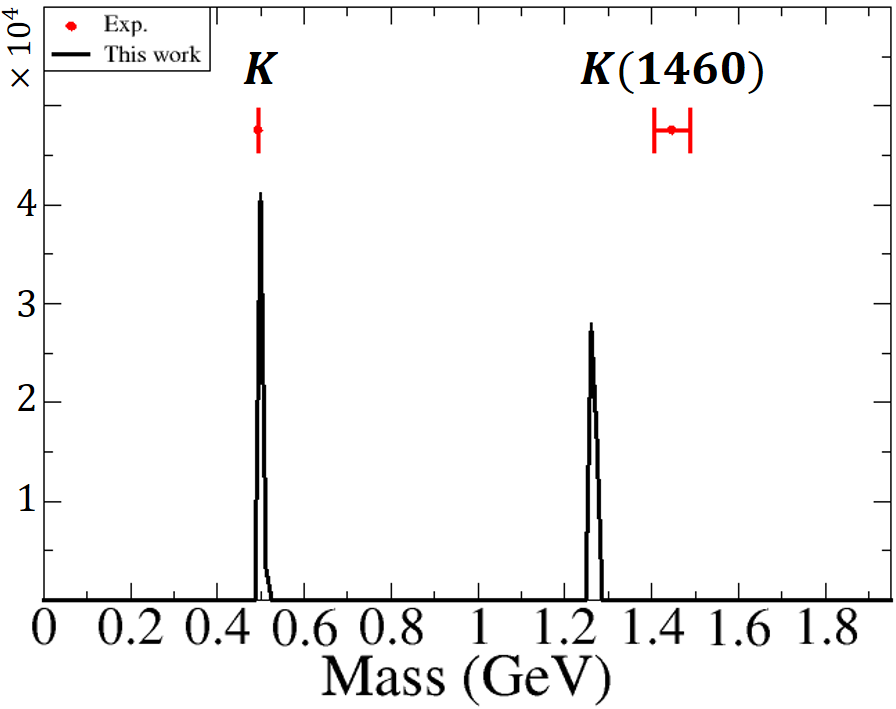}\hspace{0.02\columnwidth}
	
	\vskip .5cm
	
	\includegraphics[height=1.5in]{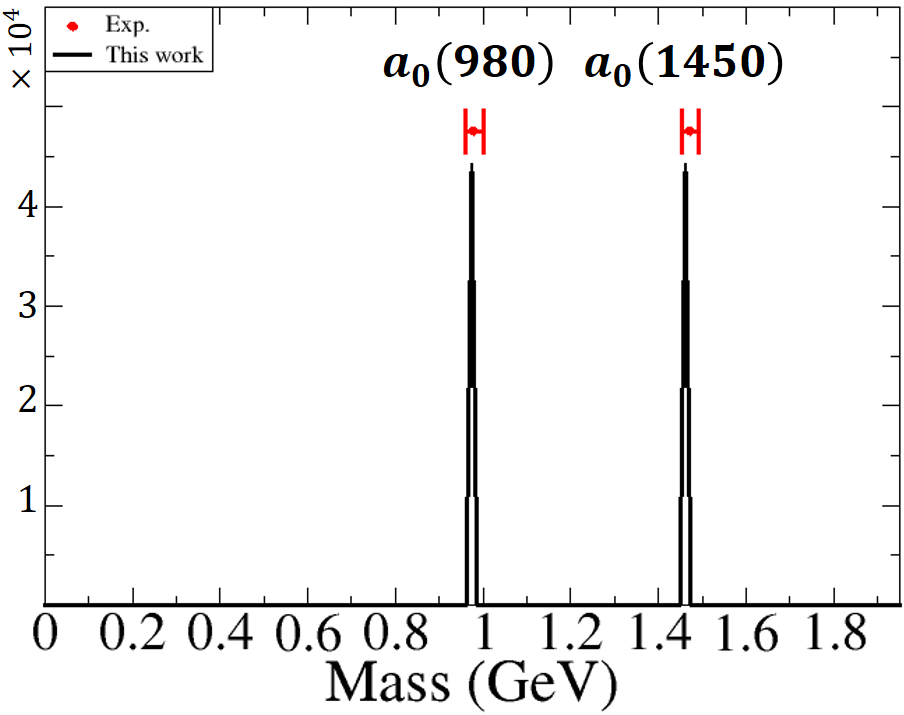}\hspace{0.02\columnwidth}\hskip .3cm
	\includegraphics[height=1.5in]{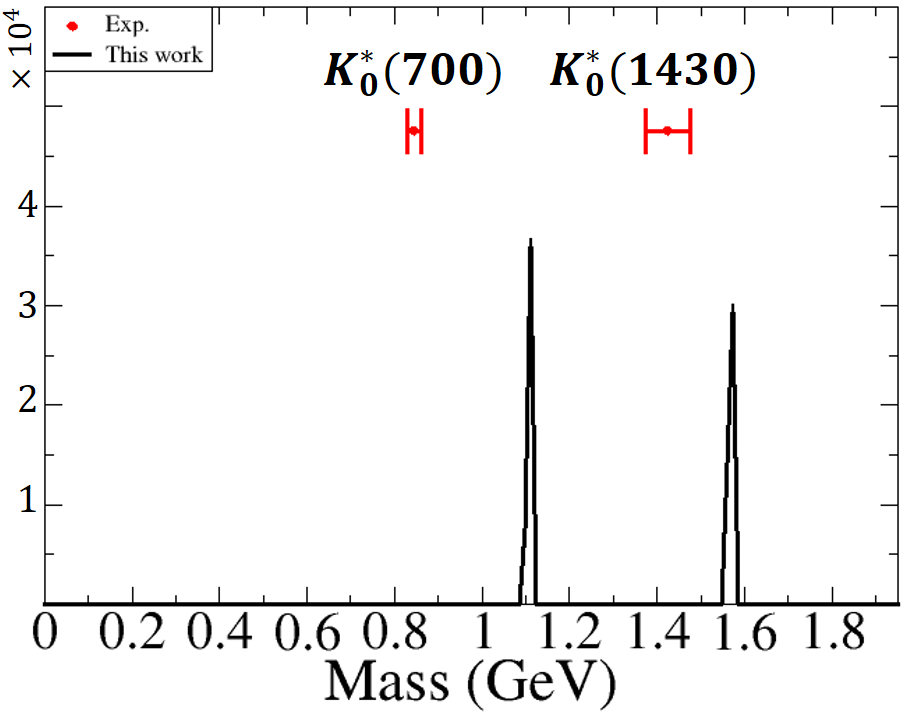}\hspace{0.02\columnwidth}
	\caption{Histograms for the masses of  isotriplet systems of $\pi$ and $a_0$, and for isodoublet systems of $K$ and $K_0^*$, produced by parameters $\mathbf p$ in set $S_{II}$ (\ref{E_SII})  with unfiltered range of $h_0 = 0.40-0.99$ GeV. The masses of $\pi$ and $a_0$ systems are inputted and the masses of $K$ and $K_0^*$ are predictions. 	}
	\label{F_piKa0kappa_masses}
\end{figure}

The histograms for the resulting masses of these states are given in Fig. \ref{F_piKa0kappa_masses} and compared with their experimental values (shown in red).   Note that, as given in (\ref{E_inputs1}), the masses of pseudoscalar and scalar  isotriplets are part of the inputs for the set $S_I$ and therefore the histograms for these masses are expected, and just double check the computation.   Although mass of $\pi(1300)$ is one of the inputs, this computation narrows down its range to 1.29-1.31 GeV.  However,  the masses of isodoubles are not part of any of the inputs and are predictions of this computation. We see in Fig. \ref{F_piKa0kappa_masses} that mass of $K(496)$ is in remarkable agreement with its experimental value.  The situation for the heavier kaon state $K(1460)$ is not well-understood and while this state is included in PDG \cite{24_PDG} as a state that is observed in $K\pi\pi$ partial-wave analysis, its reported properties in the literature are not used in any averaging by PDG.   The histogram for this state in Fig. 
\ref{F_piKa0kappa_masses} gives a lighter mass around 1.3 GeV in comparison with 1.46 GeV mentioned in PDG  \cite{24_PDG}.  The case of isodoublet scalars are more complicated.    As can be seen in Fig. \ref{F_piKa0kappa_masses}, the masses of both of $K_0^*(700)$ and $K_0^*(1430)$ are higher than the experimental ranges given in the same figure.  However, as is well known, the broad states such as $K_0^*(700)$,  and to some extent  $K_0^*(1430)$,  receive corrections to their mass and decay width due to the final-state interactions of their $\pi K$ decay products.   Specifically, in \cite{15_Fariborz_PRD92} $\pi K$ scattering is studied in the framework of the generalized linear sigma model, in which it is shown that the  physical mass of $K_0^*(700)$ which is computed from the pole of the   unitarized scattering amplitude, is 0.670-0.770 GeV   which is closer to its PDG value of 0.845 $\pm$ 0.017 GeV.   Therefore, what we see in  Fig. \ref{F_piKa0kappa_masses} for the masses of $K_0^*(700)$ and $K_0^*(1430)$ are  their masses before the unitarity corrections are taken into account, and are in fact consistent with previous study in \cite{15_Fariborz_PRD92}.

\begin{figure}
	%	\centering
	
	\includegraphics[height=1.12in]{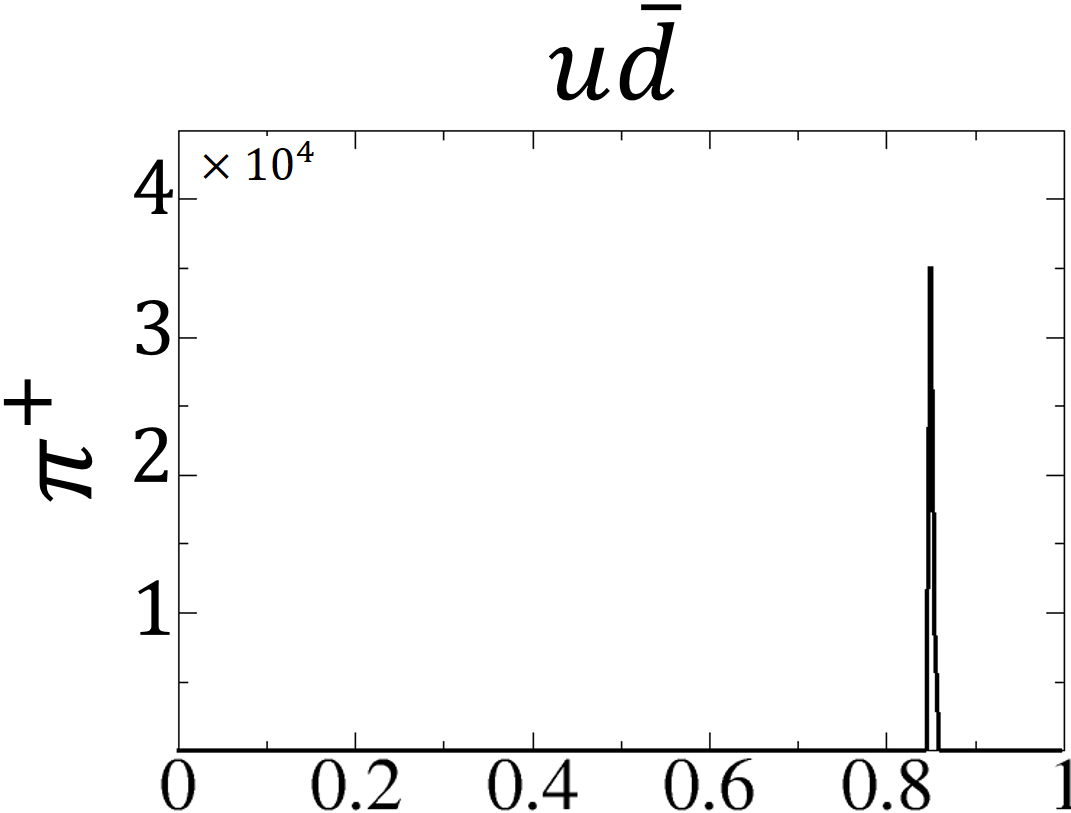}\hspace{0.02\columnwidth}\hskip -.3cm
	\includegraphics[height=1.12in]{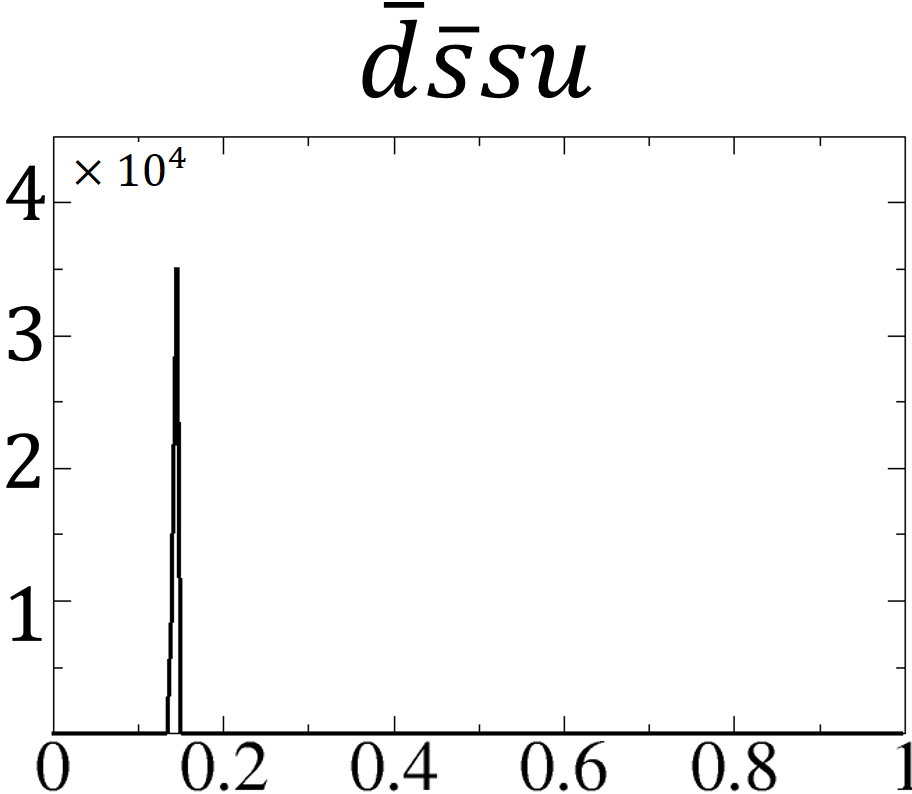}\hspace{0.02\columnwidth}\hskip .5cm
	\includegraphics[height=1.12in]{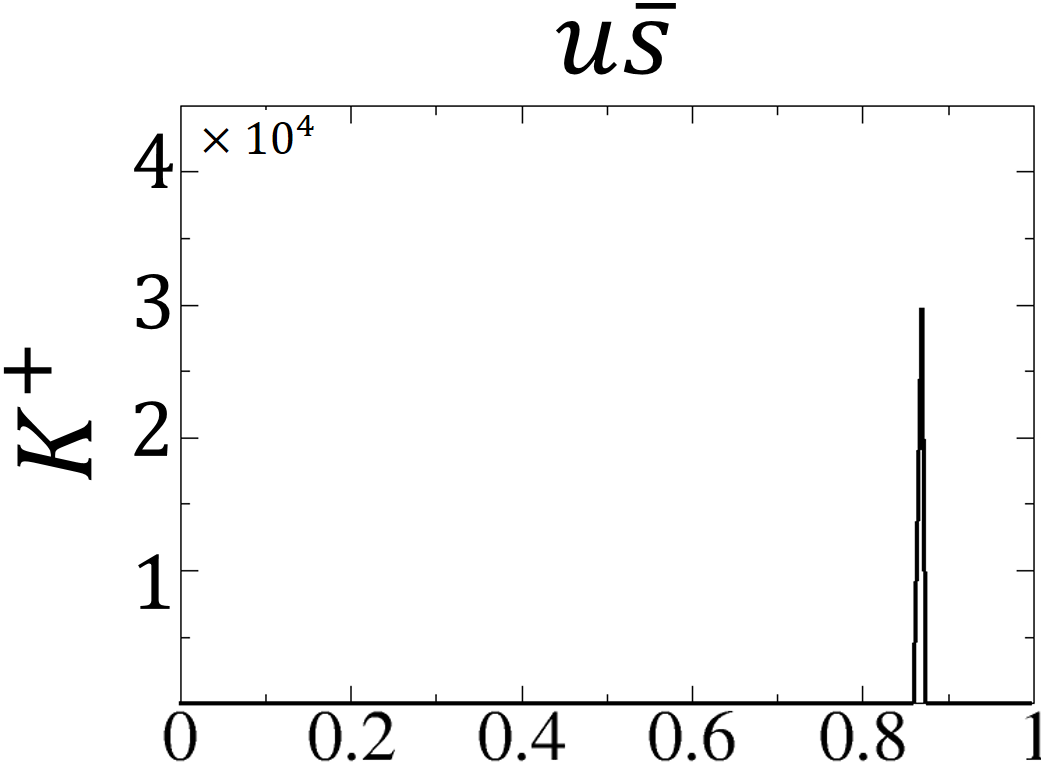}\hspace{0.02\columnwidth}\hskip -.3cm
	\includegraphics[height=1.16in]{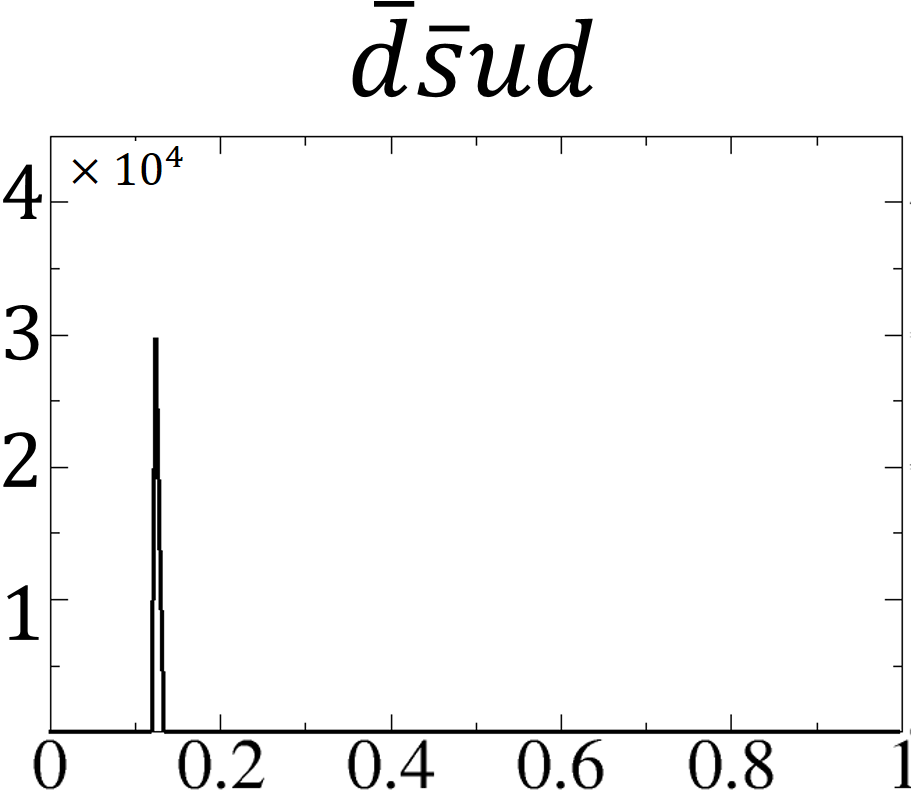}\hspace{0.02\columnwidth}	
	
	\includegraphics[height=.96in]{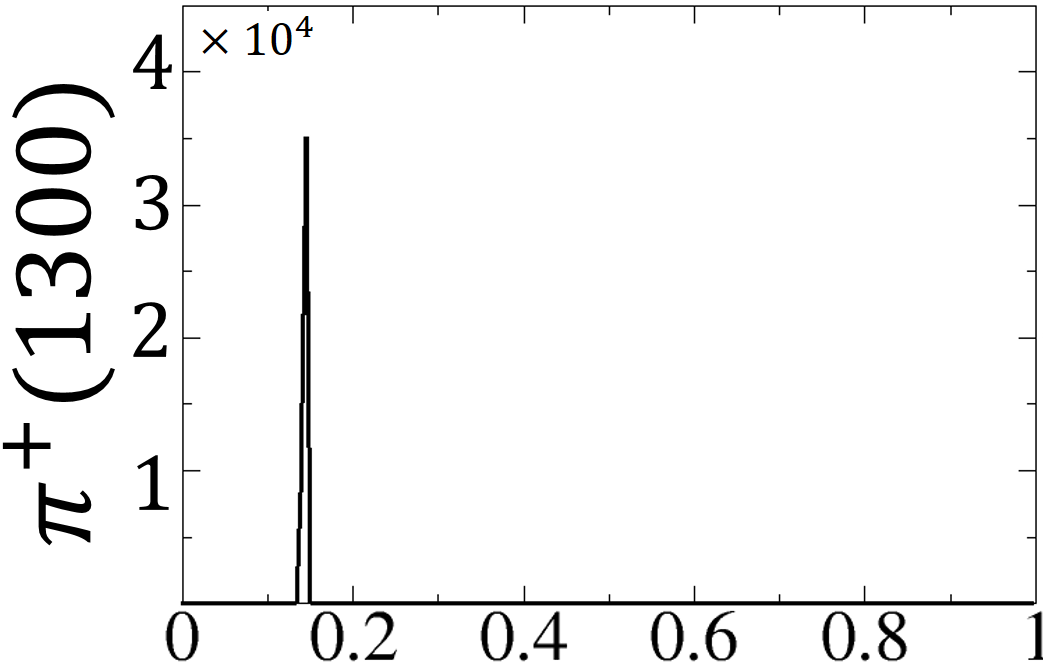}\hspace{0.02\columnwidth}\hskip -.3cm
	\includegraphics[height=.96in]{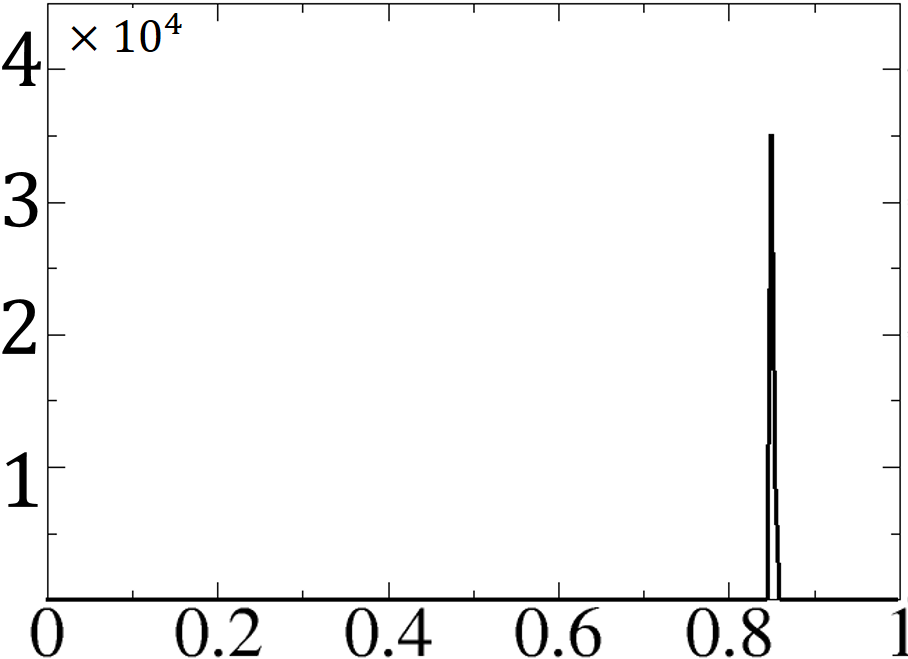}\hspace{0.02\columnwidth}\hskip .5cm
	\includegraphics[height=.96in]{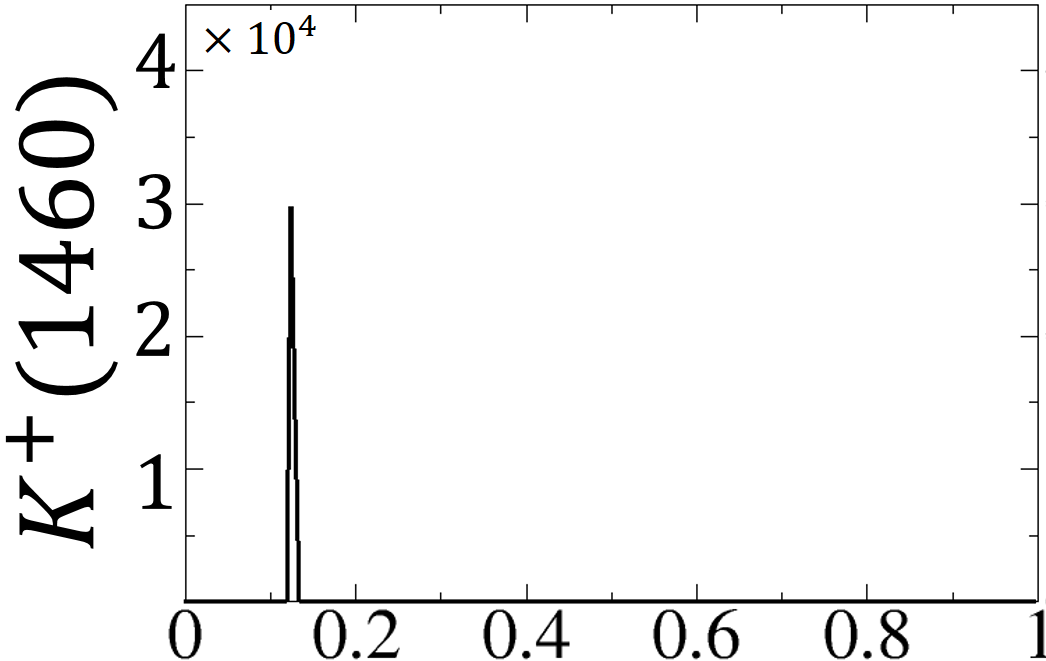}\hspace{0.02\columnwidth}\hskip -.3cm
	\includegraphics[height=.96in]{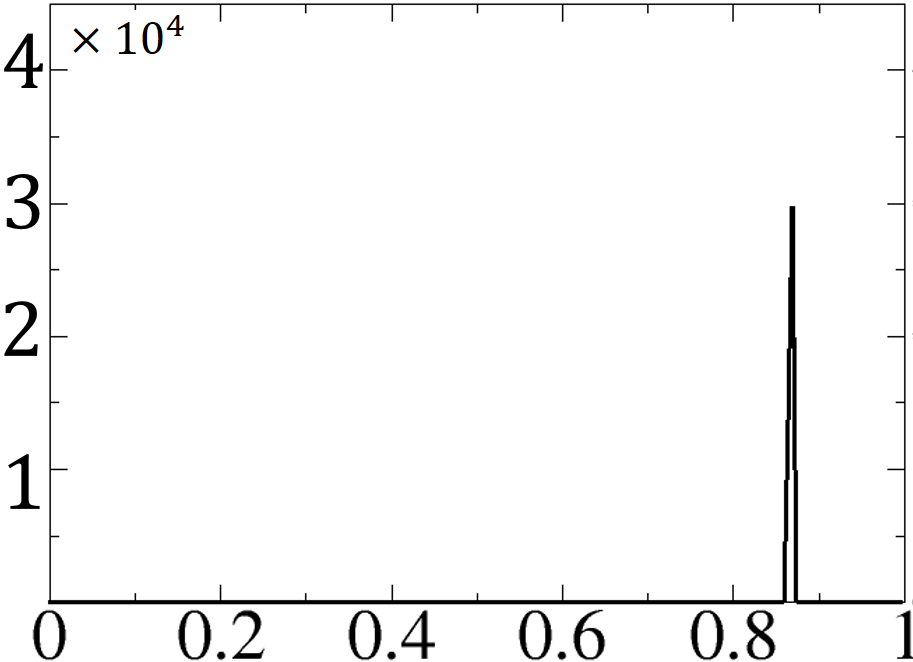}\hspace{0.02\columnwidth}
	
	\vskip .5cm
	
	\includegraphics[height=.96in]{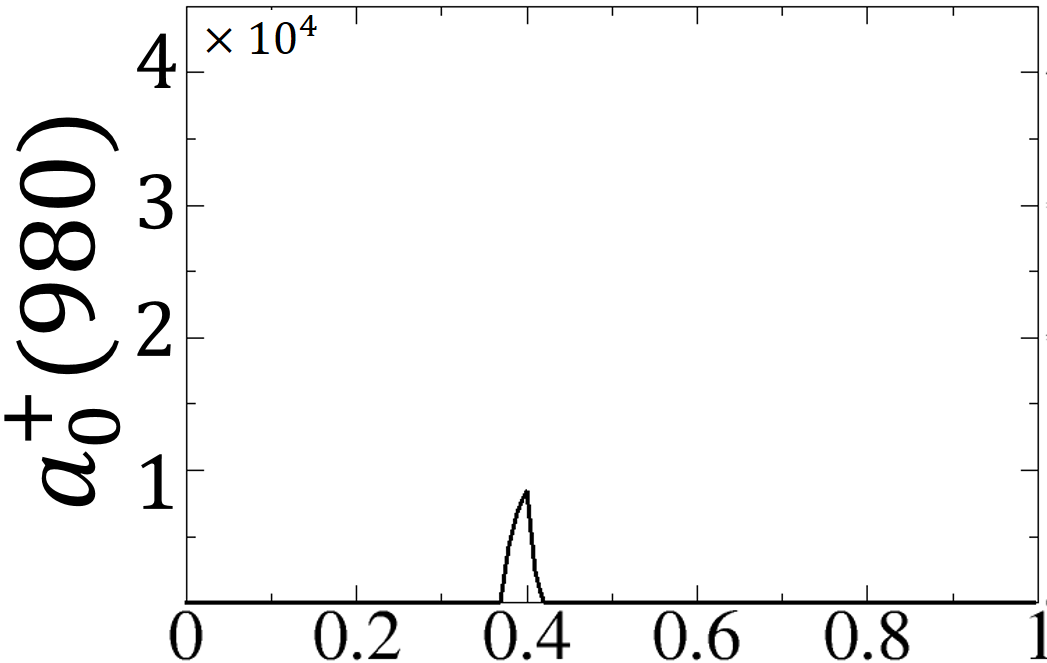}\hspace{0.02\columnwidth}\hskip -.3cm
	\includegraphics[height=.96in]{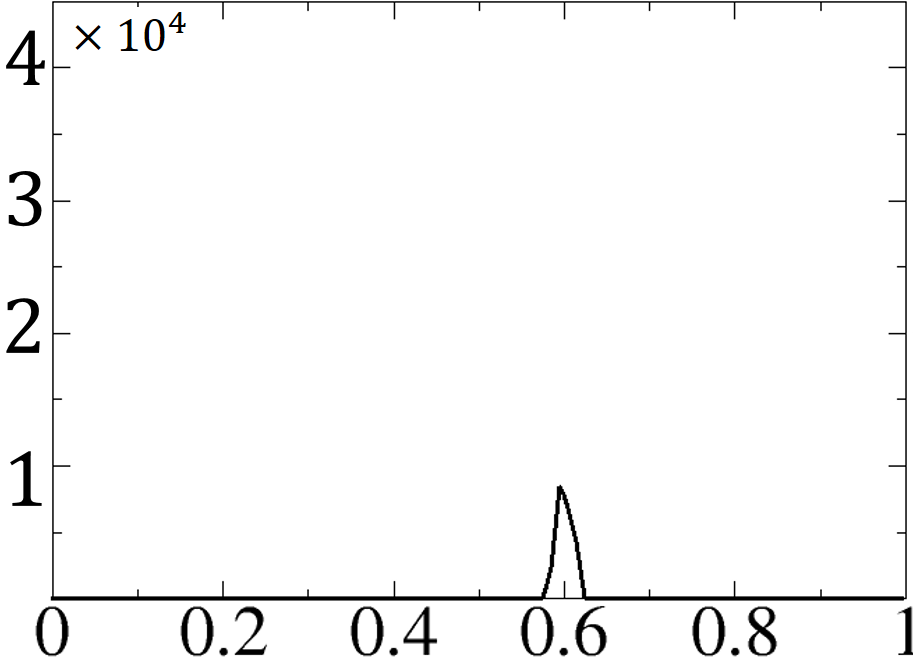}\hspace{0.02\columnwidth}\hskip .5cm
	\includegraphics[height=.96in]{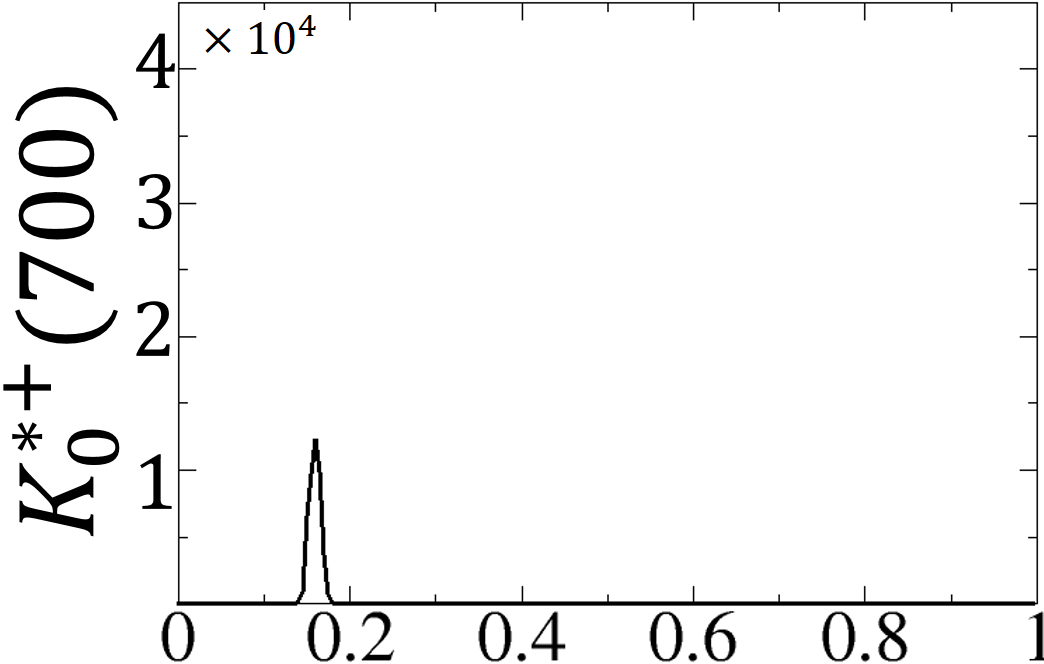}\hspace{0.02\columnwidth}\hskip -.3cm
	\includegraphics[height=.96in]{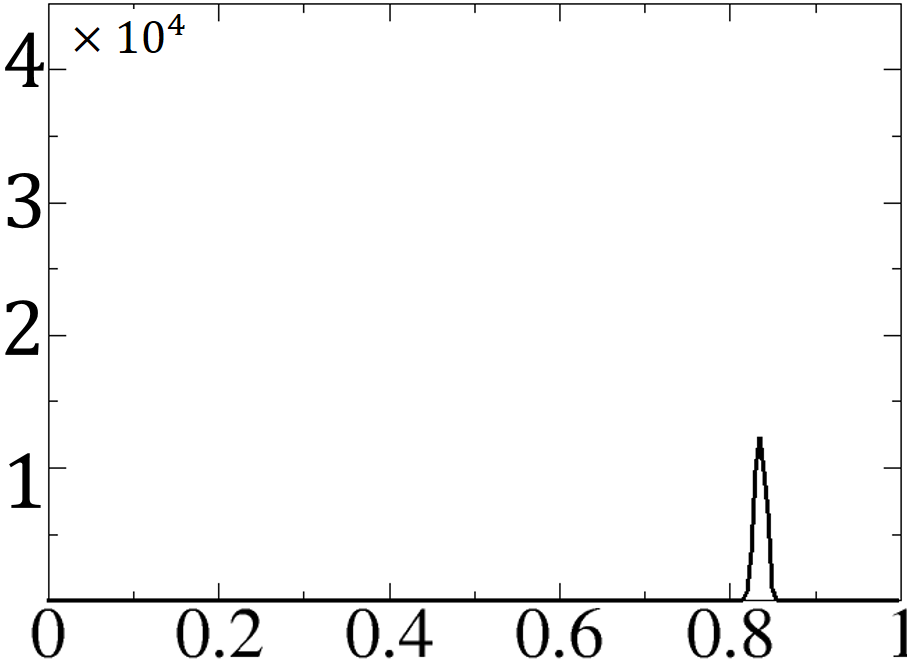}\hspace{0.02\columnwidth}
	
	\includegraphics[height=.96in]{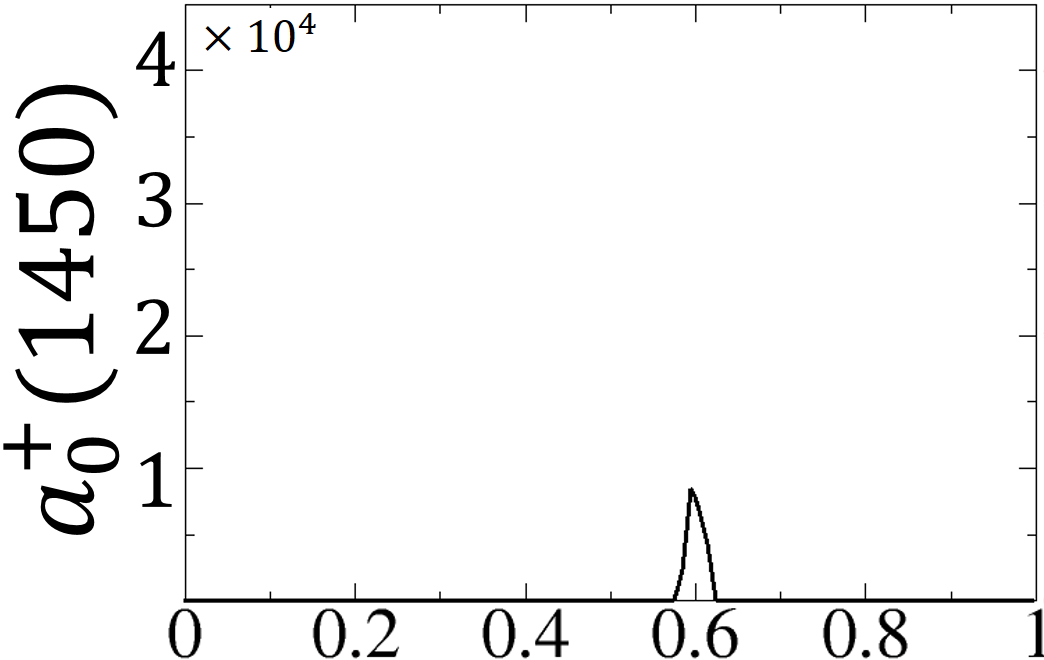}\hspace{0.02\columnwidth}\hskip -.3cm
	\includegraphics[height=.96in]{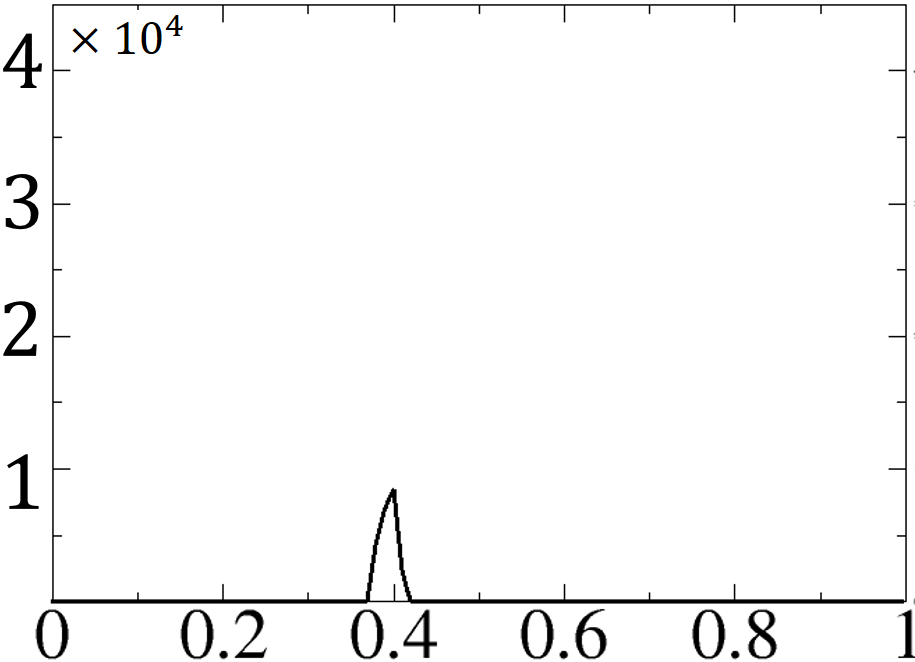}\hspace{0.02\columnwidth}\hskip .5cm
	\includegraphics[height=.96in]{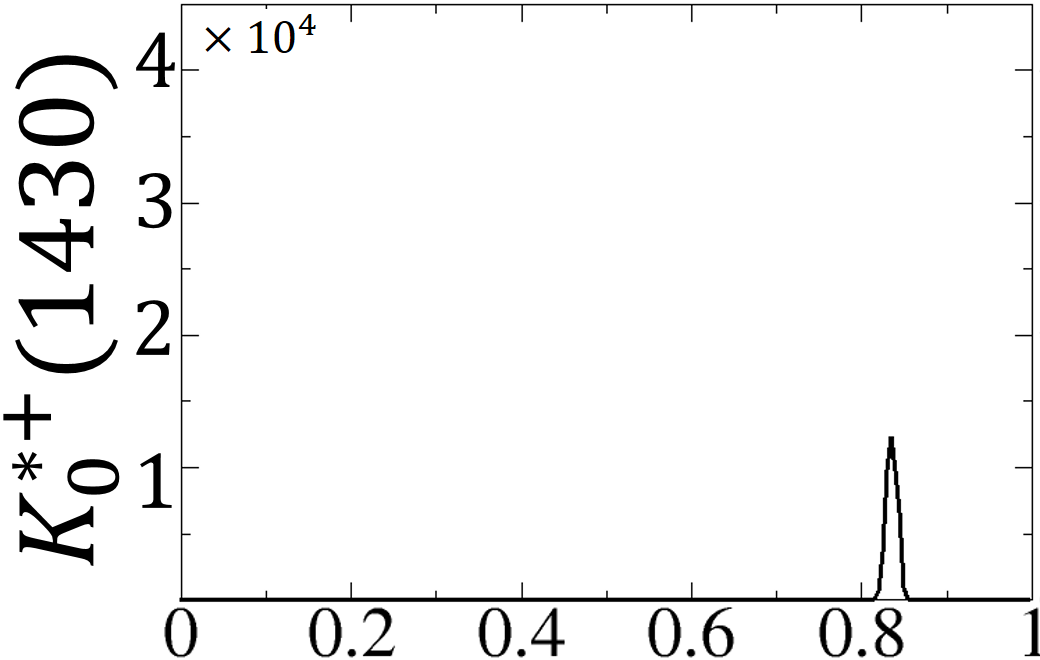}\hspace{0.02\columnwidth}\hskip -.3cm
	\includegraphics[height=.96in]{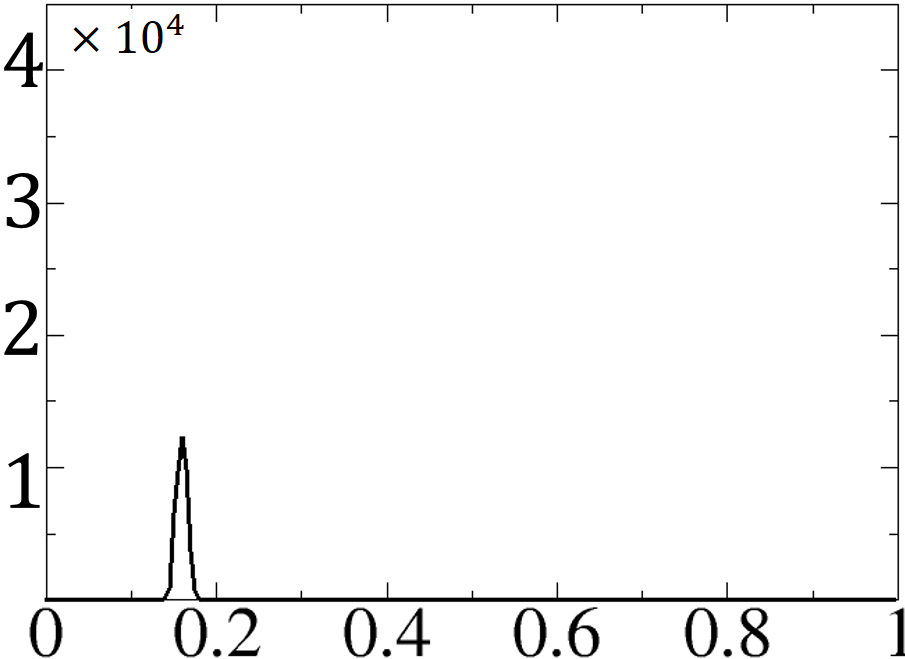}\hspace{0.02\columnwidth}
	
	\caption{Histograms for the components of isotriplets and isodoublets found in numerical simulation of this model produced by parameters $\mathbf p$ in set $S_{II}$ (\ref{E_SII})  with unfiltered range of $h_0 = 0.40-0.99$ GeV.  Each physical state is a linear combination of the quark compositions shown at the top of each column. }
	\label{F_piKa0kappa_comps}
\end{figure}

Next, the  predictions for the substructures of isotriplet and isodoublet states are presented in Fig. \ref{F_piKa0kappa_comps}.    As expected,  pion and kaon are dominantly quark-antiquark states, but the heavier pion and kaon are predicted in this model to be of four-quark type. On the other hand, as is also widely accepted, the situation with scalar states is reversed with light isotriplet and isodoublet scalars being  mostly  four-quark type,  and their heavier states being  mostly of quark-antiquark type.   These results are consistent with the predictions made in the generalized linear sigma model without glueballs in \cite{09_Fariborz_PRD79}.   The main difference between results presented in \cite{09_Fariborz_PRD79} and the present work is that in this work the constraint (\ref{E_chi_goodness}) limits the ranges of the experimental inputs [see (\ref{E_mPiA31_ranges})] which in turn leads to sharper predictions in Fig. \ref{F_piKa0kappa_comps}.

% --------------------------------------------------------------------------

\subsubsection{Isosinglet states}

As pointed out before,  the properties of the isosinglet states involve  both parameters $\mathbf p$ and $\mathbf q$ in set $S_{II}$.    Particularly, parameter $h_0$ (the glueball condensate) plays a key role in this analysis and a range of 0.75-0.825 was favored for this parameter in the analysis of this model in the SU(3) flavor limit in \cite{21_Fariborz_NPA1015}.   While we use this  SU(3) result  for comparison and for  guiding the initial ranges of parameters, we perform the  present analysis independently, and  with glueball condensate  within a more extended  range of $h_0=0.4-0.99$ GeV when performing simulations to construct $S_{II}$.  Beyond this range of $h_0$, there are no acceptable parameters.

\begin{figure}
%	\centering
	\includegraphics[height=3.5in]{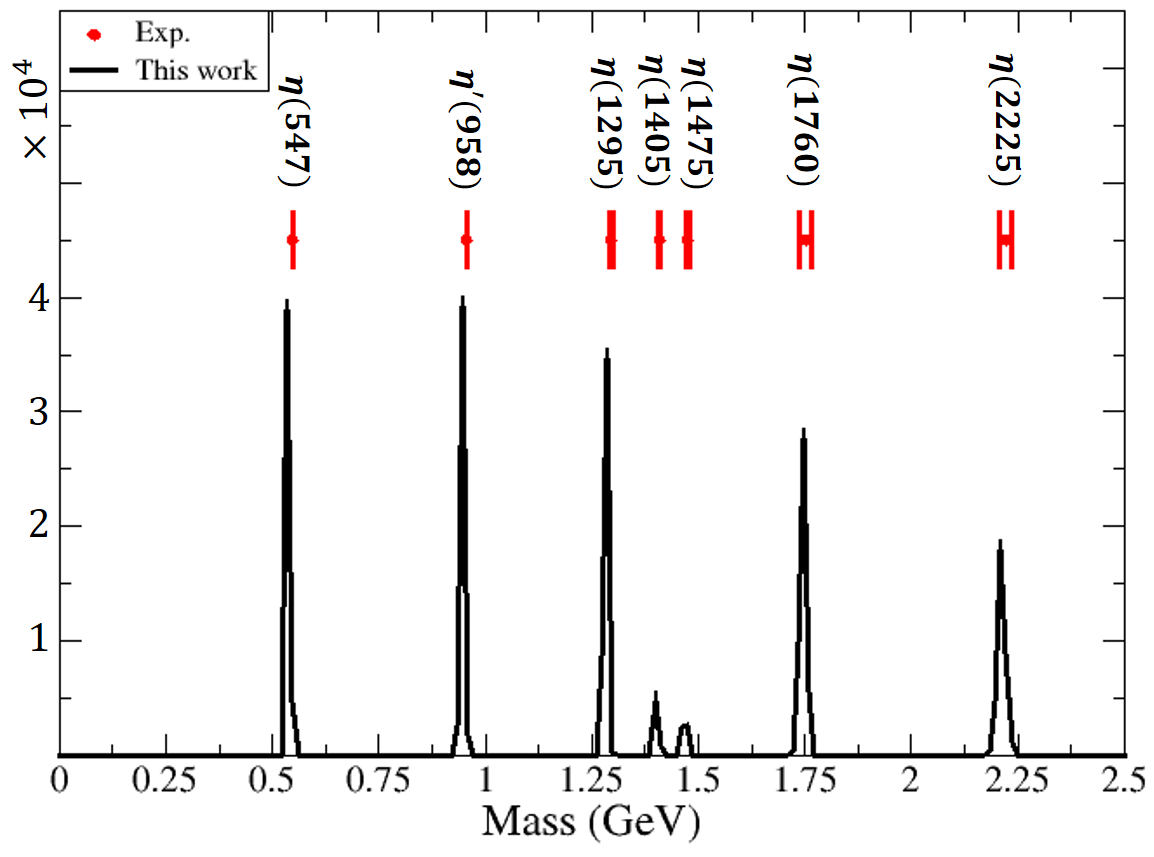}
	%\hspace{0.02\columnwidth}
	\caption{Simulation results for the eta masses predicted in  this work using set $S_{II}$ (\ref{E_SII}) with unfiltered range of $h_0=0.4-.99$ GeV are compared with experiment (in red).
	}
	\label{F_Meta_hist_h0_4-99}
\end{figure}

We use set $S_{II}$ (produced with the entire range of $h_0 =0.4-0.99$ GeV) to compute the masses and the substructure of the isosinglet states and compare with experiment.  Figure \ref{F_Meta_hist_h0_4-99} gives the predicted eta masses and shows a remarkable agreement with experiment. Note again that this is not a fit to the experimental data, but rather obtained by imposing the  condition (\ref{E_chi_goodness}) which is just a  collective constraint on the combination of the eta masses. 

The model, in its present formulation,  contains five etas (formed out of a linear combination of two etas in each of the two $M$ and $M'$ chiral nonets and a pseudoscalar glueball) and since not all simulations  favor the same five etas, the overall results in  Fig. \ref{F_Meta_hist_h0_4-99}  shows that clear signals for all seven etas are very accurately generated  by the model.   Some of these signals, such as those for $\eta(1405)$ and $\eta(1475)$,  are weak,   consistent (at least qualitatively) with the suggestions  in \cite{10_Albaladejo_PRD82}
that these two states might be dynamically generated and likely to be more complicated than being simply captured by ground state of mixed quark-antiquark, tetraquarks and a pseudoscalar glueball.   
Overall, as can be seen in Fig. \ref{F_Meta_hist_h0_4-99}, the simulations clearly favor scenario 6 defined in Table \ref{T_10_scenarios}.

\begin{figure}
%	\centering
	\includegraphics[height=3.5in]{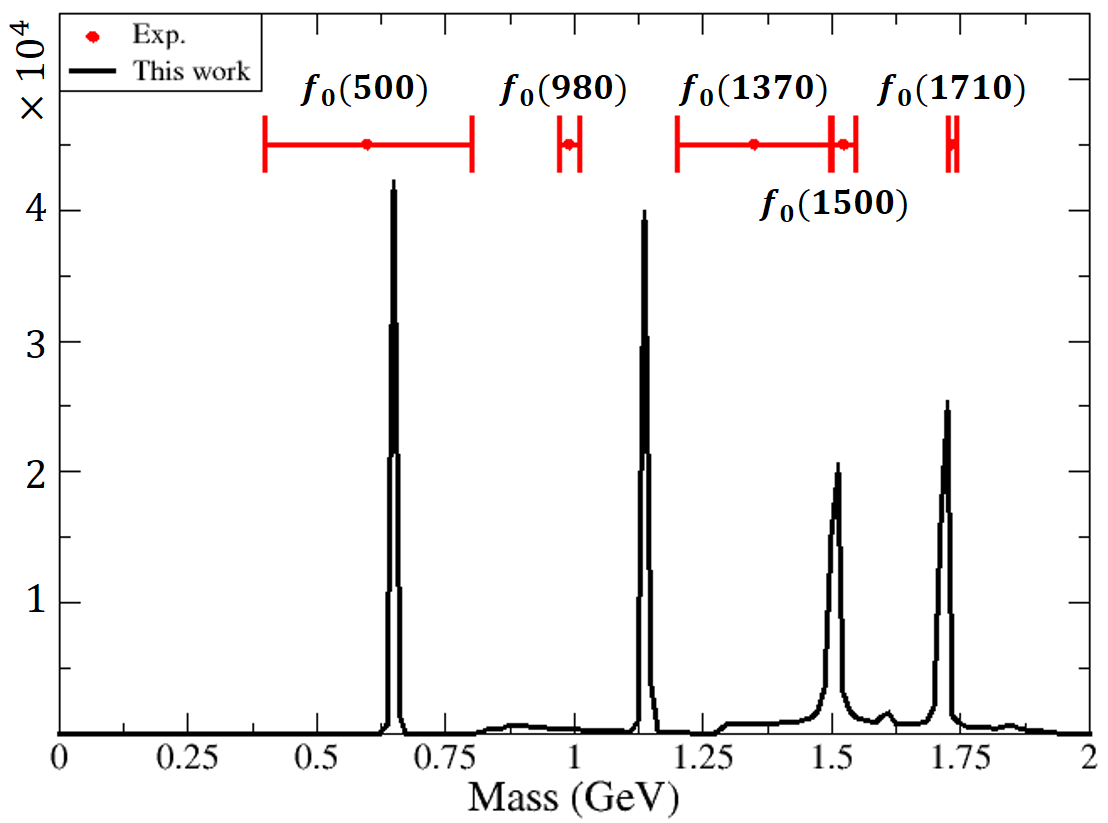}
	%\hspace{0.02\columnwidth}
\caption{Simulation results for the $f_0$ masses predicted by  this model using set $S_{II}$ (\ref{E_SII}) with unfiltered range of $h_0=0.4-0.99$ GeV  are compared with experiment.}
	\label{F_Mf0_hist_h0_4-99}
\end{figure}

The model predictions for the masses of isosinglet scalar mesons, the $f_0$ states, obtained in these simulations are given in Fig. \ref{F_Mf0_hist_h0_4-99} and compared with experiment.   The case of scalars is known to be more challenging and that is  reflected in this figure as well.   First, we again comment that due to the final-state interactions of the decay products of scalars,    the Lagrangian masses are not necessarily equal to the physical masses that are probed in the scattering analyses.   Within the generalized linear sigma model without glueballs  these were studied in \cite{11_Fariborz_PRD84,15_Fariborz_PRD92,22_Fariborz_EJPC82} in which it is shown that the physical mass of broad states [such as $f_0(500)$ or $K_0^*(700)$] probed in scattering processes are considerably lower than their Lagrangian masses.   Specifically,   it is shown in \cite{11_Fariborz_PRD84} that when the $f_0(500)$ is probed in $\pi\pi$ scattering, its mass obtained from the pole of the unitarized $\pi\pi$ scattering amplitude is reduced to $477 \pm 8$ MeV.  Similarly, the $f_0(980)$ receives some unitarity corrections for its mass and moves it closer to its experimental value \cite{11_Fariborz_PRD84}.   Secondly,   the mass of $f_0(1370)$ is not immediately seen in Fig. \ref{F_Mf0_hist_h0_4-99} due to its potential proximity to the mass of $f_0(1500)$.   This will be further discussed below when correlation with $h_0$ is examined.

\begin{figure}
%	\centering
	\includegraphics[height=1.12in]{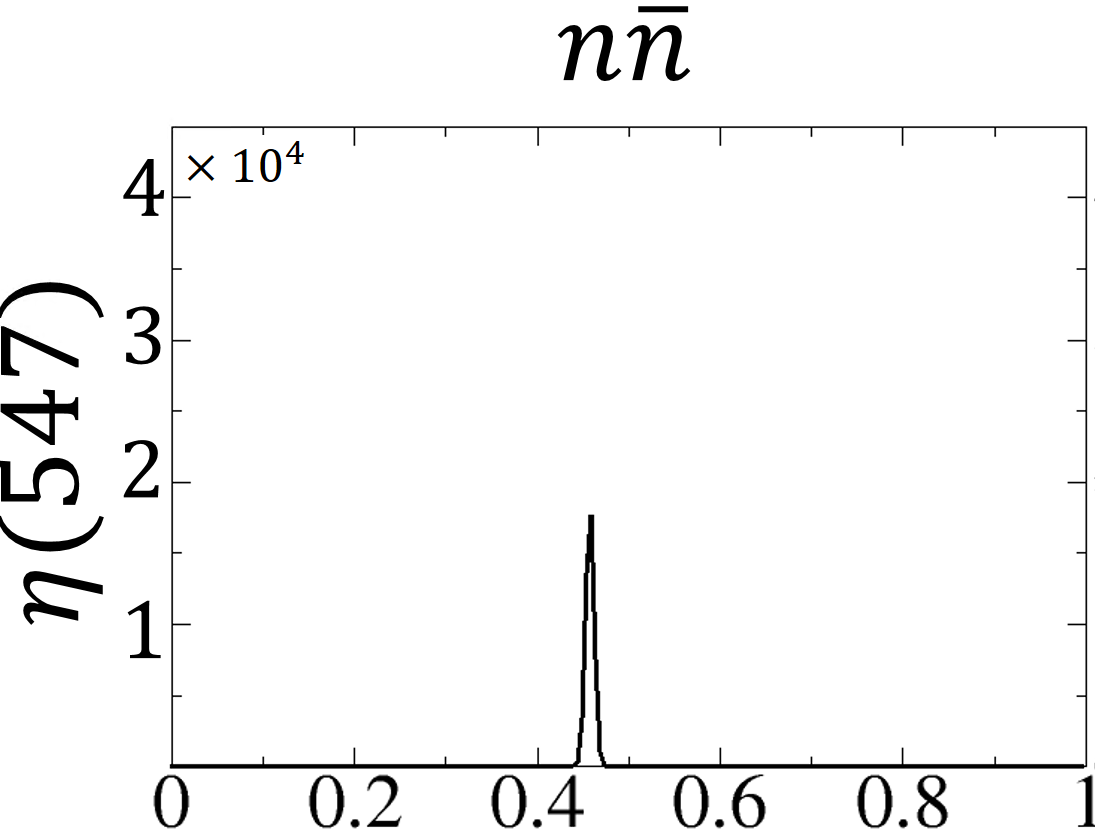}\hspace{0.02\columnwidth}\hskip -.3cm
	\includegraphics[height=1.12in]{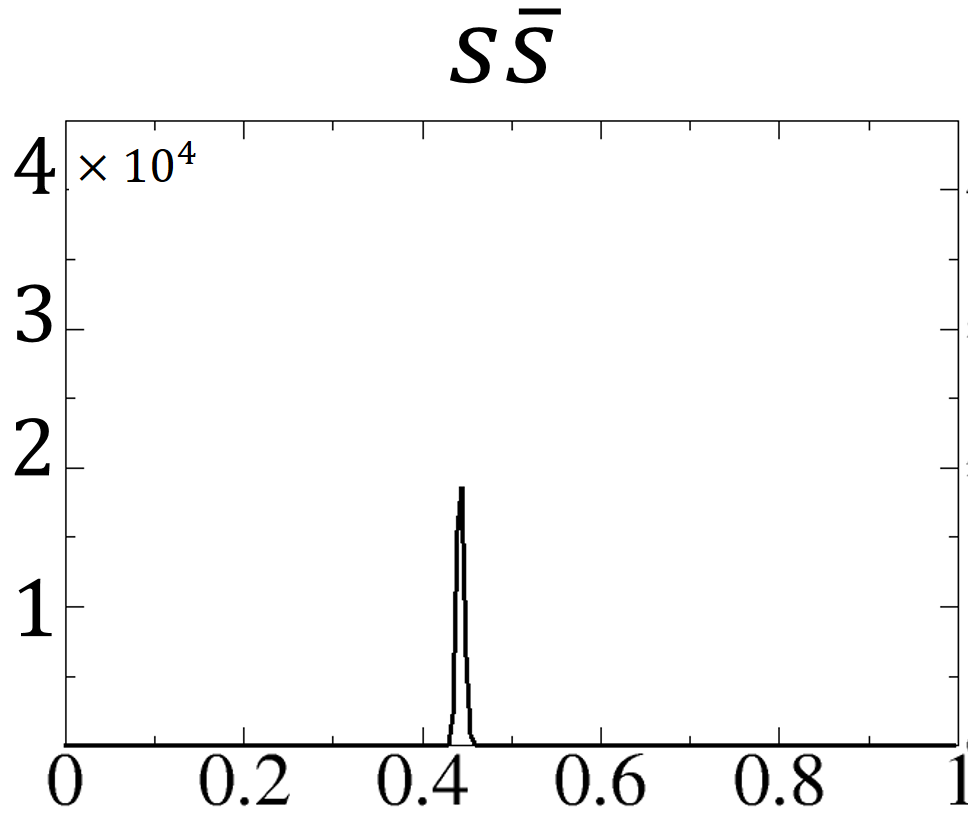}\hspace{0.02\columnwidth}\hskip -.3cm
	\includegraphics[height=1.12in]{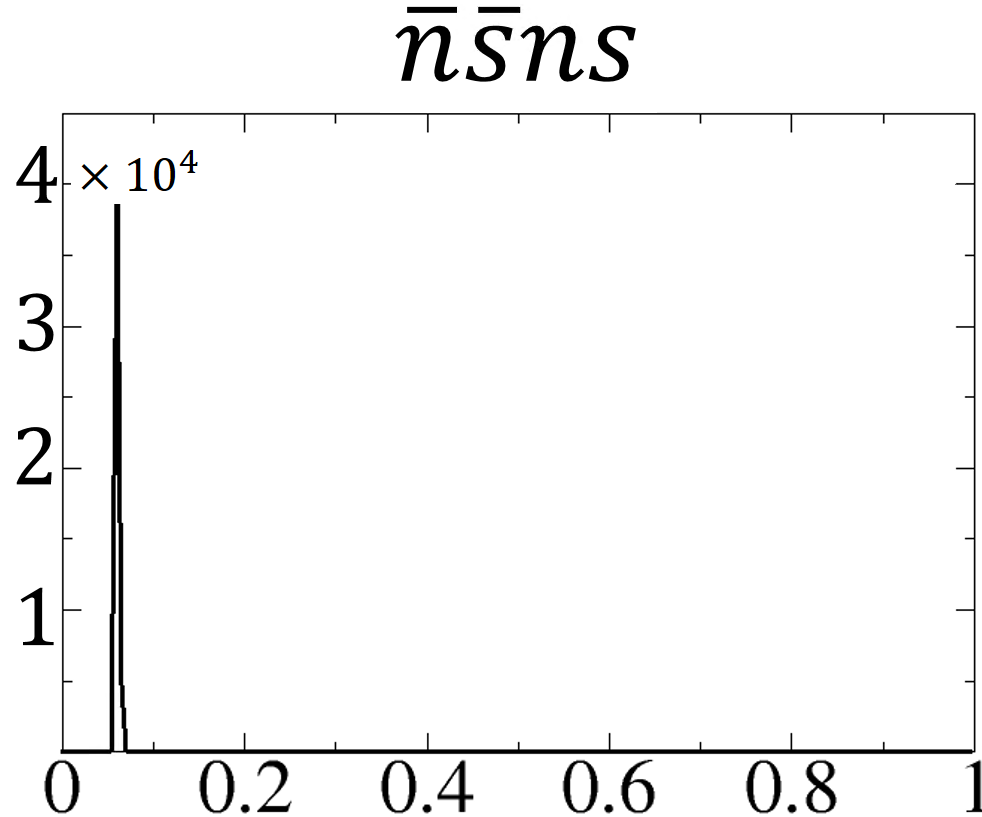}\hspace{0.02\columnwidth}\hskip -.3cm
	\includegraphics[height=1.12in]{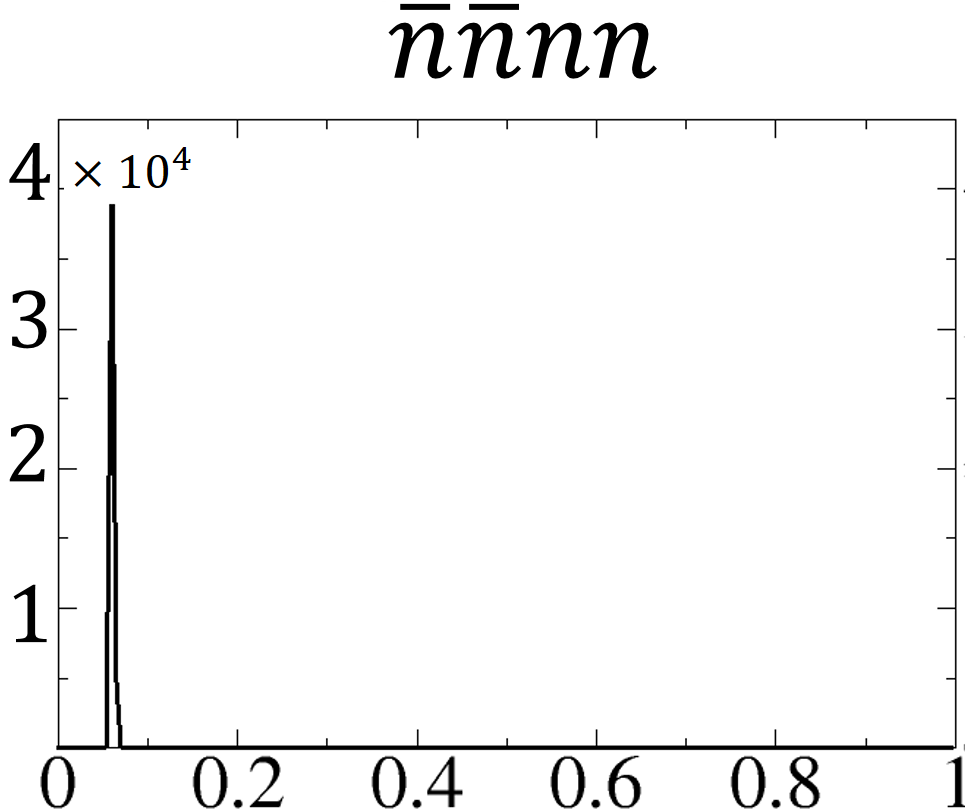}\hspace{0.02\columnwidth}\hskip -.3cm
	\includegraphics[height=1.12in]{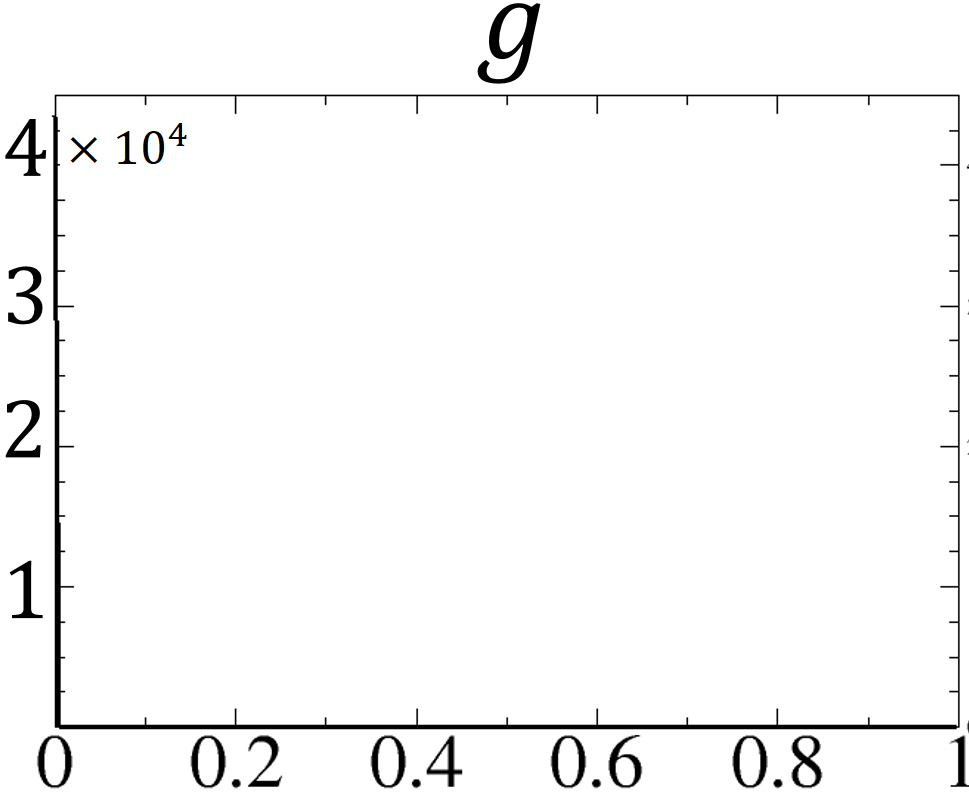}\hspace{0.02\columnwidth}
	
	\includegraphics[height=.98in]{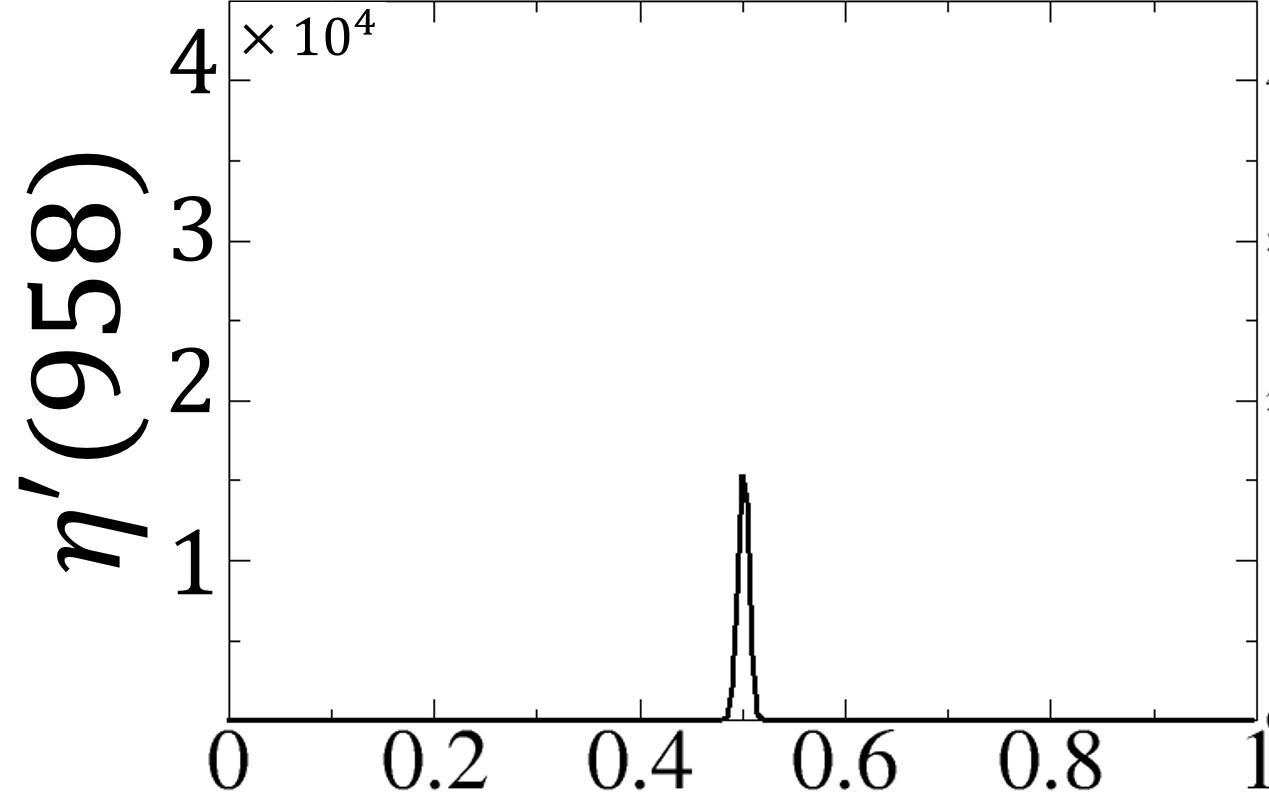}\hspace{0.02\columnwidth}\hskip -.3cm
	\includegraphics[height=.98in]{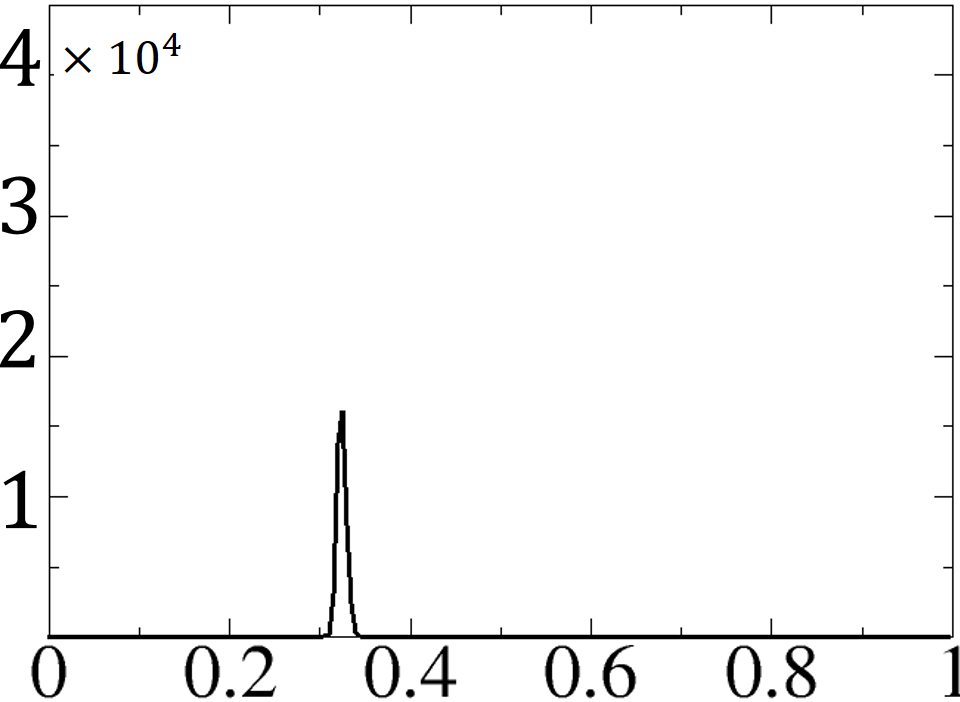}\hspace{0.02\columnwidth}\hskip -.3cm
	\includegraphics[height=.98in]{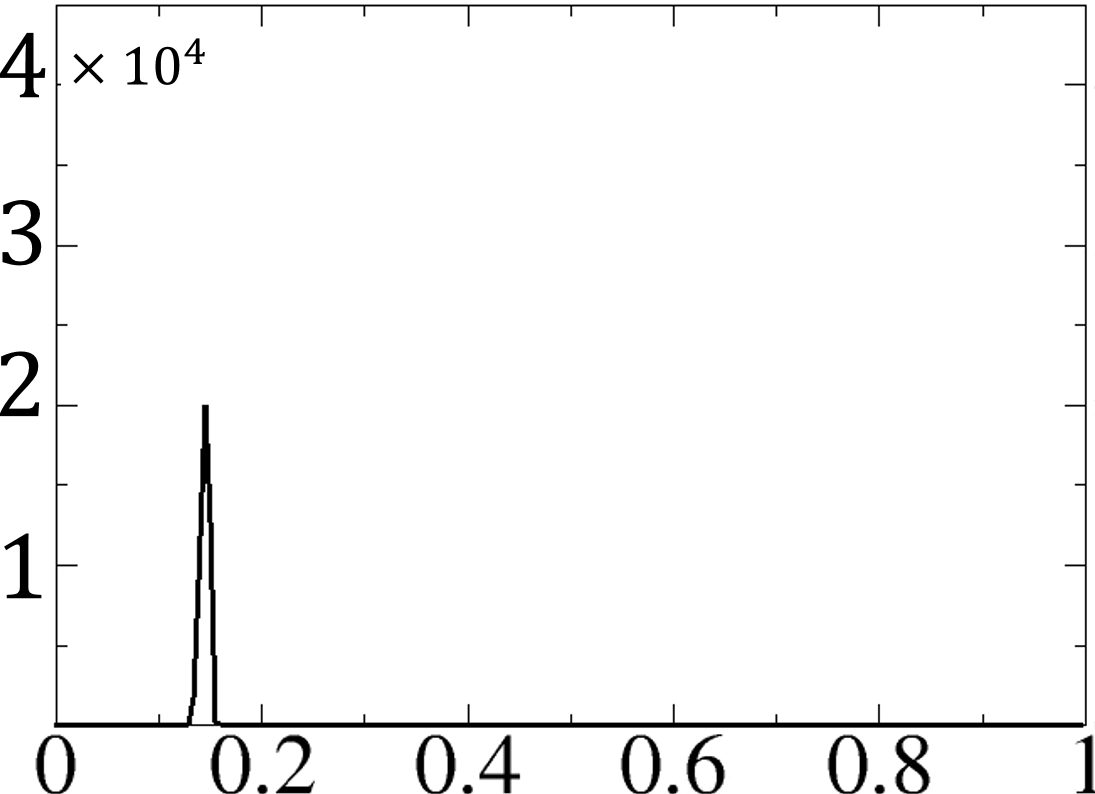}\hspace{0.02\columnwidth}\hskip -.3cm
	\includegraphics[height=.98in]{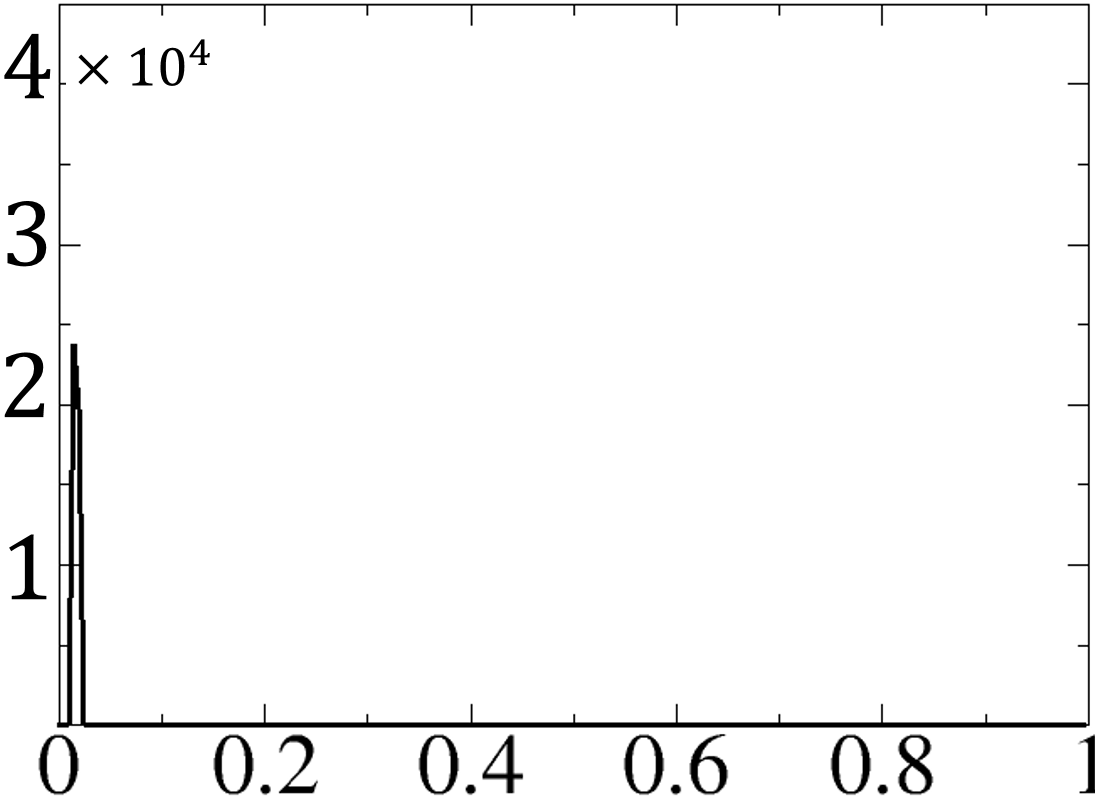}\hspace{0.02\columnwidth}\hskip -.3cm
	\includegraphics[height=.98in]{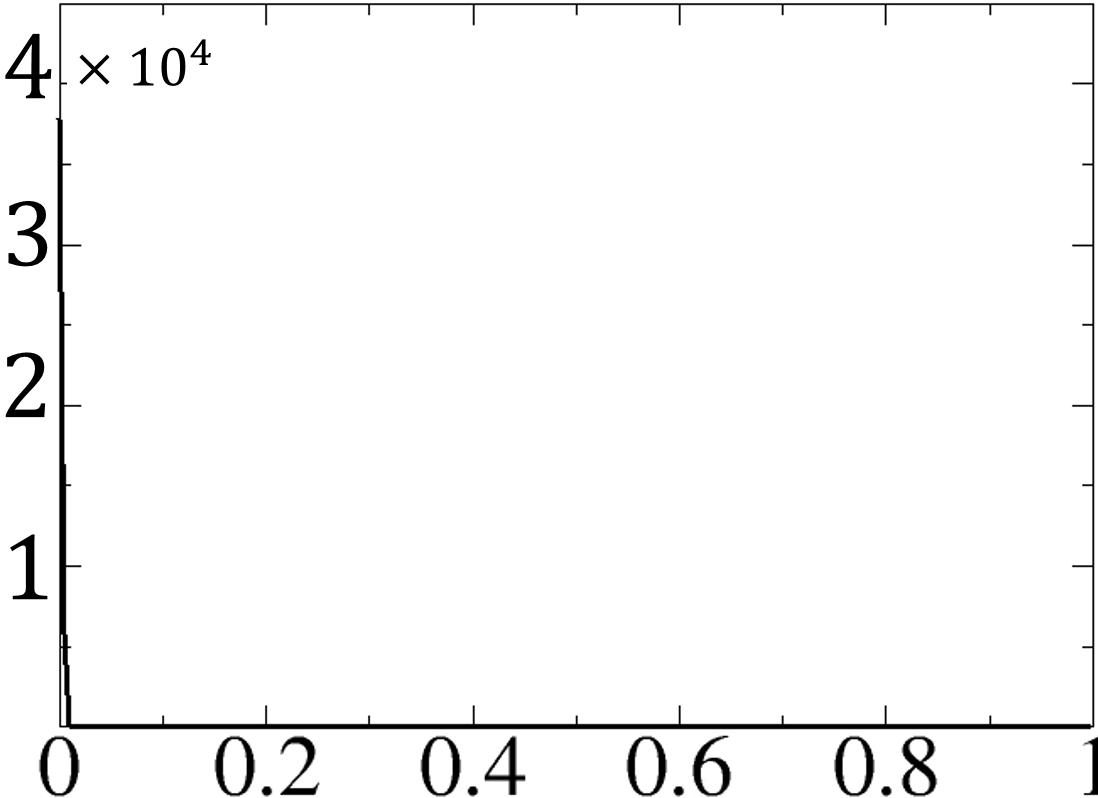}\hspace{0.02\columnwidth}
	
	\includegraphics[height=.98in]{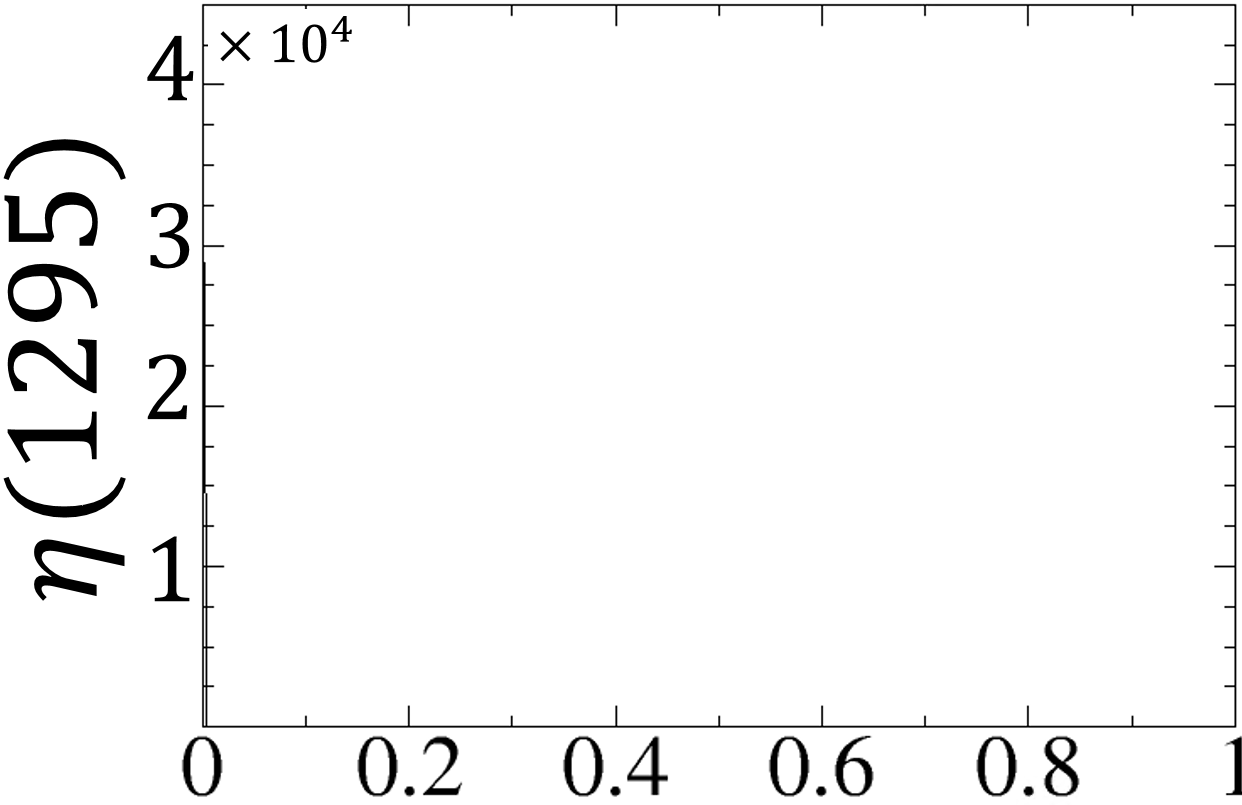}\hspace{0.02\columnwidth}\hskip -.3cm
	\includegraphics[height=.98in]{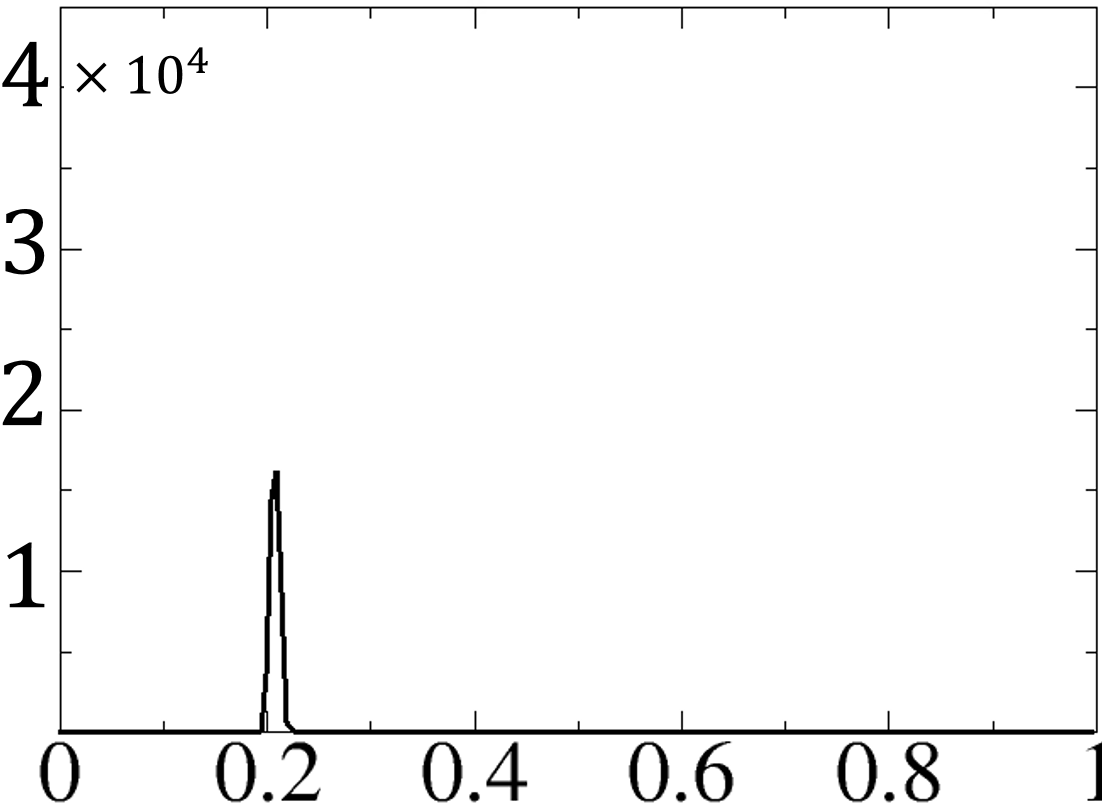}\hspace{0.02\columnwidth}\hskip -.3cm
	\includegraphics[height=.98in]{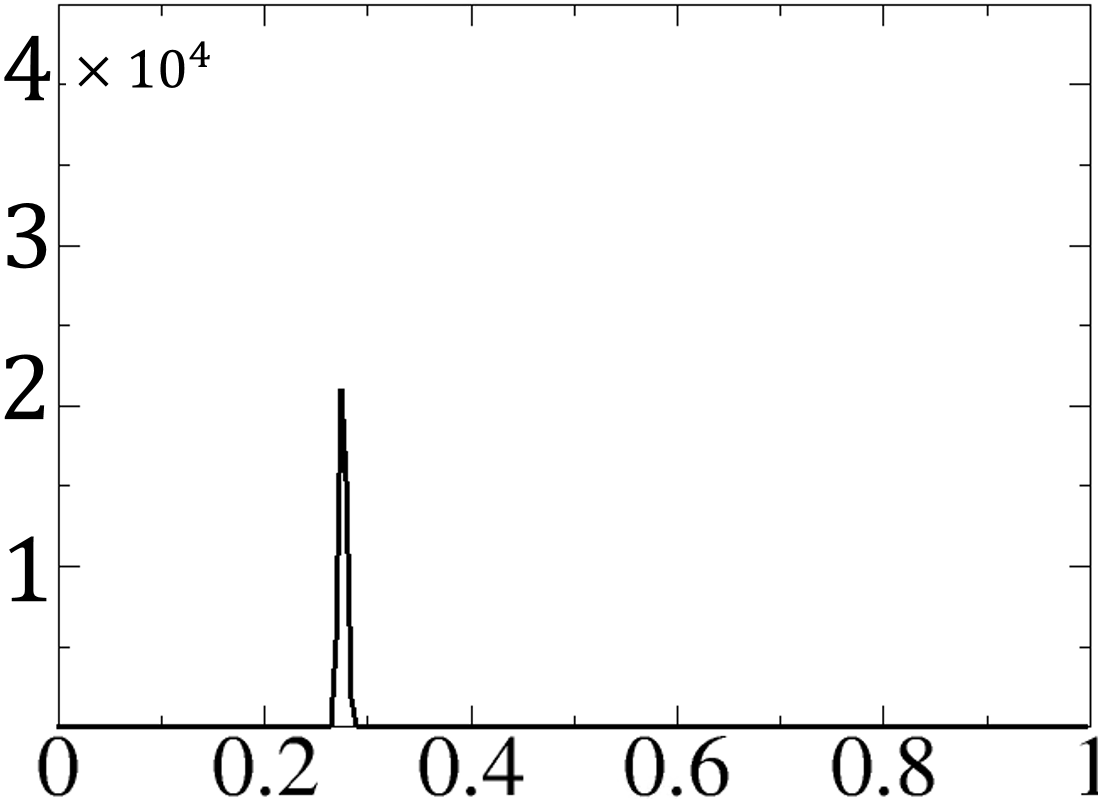}\hspace{0.02\columnwidth}\hskip -.3cm
	\includegraphics[height=.98in]{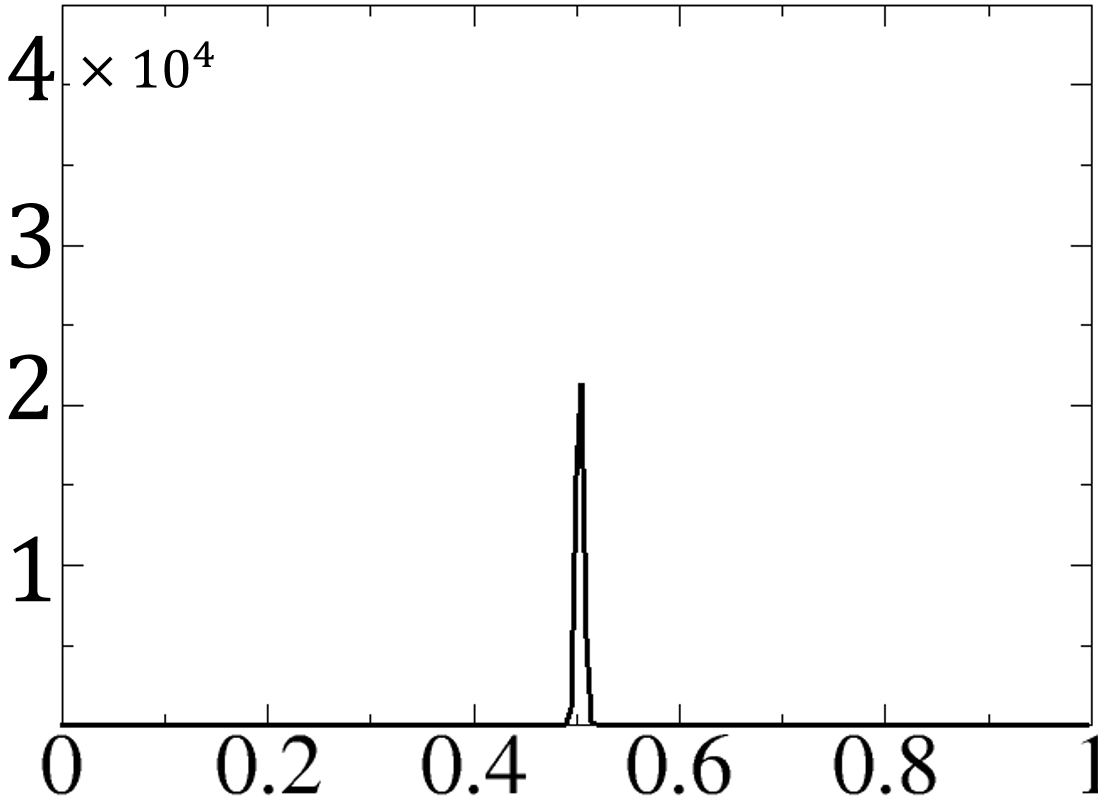}\hspace{0.02\columnwidth}\hskip -.3cm
	\includegraphics[height=.98in]{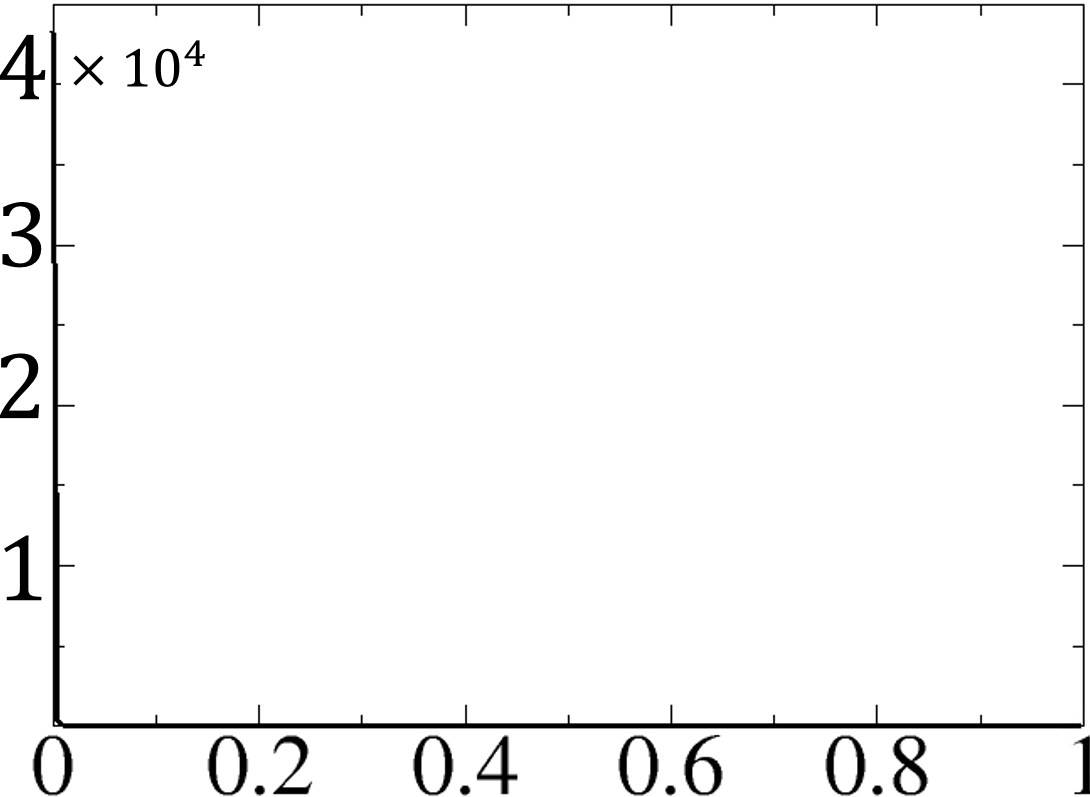}\hspace{0.02\columnwidth}
	
	\includegraphics[height=.98in]{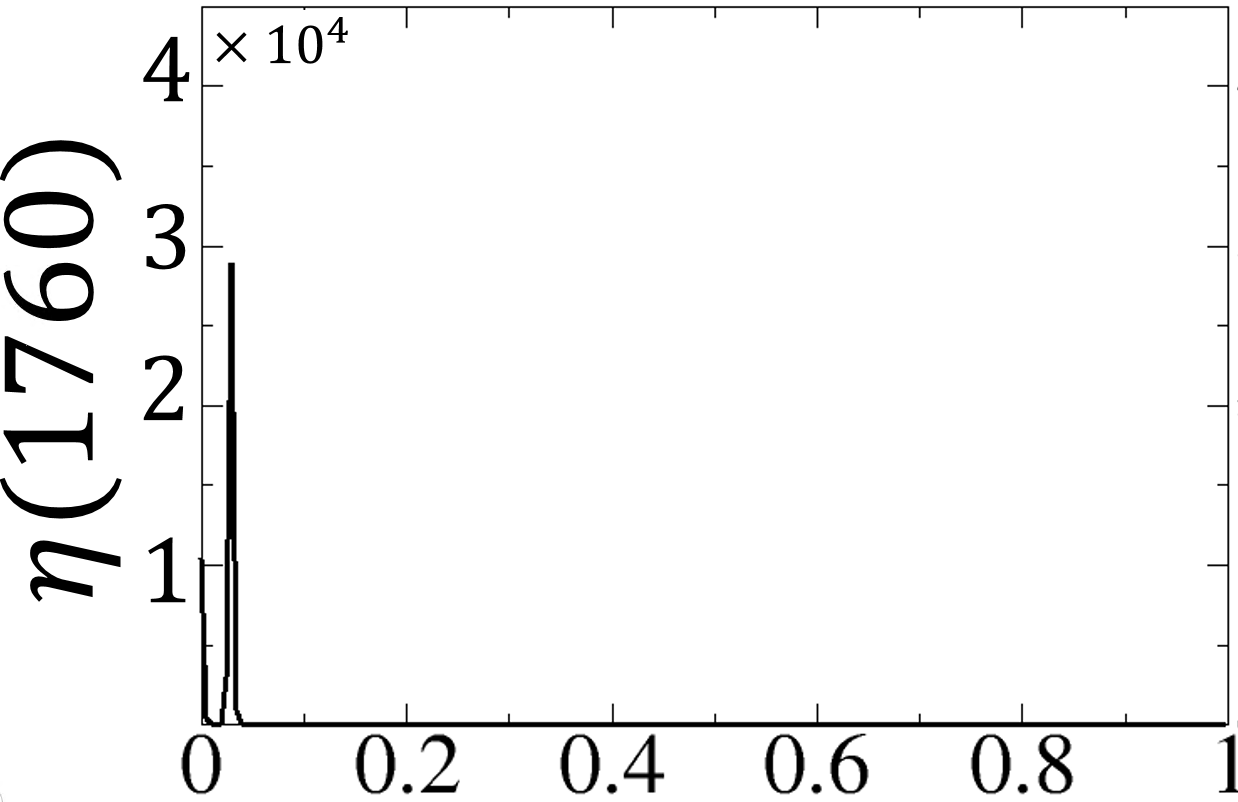}\hspace{0.02\columnwidth}\hskip -.3cm
	\includegraphics[height=.98in]{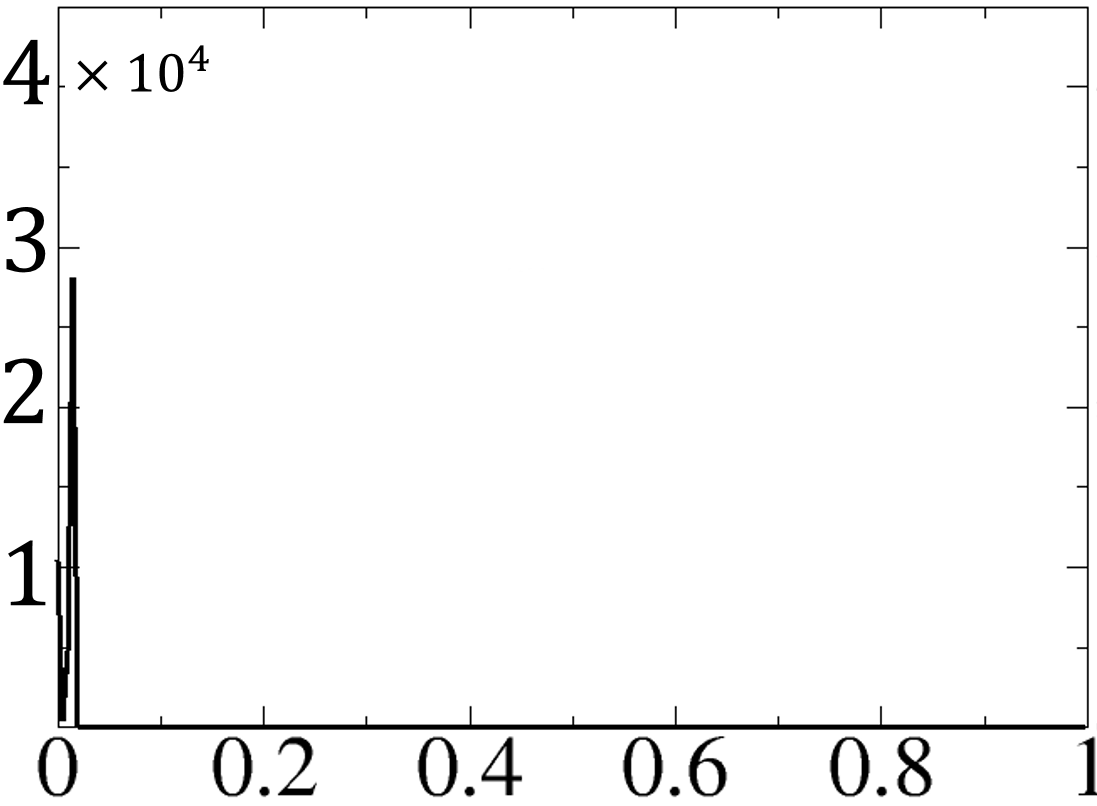}\hspace{0.02\columnwidth}\hskip -.3cm
	\includegraphics[height=.98in]{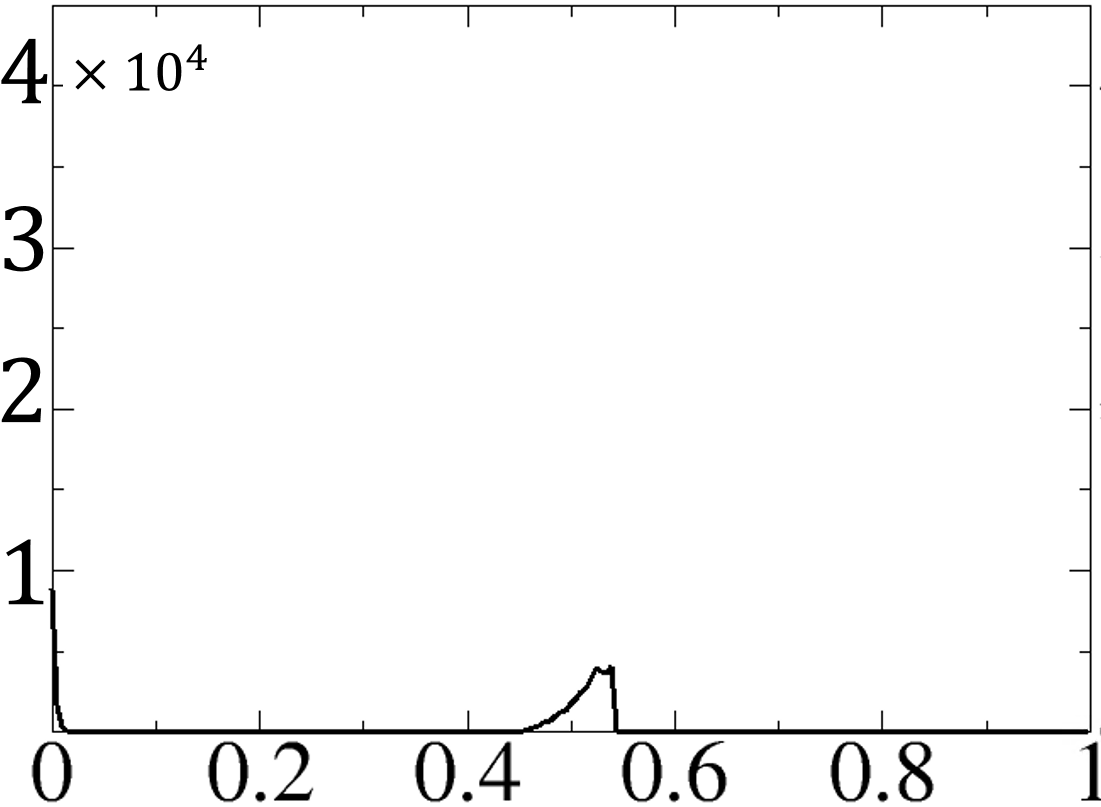}\hspace{0.02\columnwidth}\hskip -.3cm
	\includegraphics[height=.98in]{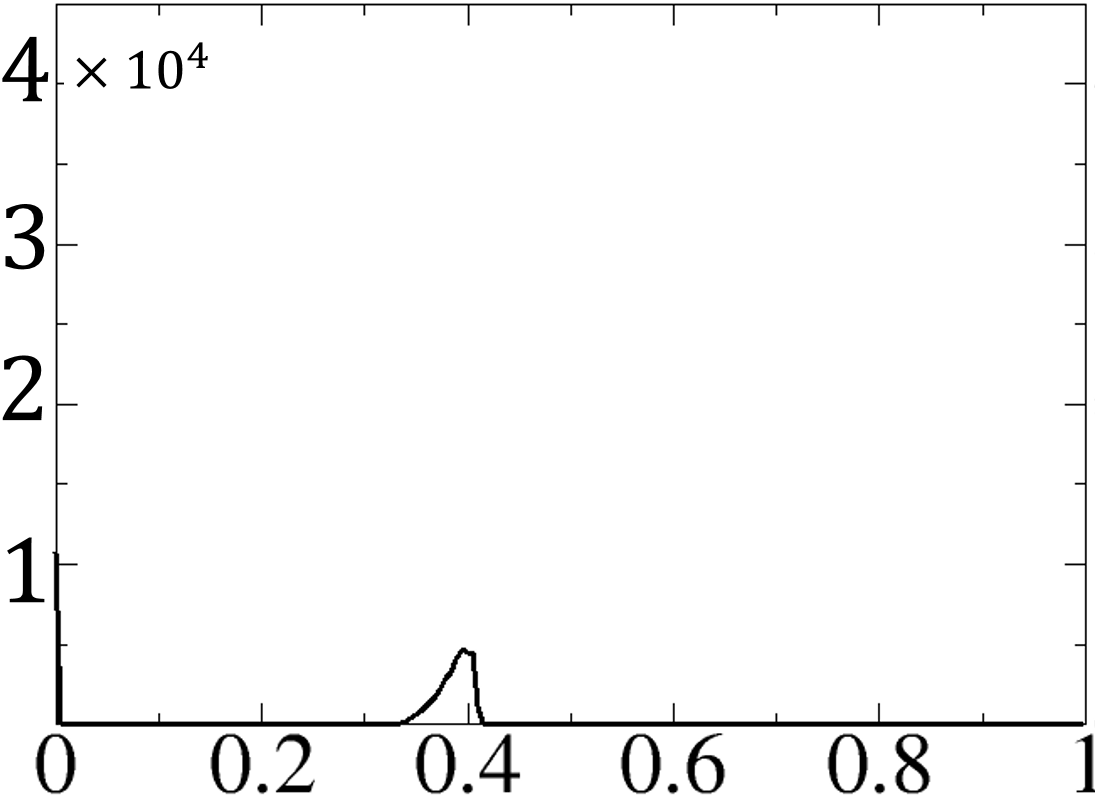}\hspace{0.02\columnwidth}\hskip -.3cm
	\includegraphics[height=.98in]{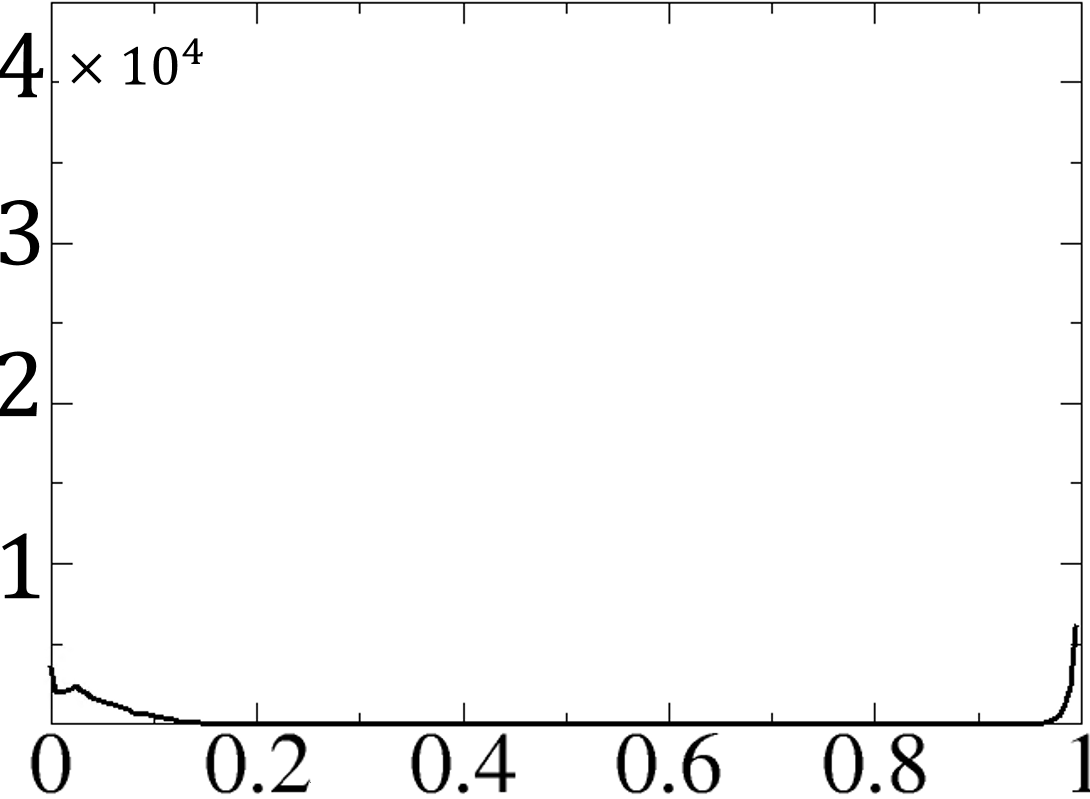}\hspace{0.02\columnwidth}
	
	\includegraphics[height=.98in]{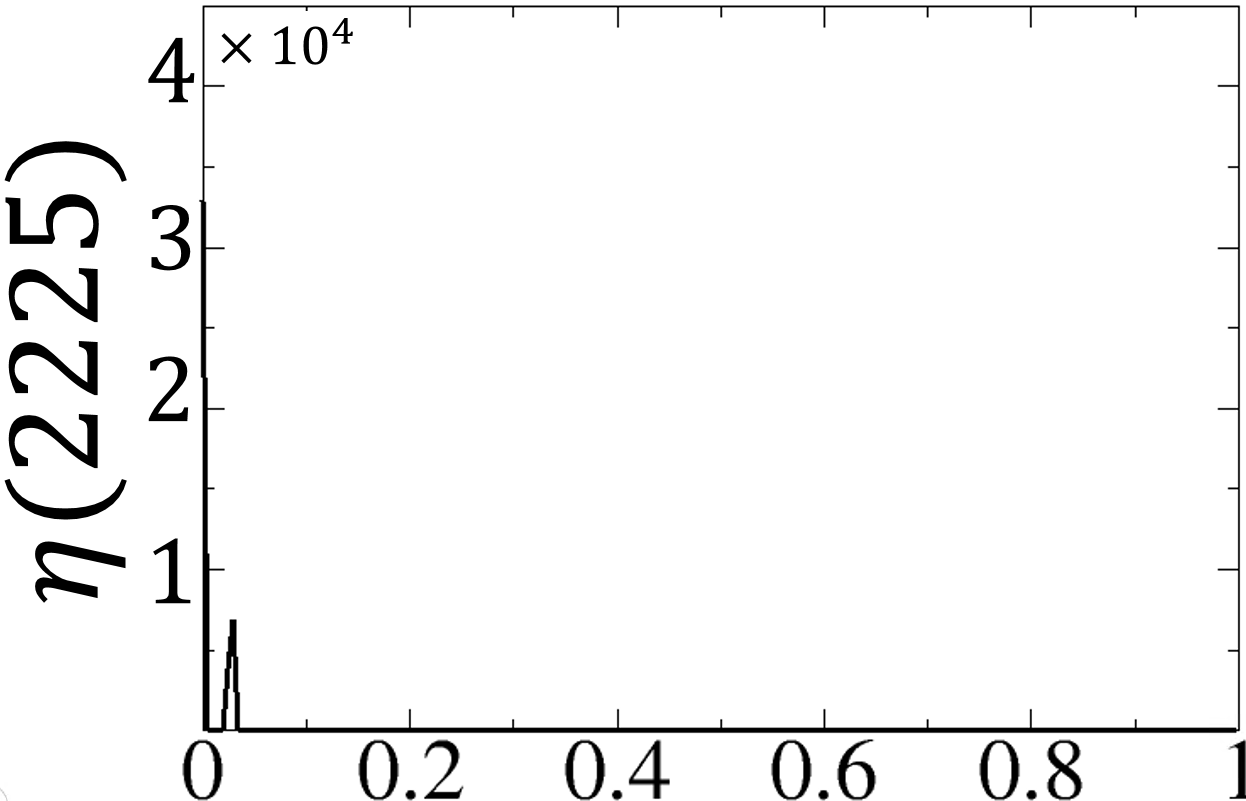}\hspace{0.02\columnwidth}\hskip -.3cm
	\includegraphics[height=.98in]{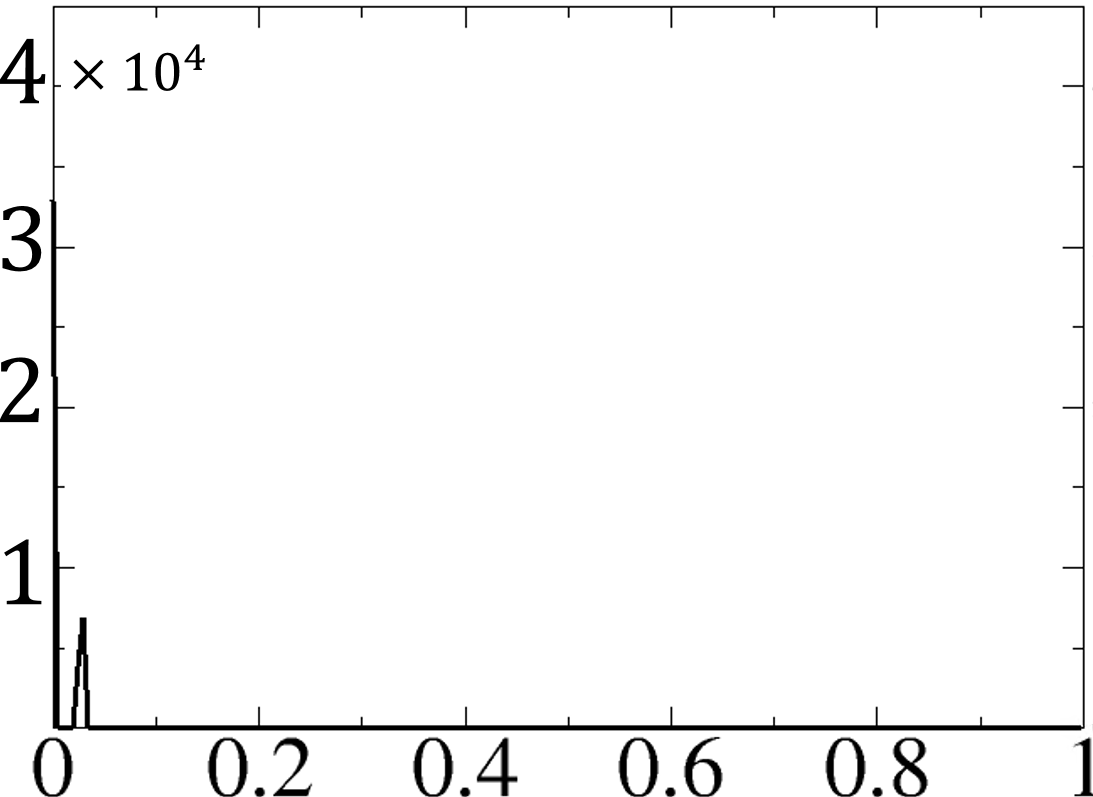}\hspace{0.02\columnwidth}\hskip -.3cm
	\includegraphics[height=.98in]{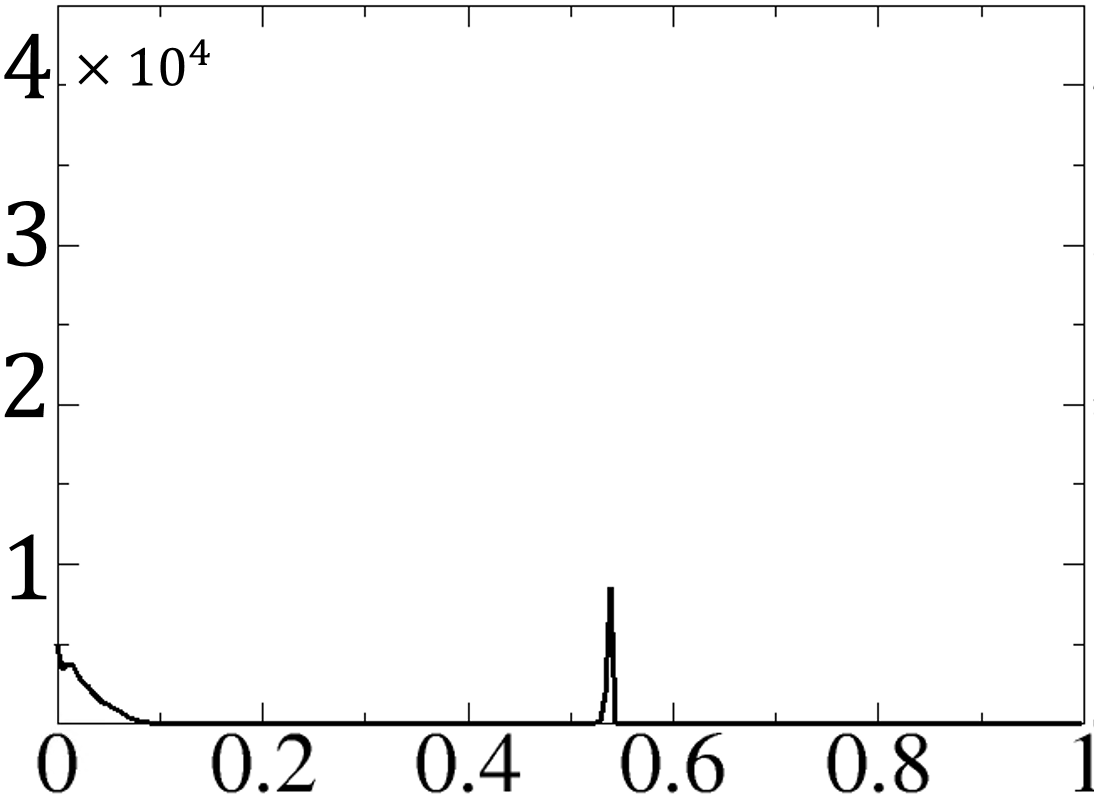}\hspace{0.02\columnwidth}\hskip -.3cm
	\includegraphics[height=.98in]{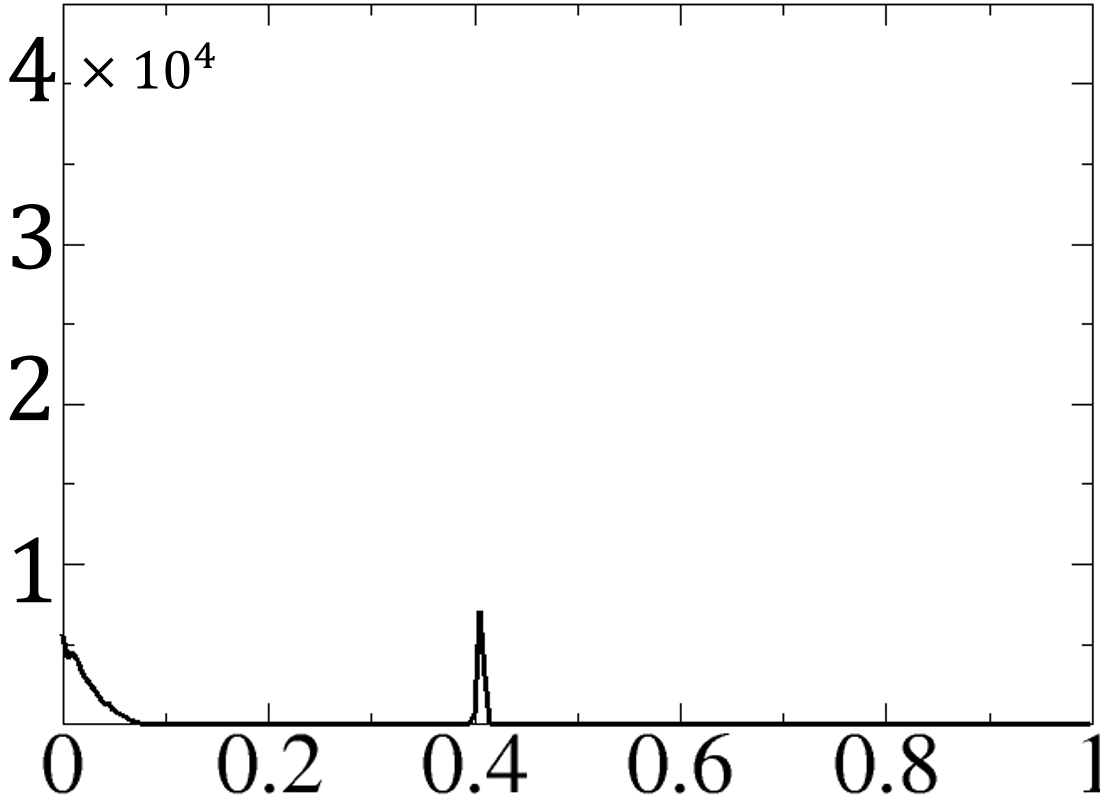}\hspace{0.02\columnwidth}\hskip -.3cm
	\includegraphics[height=.98in]{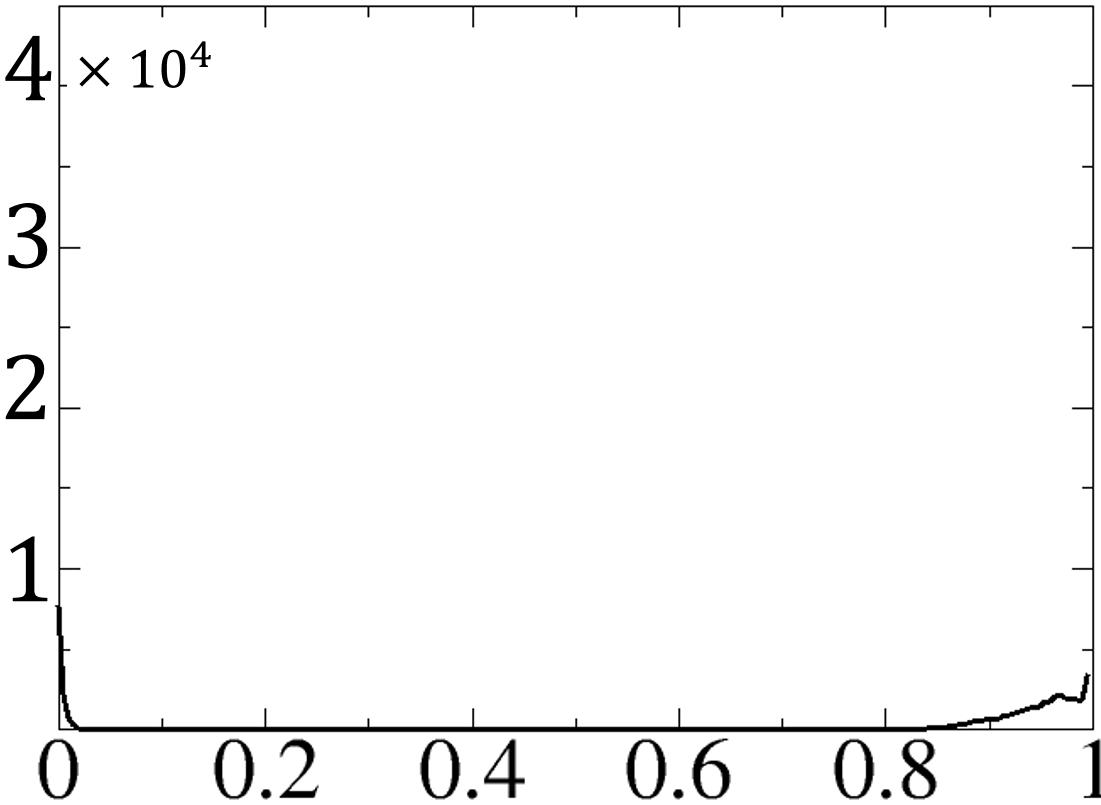}\hspace{0.02\columnwidth}
	\caption{Histograms for the components of eta states found in numerical simulation of this model using set $S_{II}$ (\ref{E_SII}) with  unfiltered range of $h_0 = 0.40-0.99$ GeV.   The five columns respectively represent non-strange quark-antiquark $\left(u {\bar u} + d {\bar d}\right) / \sqrt{2}$ (or $n {\bar n}$ in short), strange quark-antiquark $s {\bar s}$,  strange two-antiquarks two-quarks $\left({\bar d} {\bar s} d s  + {\bar u} {\bar s} u s\right) / \sqrt{2}$ (or ${\bar n} {\bar s} n s$ in short), and  non-strange two-antiquarks two-quarks ${\bar u} {\bar d} u d$ (or generically  ${\bar n} {\bar n} n n$), and the glue component.	}
	\label{F_eta_substructures_h0_4-99}
\end{figure}

The analysis also probes the substructure of the scalar and pseudoscalar states.  Fig.  \ref{F_eta_substructures_h0_4-99} gives the histograms for the substructure of the eta states consist of quark-antiquarks, four-quarks and a glue component.  The results for the two well-established states, $\eta(547)$ and $\eta'(958)$,  are quite unambiguous and in agreement with the conventional phenomenology.  Both of these states are dominantly quark-antiquark,  a small admixture with four-quarks, and no glueball component.  On the other hand, $\eta(1295)$ is dominantly a four-quark state.  The case of $\eta(1760)$ and $\eta(2225)$ are not as clear-cut as the first three states.   We see that some of their components (such as their four-quark and glue components) are not uniquely determined in these simulations with the broad range of $h_0$, as for example we see double peaks in the histograms for some of the components of these two states.  This shows the limitations in predictions when minimum constraint is imposed.  Simulations over a broad range of $h_0$ need to be further studied and narrowed down, as will be discussed later in this work.   

\begin{figure}
	\centering
	\includegraphics[height=1.12in]{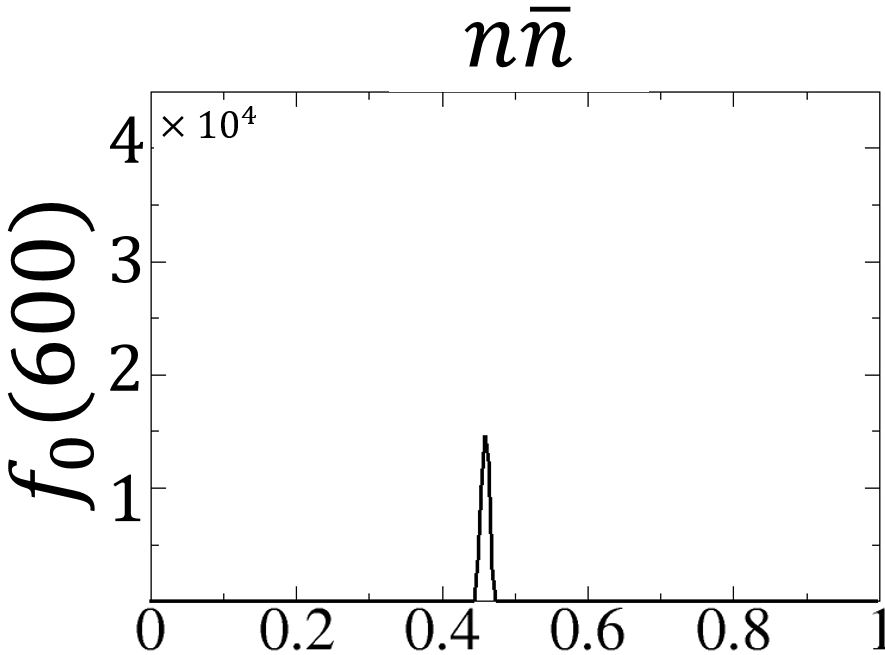}\hspace{0.02\columnwidth}\hskip -.3cm
	\includegraphics[height=1.12in]{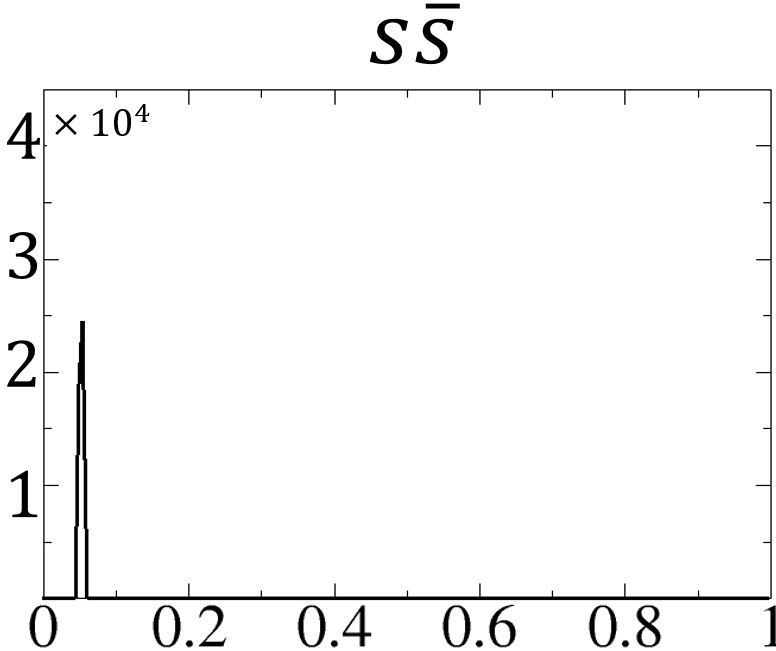}\hspace{0.02\columnwidth}\hskip -.3cm
	\includegraphics[height=1.12in]{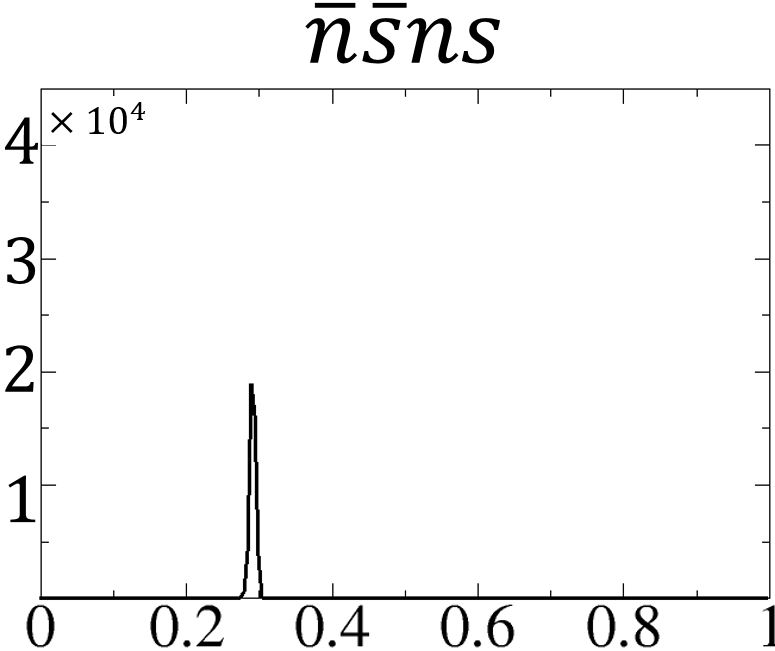}\hspace{0.02\columnwidth}\hskip -.3cm
	\includegraphics[height=1.12in]{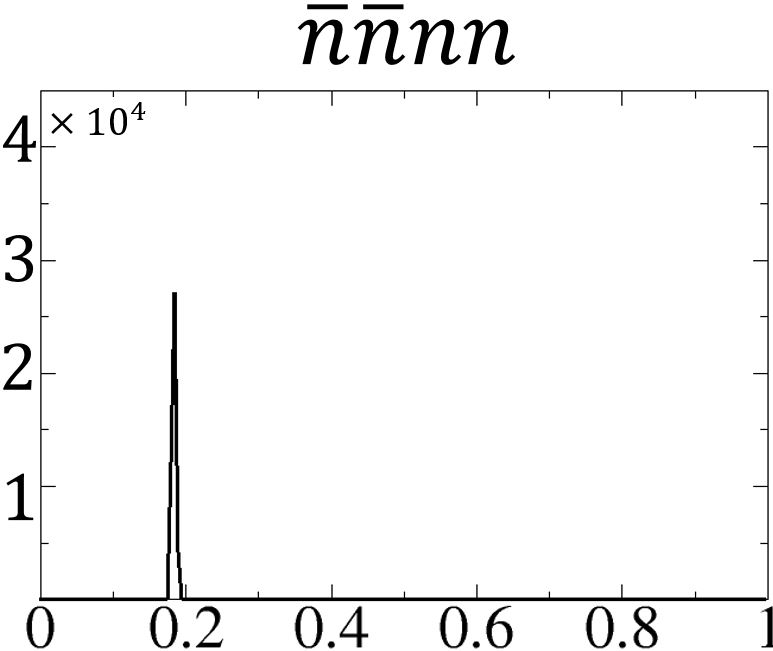}\hspace{0.02\columnwidth}\hskip -.3cm
	\includegraphics[height=1.12in]{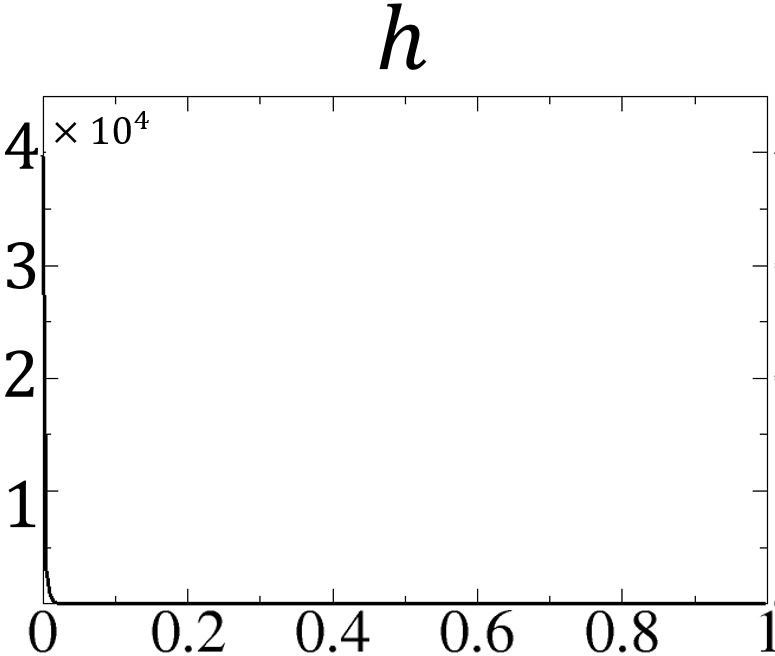}\hspace{0.02\columnwidth}
	
	\includegraphics[height=.98in]{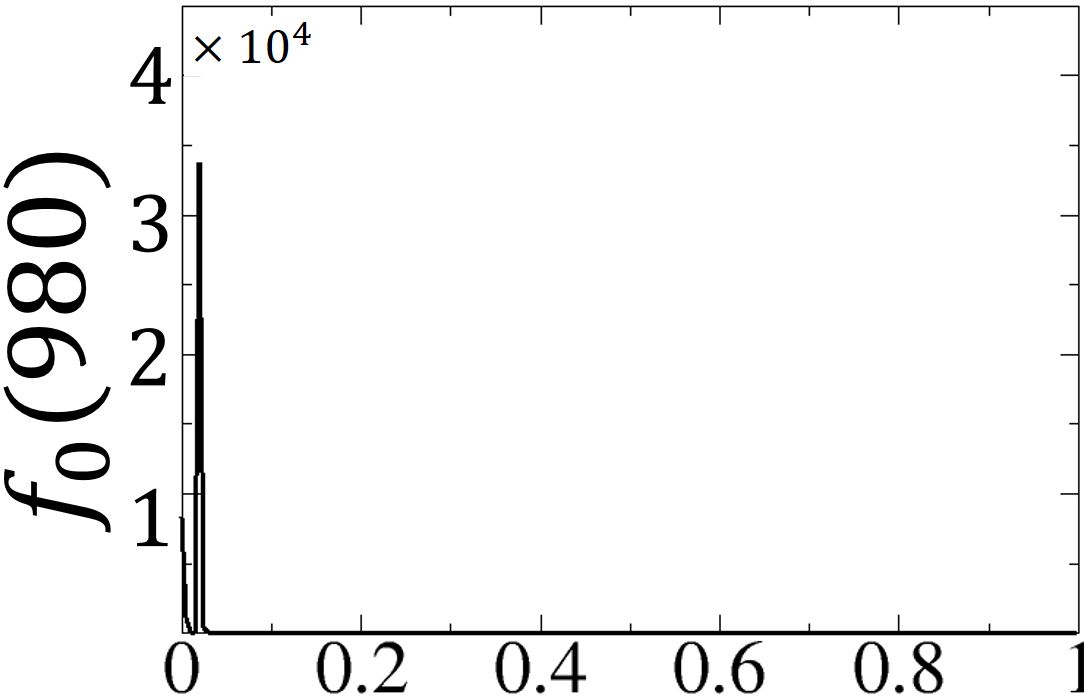}\hspace{0.02\columnwidth}\hskip -.3cm
	\includegraphics[height=.98in]{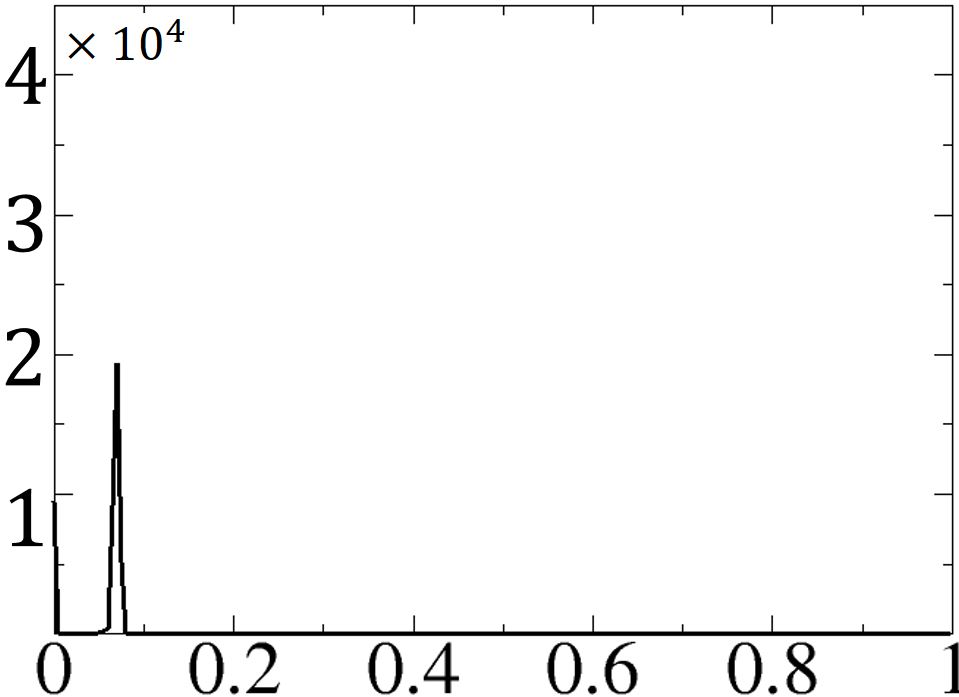}\hspace{0.02\columnwidth}\hskip -.3cm
	\includegraphics[height=.98in]{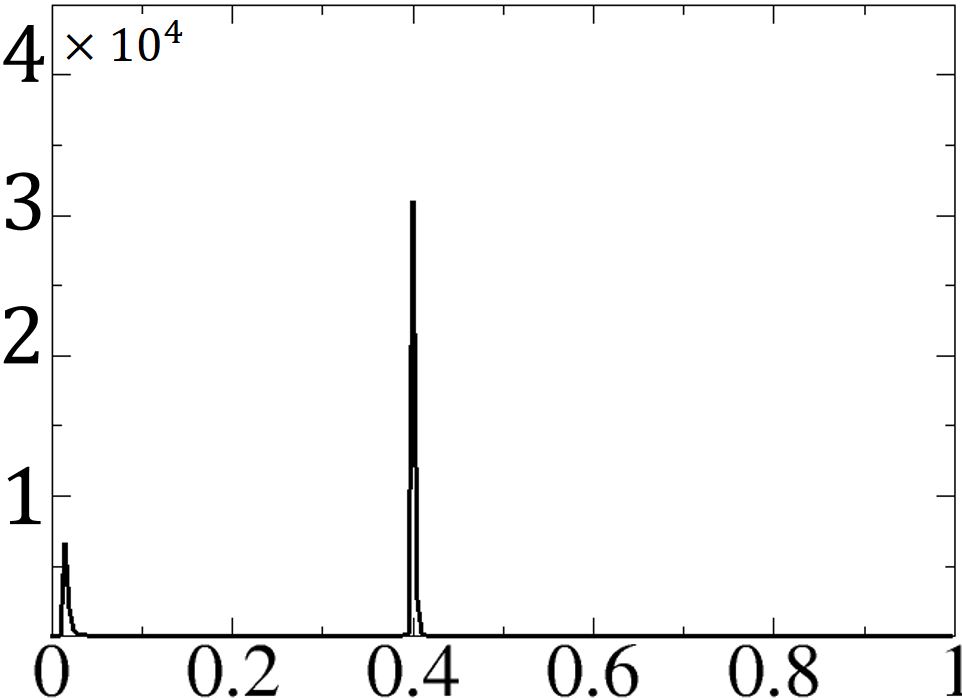}\hspace{0.02\columnwidth}\hskip -.3cm
	\includegraphics[height=.98in]{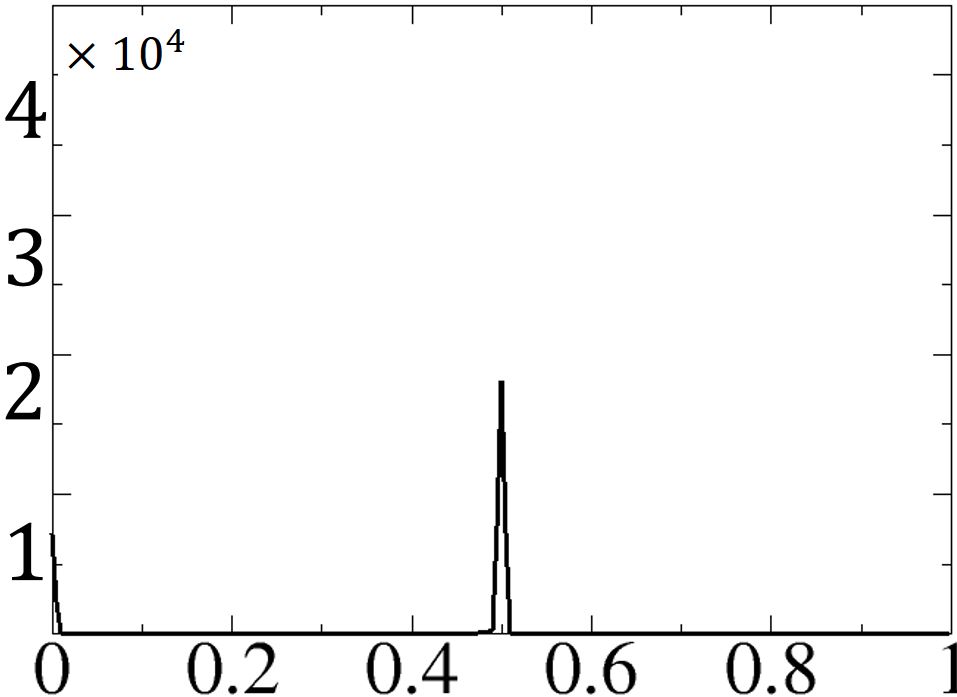}\hspace{0.02\columnwidth}\hskip -.3cm
	\includegraphics[height=.98in]{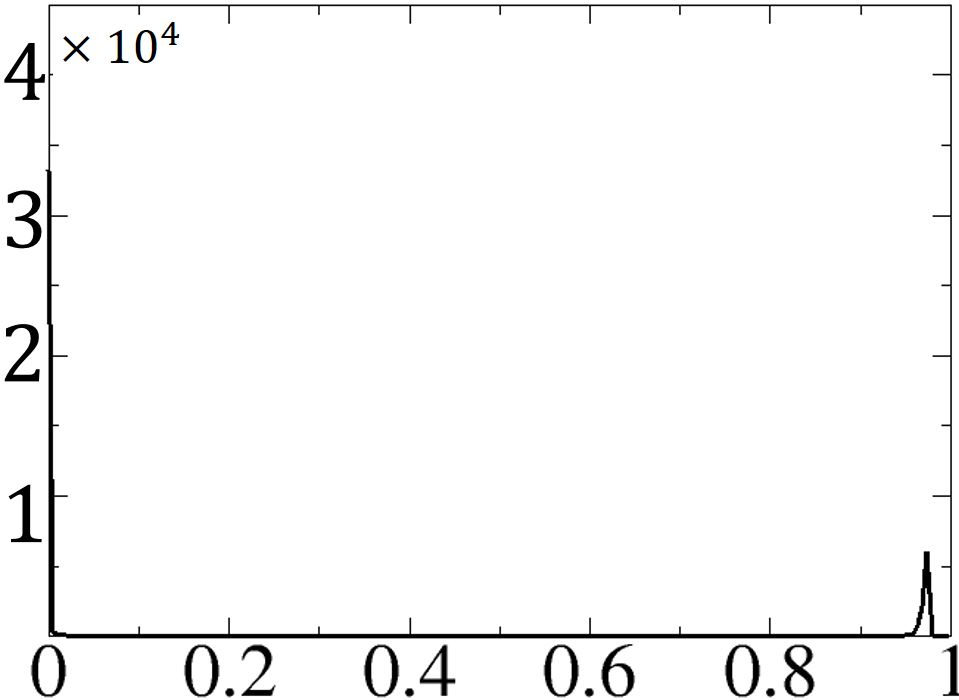}\hspace{0.02\columnwidth}
	
	\includegraphics[height=.98in]{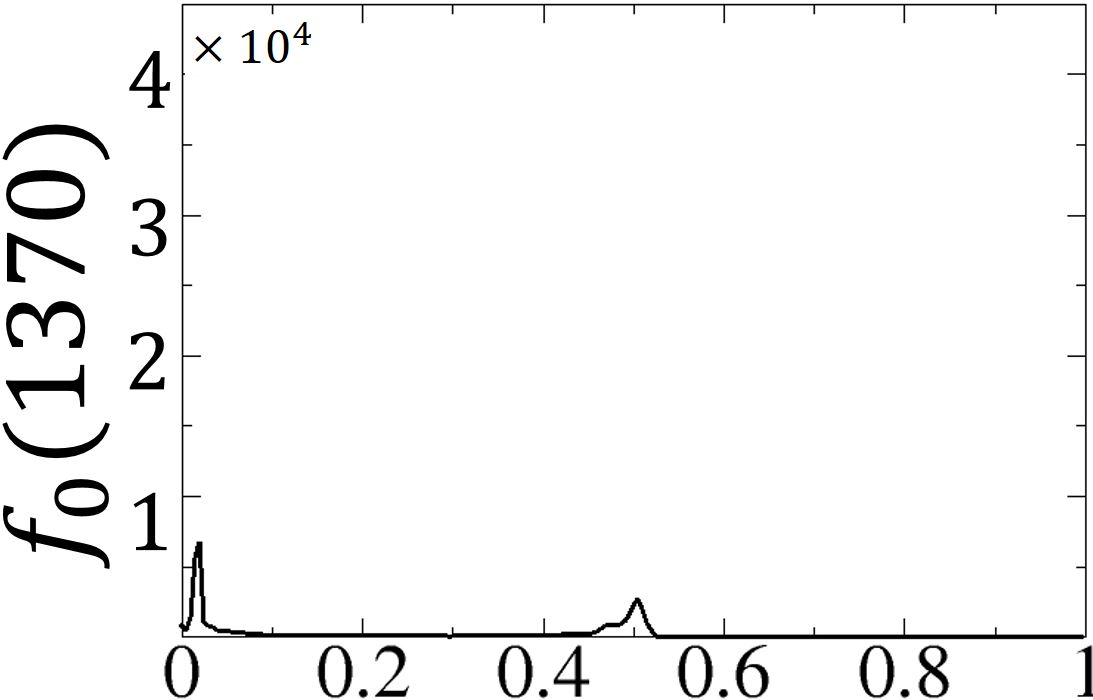}\hspace{0.02\columnwidth}\hskip -.3cm
	\includegraphics[height=.98in]{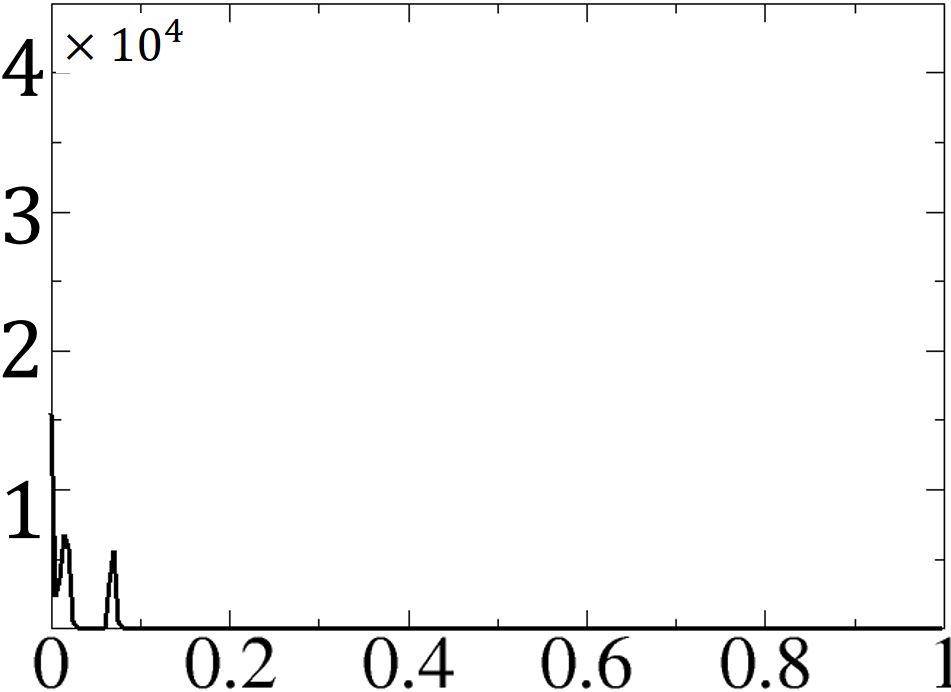}\hspace{0.02\columnwidth}\hskip -.3cm
	\includegraphics[height=.98in]{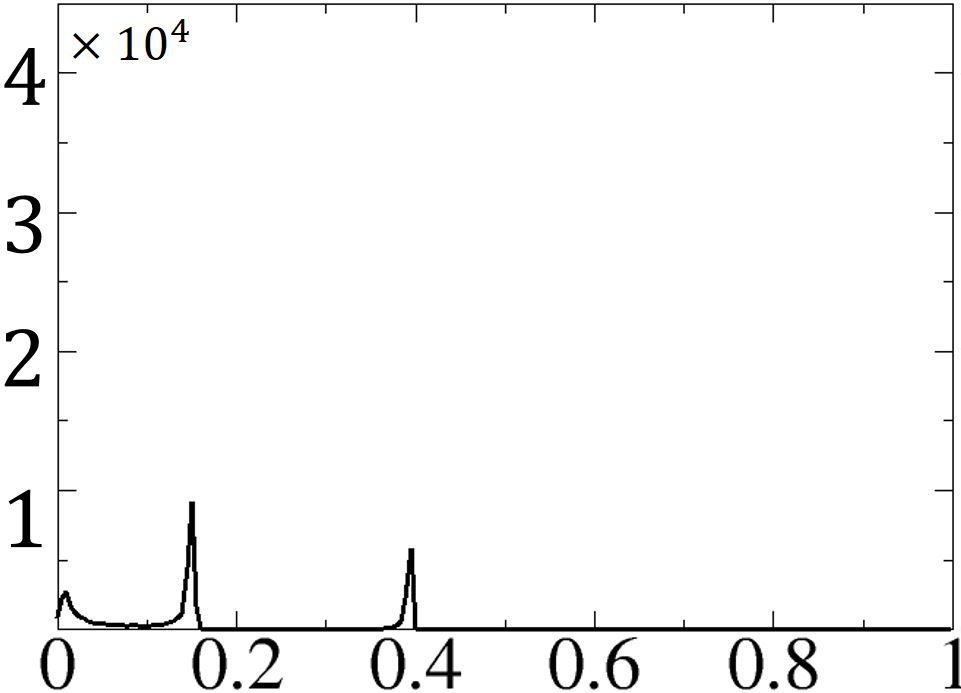}\hspace{0.02\columnwidth}\hskip -.3cm
	\includegraphics[height=.98in]{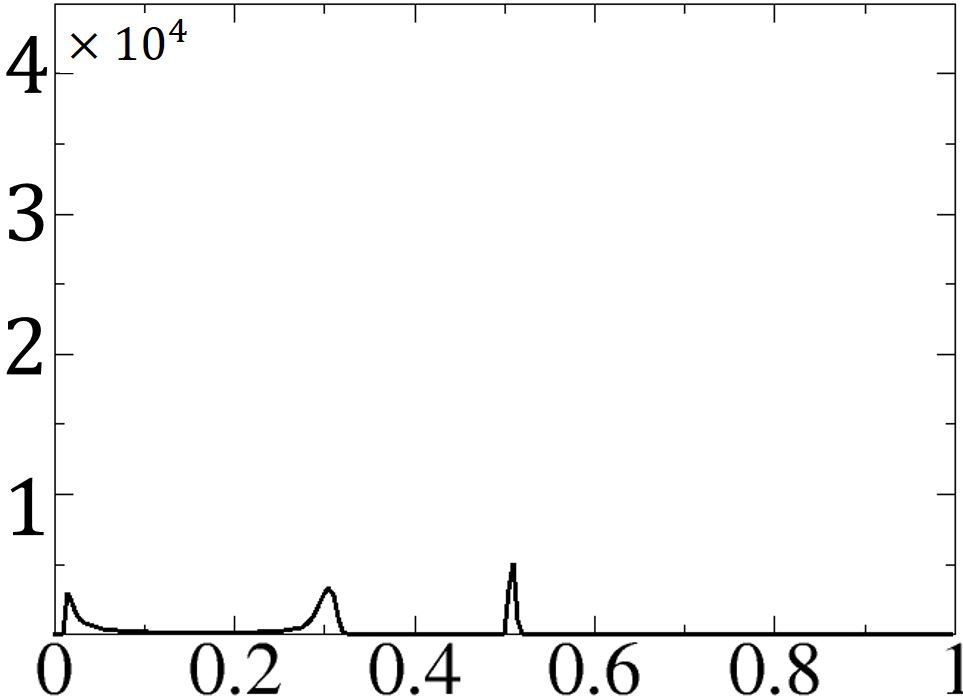}\hspace{0.02\columnwidth}\hskip -.3cm
	\includegraphics[height=.98in]{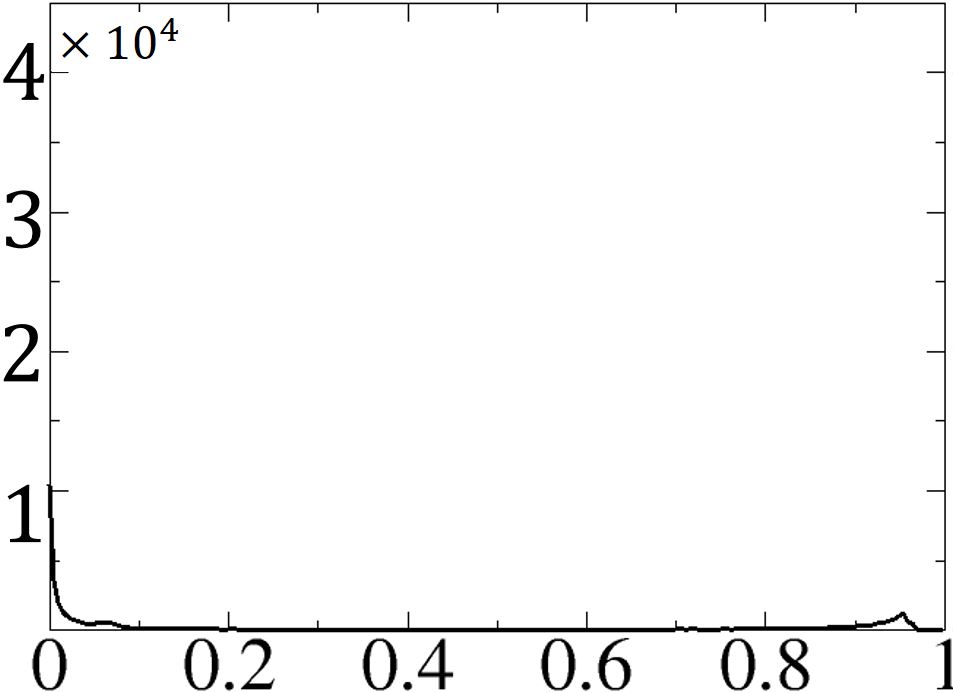}\hspace{0.02\columnwidth}
	
    \includegraphics[height=.98in]{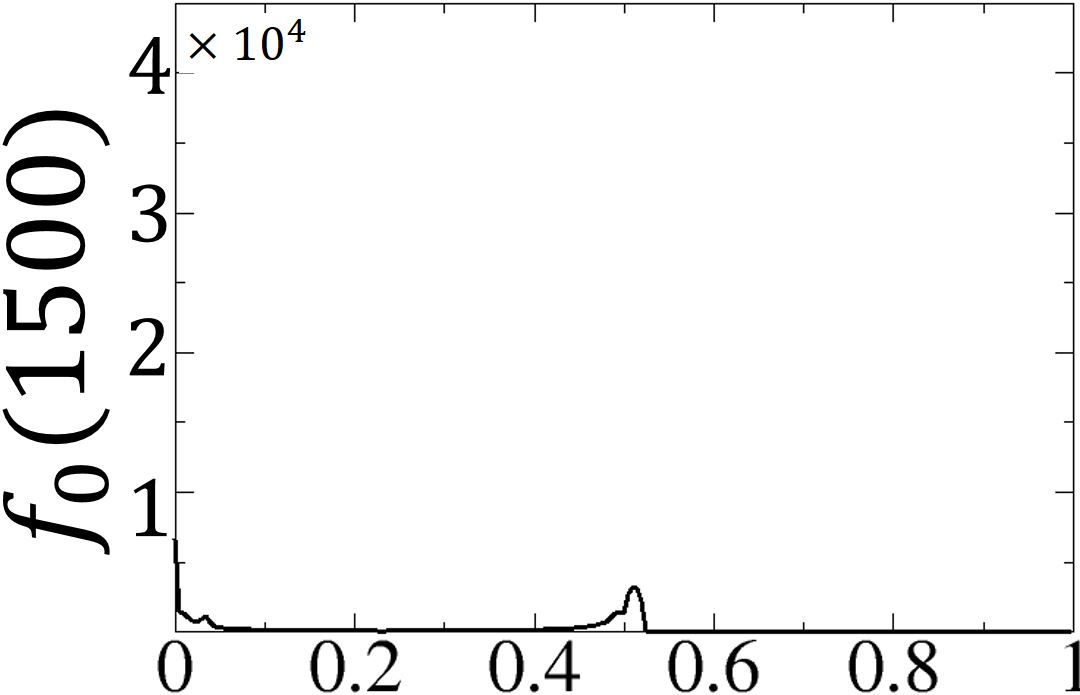}\hspace{0.02\columnwidth}\hskip -.3cm
    \includegraphics[height=.98in]{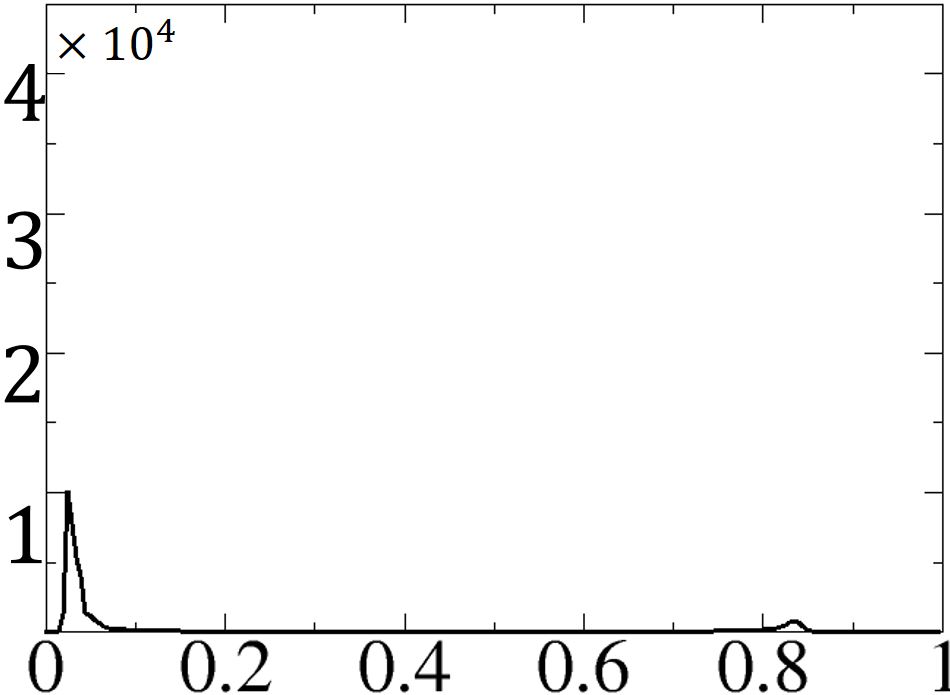}\hspace{0.02\columnwidth}\hskip -.3cm
    \includegraphics[height=.98in]{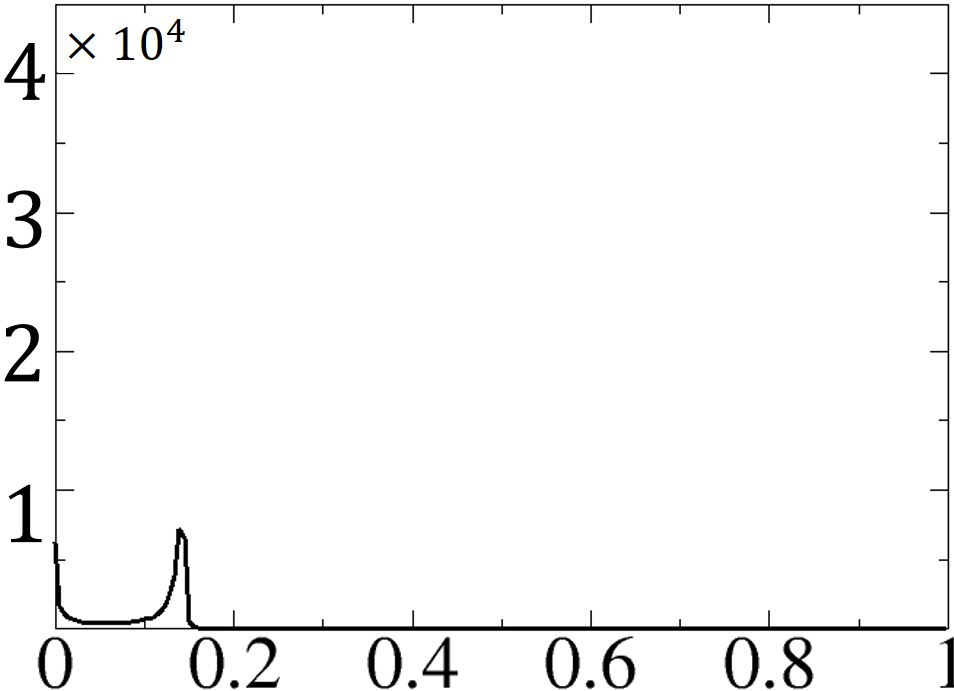}\hspace{0.02\columnwidth}\hskip -.3cm
    \includegraphics[height=.98in]{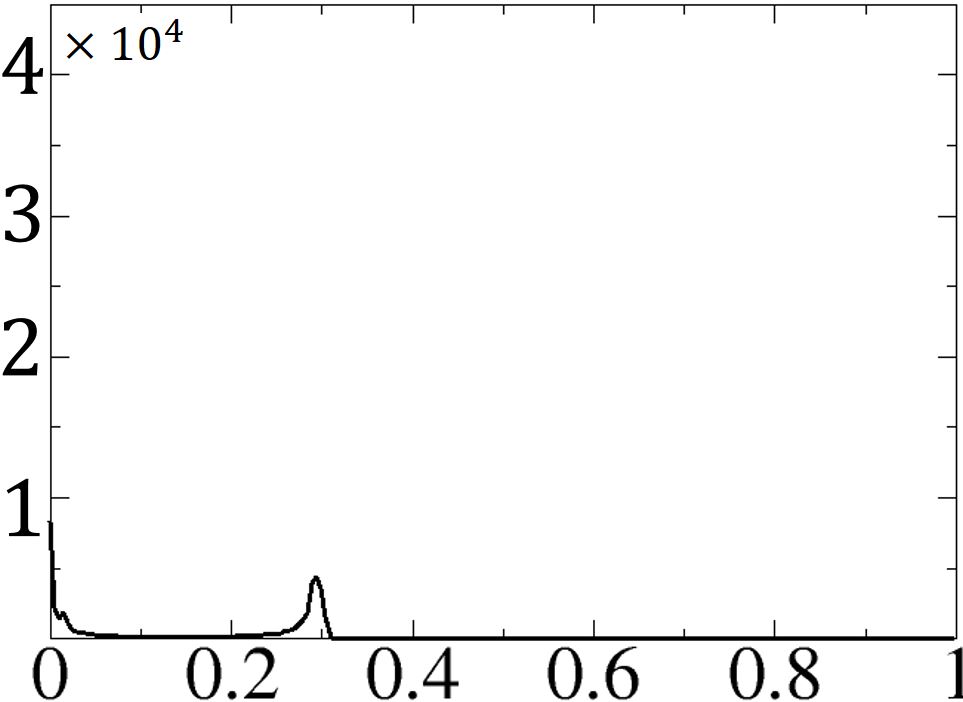}\hspace{0.02\columnwidth}\hskip -.3cm
    \includegraphics[height=.98in]{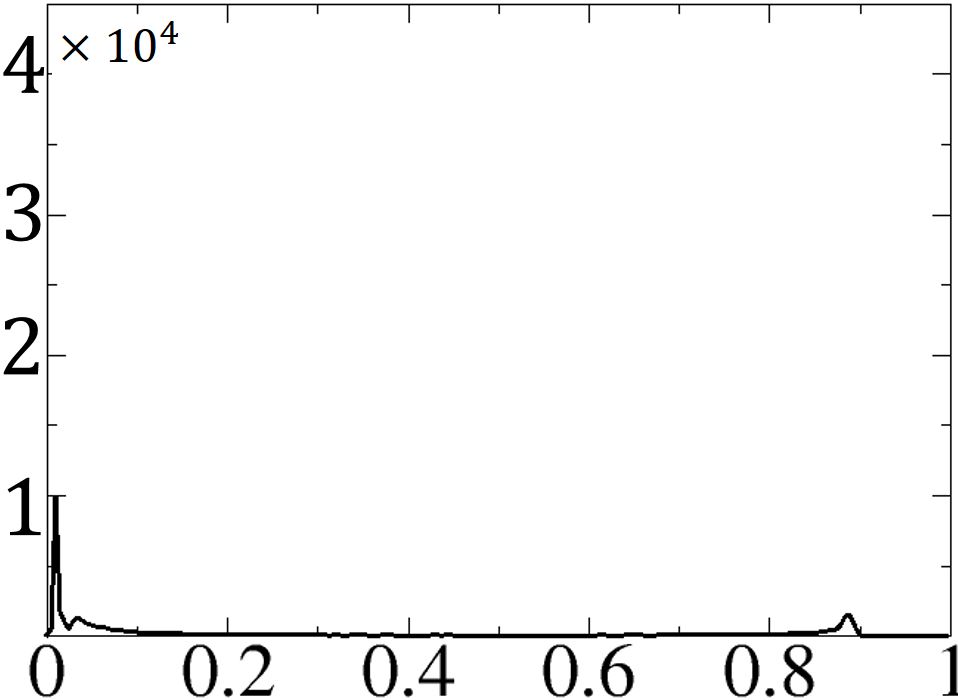}\hspace{0.02\columnwidth}
    
    \includegraphics[height=.98in]{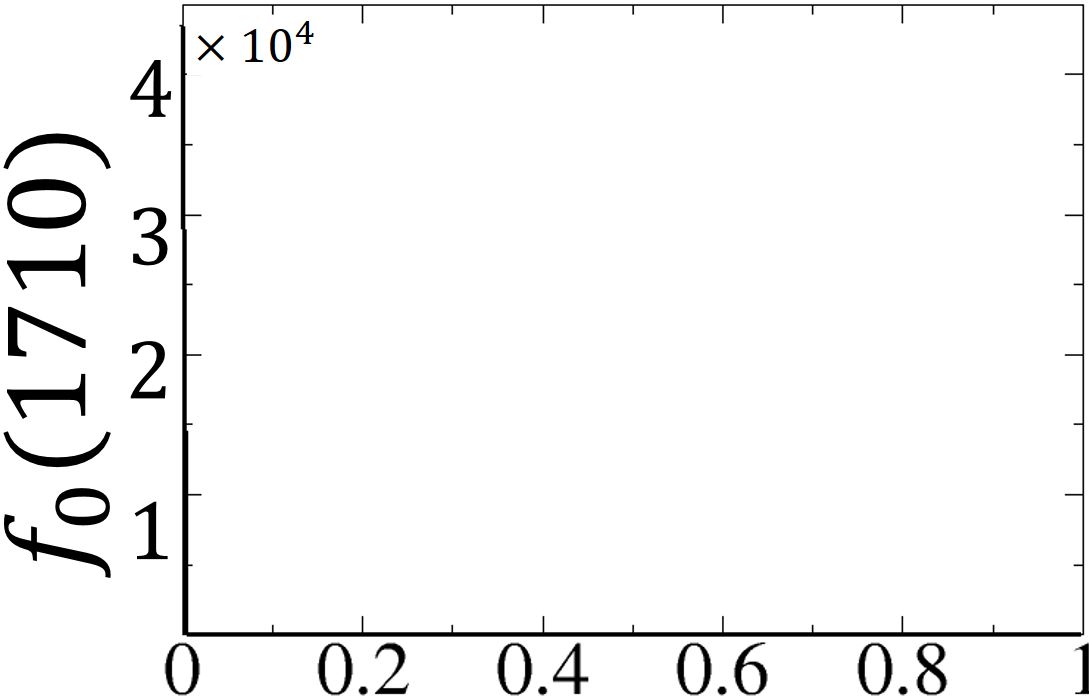}\hspace{0.02\columnwidth}\hskip -.3cm
    \includegraphics[height=.98in]{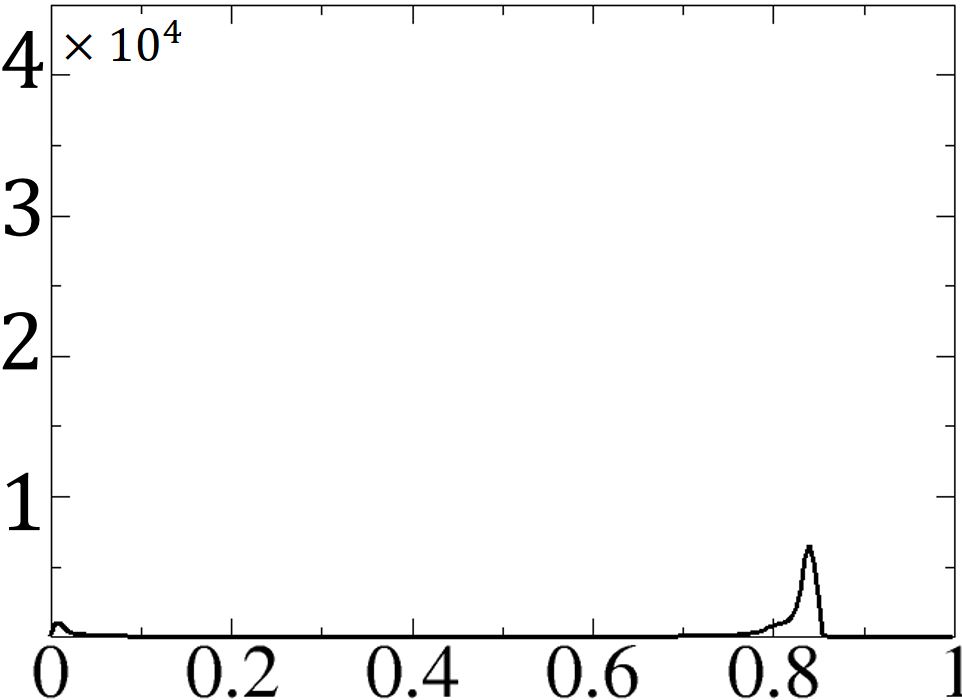}\hspace{0.02\columnwidth}\hskip -.3cm
    \includegraphics[height=.98in]{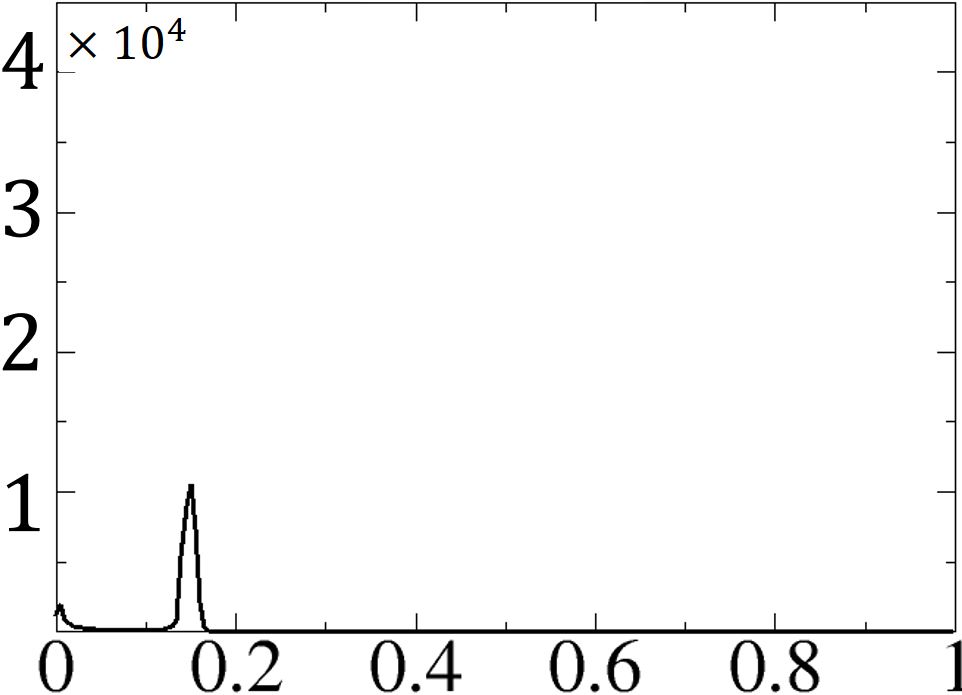}\hspace{0.02\columnwidth}\hskip -.3cm
    \includegraphics[height=.98in]{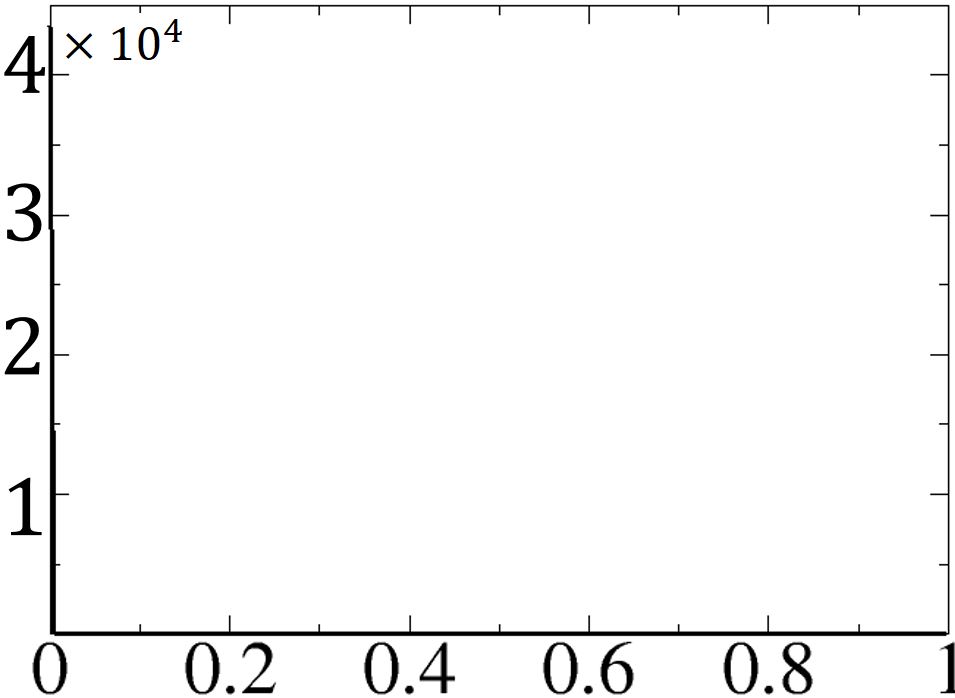}\hspace{0.02\columnwidth}\hskip -.3cm
    \includegraphics[height=.98in]{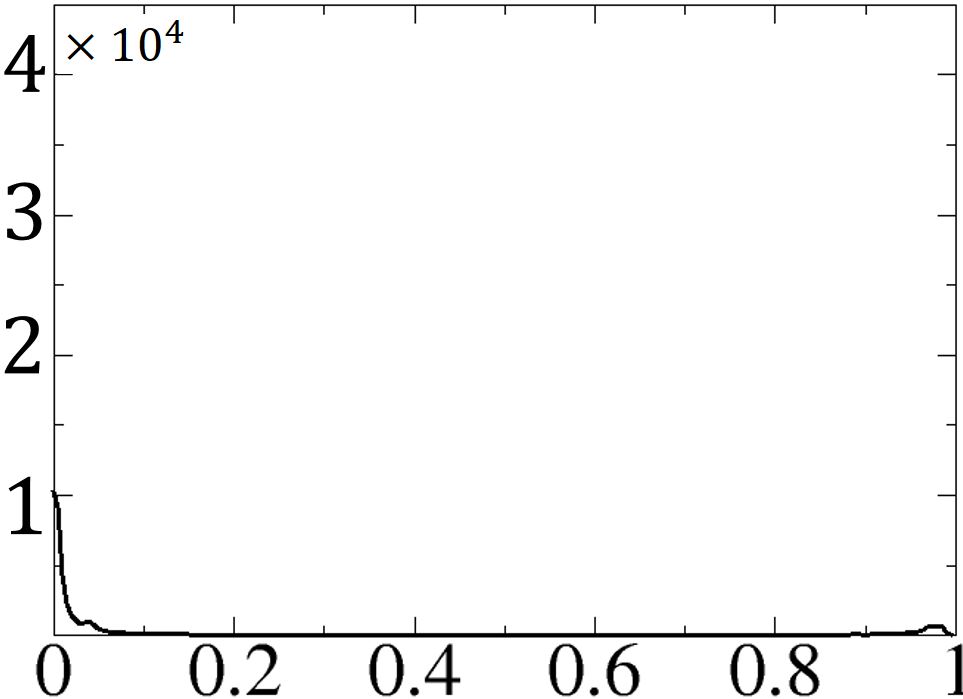}\hspace{0.02\columnwidth}

	\caption{Histograms for the components of $f_0$ states found in numerical simulation of this model  using set $S_{II}$ (\ref{E_SII}) with unfiltered range of $h_0 = 0.40-0.99$ GeV.   The five columns respectively represent non-strange quark-antiquark $\left(u {\bar u} + d {\bar d}\right) / \sqrt{2}$ (or $n {\bar n}$ in short), strange quark-antiquark $s {\bar s}$,  strange two-antiquarks two-quarks $\left({\bar d} {\bar s} d s  + {\bar u} {\bar s} u s\right) / \sqrt{2}$ (or ${\bar n} {\bar s} n s$ in short), and  non-strange two-antiquarks two-quarks ${\bar u} {\bar d} u d$  (or generically  ${\bar n} {\bar n} n n$),  and the glue component.
	}
	\label{F_f0_substructures_h0_4-99}
\end{figure}

The histograms for the scalar components are plotted in Fig.  \ref{F_f0_substructures_h0_4-99} showing  unambiguous signals for the components of the $f_0(500)$  which are consistent with the expectations for this  state to contain significant four-quark components, but the simulations for some of the components of the heavier states contain conflicting results (for example we see that the components of the $f_0(1370)$ clearly contain multiple peaks).   This observation again highlights the fact that some of the values of the glueball condensate $h_0$ in this broad scan in the minimum constraint approach lead to contradictory results, and the range of $h_0$ should be studied more carefully and narrowed down.

\subsubsection{Analysis of the glueball condensate over the range of $h_0=0.40-0.99$ GeV}

\begin{figure}
	\centering
	\includegraphics[height=3.5in]{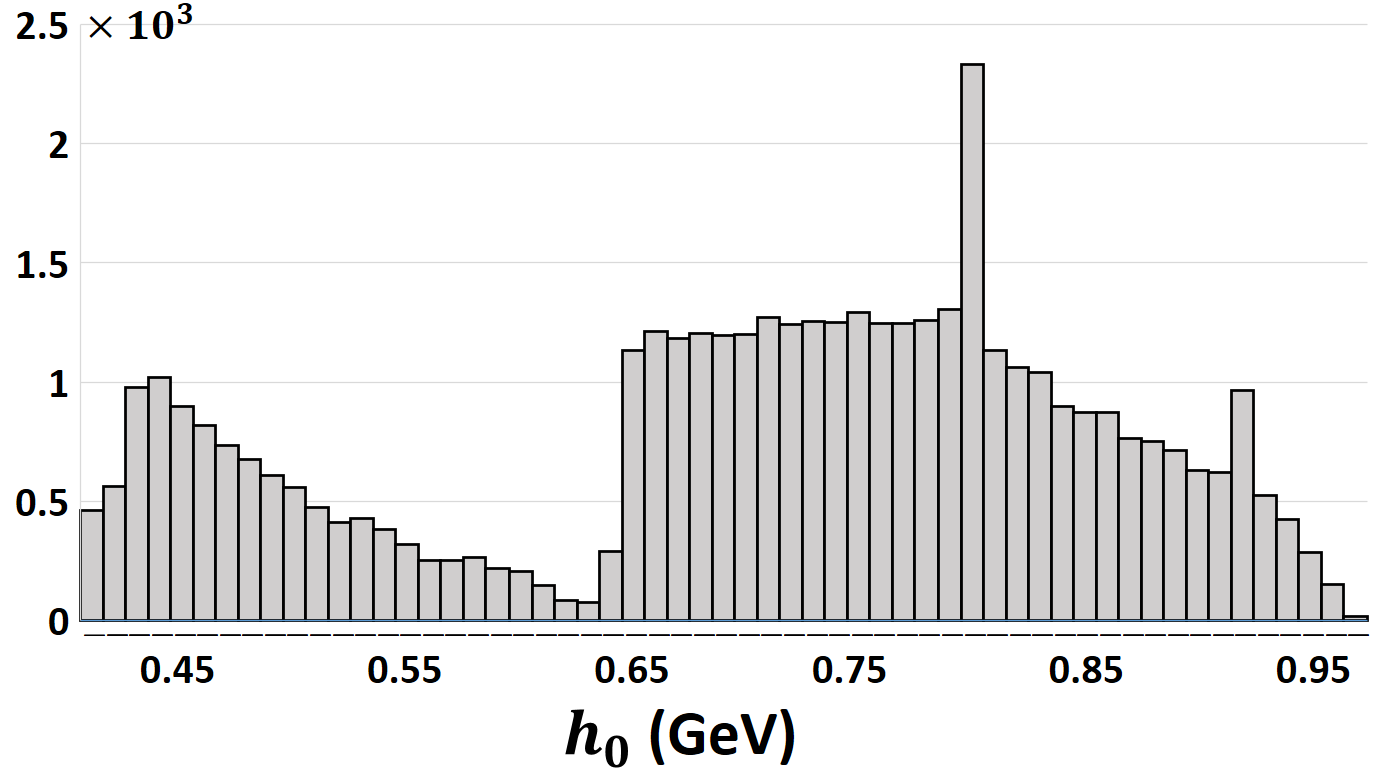}
	%\hspace{0.02\columnwidth}
	\caption{Histogram for unfiltered glueball condensate $h_0=0.40-0.99$ GeV, over set $S_{II}$ (\ref{E_SII}).   No acceptable solutions found outside this range. 
	}
	\label{F_h0_histogram}
\end{figure}

\begin{figure}
	\centering
	\includegraphics[height=3.5in]{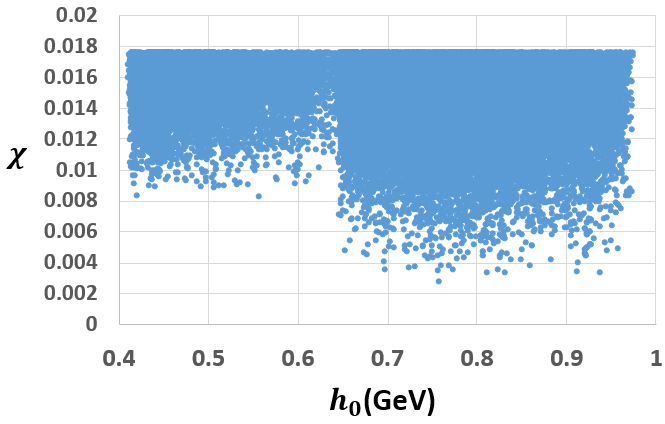}
	%\hspace{0.02\columnwidth}
	\caption{Goodness of simulations measured by function $\chi$ in Eq. (\ref{E_chi_goodness}) versus unfiltered glueball condensate $h_0=0.40-0.99$ GeV, over  set $S_{II}$ (\ref{E_SII}). 
	}
	\label{F_chi_vs_h0}
\end{figure}

The histogram for $h_0$ over its entire range of $0.40-0.99$ Gev is given in  Fig. \ref{F_h0_histogram}  (beyond this range, the model has no acceptable solutions).  The distribution is consistent with what was found in \cite{19_Fariborz_IJMPA34,21_Fariborz_NPA1015} in the SU(3) limit of this model. We see roughly two ranges, $h_0 = 0.40 - 0.65$ GeV (low range of $h_0$),  and $h_0 = 0.65-0.99$ GeV (high range of $h_0$) which give rise to results that are not quite consistent with each other. We see that more simulations favor the high range of $h_0$.   Figure \ref{F_chi_vs_h0} gives the goodness of the simulations measured by the quantity $\chi$ in Eq. (\ref{E_chi_goodness}) versus glueball condensate $h_0$,  showing a better agreement of the simulations with experiment in the high range of $h_0$.

\begin{figure}
%	\centering
	\includegraphics[height=3in]{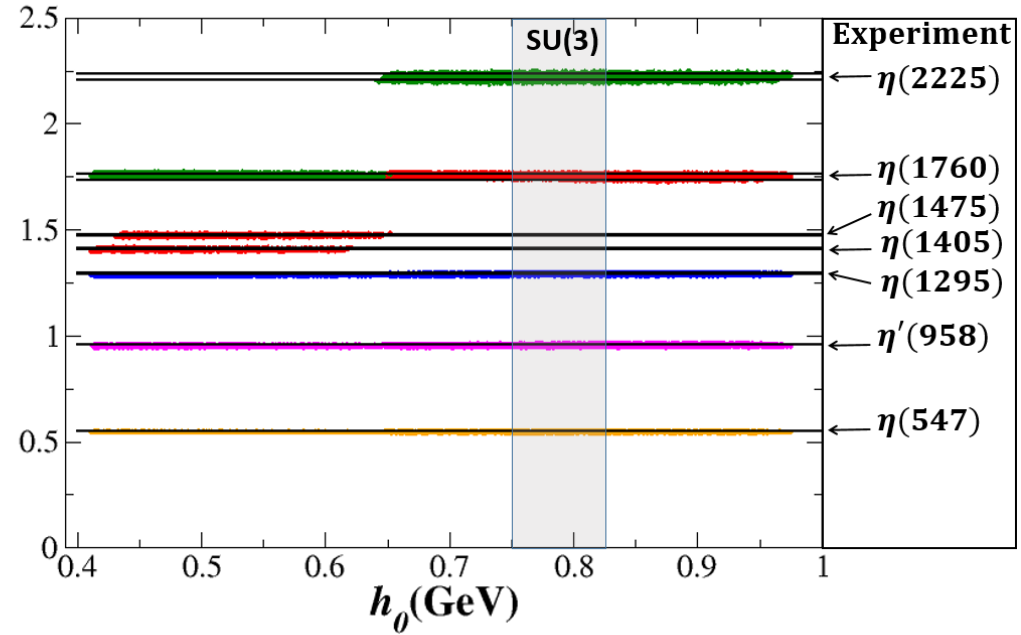}
	%\hspace{0.02\columnwidth}
	\caption{Simulation results for eta masses (dots/colors) over set $S_{II}$ (\ref{E_SII}) versus  unfiltered range of glueball condensate $h_0=0.40-0.99$ GeV.   Experimental bounds are shown in solid lines.   The three lightest etas do not depend on $h_0$ while the fourth and fifth etas found in these simulations depend on the range of $h_0$.  In the low range  $h_0 \approx 0.4-0.65$ GeV, the fourth eta is not uniquely determined and is either $\eta(1405)$ or $\eta(1475)$ and the fifth eta is $\eta(1760)$,  whereas in the high range $h_0 \approx 0.65-0.99$ GeV the fourth eta is $\eta(1760)$ and the fifth eta is $\eta(2225)$.  The gray vertical band shows the range of $h_0$ favored in the SU(3) limit of this model \cite{19_Fariborz_IJMPA34,21_Fariborz_NPA1015}.   
}
	\label{F_eta_masses_h0_4-99}
\end{figure}

Figure  Fig. \ref{F_eta_masses_h0_4-99} dissects the dependency of the eta masses on $h_0$. 
As can be seen in this figure, while  the lightest three eta masses are not sensitive to $h_0$, the remaining four eta masses are significantly affected by which range of $h_0$ is used to calculate these masses (we also note that within each range, the masses do not depend on $h_0$).   The two eta states $\eta(1405)$ and $\eta(1475)$ are only ``seen'' within the low range of $h_0$ as the fourth eta state, with some of the simulations favoring $\eta(1405)$ and some $\eta(1475)$.  Over the low range of $h_0$, the fifth eta state is clearly the  $\eta(1760)$.   On the other hand,  over the high range of $h_0$ the fourth and the fifth eta states are clearly $\eta(1760)$ and $\eta(2225)$, respectively.     Also shown in Fig. \ref{F_eta_masses_h0_4-99}, is the SU(3) limit of this model \cite{19_Fariborz_IJMPA34,21_Fariborz_NPA1015} which favors $h_0 = 0.75-0.825$ GeV, that is within the high range of $h_0$.

\begin{figure}
%	\centering
	\includegraphics[height=3in]{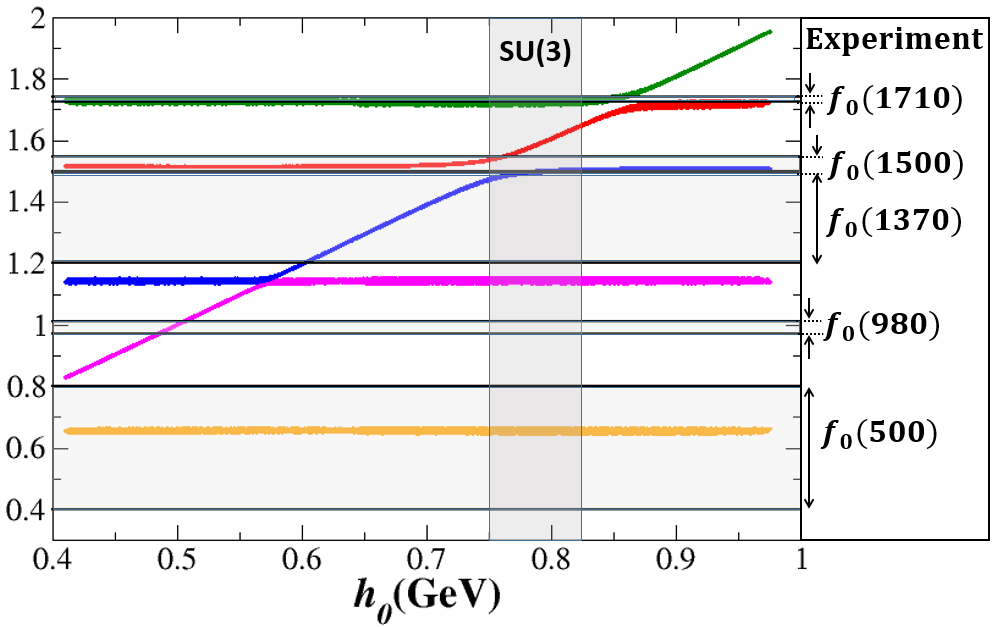}
	%\hspace{0.02\columnwidth}
	\caption{Simulation results for $f_0$ masses (dots/colors) over set $S_{II}$ (\ref{E_SII}) versus  unfiltered range of glueball condensate $h_0=0.40-0.99$ GeV.
	Experimental bounds are shown in horizontal gray bands.   The mass of $f_0(500)$ does not depend on $h_0$ while the heavier masses show some dependency on $h_0$.
	The gray band shows the range of $h_0$ favored in the SU(3) limit of this model \cite{19_Fariborz_IJMPA34,21_Fariborz_NPA1015}.  
	}
	\label{F_f0_masses_h0_4-99}
\end{figure}

A similar pattern is observed in Fig. \ref{F_f0_masses_h0_4-99} for the dependency of $f_0$ states on $h_0$.  With the exception of $f_0(500)$ mass that is completely independent of $h_0$, the rest of $f_0$ masses show some dependencies on $h_0$ ranges.    The SU(3) favored range is also shown in this figure in which the states are within or near their experimental ranges (note also that these states also receive mass corrections due to the final-state interaction of their decay products not taken into account in this graph).

\begin{figure}
	\centering
	\includegraphics[height=3in]{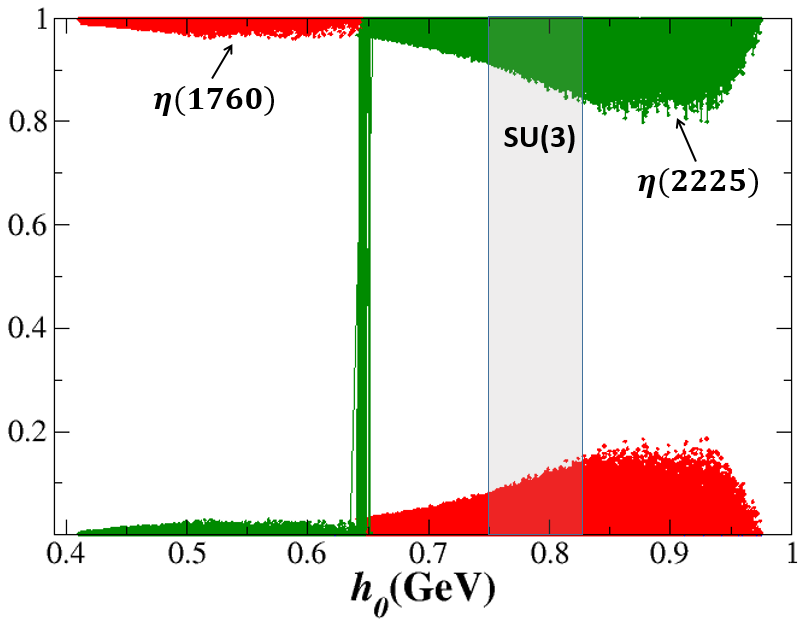}
	%\hspace{0.02\columnwidth}
\caption{The glue content of $\eta(1760)$ and $\eta(2225)$ over set $S_{II}$ (\ref{E_SII}) versus unfiltered range of glueball condensate $h_0=0.40-0.99$ GeV.  The simulations show that in the low range of  $h_0 \approx 0.4-0.65$ GeV, the $\eta(1760)$ becomes the state with maximum glue,  whereas in the high range of $h_0 \approx 0.65-0.99$ GeV,  the $\eta(2225)$ becomes the state which  
is dominantly made of glue. The gray band shows the range of $h_0$ favored in the SU(3) limit of this model \cite{19_Fariborz_IJMPA34,21_Fariborz_NPA1015}.  
	}
	\label{F_eta_glue_vs_h0}
\end{figure}

\begin{figure}
	\centering
	\includegraphics[height=3in]{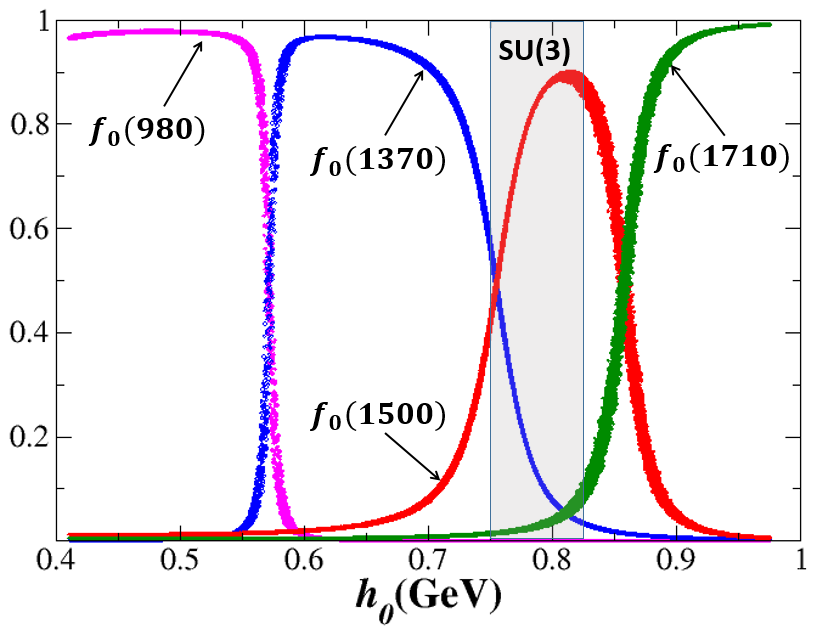}
	%\hspace{0.02\columnwidth}
	\caption{The glue content of the $f_0(980)$, $f_0(1370)$, $f_0(1500)$ and $f_0(1710)$ over  set $S_{II}$ (\ref{E_SII}) versus unfiltered range of glueball condensate $h_0=0.40-0.99$ GeV.  The simulations are sensitive to the value of $h_0$.  The gray band shows the range of $h_0$ favored in the SU(3) limit of this model \cite{19_Fariborz_IJMPA34,21_Fariborz_NPA1015}. 
	}
	\label{F_f0_glue_vs_h0}
\end{figure}

To further analyze the dependency of results on the glueball condensate,  the glueball content of $\eta(1760)$ and $\eta(2225)$ are plotted in Fig. \ref{F_eta_glue_vs_h0}
showing a distinct  dependency   on the low and the high ranges of $h_0$. Some works, including the SU(3) limit of this model favor the high energy range of $h_0$. Similarly,  in Fig. \ref{F_f0_glue_vs_h0} the glueball content of the isosinglet scalars are plotted.  Depending on the range of $h_0$, we see how each of the isosinglet states [with the exception of $f_0(500)$] can in principle contain a significant amount of glue,  and  to determine  which one is the dominant glue holder, it is imperative to further narrow down the range of $h_0$.    In the next section we will examine the correlation of the decay widths of the scalar mesons with $h_0$.

%%%%%%%%%%%%%%%%%%%%%%%%%%%%%%%%%%%%%%%%%%%%%%%%%%%%%%%%%%%%%%%%%%%%%%%%%%%%%%%%%%%%%%%%%%%%%%%%%%%%%%%%%%%%%%%%%%%%%%

\clearpage

%%%%%%%%%%%%%%%%%%%%%%%%%%%%%%%%%%%%%%%%%%%%%%%%%%%%%%%%%%%%%%%%%%%%%%%%%%%%%%%%%%%%%%%%%%%%%%%

\section{Decay widths}

To further probe $h_0$, in this section we study the tree-level partial decay widths of the isosinglet scalar mesons over the entire range of $h_0$ to see if there is a favored range that better agrees with experiment.  Although we compute the decay widths at the tree-level, and in a more comprehensive approach the effects of the final-state interactions should be taken into account, the results we will derive in this section can provide useful insights into the glueball condensate.    Details of  formulas and setup are given in Appendix \ref{A_Decay_Formulas}. 

Figure \ref{F_f1_pp} gives the decay width of $f_0(500)$ versus $h_0$ together with the experimental bounds which  are clearly in complete agreement over  the entire range of $h_0 = 0.5-0.99$ GeV. 

\begin{figure}
	\centering
	\hskip .1cm
	\includegraphics[height=2.5in]{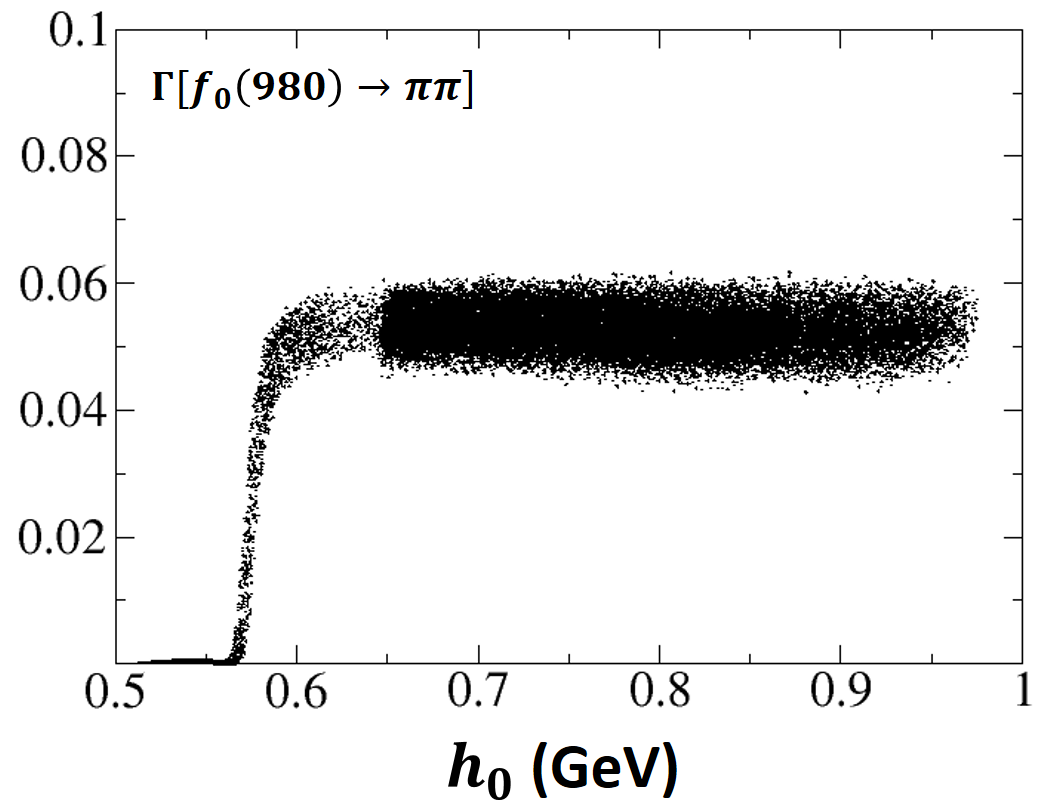}\hspace{0.02\columnwidth}\hskip -.3cm
	\includegraphics[height=2.5in]{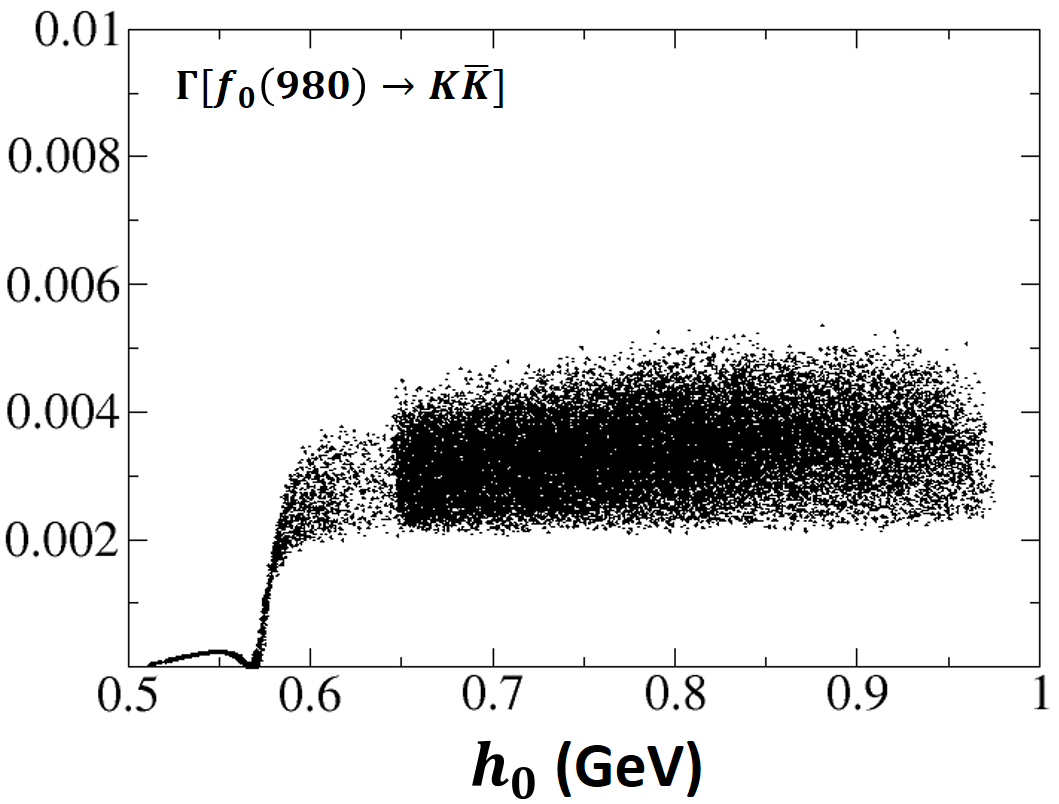}\hspace{0.02\columnwidth}

	\vskip .5cm

	\includegraphics[height=2.5in]{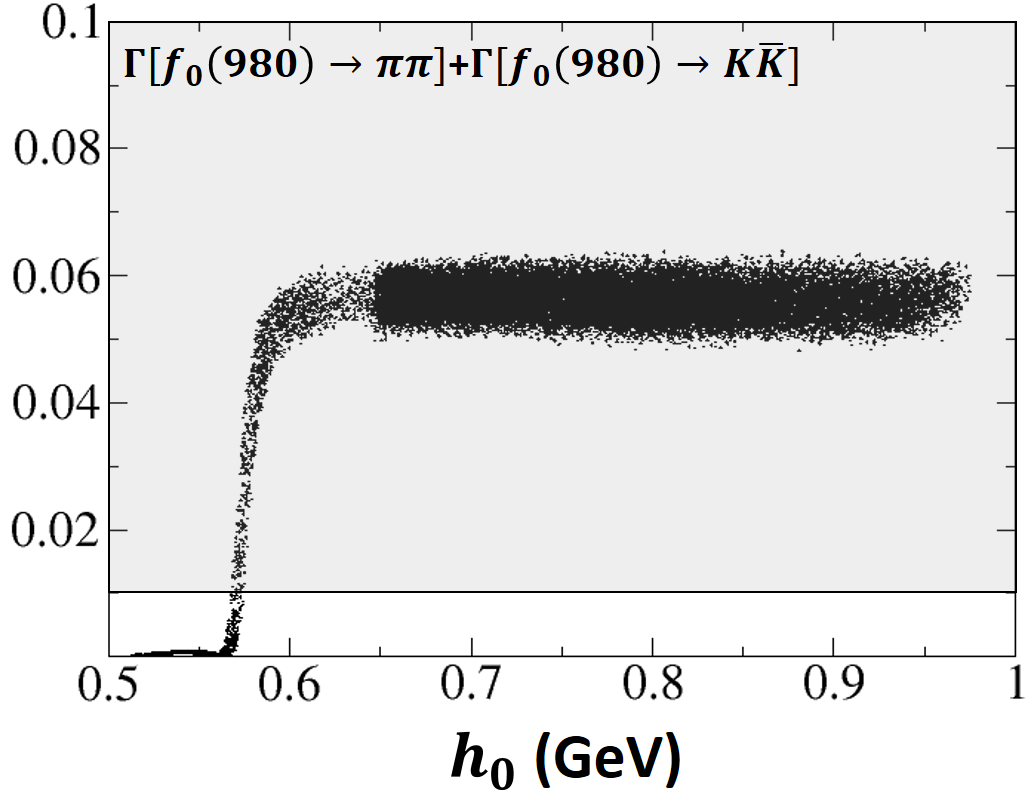}\hspace{0.02\columnwidth}\hskip -.3cm
	\includegraphics[height=2.5in]{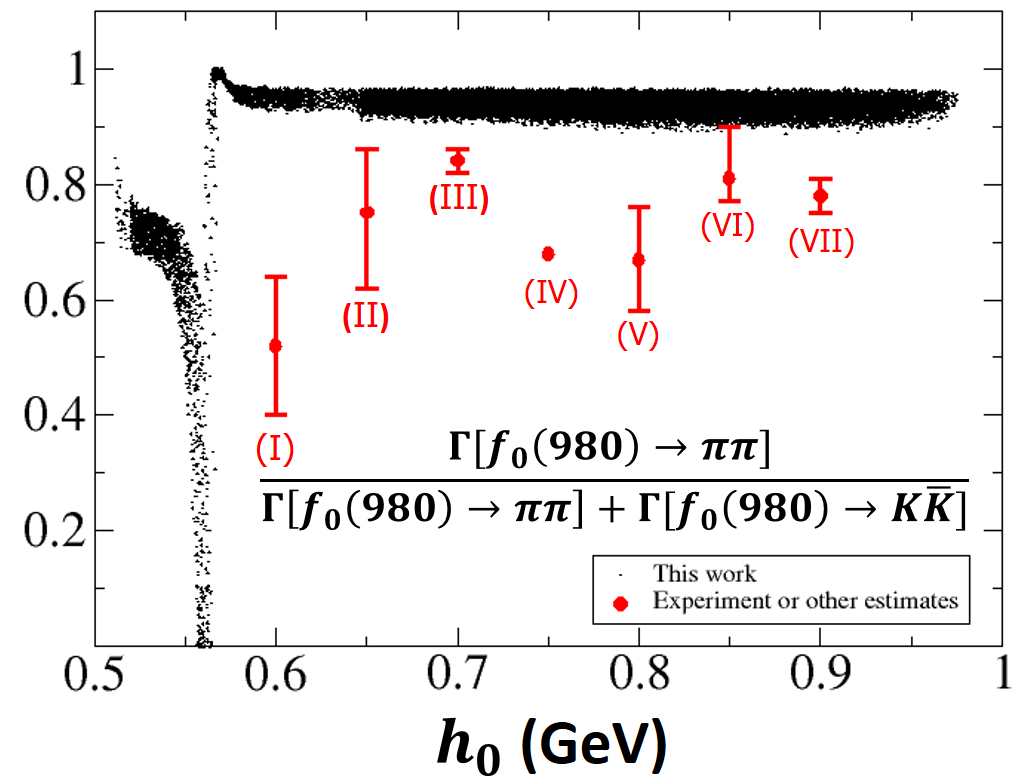}\hspace{0.02\columnwidth}

	\caption{Decay widths (in GeV) of the $f_0(980)$ to  $\pi\pi$ (top left), $K {\bar K}$ (top right), combined $\pi\pi$  and $K {\bar K}$ (bottom left) and the ratio of its decay width to $\pi\pi$ over its decay widths to $\pi\pi$ and $K {\bar K}$ (bottom right), 	
		over the range of $h_0 = 0.50-0.99$ GeV.   The gray band (bottom left) gives the experimental range for the total decay width of $f_0(980)$ \cite{24_PDG}. In bottom right figure, the error bars (I $\cdots$ VII) are estimates by other works given respectively in Refs. \cite{07_Aubert_PRD76,05_Ablikim_PRD72,02_Anisovich_PAN65,97_Oller_NPA620,80_Loverre_ZPC6,78_Cason_PRL41,76_Wetzel_NPB115}.
	}
	\label{F_f2_widths}
\end{figure}

Figure \ref{F_f2_widths} gives decay width of $f_0(980)$ to $\pi\pi$, $K {\bar K}$ together with an estimate of its total width based on these two decay channels.   Clearly, this estimate of the  total decay width is consistent with the high range of $h_0$ whereas the low range of $h_0$ does not give  a consistent estimate of the total width.   In addition,  the ratio of the decay width to $\pi\pi$ over the sum of  decay widths to $\pi\pi$ and $K{\bar K}$ is also examined in this figure and compared with experiment and other estimates and again  show more consistency with high range of $h_0$.

\begin{figure}
	\centering
	\hskip .1cm
	\includegraphics[height=2.5in]{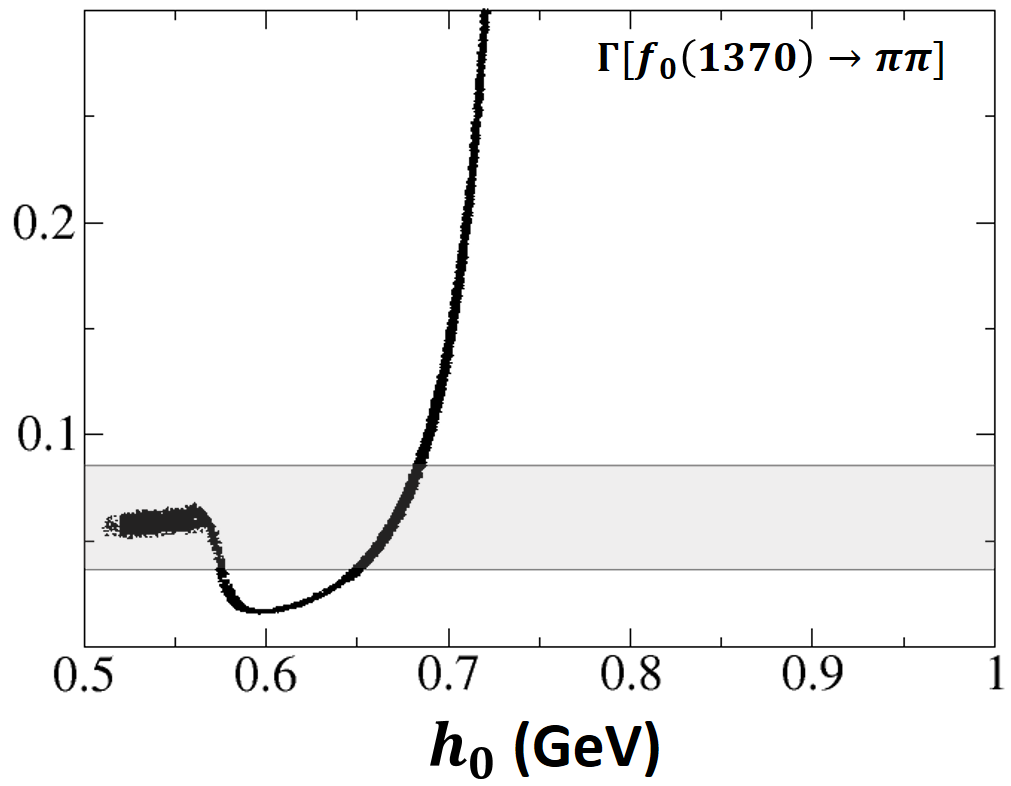}\hspace{0.02\columnwidth}\hskip -.3cm
	\includegraphics[height=2.5in]{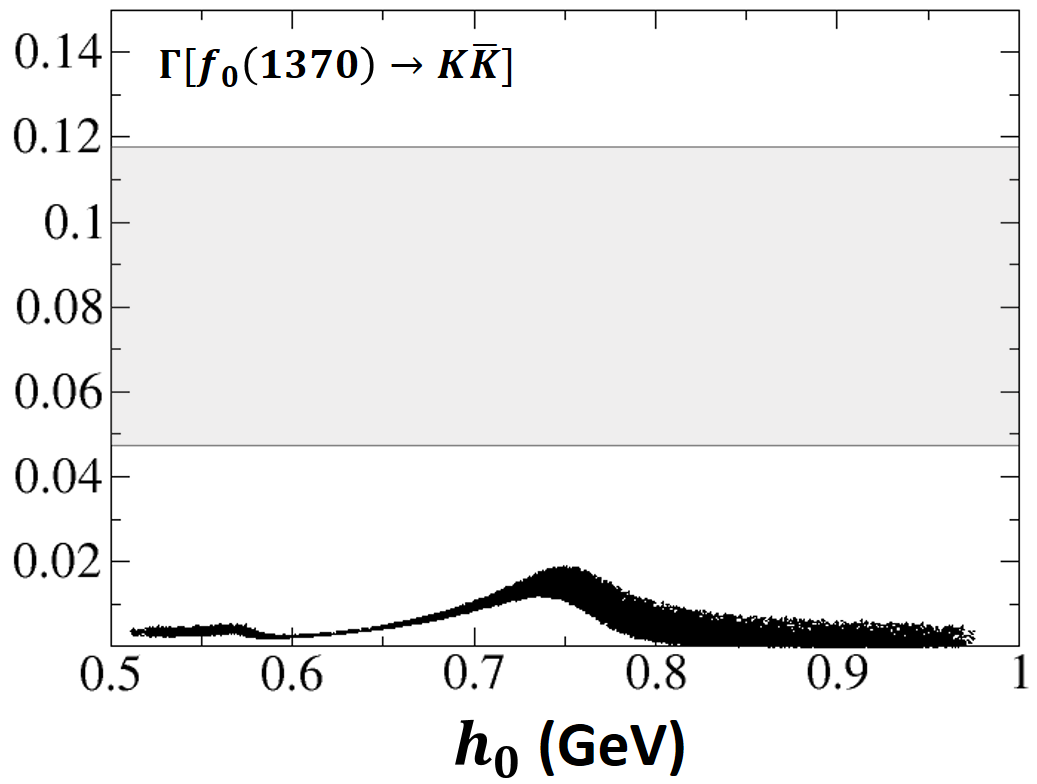}\hspace{0.02\columnwidth}
	
	\vskip .5cm
	
	\includegraphics[height=2.5in]{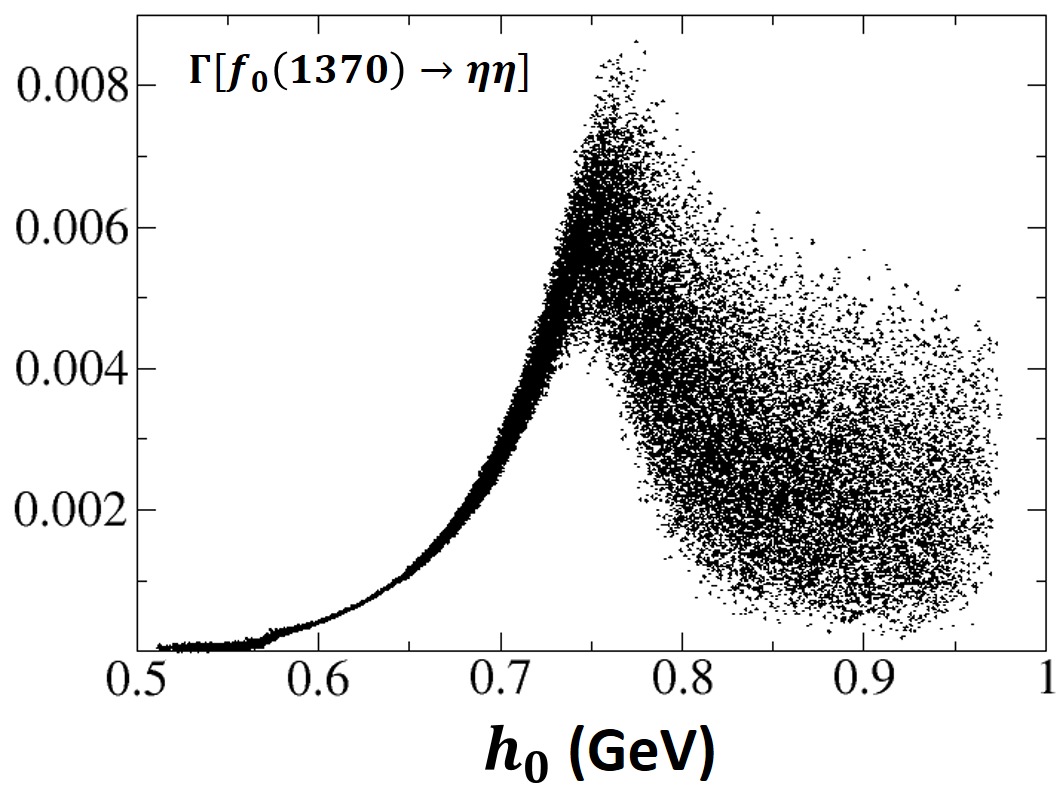}\hspace{0.02\columnwidth}\hskip -.3cm
	\includegraphics[height=2.5in]{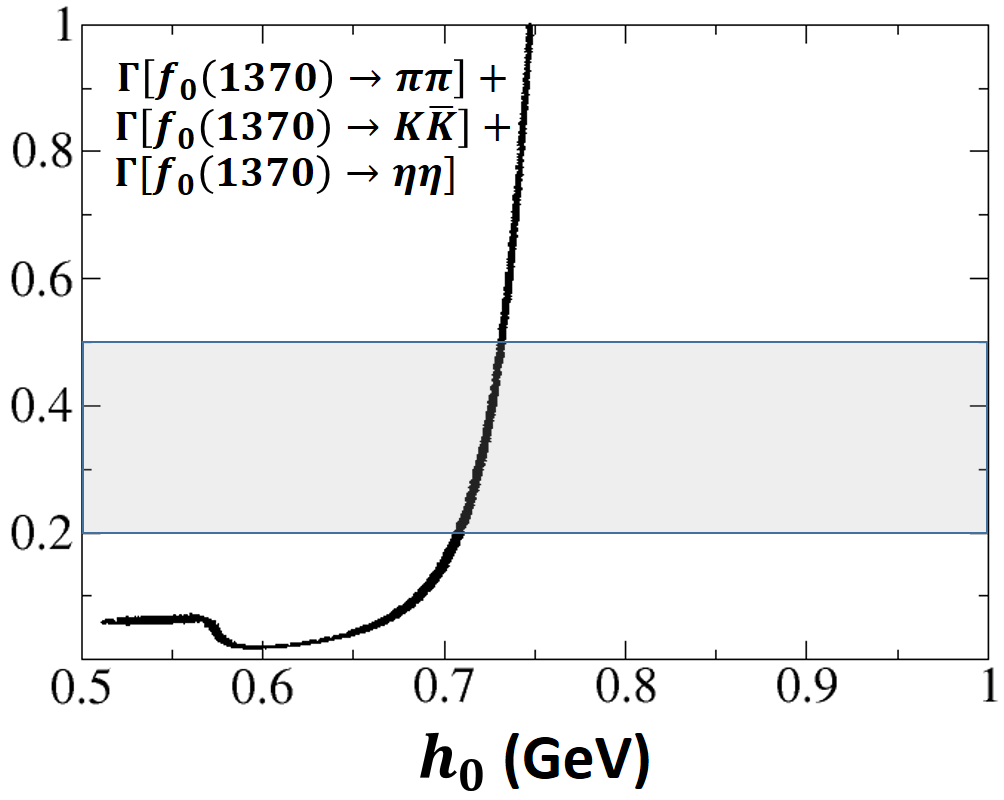}\hspace{0.02\columnwidth}

	\caption{Decay widths (in GeV) of the $f_0(1370)$ to $\pi\pi$ (top left),  $K {\bar K}$ (top right), $\eta\eta$ (bottom left), and the sum of the three (bottom right), over the range of $h_0 = 0.50-0.99$ GeV.   Gray bands (top row) are the estimates given in \cite{07_Bugg_EPJC52}.   The gray band in lower right gives the experimental bounds on the total decay width of $f_0(1370)$ \cite{24_PDG}. 
	}
	\label{F_f3_widths}
\end{figure}

The decay widths of $f_0(1370)$ to $\pi\pi$, $K{\bar K}$ and $\eta\eta$ are examined in Fig. \ref{F_f3_widths}.   We see that these decay widths  only have an order of magnitude agreement with experiment with some of them having a better  agree  with the low-range and some with the high range of $h_0$.    However, we also see in Fig. \ref{F_f3_widths}  that when the  total decay with of $f_0(1370)$ is estimated by adding the partial widths to  $\pi\pi$, $K{\bar K}$ and $\eta\eta$ (which is in principle an underestimate), the low range of $h_0$ is far below the expected range, and therefore again the high range of $h_0$ is overall favored.

\begin{figure}
	\centering
	\hskip .1cm
	\includegraphics[height=2.5in]{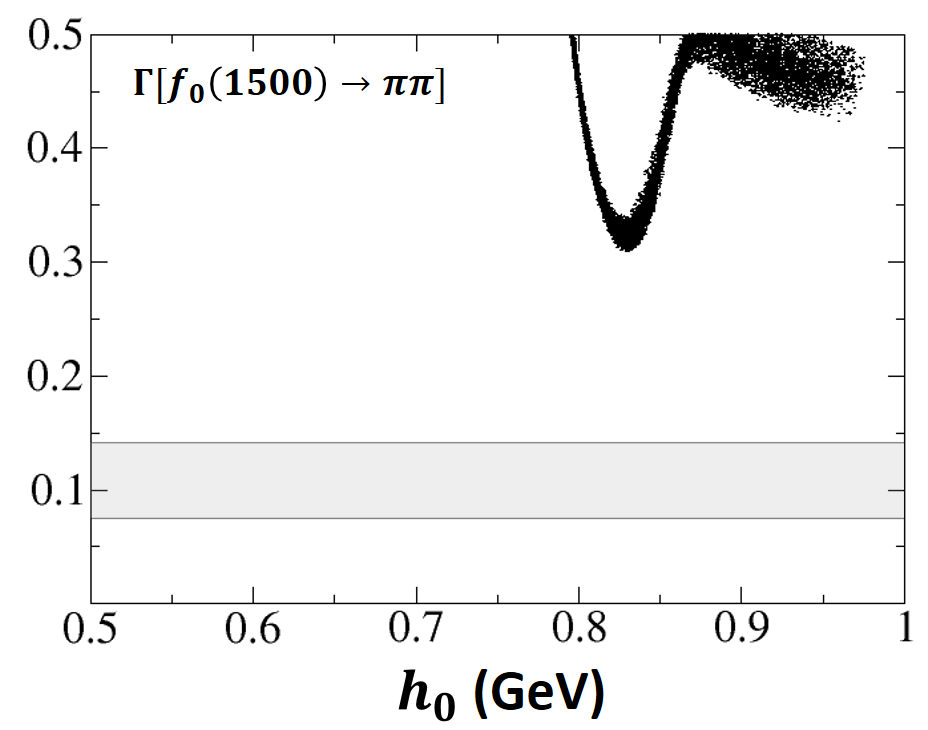}\hspace{0.02\columnwidth}\hskip -.3cm
	\includegraphics[height=2.5in]{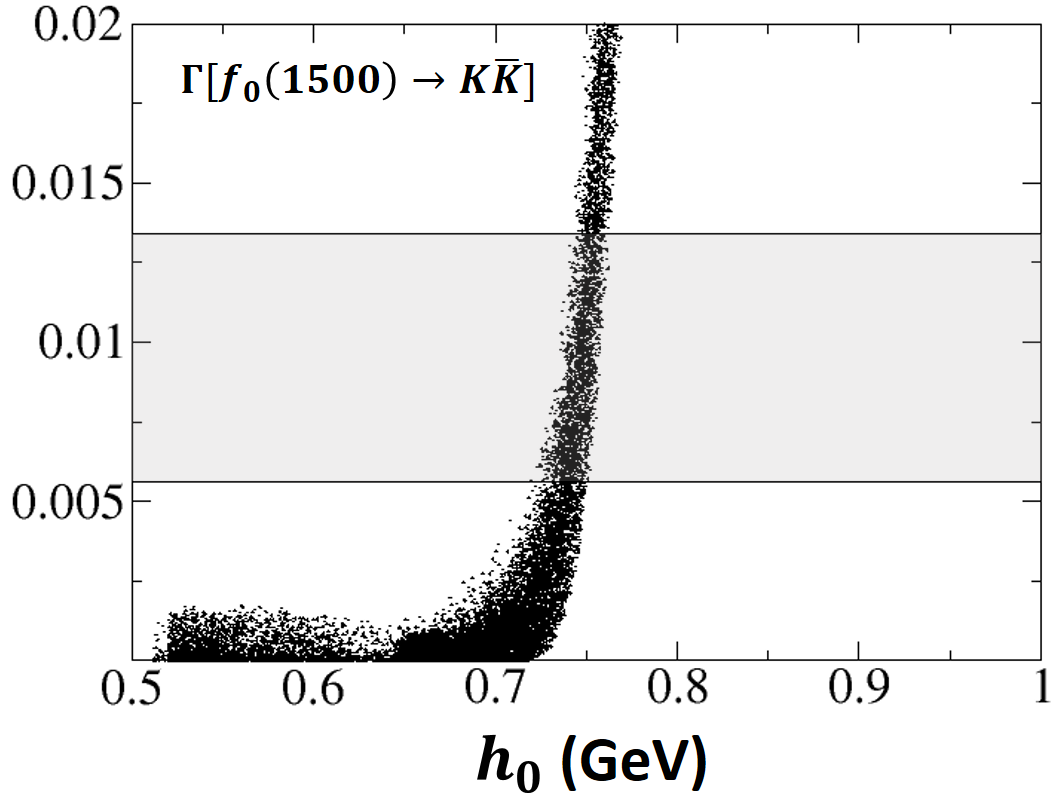}\hspace{0.02\columnwidth}
	
	\vskip .5cm
	
	\includegraphics[height=2.5in]{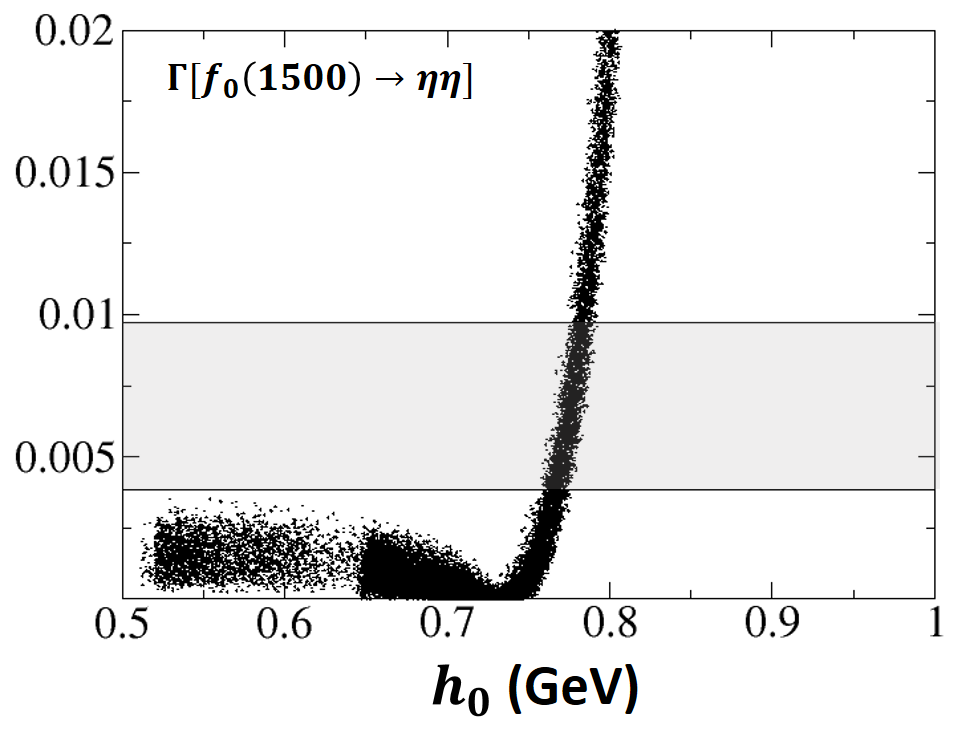}\hspace{0.02\columnwidth}\hskip -.3cm
	\includegraphics[height=2.5in]{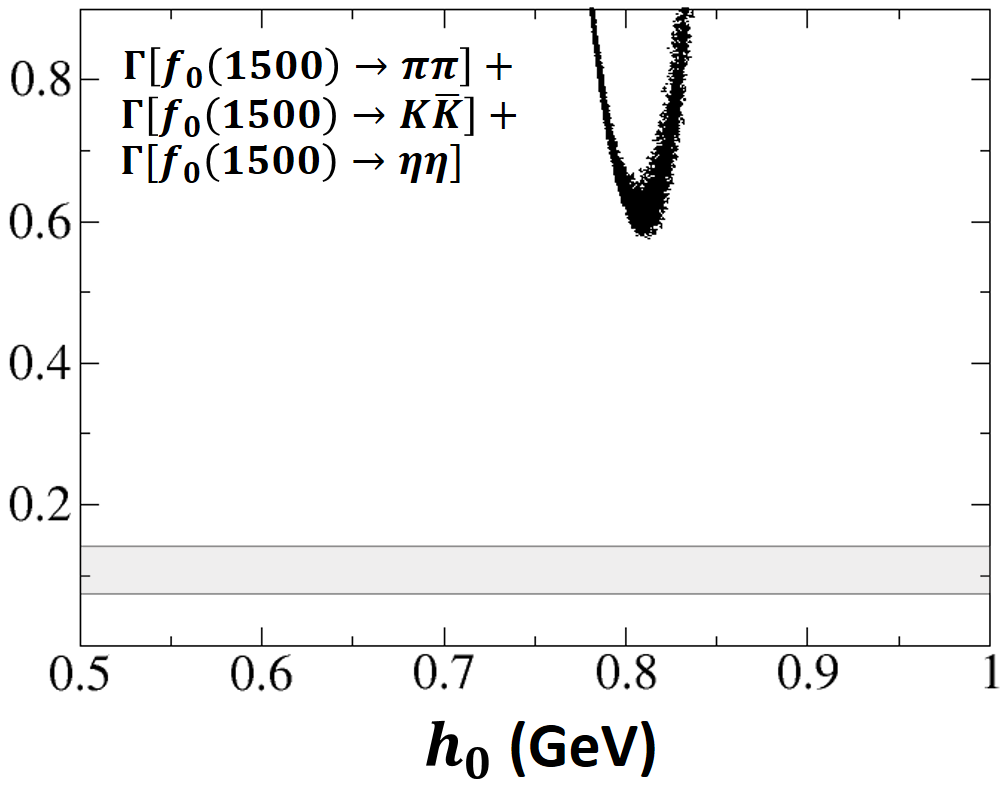}\hspace{0.02\columnwidth}

	\caption{Decay widths (in GeV) of the $f_0(1500)$ to $\pi\pi$ (top left),  $K {\bar K}$ (top right),  $\eta\eta$ (bottom left), and the sum of the three (bottom right),   over the range of $h_0 = 0.50-0.99$ GeV.   Gray bands (top row and lower left) give experimental bounds \cite{24_PDG}.  The gray band in lower right gives the experimental bounds on the total decay width of $f_0(1500)$ \cite{24_PDG}. 
	}
	\label{F_f4_widths}
\end{figure}

The predictions for the decay widths of $f_0(1500)$ are examined in Fig. \ref{F_f4_widths} and while show only an order of magnitude agreement with experiment, the high range of $h_0$ is clearly more favored.

\begin{figure}
	\centering
	\hskip .1cm
	\includegraphics[height=2.5in]{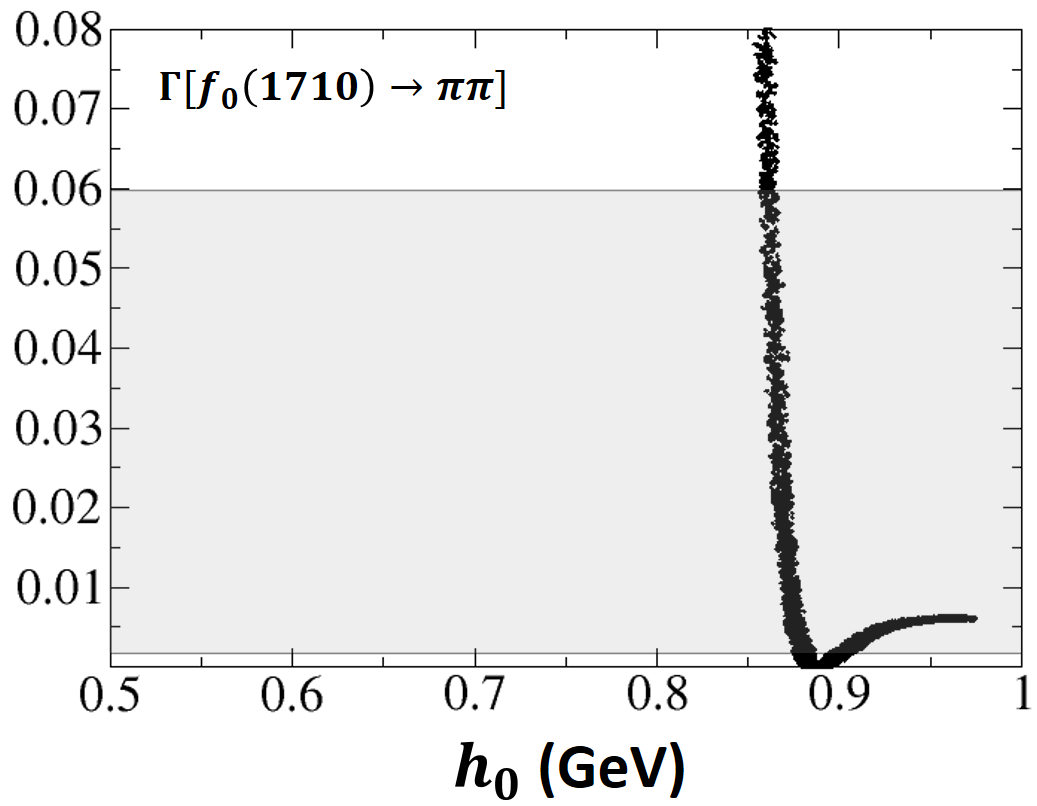}\hspace{0.02\columnwidth}\hskip -.3cm
	\includegraphics[height=2.5in]{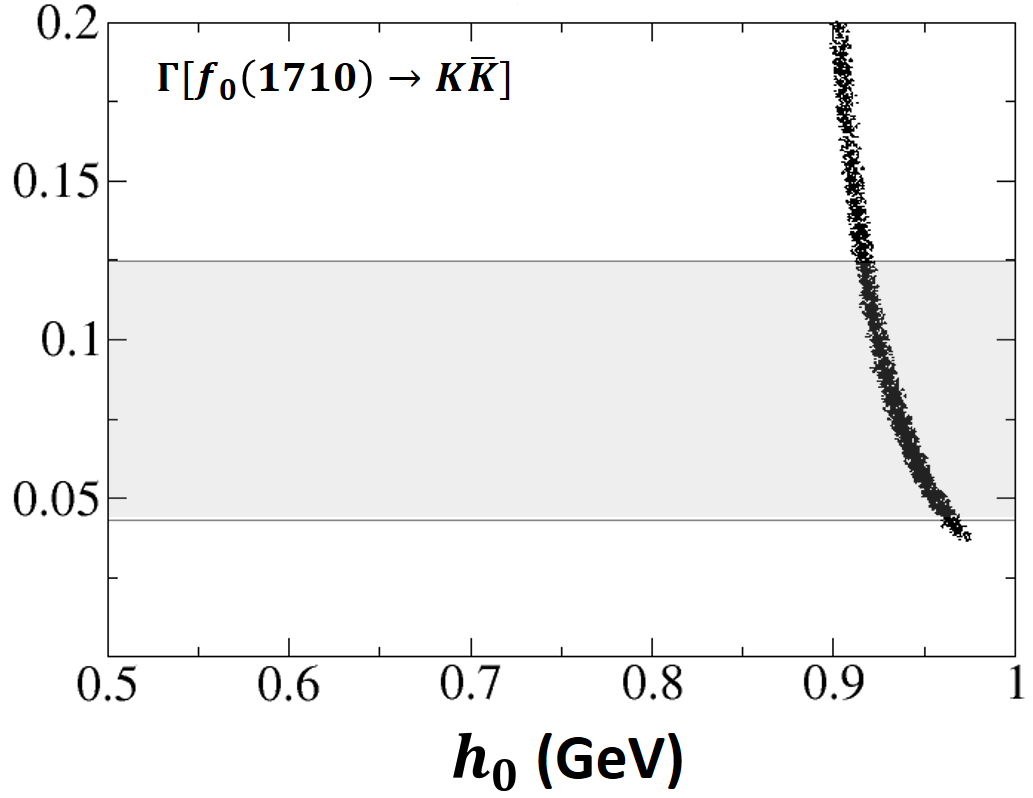}\hspace{0.02\columnwidth}
	
	\vskip .5cm
	
	\includegraphics[height=2.5in]{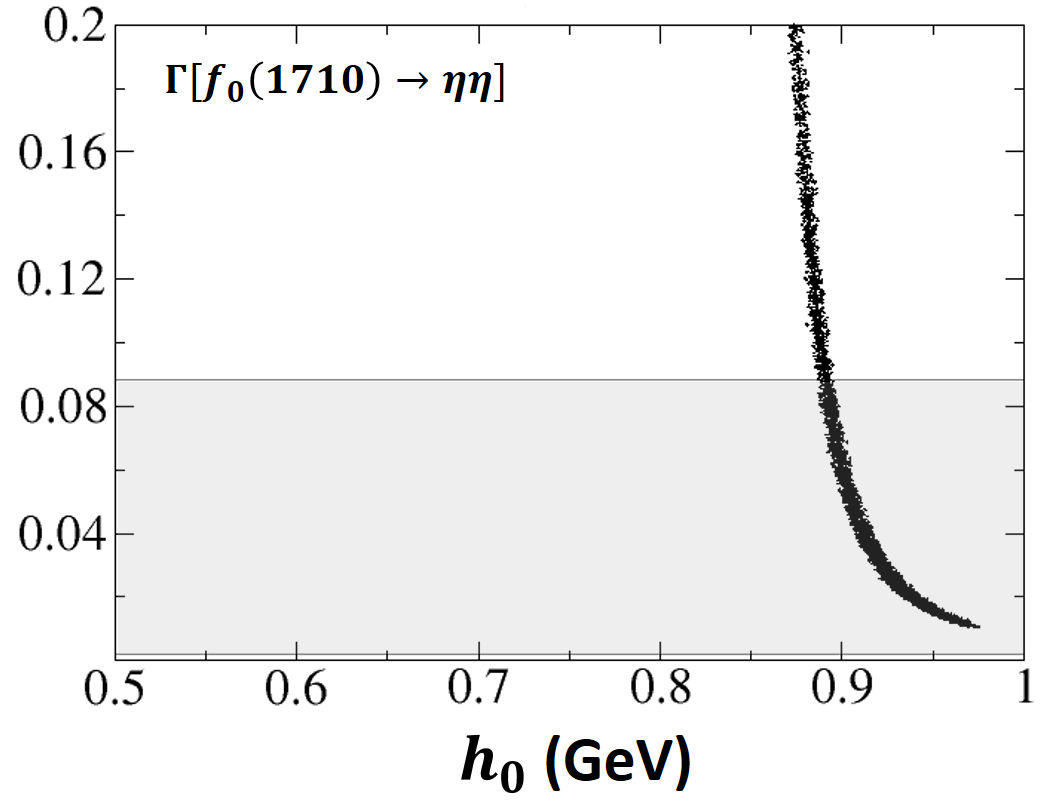}\hspace{0.02\columnwidth}\hskip -.3cm
	\includegraphics[height=2.5in]{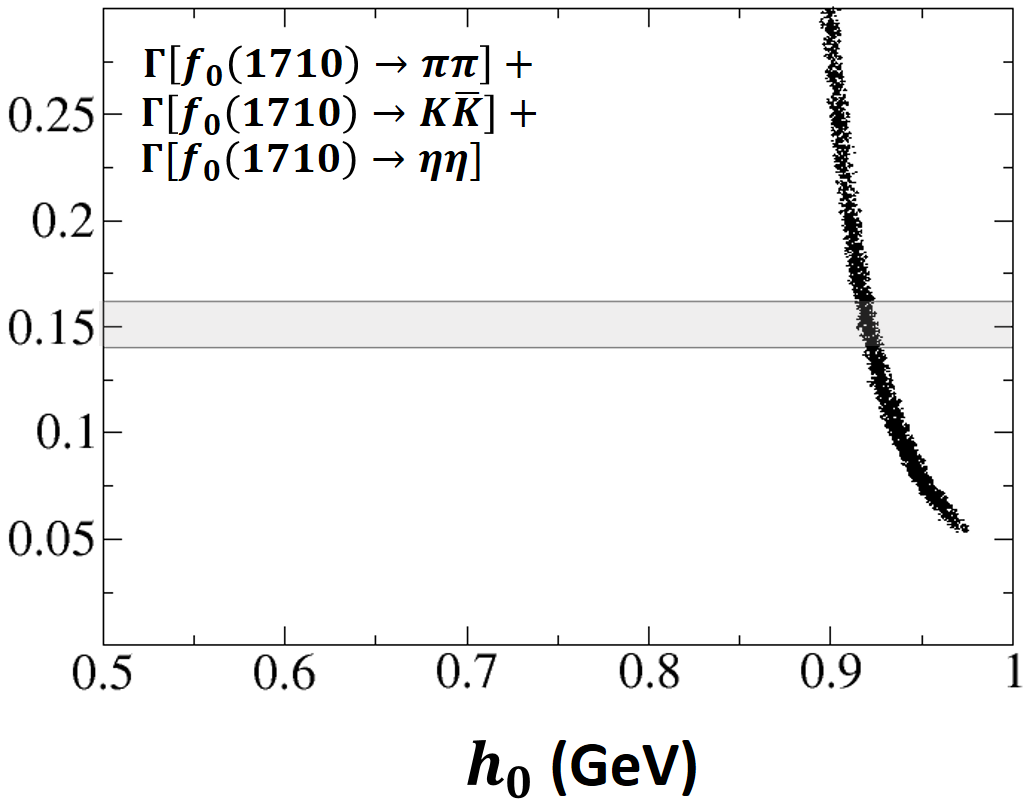}\hspace{0.02\columnwidth}

	\caption{Decay widths (in GeV)  of the $f_0(1710)$ to $\pi\pi$ (top left),  $K {\bar K}$ (top right),  $\eta\eta$ (bottom left), and the sum of the three (bottom right),   over the range of $h_0 = 0.50 - 0.99$ GeV.   Gray bands (top row and bottom left) are  the estimates given in \cite{97_Oller_NPA620}.  The gray band in lower right gives the experimental bounds on the total decay width of $f_0(1710)$ \cite{24_PDG}. 
	}
	\label{F_f5_widths}
\end{figure}

Figure \ref{F_f5_widths} gives the decay widths of $f_0(1710)$ to $\pi\pi$, $K{\bar K}$ and $\eta\eta$ and clearly the high values of $h_0$ agree with experiment.  

\begin{figure}
	\centering
	\hskip .1cm
	\includegraphics[height=2.5in]{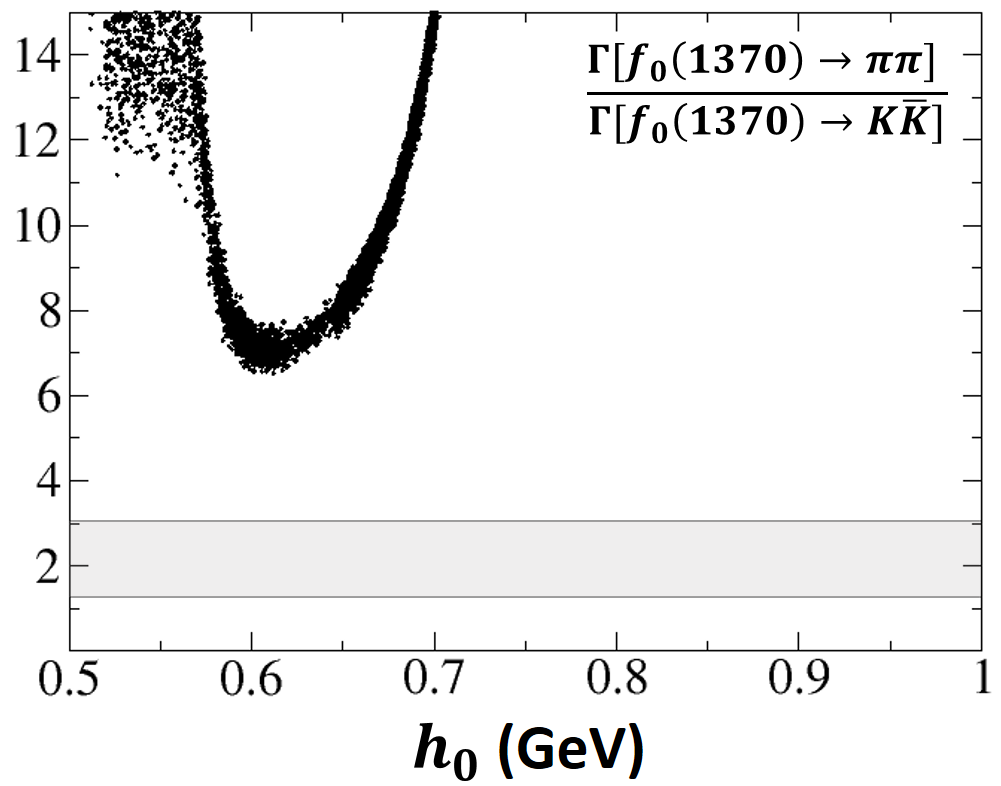}\hspace{0.02\columnwidth}\hskip -.3cm
	\includegraphics[height=2.5in]{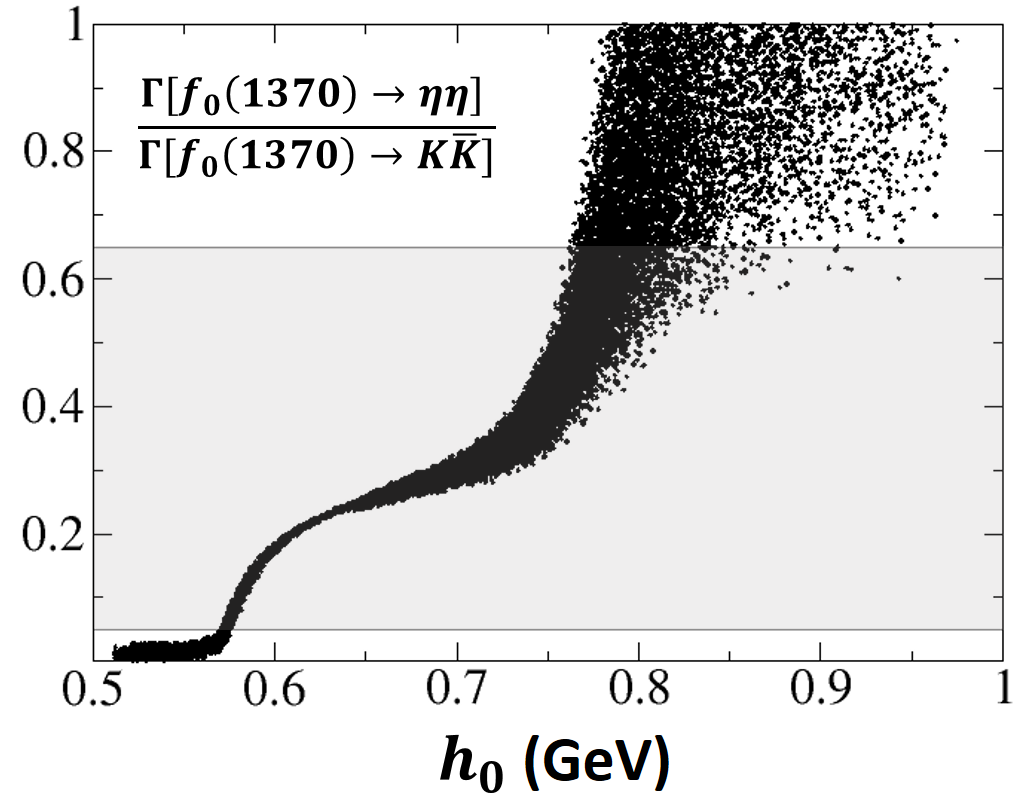}\hspace{0.02\columnwidth}
	
	\caption{Decay ratios of the $f_0(1370)$ computed  over the range of $h_0 = 0.5-0.99$ GeV.   Gray bands give experimental bounds by WA102 Collaboration \cite{WA102}.
	}
	\label{F_f3_DR}
\end{figure}

Next, we examine several decay ratios for $f_0(1370)$, $f_0(1500)$ and $f_0(1710)$ and compare with the experimental values given by WA102 collaboration \cite{WA102}.  Figure \ref{F_f3_DR} gives two decay ratios of $f_0(1370)$ and compares with experiment (gray bands).  For  $\Gamma [f_0(1370) \rightarrow \pi\pi] /\Gamma [f_0(1370) \rightarrow K{\bar K}]$, we see that the closest agreement is with $h_0 = 0.6$ GeV which is close to the middle of the two regions of $h_0$.   The second figure, gives $\Gamma [f_0(1370) \rightarrow \eta\eta] /\Gamma [f_0(1370) \rightarrow K{\bar K}]$  showing a clear agreement with experiment mostly within the high range of $h_0$.  

\begin{figure}
	\centering
	\includegraphics[height=2.5in]{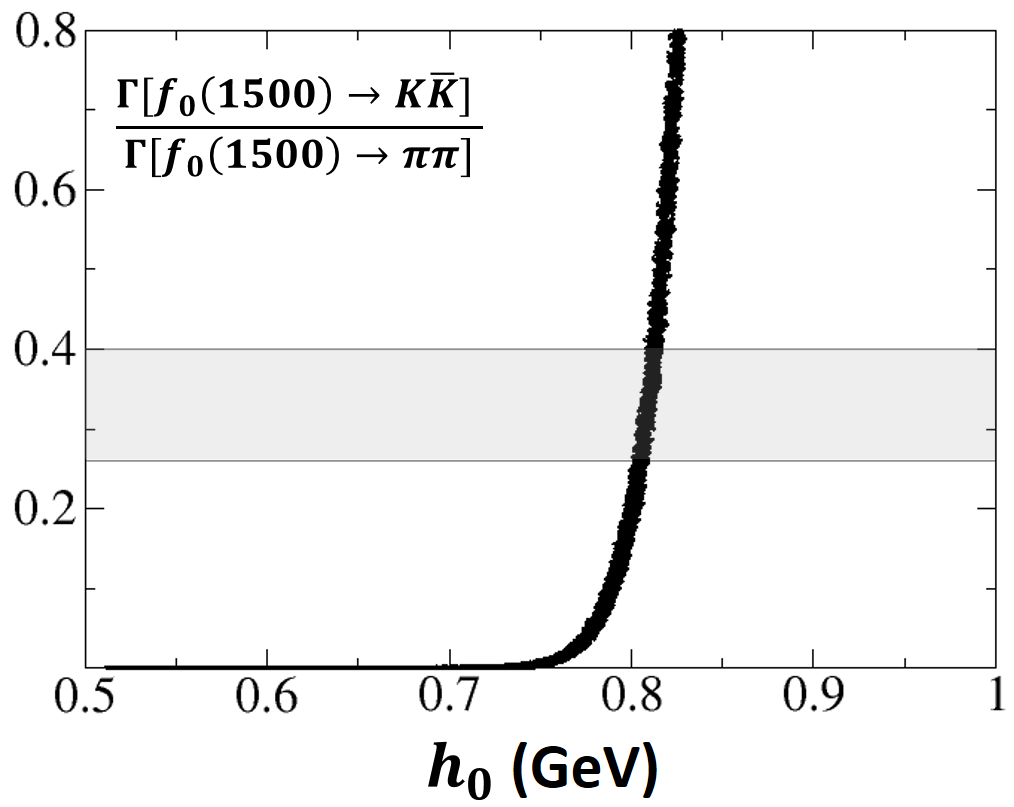}\hspace{0.02\columnwidth}\hskip -.3cm
	\includegraphics[height=2.5in]{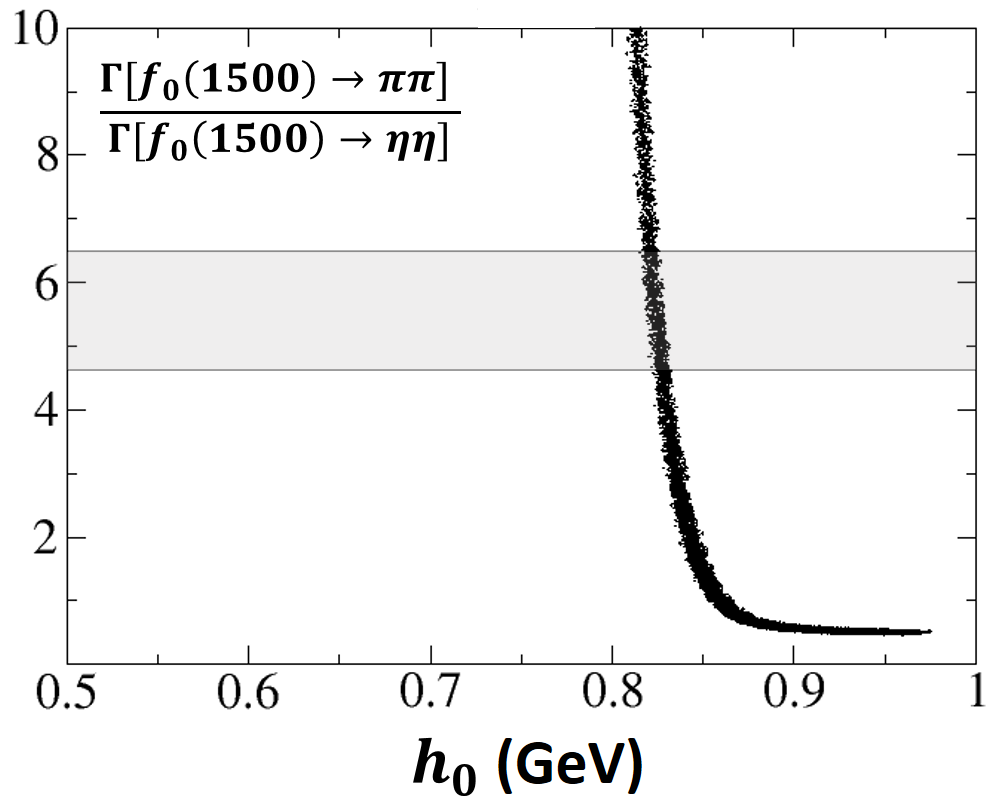}\hspace{0.02\columnwidth}

\vskip .5cm

	\includegraphics[height=2.5in]{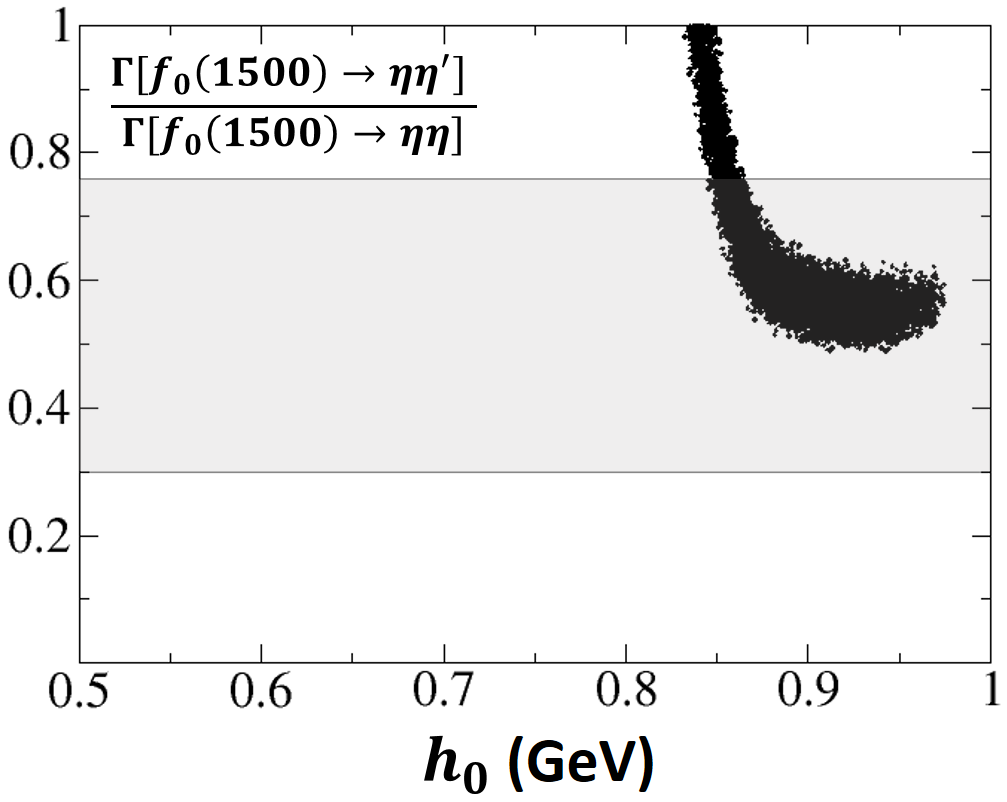}\hspace{0.02\columnwidth}

	\caption{Decay ratios of the $f_0(1500)$ computed  over the range of $h_0 = 0.5-0.99$ GeV.   Gray bands give experimental bounds by WA102 Collaboration \cite{WA102}.
	}
	\label{F_f4_DR}
\end{figure}

Figure \ref{F_f4_DR} gives three decay ratios of $f_0(1500)$, which are $\Gamma [f_0(1500) \rightarrow  K{\bar K}] /\Gamma [f_0(1500) \rightarrow \pi\pi]$, 
$\Gamma [f_0(1500) \rightarrow \pi\pi] /\Gamma [f_0(1500) \rightarrow \eta \eta]$ and 
$\Gamma [f_0(1500) \rightarrow \eta\eta'] /\Gamma [f_0(150) \rightarrow \eta\eta]$.
All in agreement with the experiment in the high region of $h_0$.

\begin{figure}
	\centering

	\includegraphics[height=2.5in]{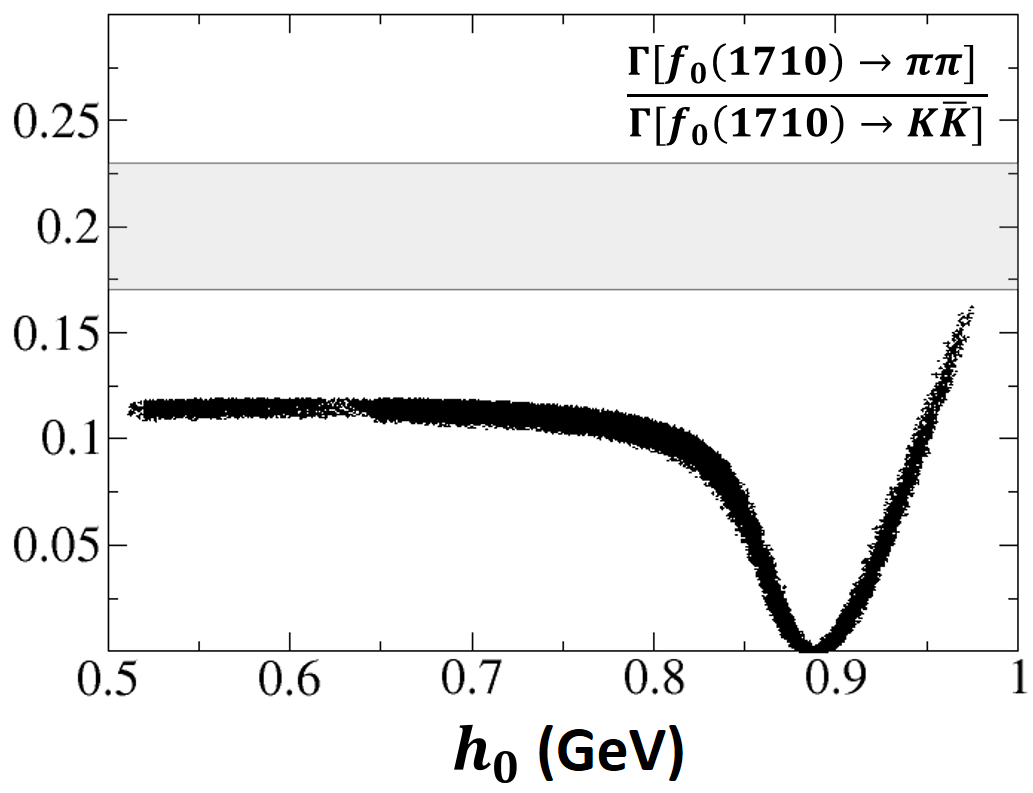}\hspace{0.02\columnwidth}\hskip -.3cm
	\includegraphics[height=2.5in]{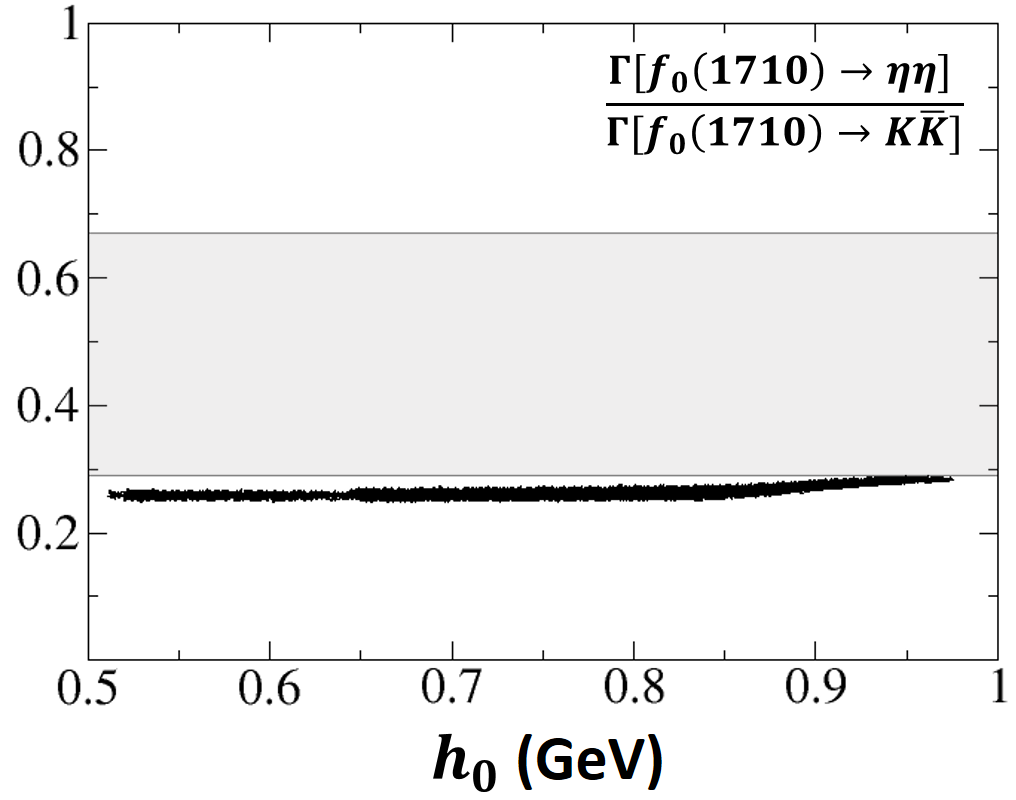}\hspace{0.02\columnwidth}

	\caption{Decay ratios of the $f_0(1710)$ computed  over the range of $h_0 = 0.5-1.0$ GeV.   Gray bands give experimental bounds by WA102 Collaboration \cite{WA102}.
	}
	\label{F_f5_DR}
\end{figure}

Figure \ref{F_f5_DR} gives two decay ratios of $f_0(1710)$.  The first one is $\Gamma [f_0(1710) \rightarrow \pi\pi] /\Gamma [f_0(1710) \rightarrow K{\bar K}]$ which is in an order of magnitude agreement mostly with 
$h_0 = 0.5-0.8$ GeV.    The second figure gives   $\Gamma [f_0(1710) \rightarrow \eta\eta] /\Gamma [f_0(1710) \rightarrow K{\bar K}]$ which is in close agreement with experiment over  the entire range of $h_0 = 0.50-0.99$ GeV.

\begin{figure}
	\centering

	\includegraphics[height=2.5in]{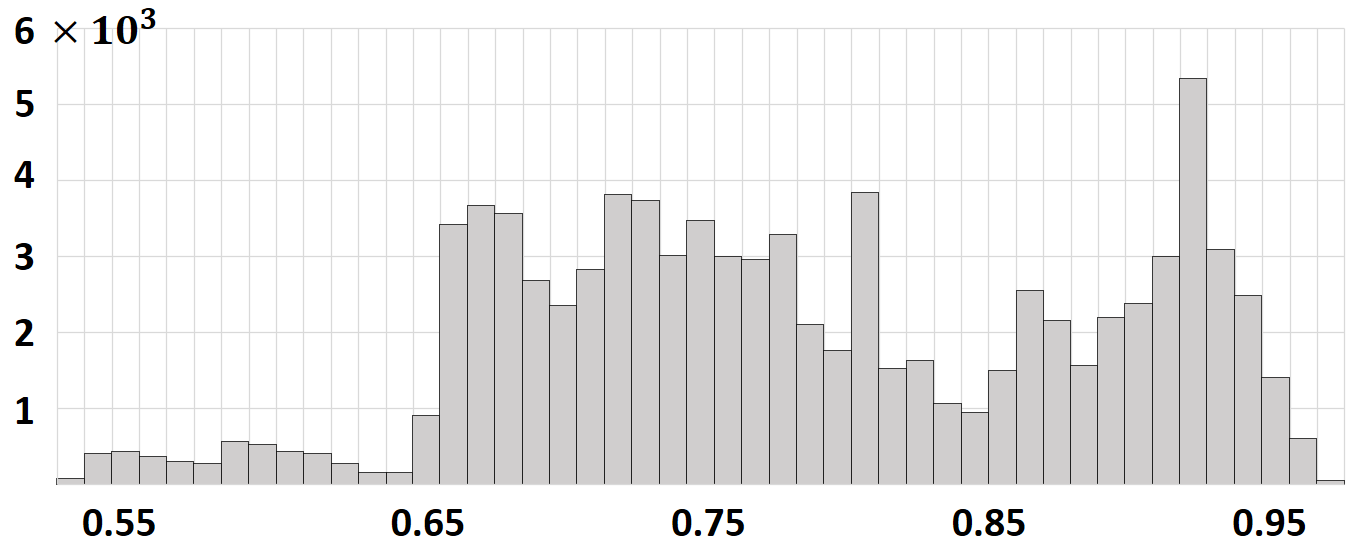}\hspace{0.02\columnwidth}\hskip -.3cm

	\caption{Histogram representing the number of decay widths and decay ratios predicted by the model computed over the range of $h_0 = 0.50-0.99$ GeV that are in agreement with the experimental bounds.   
		}
	\label{F_hist_f_decays}
\end{figure}

Therefore, overall, most of these decay widths and decay ratios favor the high range of $h_0 = 0.65 - 0.99$ GeV.  To quantify this observation,  for each narrow range of $h_0$ we count the number of decay widths and decay ratios that are in agreement with  their corresponding experimental data (or with other estimates). This results in a histogram given in    Fig. \ref{F_hist_f_decays} which clearly shows that the high range of $h_0$ is favored by the available data on the decay widths or decay ratios of the isosinglet scalars.

%%%%%%%%%%%%%%%%%%%%%%%%%%%%%%%%%%%%%%%%%%%%%%%%%%%%%%%%%%%%%%%%%%%%%%%%%%%%%%%%%%%%%%%%%

\begin{figure}
	\centering
	\hskip .1cm
	\includegraphics[height=2.5in]{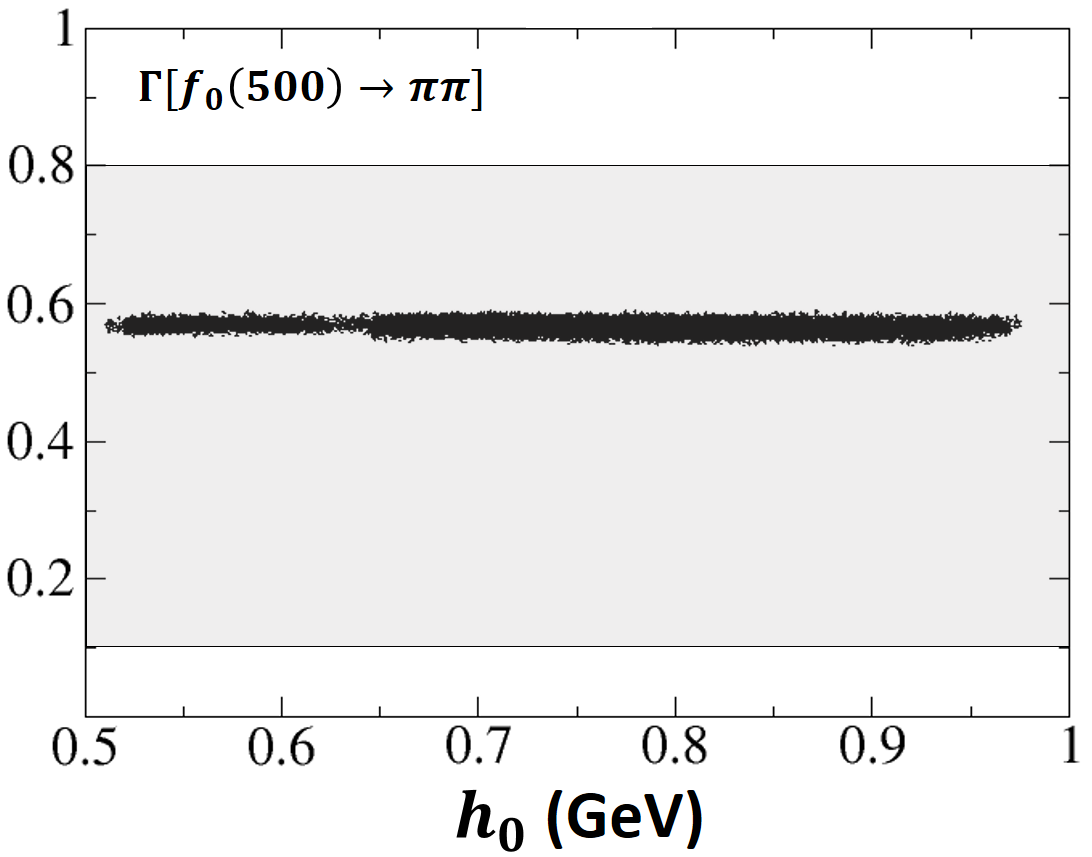}\hspace{0.2\columnwidth}
	
	\caption{Decay width (in GeV) of the $f_0(500)$ to two pions over the range of $h_0 = 0.50-0.99$ GeV.  The gray band gives the experimental bound on the total decay width of $f_0(500)$ \cite{24_PDG}.
	}
	\label{F_f1_pp}
\end{figure}

\clearpage

%%%%%%%%%%%%%%%%%%%%%%%%%%%%%%%%%%%%%%%%%%%%%%%%%%%%%%%%%%%%%%%%%%%%%%%%%%%%%%%%%%%%%%%%%

\section{Filtered results}

%%%%%%%%%%%%%%%%%%%%%%%%%%%%%%%%%%%%%%%%%%%%%%%%%%%%%%%%%%%%%%%%%%%%%%%%%%%%%%%%%%%%%%%%%

In previous section we determined the model parameters and their correlations with the experimental inputs, together with the subsequent model predictions such as masses and quark and glue substructures of various scalar and pseudoscalar states.  We noted that one of the model parameters is the glueball condensate $h_0$ which plays a key role in the predictions of this model.  It can, in principle,  have a wide range of $h_0 = 0.40-0.99$ GeV, however, it was noticed that the resulting predictions stemming from $h_0 = 0.40-0.65$ GeV (the low range of $h_0$) and those from $h_0 = 0.65-0.99$ GeV (the high range of $h_0$), are  contradictory, and therefore, one of these ranges should be ruled out.  We made a number of observations  that favored the high range of $h_0$:  
\begin{itemize}
	\item Fig. \ref{F_h0_histogram} shows that most of the acceptable parameters that satisfy condition (\ref{E_chi_goodness}) are within the high range of $h_0$, including the highest histogram in this region around $h_0 = 0.80$ GeV.
	
	\item Fig. \ref{F_chi_vs_h0} shows improvement of fits in the high range of $h_0$.
	
	\item SU(3) flavor limit of this model studied in \cite{19_Fariborz_IJMPA34,21_Fariborz_NPA1015} favors the range of $h_0=0.75-0.825$ GeV which is in the high range of $h_0$.
	
	\item The decay widths and decay ratios of isosinglet states investigated in previous section, overall,  favored the high range of $h_0$.
	
	\item Previous investigations of glueball characteristics within nonlienar sigme model \cite{04_Fariborz_IJMPA19,06_Fariborz_PRD74,15_Fariborz_PRD91,15_Fariborz_PRD92A}, as well as most (if not all) studies of glueballs favor a combination of $f_0(1370)$, $f_0(1500)$ and $f_0(1710)$ as states that contain a significant  glue component \cite{96_Amsler_PRD53, 97_Anisovich_PLB395, 97_Close_PLB397, 98_Narison_PLB509, 01_Close_EPJC21, 02_Amsler_PLB541,05_Close_PRD71,06_Cheng, 00_Lee_PRD61, 14_Janowski_PRD90, 15_Brunner_PRL115, 21_Sar_PLB816, 22_Rodaz_EPJ82}. 
\end{itemize}

The above rather robust (albeit circumstantial) evidences  seem to warrant a focused look at the predictions of this model by filtering out the low range of $h_0$ and focusing on the high range of this parameter.   This section is dedicated to that objective.   In addition to the high range of $h_0 = 0.65-0.99$ GeV, for comparison, we also analyze the predictions in the range of $h_0=0.75-0.825$ GeV determined in the flavor SU(3) limit of this model \cite{19_Fariborz_IJMPA34,21_Fariborz_NPA1015} that can be used to study the  SU(3) breaking effects.  All the results presented in this section are given for both of these two ranges. 

Table \ref{T_Lag_Par_h0_75-825} gives the averages of model parameters (together with standard deviations around the averages).   Table 
\ref{T_eta_f0_masses} gives the predicted masses of eta and $f_0$ systems and compares them with the PDG \cite{24_PDG}.  

Table \ref{T_eta_comps}  gives the quark and glue components of the eta system, which are practically constant throughout the high range of $h_0$.  We see that the  $\eta(547)$ and the $\eta'(958)$, expectedly,  are quark-antiquark states; the $\eta(1295)$ and the $\eta(1760)$ are  dominantly four-quark states, while the heaviest eta, the $\eta(2225)$,  is almost entirely made of glue [see also comment about $X(2370)$ after (\ref{E_eta_exp})].  To further examine these predictions, in Table \ref{T_eta_comps_OS} the quark and glue components of these states are given in the octet-singlet language showing that $\eta(547)$ and $\eta'(958)$ are, as we expect, the quark-antiquark octet and singlet states, respectively.   Also we see  similar characteristics   for the next two eta states with the $\eta(1295)$ and the $\eta(1760)$ being respectively the four-quark octet and singlet states.

The components of the isosinglet scalars are given in \ref{T_f0_comps} and are less clear-cut than those of etas discussed above.  The properties of the  $f_0(500)$ and the $f_0(980)$ are independent of the specific value of $h_0$ within its high range,  whereas the properties of the $f_0(1370)$, $f_0(1500)$ and $f_0(1710)$ are  affected by the value of  $h_0$ within its high range.   The components averages  over the entire high range of $h_0$ show that these three states are nearly equal contenders for the glue.   In the range of $h_0$ favored by the flavor SU(3) limit,  clearly $f_0(1500)$ becomes the dominant glue holder.   For further insight, the quark components in the octet-singlet basis are given in  
Table \ref{T_f0_comps_OS}.   The $f_0(500)$ is predicted to be a maximally mixed two and four-quark singlet state whereas  the $f_0(980)$ is dominantly a four-quark octet state.    This singlet octet nature of $f_0(500)$ and $f_0(980)$ is clearly consistent with the findings of  \cite{03_Oller_NPA727}. The $f_0(1370)$ and the $f_0(1500)$ contain high admixtures of octet and singlet (of both two and four-quark types); and the quark component of the $f_0(1710)$ is predicted to be mainly of the quark-antiquark octet type.

\begin{table}[!htbp]
	\centering
	\caption{Lagrangian parameters (Ave. $\pm \sigma$) determined over the high range of $h_0$ (first column) and over the range of $h_0$ favored by the SU(3) limit of this model (second column). 
}
	\renewcommand{\tabcolsep}{0.4pc} % enlarge column spacing
	\renewcommand{\arraystretch}{1.5} % enlarge line spacing
	\begin{tabular}{c|c|c}
		\noalign{\hrule height 1pt}
		\noalign{\hrule height 1pt}	
		Parameter & $h_0 = 0.65-0.99$ GeV &  $ h_0 = 0.75-0.825$ GeV   \\
		\noalign{\hrule height 1pt}		
		\noalign{\hrule height 1pt}	
		$u_1 \times 10$  &  $-1.375 \pm 0.563$  &  $-1.320 \pm 0.128$ \\  
		\noalign{\hrule height 1pt}		
		$u_2 \times 10^{-1}$  & $3.880 \pm 0.011$ & $3.876 \pm 0.023$ \\  
		\noalign{\hrule height 1pt}		
		$u_3$  & $1.209 \pm 0.480 $ & $1.167 \pm 0.093 $ \\  
		\noalign{\hrule height 1pt}		
		$u_4$  & $-2.422 \pm 0.454$ & $-2.390 \pm 0.104$\\  
		\noalign{\hrule height 1pt}		
		$u_6$  & $4.049 \pm 1.706$& $3.886 \pm 0.274$\\  
		\noalign{\hrule height 1pt}		
		$\alpha_1\times 10^{2}$ & $6.065\pm 0.001  $ & $6.066 \pm 0.002$ \\  
		\noalign{\hrule height 1pt}		
		$\alpha_3 \times 10^{2}  $    & $7.832\pm 0.003$ & $7.833 \pm 0.028 $\\  
		\noalign{\hrule height 1pt}		
		$\beta_1 \times 10^{2}$  & $2.473 \pm 0.002 $ & $2.472\pm 0.005$\\  
		\noalign{\hrule height 1pt}		
		$\beta_3 \times 10^{2}$  & $1.915\pm 0.001 $ & $1.914\pm 0.010$ \\  
		\noalign{\hrule height 1pt}		
		$h_0 \times 10$    & $7.858 \pm 1.318$ &  $7.887 \pm 0.347$ \\  
		\noalign{\hrule height 1pt}
		$A_1 \times  10^{4}$   & $6.651 \pm 0.001 $&  $6.650 \pm 0.003$\\  
		\noalign{\hrule height 1pt}
		$A_3 \times 10^{2}$    & $1.955 \pm 0.010$    & $1.955\pm 0.026$\\  
		\noalign{\hrule height 1pt}
		$\xi \times 10$  & $8.166 \pm 1.000$& $8.025 \pm 0.505$\\  
		\noalign{\hrule height 1pt}
		$\gamma_1 \times 10^{3}$ & $6.551 \pm 1.882$ & $6.402 \pm 2.243$ \\  
		\noalign{\hrule height 1pt}
		$c_3 \times 10^{4}$  & $-4.757 \pm 1.130$&  $-4.952 \pm 0.902$ \\  
		\noalign{\hrule height 1pt}
	\end{tabular}\\[2pt]
	\label{T_Lag_Par_h0_75-825}
\end{table}

\begin{table}[!htbp]
	\centering
	\caption{Predicted masses  (Ave. $\pm \sigma$) of $\eta$'s and $f_0$'s (in GeV) determined over the high range of $h_0$ (first column) and over the range of $h_0$ favored by the SU(3) limit of this model (second column).   The corresponding experimental values by PDG \cite{24_PDG} are given in the last column.  The underlined quantities have been inputted in the computations, all other estimates are predictions.
	}
	\renewcommand{\tabcolsep}{0.4pc} % enlarge column spacing
	\renewcommand{\arraystretch}{1.5} % enlarge line spacing
	\begin{tabular}{c|c|c|c}
		\noalign{\hrule height 1pt}
		\noalign{\hrule height 1pt}	
		State & $h_0 = 0.65-0.99$ GeV &  $h_0 = 0.75-0.825$ GeV & PDG  \\
		\noalign{\hrule height 1pt}		
		\noalign{\hrule height 1pt}
		$m\left[\pi\right]$ & \underline{0.137} & \underline{0.137} & 0.137 (Ave.)\\
			\noalign{\hrule height 1pt}	
			$m\left[\pi(1300)\right]$ &$1.300 \pm 0.004$ & $1.301 \pm 0.004$ & $1.300 \pm 0.100$ \\
	\noalign{\hrule height 1pt}			
		$m\left[K\right]$ & $0.509 \pm 0.002$& $0.509 \pm 0.002$ & 0.496 (Ave.)\\
			\noalign{\hrule height 1pt}		
		$m\left[K(1460)\right]$ & $1.274 \pm 0.004$ &$1.274 \pm 0.004$ & $\approx 1.4-1.5$\\
			\noalign{\hrule height 1pt}		
		$m\left[\eta(547)\right]$ &  $ 0.548 \pm 0.002 $  &  $0.548 \pm 0.002$ & $0.547862 \pm 0.000017$\\  
		\noalign{\hrule height 1pt}		
     	$m\left[\eta'(958)\right]$  & $ 0.957\pm 0.004$ & $0.957 \pm 0.004 $ & $0.95778 \pm 0.00006$ \\  
		\noalign{\hrule height 1pt}		
      	$m\left[\eta(1295)\right]$  & $1.292 \pm 0.004$ & $1.292 \pm 0.004 $ & $1.294 \pm 0.004$\\  
		\noalign{\hrule height 1pt}		
		$m\left[\eta(1760)\right]$  & $1.753 \pm 0.007$ & $1.753 \pm 0.006$ & $1.751 \pm 0.015$\\  
		\noalign{\hrule height 1pt}		
		$m\left[\eta(2225)\right]$ & $ 2.221\pm 0.009$ & $2.221 \pm 0.009$ & $2.221 ^{+0.013}_{- 0.010}$\\
	\noalign{\hrule height 1pt}		  
	$m\left[a_0(980)\right]$ & $\underline{0.9835}$&
	$\underline{0.9835}$ &
	$0.980 \pm 0.020$
	 \\
		\noalign{\hrule height 1pt}		
	$m\left[a_0(1450)\right]$ &
	$\underline{1.474}$ & $\underline{1.474}$ &
	$1.439 \pm 0.034$ \\
		\noalign{\hrule height 1pt}		
		\noalign{\hrule height 1pt}		
		$m\left[K_0^*(700)\right]$ &
		$1.116 \pm 0.003$ & $1.116 \pm 0.003$ & $0.838 \pm 0.011$\\
\noalign{\hrule height 1pt}		
		$m\left[K_0^*(1430)\right]$ &$1.576 \pm 0.003$ &$1.576 \pm 0.003$ & $1.425 \pm 0.050$\\
\noalign{\hrule height 1pt}		
		$m\left[f_0(500)\right]$   & $ 0.656\pm 0.003$ & $0.656 \pm 0.003  $ &$0.400 \rightarrow 0.800$\\  
		\noalign{\hrule height 1pt}		
		$m\left[f_0(980)\right]$ & $1.145 \pm 0.004$ & $1.146\pm 0.004$ &$0.990 \pm 0.020$\\  
		\noalign{\hrule height 1pt}		
		$m\left[f_0(1370)\right]$ & $ 1.461 \pm 0.063$ & $1.496\pm 0.009 $ &$ 1.200 - 1.500$\\  
		\noalign{\hrule height 1pt}		
		$m\left[f_0(1500)\right]$   & $1.600 \pm 0.078$ &  $1.590 \pm 0.033 $ & $1.522 \pm 0.025$ \\  
		\noalign{\hrule height 1pt}
		$m\left[f_0(1710)\right]$  & $1.747 \pm 0.045$ &  $1.726 \pm 0.004 $ & $1.733^{+0.008}_{-0.007}  $\\  
		\noalign{\hrule height 1pt}
		\end{tabular}\\[2pt]
	\label{T_eta_f0_masses}
\end{table}

\begin{table}[!htbp]
	\centering
	\caption{Predicted components (Ave. $\pm \sigma$) of $\eta$'s in the strange-nonstrange basis.  For each component, the top number is determined over the high range of glueball condensate ($h_0 = 0.65-0.99$ GeV) and the bottom number in parentheses is obtained in the range of $h_0 = 0.75-0.825$ GeV favored by the SU(3) limit of this model. 
			}
	\renewcommand{\tabcolsep}{0.4pc} % enlarge column spacing
	\renewcommand{\arraystretch}{1.5} % enlarge line spacing
	\begin{tabular}{c|c|c|c|c|c||c|c}
		\noalign{\hrule height 1pt}
		\noalign{\hrule height 1pt}	
		State & 
		$\frac{1}{\sqrt{2}} \left( u{\bar u} + d{\bar d}\right)$ &  
		$s{\bar s}$ & 
		%$\frac{1}{\sqrt{2}} \left({\bar d} {\bar s} d s + {\bar s} {\bar u} s %u\right)$ 
		$\frac{\left({\bar d} {\bar s} d s + {\bar s} {\bar u} s u\right)}{\sqrt{2}}$
		& 
		${\bar u} {\bar d} u d$
		&
		$g$ & Total & Total 
		 \\
		 &&&&&& two quark & four quark\\
		\noalign{\hrule height 1pt}		
		\noalign{\hrule height 1pt}	
%==========================================================
		$\eta(547)$ & 
		$0.461 \pm 0.005$     &  
		$0.447 \pm 0.004$     & 
		$0.029 \pm 0.001$      &      
		$0.063 \pm 0.001$      &  
		$\approx 0 $ &
		$0.907 \pm 0.001$    &
		$0.092 \pm 0.002$
		\\
		&
		($0.460 \pm 0.004$)
		&
    	($0.447 \pm 0.004$)
		&
		($0.029 \pm 0.001$)
		&
		($0.063 \pm 0.001$)
		&
		($ \approx 0 $) 
		&  
		($0.907 \pm 0.001$) 
		&
		($0.092 \pm 0.002$) 
		\\  
		\noalign{\hrule height 1pt}		
%==========================================================
		$\eta'(958)$  & 
        $0.504 \pm 0.005$     &  
		$0.326 \pm 0.005$     & 
		$0.147 \pm 0.004$      &      
		$0.020 \pm 0.002$      &  
		$0.003 \pm 0.002$ &
		$0.831 \pm 0.003$ 
		&
		$0.166 \pm 0.004$
		\\    
				&
		($0.504 \pm 0.005$)
		&
		($0.326 \pm 0.005$)
		&
		($0.146 \pm 0.004$)
		&
		($0.020 \pm 0.002$)
		&
		($0.003 \pm 0.002$) 
		&
		($0.831 \pm 0.003$)
		&
		($0.166 \pm 0.004$)
		\\  
		\noalign{\hrule height 1pt}		
%========================================================
		$\eta(1295)$  & 
	    $\approx 0.003  $     &  
		$0.211 \pm 0.004$     & 
		$0.280 \pm 0.003$      &      
		$0.506 \pm 0.003$      &  
		$\approx 0.001$    
		&
		$0.214 \pm 0.005$
		&
		$0.786 \pm 0.005$
		\\  
				&
		($\approx 0.003$)
		&
		($0.210 \pm 0.004$)
		&
		($0.280 \pm 0.003$)
		&
		($0.506 \pm 0.003$)
		&
		($\approx 0.001$) 
		&
		($0.213 \pm 0.005$)
		&
		($0.786 \pm 0.005$)
		\\     
		\noalign{\hrule height 1pt}		
%======================================================
        $\eta(1760)$  & 
	    $0.032 \pm 0.002$     &  
		$0.016 \pm 0.001$     & 
		$0.519 \pm 0.019$      &      
		$0.389 \pm 0.016$      &  
		$0.044 \pm 0.034$  
		&
		$0.048 \pm 0.002$
		&
		$0.908 \pm 0.035$
		\\
				&
		($0.032 \pm 0.002$)
		&
		($0.016 \pm 0.001$)
		&
		($0.515 \pm 0.018$)
		&
		($0.386 \pm 0.015$)
		&
		($0.050 \pm 0.031$) 
		&
		($0.048 \pm 0.002$)
		&
		($0.902 \pm 0.032$)
		\\  
		\noalign{\hrule height 1pt}		
%==========================================================	
	$\eta(2225)$ & 
		$\approx 0 $     &  
		$\approx  0 $     & 
		$0.026 \pm 0.020$      &      
		$0.022 \pm 0.017$      &  
		$0.953 \pm 0.037$ 
		&
		$\approx 0$
		&
		$0.047 \pm 0.037$
		\\ 
				&
		($\approx 0 $)
		&
		($\approx 0 $)
		&
		($0.029 \pm 0.018$)
		&
		($0.025 \pm 0.015$)
		&
		($0.946 \pm 0.034$) 
		&
		($\approx 0$)
		&
		($0.054 \pm 0.034$)
		\\  
		\noalign{\hrule height 1pt}		
	\end{tabular}\\[2pt]
	\label{T_eta_comps}
\end{table}

\begin{table}[!htbp]
	\centering
	\caption{Predicted components (Ave. $\pm \sigma$) of $\eta$'s in the octet-singlet basis.  The four columns respectively give the quark-antiquark octet, the four-quark octet, the quark-antiquark singlet and the four-quark singlet. 	For each component, the top number is determined over the high range of glueball condensate ($h_0=0.65-0.99$ GeV) and the bottom number in parentheses is obtained in the range of $h_0 = 0.75-0.825$ GeV favored by the SU(3) limit of this model.}
	\renewcommand{\tabcolsep}{0.4pc} % enlarge column spacing
	\renewcommand{\arraystretch}{1.5} % enlarge line spacing
	\begin{tabular}{c|c|c|c|c}
		\noalign{\hrule height 1pt}
		\noalign{\hrule height 1pt}	
		State & 
		$\phi_8$ &  
		$\phi'_8$ & 
		$\phi_0$ &
		$\phi'_0$
		\\ 
		\noalign{\hrule height 1pt}		
		\noalign{\hrule height 1pt}	
		%==========================================================
		$\eta(547)$        & 
		$0.879 \pm 0.002$ &  
		$ 0.092\pm 0.002$ & 
		$0.028 \pm 0.001$ &      
		$\approx 0$   
		\\
		&
		($0.879 \pm 0.002$ )
		&
		($ 0.092 \pm 0.002 $)
		&
		($ 0.028 \pm 0.001 $)
		&
		($\approx 0$)
		\\    
		\noalign{\hrule height 1pt}		
		%==========================================================		
		$\eta'(958)$  & 
		$0.003 \pm 0.001$     &  
		$0.113 \pm 0.003$     & 
		$0.827 \pm 0.003$      &      
		$0.053 \pm 0.003$        
		\\    
		&
		($0.003 \pm 0.001$)
		&
		($0.113 \pm 0.003$)
		&
		($0.828 \pm 0.003$)
		&
		($0.053 \pm 0.003$)
		\\    
		\noalign{\hrule height 1pt}		
		%==========================================================
		$\eta(1295)$  & 
		$0.118 \pm 0.002$     &  
		$0.785 \pm 0.005$     & 
		$0.096 \pm 0.003$      &      
		$\approx 0 $      
		\\    
		&
		($0.118 \pm 0.002$)
		&
		($0.785 \pm 0.005$)
		&
		($0.096 \pm 0.003$)
		&
		($\approx 0$)
		\\     
		\noalign{\hrule height 1pt}		
		%==========================================================		
		$\eta(1760)$  & 
		$\approx 0  $     &  
		$0.009 \pm 0.001$     & 
		$0.048 \pm 0.002$      &      
		$0.900 \pm 0.034$     
		\\
		&
		($\approx 0$)
		&
		($0.009 \pm 0.001$)
		&
		($0.048 \pm 0.002 $)
		&
		($0.893\pm 0.032$)
		\\
		\noalign{\hrule height 1pt}		                                                                                                                                                                                                               	%==========================================================		
		$\eta(2225)$ & 
		$\approx 0$     &  
		$0.001 \pm 0.001$     & 
		$\approx 0$      &      
		$0.046 \pm 0.036$
		\\    
		&
		($\approx 0$)
		&
		$(0.001 \pm 0.001)$
		&
		($\approx 0$)
		&
		($0.053 \pm 0.033$)\\
		\noalign{\hrule height 1pt}		
	\end{tabular}\\[2pt]
	\label{T_eta_comps_OS}
\end{table}

\begin{table}[!htbp]
	\centering
	\caption{Predicted components (Ave. $\pm \sigma$) of $f_0$'s  in the strange-nonstrange basis.  For each component, the top number is determined over the high range of glueball condensate ($h_0 = 0.65-0.99$ GeV) and the bottom number in parentheses is obtained in the range of $h_0 = 0.75-0.825$ GeV favored by the SU(3) limit of this model. 
	}
	\renewcommand{\tabcolsep}{0.4pc} % enlarge column spacing
	\renewcommand{\arraystretch}{1.5} % enlarge line spacing
	\begin{tabular}{c|c|c|c|c|c||c|c}
		\noalign{\hrule height 1pt}
		\noalign{\hrule height 1pt}	
		State & 
		$\frac{1}{\sqrt{2}} \left( u{\bar u} + d{\bar d}\right)$ &  
		$s{\bar s}$ & 
		%$\frac{1}{\sqrt{2}} \left({\bar d} {\bar s} d s + {\bar s} {\bar u} s u\right)$ & 
		%${\bar u} {\bar d} u d$
		$\frac{\left({\bar d} {\bar s} d s + {\bar s} {\bar u} s u\right)}{\sqrt{2}}$ 
		& 
		${\bar u} {\bar d} u d$
		&
		$h$
		&
		Total
		&
		Total
		\\
		&&&&&&
		two quark 
		&
		four quark
		\\ 
		\noalign{\hrule height 1pt}		
		\noalign{\hrule height 1pt}	
		%==========================================================
		$f_0(500)$ & 
		$0.463 \pm 0.005$ &  
		$0.055 \pm 0.001$ & 
		$0.295 \pm 0.003$ &      
		$0.187 \pm 0.002$ &  
		$\approx 0 $ 
		&
		$0.518 \pm 0.006$
		&
		$0.482 \pm 0.006$
		\\
						&
		($0.463 \pm 0.005$)
		&
		($0.055 \pm 0.001$)
		&
		($0.295 \pm 0.003$)
		&
		($0.186 \pm 0.002$)
		&
		($ \approx 0 $) 
		& ($ 0.518 \pm 0.006$)
		
		& ($ 0.482 \pm 0.006$)
		\\    
		\noalign{\hrule height 1pt}		
		%==========================================================		
		$f_0(980)$  & 
		$0.022 \pm 0.001$     &  
		$0.072 \pm 0.003$     & 
		$0.403 \pm 0.001$      &      
		$0.502 \pm 0.003$      &  
		$ \approx 0$   
		&
		$0.094 \pm 0.004$
		&
		$0.906 \pm 0.004$
		\\    
								&
		($0.022 \pm 0.001$)
		&
		($0.072 \pm 0.003$)
		&
		($0.402 \pm 0.001$)
		&
		($0.503 \pm 0.003$)
		&
		($ \approx 0$) 
		&
		($0.095 \pm 0.004$)  
		&
		($0.905 \pm 0.004 $)
		\\    
		\noalign{\hrule height 1pt}		
		%==========================================================
		$f_0(1370)$  & 
		$0.312 \pm 0.205$     &  
		$0.010 \pm 0.009$     & 
		$0.100 \pm 0.059$      &      
		$0.195 \pm 0.121$      &  
		$0.383_{-0.383}^{+0.392}$  
		&
		$0.322 \pm 0.212$
		& 
		$0.295 \pm 0.180$ 
		\\    
								&
		($0.416 \pm 0.081$)
		&
		($0.011 \pm 0.006$)
		&
		($0.135 \pm 0.020$)
		&
		($0.258 \pm 0.047$)
		&
		($0.181 \pm 0.152$) 
		&
		($0.427 \pm 0.086$)
		&
		($ 0.393 \pm 0.067$)
		\\     
		\noalign{\hrule height 1pt}		
		%==========================================================		
		$f_0(1500)$  & 
		$0.202_{-0.202}^{+206}  $     &  
		$0.212_{-0.212}^{+0.298} $     & 
		$0.079 \pm 0.057$      &      
		$0.116^{+0.121}_{-0.116}$      &  
		$0.391 \pm 0.347$ 
		&
		$0.414 \pm 0.246$
		&
		$0.195 \pm 0.158$
		\\
		&
		($0.098 \pm 0.081$)
		&
		($0.050 \pm 0.013$)
		&
		($0.015_{-0.015}^{+0.019} $)
		&
		($0.053 \pm 0.046$)
		&
		($0.784 \pm 0.139$) 
		&
		($0.148 \pm 0.074$)
		&
		($0.068 \pm 0.066$)
		\\    
		\noalign{\hrule height 1pt}		
		%==========================================================		
		$f_0(1710)$ & 
		$0.001 \pm 0.001$     &  
		$0.651 \pm 0.305$     & 
		$0.123\pm 0.053$      &      
		$\approx 0$    
		&  
		$0.225_{-0.225}^{+0.357} $ 
		&
		$0.652 \pm 0.304$
		&
	    $0.123 \pm 0.053$
		\\
		&
		($\approx 0$)
		&
		($0.812 \pm 0.019$)
		&
		($0.153 \pm 0.006$)
		&
		($\approx 0$)
		&
		($0.035\pm 0.018$)
		&
		($0.812 \pm 0.019$)
				& 
		($0.153 \pm 0.006$)
		\\    
		\noalign{\hrule height 1pt}		
	\end{tabular}\\[2pt]
	\label{T_f0_comps}
\end{table}

\begin{table}[!htbp]
	\centering
	\caption{Predicted octet and singlet components (Ave. $\pm \sigma$) of $f_0$'s in the octet-singlet basis.  The four columns respectively give the quark-antiquark octet, the four-quark octet, the quark-antiquark singlet and the four-quark singlet. 	For each component, the top number is determined over the high range of glueball condensate ($h_0=0.65-0.99$ GeV) and the bottom number in parentheses is obtained in the range of $h_0 = 0.75-0.825$ GeV favored by the SU(3) limit of this model.	}
	\renewcommand{\tabcolsep}{0.4pc} % enlarge column spacing
	\renewcommand{\arraystretch}{1.5} % enlarge line spacing
	\begin{tabular}{c|c|c|c|c}
		\noalign{\hrule height 1pt}
		\noalign{\hrule height 1pt}	
		State & 
		$S_8$ &  
		$S'_8$ & 
		$S_0$ &
		$S'_0$
		\\ 
		\noalign{\hrule height 1pt}		
		\noalign{\hrule height 1pt}	
		%==========================================================
		$f_0(500)$        & 
		$0.040 \pm 0.001$ &  
		$\approx 0.002 $ & 
		$0.477 \pm 0.005$ &      
		$0.481 \pm 0.006$   
		\\
		&
		($0.040 \pm 0.001$)
		&
		($ \approx 0.002$)
		&
		($0.478 \pm 0.005$)
		&
		($0.480 \pm 0.006$)
		\\    
		\noalign{\hrule height 1pt}		
		%==========================================================		
		$f_0(980)$  & 
		$0.093 \pm 0.004$     &  
		$0.894 \pm 0.004$     & 
		$\approx 0.001 $      &      
		$\approx 0.012 $        
		\\    
		&
		($0.094 \pm 0.004$)
		&
		($0.893 \pm 0.004$)
		&
		($\approx 0.001$)
		&
		($\approx 0.012$)
		\\    
		\noalign{\hrule height 1pt}		
		%==========================================================
		$f_0(1370)$  & 
		$0.158 \pm 0.117$     &  
		$0.032 \pm 0.021$     & 
		$0.164 \pm 0.096$      &      
		$0.263 \pm 0.159$      
		\\    
		&
		($0.207 \pm 0.053$)
		&
		($0.041 \pm 0.009$)
		&
		($0.220 \pm 0.033$)
		&
		($0.351 \pm 0.058$)
		\\     
		\noalign{\hrule height 1pt}		
		%==========================================================		
		$f_0(1500)$  & 
		$0.280 \pm 0.157  $     &  
		$0.035 \pm 0.017 $     & 
		$0.134 \pm 0.098$      &      
		$0.159 \pm 0.146$     
		\\
		&
		($0.124 \pm 0.042$)
		&
		($0.016 \pm 0.008$)
		&
		($0.024^{+0.032}_{-0.024}$)
		&
		($0.052^{+0.058}_{-0.052}$)
		\\    
		\noalign{\hrule height 1pt}		
		%==========================================================		
		$f_0(1710)$ & 
		$0.427 \pm 0.208$     &  
		$0.038 \pm 0.019$     & 
		$0.224\pm 0.097$      &      
		$0.085 \pm 0.035$
				\\    
				&
				($0.535 \pm 0.018$)
				&
				($0.048 \pm 0.002$)
				&
				($0.277 \pm 0.003$)
				&
				($0.105 \pm 0.005$)
				\\
		\noalign{\hrule height 1pt}		
	\end{tabular}\\[2pt]
	\label{T_f0_comps_OS}
\end{table}

\clearpage

\section{Summary and Discussion}

In this work we presented a comprehensive treatment of the generalized linear sigma model with glueballs in the isospin invariant limit.    The connection to pseudoscalar and scalar glueballs were made by inclusion  of effective terms in the Lagrangian that exactly model the U(1)$_{\rm A}$ axial anomaly and the trace anomaly, respectively.  One important aspect of modeling the axial anomaly was the need for inclusion of two peseudoscalar glueballs, a physical one together with a second one with an unphysical mass which gets integrated out and  results in an effective instanton-type term that supplies the additional mass to the etas.  The essential and intricate interplay between these two pseudoscalar glueballs allows a description of all eta masses in a remarkable agreement with experiment. 

In the leading order of this framework which corresponds to including effective terms in the Lagrangian with eight or fewer underlying quark and antiquark lines, together with the simplifying assumption about the modeling of trace anomaly in terms of  the scalar glueball field only, we performed an extensive numerical analysis to probe the 20 free parameters of the model.   It was shown that the magnitude of the glueball condensate plays a key role in characterization of the isosinglet pseudoscalar mesons $\eta(1760)$ and $\eta(2225)$, and the scalar mesons above 1 GeV.   While a broad range of $h_0 = 0.4-0.99$ Gev can give a good  description of the eta and $f_0$ mass spectra,  the substructure of these states depend on the specific range of $h_0$.  We showed that several factors favor the high range of $h_0 = 0.65-0.99$ GeV. After filtering out the low range from the simulations, we worked out in  detail the  model parameters, mass spectra, and the substructures of the isosinglet scalar and pseudoscalar  (the substructures of the isodoublets and isotriplet states were also presented in this work, but they are independent of the $h_0$).   The simulations clearly support identification of $\eta(2225)$ [or $X(2370)$ as pointed out after Eq. (\ref{E_eta_exp})] as a state with dominant glue component.   However, the exact identification of  which scalar meson above 1 GeV is dominantly made of glue still sharply depends on $h_0$,  even within its  high range (see Fig. \ref{F_f0_glue_vs_h0}), and therefore requires further investigation.   We see in Fig. \ref{F_f0_glue_vs_h0} that the  $f_0(1370)$, $f_0(1500)$ and $f_0(1710)$ can each (or a combination of them) be the dominant glue holder(s), and the question of which of these state(s) wins the match, in this framework solely hinges upon the exact value of the glueball condensate $h_0$.  Unlike the case of pseudoscalars which, within the high range of $h_0$, there remains no major uncertainty (see Fig. \ref{F_eta_glue_vs_h0}) that the main glue holder is the eta above 2 GeV [$\eta(2225)$ or $X(2370)$],  the three scalar states $f_0(1370)$, $f_0(1500)$ and $f_0(1710)$ can all have noticeable glue content.   The SU(3) flavor limit of this model favors $h_0=0.75-0.825$, which would imply that $f_0(1500)$ can be the dominant glue holder.  Nevertheless, more study is needed before the convoluted case of scalar mesons and their mixing with glueballs is fully understood.    The generalized linear sigma model is one possible framework that can further our understanding of the complex issue of glueball mixing with mesons.  The present work showed that the glueball condensate is the key quantity  that perhaps holds all (or most) of the secrets and we intend to chase it in future works.

There are several specific directions for future works.   Extending the simulations beyond the simplifying limit (\ref{E_LTA_def}) can improve the predictions of mixing.   Relaxing this condition will introduce   effective terms in the   trace anomaly that contain mixing between the quark-antiquark and four-quark chiral nonets, and  would be interesting to examine their impacts.    

\begin{figure}[!htb]
	\centering
	\includegraphics[scale=0.6]{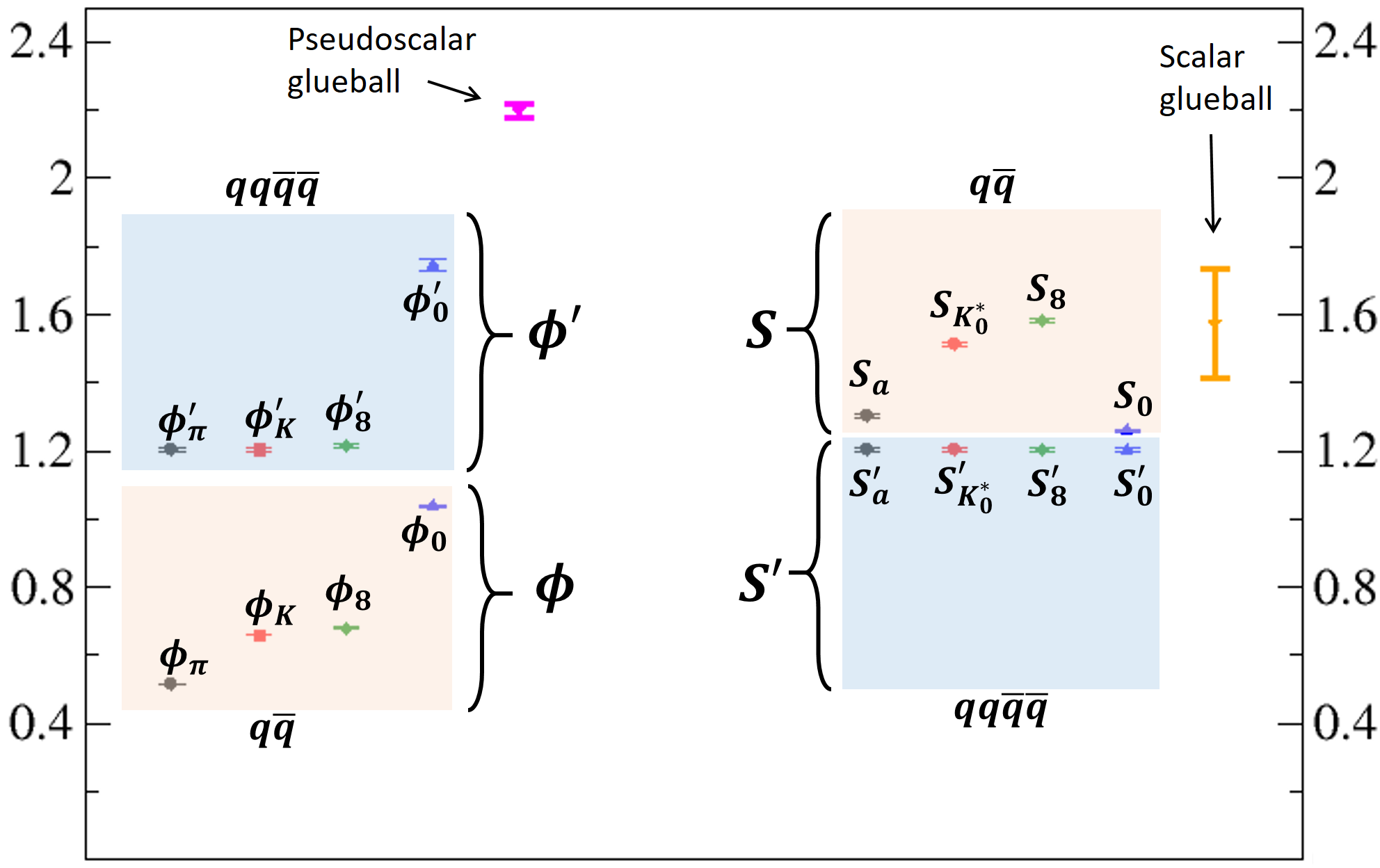}
	\caption{Masses of bare (unmixed) states (in GeV) in the scalar and  pseudoscalar nonets together with the masses of the scalar and  pseudoscalar glueballs.  The simulations show that the quark-antiquark pseudoscalar nonet is lighter than the four-quark pseudoscalar nonet (left), whereas in the scalar case this is reversed (right).  In each of the four nonets, the subscripts specify the unmixed states (for example, $\phi_\pi$ is the pure quark-antiquark pion in nonet $\phi$).  }
	\label{F_Bare_Masses}
\end{figure}

Extending the model beyond its leading order (which contains terms with only eight or fewer quark and antiquark lines) can reduce the uncertainties of  predictions.   However, this will also lead to the extension of the parameter space that poses new challenges.  Such extensions get closely related to  the spectroscopy of the bare (unmixed) states.       At the present order of the model,  the bare (unmixed) masses are given in  Table \ref{T_Bare_masses} and summarized in Fig. \ref{F_Bare_Masses}.   These results show several characteristics of the bare masses expected from schematic spectroscopy.   We see that the quark-antiquark  pseudoscalar nonet is seen to be lighter than the four-quark nonet of these states. This in contrast to the reversed case of scalars where the four-quark underlying bare nonet is lighter than the quark-antiquark one.    When only effective terms with eight or fewer quark  and antiquark lines are considered, the mass term for $M'$ chiral nonet is only limited to $u_3$ term in (\ref{E_f_terms}), and that results in bare four-quark states being degenerate in mass [with the exception of $\phi'_0$ which also receives mass through the axial anomaly breaking term (\ref{E_fA_fianl})].   To further refine this underlying spectroscopy which holds  useful information about quark model and  low-energy QCD dynamics, it is necessary to introduce more mass terms for $M'$ chiral nonet, and that in turn means we need to go beyond the leading order.

There are several additional isosinglet scalars, $f_0(1770)$, $f_0(2020)$, $f_0(2200)$, $f_0(2330)$ and $f_0(2470)$ listed in PDG \cite{24_PDG}, with different degrees of confidence and perhaps more experimental data is needed before they are fully characterized.  Nevertheless, it is interesting to examine whether  additional scalar glueballs in the framework of the generalized linear sigma model can describe some of their properties.

The property of the pure scalar glueball is another interesting topic for future studies.  In the present framework, the scalar glueball is introduced through its various interaction terms with matter fields, and as a result, its mass  generation is not only through its own condensate [$u_5$ in (\ref{E_f_terms})], it is also  through quark-antiquark and four-quark condensates (\ref{E_S_Sp_VEVs}), together with  the interactions of scalar glueball with the two chiral nonets [see matrix element 55 in (\ref{E_X02_LTA})]. In other words, the present framework is primarily designed to study mesons and their mixing patterns  with glueballs, and not inherently meant for studies  of glueballs isolated from mesons.  Therefore, the mass of the pure scalar glueball is not directly provided as can be seen in $\left(X_0^2\right)_{55}$ of (\ref{E_X02_LTA}).  However, one can make indirect inferences.  For example, as Fig. \ref{F_f0_glue_vs_h0} shows,  as the glueball condensate approaches 1.0 GeV, the fifth isosinglet state approaches a pure glueball, and since in this model the mass of scalar glueball is around 2 $h_0$ (see \cite{19_Fariborz_PLB790}),  it is possible that {\it pure} scalar glueball has a mass around 2.0 GeV (or slightly higher).   This indirect observation seems consistent with a lattice QCD study in \cite{24_Morningstar_PoS004}.  It would be interesting to examine the characteristics of the pure scalar glueball further.

A more detailed analysis of the partial decay widths of isosinglet states including the effects of the final-state interactions of the decay products is another important future plan which can bring further refined  insights into the mixing patterns of the glueballs and mesons.

\begin{table}[!htbp]
	\centering
	\caption{Predicted bare masses (Ave. $\pm \sigma$) in GeV. 
The top number is determined over the high range of glueball condensate ($h_0 = 0.65-0.99$ GeV) and the bottom number in parentheses is obtained in the range of $h_0 = 0.75-0.825$ GeV favored by the SU(3) limit of this model.		
	}
	\renewcommand{\tabcolsep}{0.4pc} % enlarge column spacing
	\renewcommand{\arraystretch}{1.5} % enlarge line spacing
	\begin{tabular}{c|c||c|c}
		\noalign{\hrule height 1pt}
		\noalign{\hrule height 1pt}	
		Bare state & Mass &  Bare state & Mass  \\
		\noalign{\hrule height 1pt}		
		\noalign{\hrule height 1pt}	
		$\phi_\pi$ &  $ \approx 0.512  $  & 	$S_a $ & $1.301 \pm 0.004$\\  
		      &	($\approx 0.512$)    &         & ($1.301 \pm 0.004$)\\
%----------------------------------------------------------------------------
		\noalign{\hrule height 1pt}		
		$\phi'_\pi$  & $1.203 \pm 0.005 $ & $S'_a$ & $1.203 \pm 0.005$ \\
	        	&($1.203\pm 0.005$)  &         & ($1.203 \pm 0.005$)\\  
%----------------------------------------------------------------------------
		\noalign{\hrule height 1pt}		
		$\phi_K$     & $0.660 \pm 0.002$  & $S_{K_0^*}$ & $1.511\pm 0.005$\\  
		        & ($0.660\pm 0.002$) &         & ($1.510 \pm 0.005$) \\
%----------------------------------------------------------------------------	   
  		\noalign{\hrule height 1pt}		
		$\phi'_K$    & $1.203 \pm 0.005$  &$S'_{K_0^*}$& $1.203 \pm 0.005$\\
	        	& ($1.203 \pm 0.005$)&   		& ($1.203 \pm 0.005$)\\ 
%----------------------------------------------------------------------------      
     	\noalign{\hrule height 1pt}		
		$\phi_8$& $0.680 \pm 0.002$  & $S_8$    & $1.585 \pm 0.005$\\
	        	 &($0.680 \pm 0.002$) &          & ($1.584 \pm 0.005$)\\  
%---------------------------------------------------------------------------- 
        \noalign{\hrule height 1pt}		
		$\phi'_8$ & $1.213 \pm 0.005$  & $S'_8  $ & $1.203 \pm 0.005$\\ 
		          & ($1.214 \pm 0.005$)&          & ($1.203\pm 0.005$) \\ 
%----------------------------------------------------------------------------
		\noalign{\hrule height 1pt}		
		$\phi_0$  & $1.036 \pm 0.003$  &  $S_0$   & $1.259 \pm 0.005$\\  
		          & ($1.037 \pm 0.003$)&          & ($1.259 \pm 0.005$) \\
%----------------------------------------------------------------------------
		\noalign{\hrule height 1pt}		
		$\phi'_0$ & $1.745  \pm 0.019$ &  $S'_0 $ &$1.203 \pm 0.005$\\  
		          & ($1.748 \pm 0.017$)&          & ($1.203 \pm 0.005$) \\
%----------------------------------------------------------------------------     
		\noalign{\hrule height 1pt}		
		$g$       & $2.199 \pm 0.019$  & 	$h$   &  $1.575\pm 0.161$ \\
	           	  &($2.196 \pm 0.018$) &          &  ($1.581 \pm 0.042$)\\  
%----------------------------------------------------------------------------
		\noalign{\hrule height 1pt}
	\end{tabular}\\[2pt]
	\label{T_Bare_masses}
\end{table}

\section*{Acknowledgement}
The author gratefully acknowledges invaluable communications with R.M. Corless  on the application of Maple software in big data computing.   

\clearpage

\appendix

% A A A A A A A A  A A A A A A A A  A A A A A A A A  A A A A A A A A  A A A A A A A A 

\section{Parameter determination} \label{A_par_determ}
   
In this appendix we show how model parameters $\alpha_1, \beta_1, \alpha_3, \beta_3, A_1, A_3, u_2$ and the combined parameters
$u_1 h_0^2$, $u_3 h_0^2$ and $u_4 h_0$ can be determined from experimental inputs (\ref{E_inputs1}).   We keep the mass of $a_0(980)$ fixed to its central value of $0.980$ GeV, and  mass of $a_0(1450)$ to 1.474 GeV to remain numerically consistent with the prior analysis of this model without glueballs in \cite{09_Fariborz_PRD79} as well as with the study of this model with inclusion  of glueballs in the SU(3) flavor limit in \cite{19_Fariborz_IJMPA34,21_Fariborz_NPA1015}.   Using $M_\pi^2$ in (\ref{E_Mpi2_LTA}) and $X_a^2$ in (\ref{E_Xa2_LTA}), we can write down  relations:
\begin{eqnarray}
{\rm Tr} \left( M_\pi^2 \right) &=& 4 u_2 \alpha_1^2 + 4 u_4 h_0 \beta_3  + 2 u_1 h_0^2 + 2 u_3 h_0^2 = m_\pi^2 + m_{\pi'}^2
\nonumber \\
{\rm det} \left( M_\pi^2 \right) &=& 
8 \alpha_1^2 h_0^2 u_2 u_3 
-16 u_4^2h_0^2 \alpha_3^2 
+ 8 \beta_3 h_0^3 u_3 u_4 
+ 4 h_0^4 u_1 u_3
 = m_\pi^2 m_{\pi'}^2  
\nonumber \\
{\rm Tr} \left( X_a^2 \right) &=& 12 u_2 \alpha_1^2 - 4 u_4 h_0 \beta_3  + 2 u_1 h_0^2 + 2 u_3 h_0^2 = m_a^2 + m_{a'}^2
\nonumber \\
{\rm det} \left( X_a^2 \right) &=& 
24 \alpha_1^2 h_0^2 u_2 u_3 
-16 u_4^2h_0^2 \alpha_3^2 
- 8 \beta_3 h_0^3 u_3 u_4 
+ 4 h_0^4 u_1 u_3 = m_a^2 m_{a'}^2.
\label{E_Trdet_relations}
\end{eqnarray}  

Then note that:
\begin{eqnarray}
{\rm det} \left( X_a^2 \right) - {\rm det} \left( M_\pi^2 \right) &=& 
\left(8 \alpha_1^2 u_2 - 8 \beta_3 h_0 u_4\right) 
\left( 2 u_3 h_0^2\right)  
\nonumber\\
{\rm Tr} \left( X_a^2 \right) - {\rm Tr} \left( M_\pi^2 \right) &=& 
8 \alpha_1^2 u_2 - 8 \beta_3 h_0 u_4
\label{E_TrXM_xap}
\end{eqnarray}
which determines the combined quantity $u_3 h_0^2$  in terms of the experimental masses for isotriplets:
\begin{equation}
u_3 h_0^2  = 
\frac{1}{2}
\left( 
\frac{ {\rm det} \left( X_a^2   \right) - 
	   {\rm det} \left( M_\pi^2 \right)
     }
       { {\rm Tr} \left( X_a^2   \right) - 
 	   {\rm Tr} \left( M_\pi^2 \right)
     }
\right)
=
\frac{1}{2}
\left(
 	{  {m_a^2 m_{a'}^2  - m_\pi^2\, m_{\pi'}^2} \over
 		{m_a^2 +m_{a'}^2 - m_\pi^2 - m_{\pi'}^2} }
\right).
\label{E_u3h02}
 \end{equation}
At the same time,  it is easy to show:
\begin{equation}
\frac{1}{\left(2u_3 h_0^2\right)} {\rm Tr} \left( X_a^2   \right) - 
\frac{1}{\left(2u_3 h_0^2\right)^2} {\rm det} \left( X_a^2   \right) - 1
= \frac{4 u_4^2 \alpha_3^2 h_0^2 }{\left( u_3 h_0^2 \right)^2}
\label{E_za2}
\end{equation}
which would then allow determination of  the combined parameters $\alpha_3 u_4 h_0$ in terms of the combined parameters $u_3 h_0^2$ in (\ref{E_u3h02}) and the experimental masses:
\begin{equation}
\left(\alpha_3 u_4  h_0 \right)^2 =\frac{1}{64}\Biggl( (m_a^2 -
m_{a'}^2)^2  - \left[4 \d_2 - (m_a^2
+m_{a'}^2)\right]^2 \Biggr).
\label{E_alpha3u4h0}
\end{equation}

Then, the two combined quantities (\ref{E_u3h02}) and (\ref{E_alpha3u4h0}), together with the  third minimum equation in (\ref{E_MinEq_LTA}), determine the ratio
\begin{equation}
 	{\beta_1\over \alpha_1} = -{{2 \left(\alpha_3
 			u_4 h_0\right)} \over \d_2}.
 	\label{E_beta1_over_alpha1}
 \end{equation}
 
 Next, the $\pi-\pi'$ mixing angle is found
 from the diagonalization of $M_\pi^2$: 
 
 \begin{equation} 
 	{\rm cos}\, 2\theta_\pi =
 	{
 		{4\, \d_2 - m_\pi^2 - m_{\pi'}^2}
 		\over
 		{
 			\sqrt{
 				64\, \left(\alpha_3 u_4 h_0 \right)^2
 				+ 16\, \d_2^2
 				- 8\, \d_2 \, (m_\pi^2 + m_{\pi'}^2)
 				+ (m_\pi^2 + m_{\pi'}^2)^2
 			}
 		}
 	}.
 \label{E_cos_2theta_pi}
 \end{equation}
The axial current relates the pion decay constant to the mixing angle:
\begin{equation}
F_\pi = 2 \alpha_1 \cos \theta_\pi - 2 \beta_1 \sin \theta_\pi
\end{equation} 
which can then be used to compute $\alpha_1$ in terms of the known quantities determined above:
\begin{equation} 
 	\alpha_1 =
 	{1\over 2}\,
 	{
 		{F_\pi}
 		\over
 		{ {\rm cos}\, \theta_\pi -
 			\left({\beta_1\over \alpha_1}\right) \,
 			{\rm sin}\, \theta_\pi
 		}
 	},  
 \label{E_alpha1}
 \end{equation}
which in turn determines $\beta_1$ using  the relationship (\ref{E_beta1_over_alpha1}) above.

Next, we compute the combined quantity $\alpha_1 u_4 h_0$ for an input of  $A_3/A_1$ (the ratio of 
 strange to non-strange quark mass).
 We use the first two minimum equations in (\ref{E_MinEq_LTA}), 
 \begin{eqnarray}
 	A_1 &=&
 	2\, \beta_3 \, \left( \alpha_1 u_4 h_0 \right)  + 2\, \beta_1 \, 
 	\left(\alpha_3 u_4 h_0 \right) + u_1 h_0^2 \alpha_1 +
 	2 \, \C4 \, \alpha_1^3,
 	\nonumber \\  
 	A_3 &=&
 	4 \, \beta_1 \, \left( \alpha_1 u_4 h_0 \right) +  u_1 h_0^2
\alpha_3 + 2 \, \C4 \, \alpha_3^3,
 	\label{E_A31}
 \end{eqnarray}
to construct  $A_3/A_1$ in this model.  We see that in addition to our target quantity $\alpha_1 u_4 h_0$, the model prediction for $A_3/A_1$ also depends on 
quantities $\alpha_3$, $\beta_3$,  $u_2$ and  $u_1 h_0^2$.  However, the first  three quantities can be in turn  expressed in terms of the target quantity $\alpha_1 u_4 h_0$.  Using $\alpha_3 u_4 h_0$ from (\ref{E_alpha3u4h0}) and $\alpha_1$ from (\ref{E_alpha1}), we can express $\alpha_3$ in terms of $\alpha_1 u_4 h_0$:
\begin{equation}
	\alpha_3 = \frac{\left(\alpha_3 u_4 h_0\right)}{\left(\alpha_1 u_4 h_0\right)} \, \alpha_1 
\label{E_alpha3}
\end{equation}
Using  now known quantity $u_3 h_0^2$ from (\ref{E_u3h02}), $\alpha_1$ from (\ref{E_alpha1}), and the fourth minimum equation in (\ref{E_MinEq_LTA}), we can express $\beta_3$ in terms of $\alpha_1 u_4 h_0$:
\begin{equation}
\beta_3 = - \frac{2\left(\alpha_1 u_4 h_0\right)}{\left(u_3 h_0^2\right)}\, \alpha_1.
\label{E_beta3}	
\end{equation}
Substituting $\beta_3$ from (\ref{E_beta3}) into the second equation in (\ref{E_TrXM_xap}),  and solving for $u_2$ we find
\begin{equation}
 	\C4 = {1 \over {8\, \alpha_1^2}}
\left[
m_a^2 + m_{a'}^2
- m_\pi^2 - m_{\pi'}^2
- 16 \,{ {\left( \alpha_1 u_4 h_0\right)^2} \over \d_2}
\right].   
\label{E_u2}
\end{equation}
The quantity $u_1 h_0^2$ can be expressed in terms of $u_2$ in (\ref{E_u2}), $\alpha_1$ in (\ref{E_alpha1}), and $u_3 h_0^2$ in (\ref{E_u3h02}).  Starting from 
\begin{eqnarray}
{\rm Tr} \left( X_a^2\right) + {\rm Tr} \left( M_\pi^2\right) = 16 u_2 \alpha_1^2 + 4 u_1 h_0^2 + 4 u_3 h_0^2  
\label{E_TrXM_u2u1u3}
\end{eqnarray} 
and solving for $u_1 h_0^2$ we find:
\begin{equation}
u_1 h_0^2  = \frac{1}{4} \left( m_a^2 + m_{a'}^2 + m_\pi^2 + m_{\pi'}^2 - 16 u_2 \alpha_1^2 - 4 u_3 h_0^2\right).
\label{E_u1h02}
\end{equation}
Therefore,  substituting (\ref{E_alpha3}), (\ref{E_beta3}),  (\ref{E_u2}) and (\ref{E_u1h02}) into $A_3/A_1$ constructed from (\ref{E_A31}) yields one equation in the combined unknown $h_0 u_4 \alpha_1$ which we can solve for, and subsequently substitute it back into (\ref{E_alpha3}), (\ref{E_beta3}),  (\ref{E_u2}) and (\ref{E_u1h02}) and determine $\alpha_3$, $\beta_3$, $u_2$ and $u_1 h_0^2$.   We can also substitute the necessary inputs into (\ref{E_A31}) and find $A_1$ and $A_3$.     Furthermore, we can use these combined values to also compute:
 \begin{equation}
 	u_4 h_0 = { {\left( u_4 h_0 \alpha_1\right)} \over
 		\alpha_1}.
\label{E_u4h0alpha3beta3}
 \end{equation}
 
In summary,  in this appendix we determined parameters $\alpha_1, \beta_1, \alpha_3, \beta_3, A_1, A_3, u_2$ and the combined parameters
$u_1 h_0^2$, $u_3 h_0^2$ and $u_4 h_0$ from experimental inputs (\ref{E_inputs1}).

% A A A A A A A A  A A A A A A A A  A A A A A A A A  A A A A A A A A  A A A A A A A A 

% A A A A A A A A  A A A A A A A A  A A A A A A A A  A A A A A A A A  A A A A A A A A 

\section{Decay widths} \label{A_Decay_Formulas}

\subsection{The interaction Lagrangian and the fields}

In this appendix we give the notation and formulas for the two-body decay widths of isosinglet scalars into two pseudoscalars.   The relevant part of the Lagrangian can be written in the form

\def\d_2{\left(u_3 h_0^2\right)}

\begin{equation}
-{\cal L} = 
{ {1} \over \sqrt{2} } 
\sum_{R,A,B}
\gamma{_{_{_{F_{_R} \Pi_{_A} \Pi_{_B} }}}} 
{\boldsymbol F}_{_R}{\boldsymbol \Pi}_{_A} \cdot {\boldsymbol \Pi}_{_B}
+
{ {1} \over \sqrt{2} } 
\sum_{R,A,B}
\gamma{_{_{_{F_{_R} K_{_A} K_{_B} }}}} 
{\boldsymbol F}_{_R}{\boldsymbol {\bar K}}_{_A} {\boldsymbol K}_{_B}
+
\sum_{R,S,T}
\gamma{_{_{_{F_{_R} \eta_{_S} \eta_{_T} }}}} 
{\boldsymbol F}_{_R}{\boldsymbol \eta}_{_S} {\boldsymbol \eta}_{_T}
+ \cdots
\label{E_L_Decay}
\end{equation}
where fields $\boldsymbol F$ and $\boldsymbol \eta$  respectively contain five isoninglet scalars and five isosinglet pseudoscalars.   Each of these two fields has five components labeled by indices $R, S, T = 1\cdots 5$ and each component represents a physical state,  as further shown below.    The indices on fields  $\boldsymbol \Pi$   and $\boldsymbol K$ are different, and label the fields themselves and not their components.   There are two distinct  $\boldsymbol \Pi$ fields, each labeled by $A,B = 1, 2$, as well as two  distinct  $\boldsymbol K$ fields, labeled by $A,B = 1, 2$.   Each $\boldsymbol \Pi$ field has three components (representing an isotriplet pseudoscalar) and each $\boldsymbol K$ field has two components (representing an isodoublet pseudoscalar).  This Lagrangian can  be rewritten in terms of the physical (mixed)  isotriplet and isodoublet states.

Specifically, field  $\boldsymbol F$ contains the five isosinglet scalars, and  is related to the bare (unmixed) states $\boldsymbol F_0$ through rotation matrix $L_0$: 
\begin{equation}
	\boldsymbol{F} =  
	\left[
\begin{array}{c}
	f1\\
	f_2\\
	f_3\\
	f_4\\
	f_5
\end{array}
\right] =
	\left[
	\begin{array}{c}
		f_0(500)\\
		f_0(980)\\
		f_0(1370)\\
		f_0(1500)\\
		f_0(1710)
	\end{array}
	\right]
	= L_0^{-1} 		\boldsymbol{F_0}   
= L_0^{-1}
\left[
\begin{array}{c}
	f_a\\
	f_b\\
	f_c\\
	f_d\\
	h
\end{array}
\right]
= L_0^{-1}
\left[
\begin{array}{c}
	{1\over \sqrt{2}} \left(S_1^1 + S_2^2 \right) \\
	S_3^3 \\
	{1\over \sqrt{2}} \left({S'}_1^1 + {S'}_2^2\right)\\
	{S'}_3^3\\
	h
\end{array}
\right].    
\label{E_Fvec_def}
\end{equation}
 
The quark content of the unmixed states are:
\begin{eqnarray}
	f_a&=&\frac{S^1_1+S^2_2}{\sqrt{2}} \hskip .7cm
	\propto \hskip .5cm n{\bar n},
	\nonumber  \\
	f_b&=&S^3_3 \hskip 1.6 cm \propto \hskip .5cm s{\bar s},
	\nonumber    \\
	f_c&=&  \frac{S'^1_1+S'^2_2}{\sqrt{2}}
	\hskip .5 cm \propto \hskip .5cm ns{\bar n}{\bar s},
	\nonumber   \\
	f_d&=& S'^3_3
	\hskip 1.5 cm \propto \hskip .5cm nn{\bar n}{\bar n}.
	\label{E_f_basis}
\end{eqnarray}

Similarly,  the field $\boldsymbol \eta$ contains the physical isosinglet pseudoscalars and is related to the unmixed states $\boldsymbol \eta_0$ by rotation matrix $R_0$:
\begin{equation}
	\boldsymbol{\eta} = 
		\left[
\begin{array}{c}
	\eta_1\\
	\eta_2\\
	\eta_3\\
	\eta_4\\
	\eta_5
\end{array}
\right]
=
	\left[
	\begin{array}{c}
		\eta(547)\\
		\eta'(958)\\
		\eta(1295)\\
		\eta(1760)\\
		\eta(2225)
	\end{array}
	\right] = R_0^{-1} 		\boldsymbol{\eta_0} 
	= R_0^{-1}
		\left[
\begin{array}{c}
	\eta_a\\
	\eta_b\\
	\eta_c\\
	\eta_d\\
	g
\end{array}
\right]
= 	R_0^{-1}
	\left[
	\begin{array}{c}	
		{1\over \sqrt{2}} \left(\phi_1^1 + \phi_2^2\right)\\
		\phi_3^3 \\
		{1\over \sqrt{2}} \left({\phi'}_1^1 + {\phi'}_2^2 \right)\\
		{\phi'}_3^3\\
		g
	\end{array}
	\right].  
\label{E_Etavec_Eta0vec_def}  
\end{equation}
The quark composition of the unmixed states are:
\begin{eqnarray}
	\eta_a&=&\frac{\phi^1_1+\phi^2_2}{\sqrt{2}} \hskip .7cm
	\propto \hskip .5cm n{\bar n},
	\nonumber  \\
	\eta_b&=&\phi^3_3 \hskip 1.6 cm \propto \hskip .5cm s{\bar s},
	\nonumber    \\
	\eta_c&=&  \frac{\phi'^1_1+\phi'^2_2}{\sqrt{2}}
	\hskip .5 cm \propto \hskip .5cm {\bar n}{\bar s}ns,
	\nonumber   \\
	\eta_d&=& \phi'^3_3
	\hskip 1.5 cm \propto \hskip .5cm {\bar n}{\bar n}nn.
	\label{E_eta_basis}
\end{eqnarray}

The two fields $\boldsymbol \Pi_1$ and $\boldsymbol \Pi_2$ respectively represent the isotriplet pions $\pi(137)$ and $\pi(1300)$: 
\begin{equation}
	{\boldsymbol \Pi}_1
	= 
	\left[
	\begin{array}{c}
		\pi_1\\
		\pi_2\\
		\pi_3
	\end{array}
	\right],
\end{equation}
with
\begin{eqnarray}
	\pi^{\pm}(137) 	& = & {1\over \sqrt{2}} \left(\pi_1 \mp i \pi_2 \right), \nonumber \\
	\pi^0(137)	& = &  \pi_3.	
\end{eqnarray}
Similarly
\begin{equation}
	{\boldsymbol \Pi}_2
	= 
	\left[
	\begin{array}{c}
		\pi'_1\\
		\pi'_2\\
		\pi'_3
	\end{array}
	\right],
\end{equation}
with
\begin{eqnarray}
	\pi^{\pm}(1300) 	& = & {1\over \sqrt{2}} \left(\pi'_1 \mp i \pi'_2 \right), \nonumber \\
	\pi^0(1300)	& = &  \pi'_3.	
\end{eqnarray}
The physical states are related to the bare states by rotation matrix $R_\pi$:
\begin{equation}
	{\boldsymbol \Pi}^+ = 
	\left[
\begin{array}{c}
	\pi^+(137)\\
	\pi^+(1300)
\end{array}
\right]
=
R_\pi^{-1} {\boldsymbol \Pi}_0^+
=
R_\pi^{-1}
	\left[
\begin{array}{c}
	\phi_1^2\\
	{\phi'}_1^2
\end{array}
\right],
\end{equation}
where the unmixed states have quark contents:
\begin{eqnarray}
	\phi_1^2 \hskip .4cm
	&\propto& \hskip .4cm n{\bar n},
	\nonumber  \\
	{\phi'}_1^2 \hskip .4cm
    &\propto& \hskip .4cm {\bar n}{\bar s}sn.
	\label{E_pi_basis}
\end{eqnarray}

The fields $\boldsymbol K_1$ and $\boldsymbol K_2$ respectively represent $K(496)$ and $K(1460)$:
\begin{equation}
	{\boldsymbol K}_1
	= 
	\left[
	\begin{array}{c}
		K^+(496)\\
		K^0(496)
	\end{array}
	\right],
\hskip 2cm 
	{\boldsymbol K}_2
= 
\left[
\begin{array}{c}
	K^+(1460)\\
	K^0(1460)
\end{array}
\right].
\end{equation}
These physical states are related to the bare states by rotation matrix $R_K$:
\begin{equation}
{\boldsymbol K}^+ =
	\left[
	\begin{array}{c}
		K^+(496)\\
		K^+(1460)
	\end{array}
	\right]
	=
	R_K^{-1}
{\boldsymbol K}_0^+ =
	R_K^{-1}
	\left[
	\begin{array}{c}
		\phi_1^3\\
		{\phi'}_1^3
	\end{array}
	\right],
\end{equation}
and the quark content of these unmixed states are:
\begin{eqnarray}
	\phi_1^3 \hskip .4cm
	&\propto& \hskip .4cm n{\bar s},
	\nonumber  \\
	{\phi'}_1^3 \hskip .4cm
	&\propto& \hskip .4cm {\bar n}{\bar s}nn.
	\label{E_K_basis}
\end{eqnarray}

\subsection{Decay width formulas}

The decay widths of isosinglet scalars to two-pseudoscalars are:

\begin{eqnarray}
	\Gamma\left[ {\boldsymbol F}_{_R} \rightarrow {\boldsymbol \Pi}_{_A} {\boldsymbol \Pi}_{_B} \right] & = & 
	{{3 \left(2 - \delta_{_{AB}} \right) \gamma^2_{_{_{F_{_R} \Pi_{_A} \Pi_{_B} }}} q_{_{_{\Pi_{_A} \Pi_{_B}} }}} \over {8 \pi\,  m^2_{_{F_{_R}} }} },
 		\nonumber \\
	\Gamma\left[ {\boldsymbol F}_{_R} \rightarrow {\boldsymbol K}_{_A} {\boldsymbol K}_{_B} \right] & = & 
{{ \left(2 - \delta_{_{AB}} \right) \gamma^2_{_{_{F_{_R} K_{_A} K_{_B} }}} q_{_{_{K_{_A}K_{_B}} }}} \over {8 \pi\,  m^2_{_{F_{_R}} }} },
\nonumber \\
	\Gamma\left[ {\boldsymbol F}_{_R} \rightarrow {\boldsymbol \eta}_{_S} {\boldsymbol \eta}_{_T} \right] & = & 
{{ \left(1 + \delta_{_{ST}} \right) \gamma^2_{_{_{F_{_R} \eta_{_S} \eta_{_T} }}} q_{_{_{\eta_{_S} \eta_{_T}} }}} \over {8 \pi\,  m^2_{_{F_{_R}} }} },
\label{E_Decay_formulas}
\end{eqnarray}
where $q$ is the center of mass momentum in a general two-body decay $R\rightarrow AB$ given by 
$q = \sqrt{
	[m_R^2 - (m_A + m_B)^2 ]
	[m_R^2 - (m_A - m_B)^2 ] 
}/(2 m_R)
$.
The physical coupling constants are calculated in terms of the bare (unmixed) coupling constants:
\begin{eqnarray}
	\gamma{_{_{_{F_{_R} \Pi^+_{_A} \Pi^-_{_B} }}}} & = & 
		{1\over \sqrt{2}} 
	\left\langle
	{ {\partial^3 V}
		\over 
		{
			\partial  {\boldsymbol F}_{_R} 
			\partial  \left( {\boldsymbol \Pi^+}\right)_A  
			\partial \left( {\boldsymbol \Pi^-}\right)_B }} 
	\right\rangle_0
	\nonumber \\
&=&	{1\over \sqrt{2}} \sum_{r,a,b} 
    \left\langle
	{ {\partial^3 V}
		\over 
	  {
	  	\partial \left( {\boldsymbol F_0}\right)_r 
	  	\partial  \left( {\boldsymbol \Pi_0^+}\right)_a  
		\partial \left( {\boldsymbol \Pi_0^-}\right)_b }} 
		  \right\rangle_0 
		  \left(L_0\right)_{rR}
          \left(R_\pi\right)_{aA}
            \left(R_\pi \right)_{bB}.
\label{E_gP_FR_PA_PB}
\end{eqnarray}

\begin{eqnarray}
	\gamma{_{_{_{F_{_R} K^+_{_A} K^-_{_B} }}}} & = & 
	\sqrt{2} 
	\left\langle
	{ {\partial^3 V}
		\over 
		{
			\partial  {\boldsymbol F}_{_R} 
			\partial  \left( {\boldsymbol K^+}\right)_A  
			\partial \left( {\boldsymbol K^-}\right)_B }} 
	\right\rangle_0
	\nonumber \\
	&=&	\sqrt{2} \sum_{r,a,b} 
	\left\langle
	{ {\partial^3 V}
		\over 
		{
			\partial \left( {\boldsymbol F_0}\right)_r 
			\partial  \left( {\boldsymbol K_0^+}\right)_a  
			\partial \left( {\boldsymbol K_0^-}\right)_b }} 
	\right\rangle_0 
	\left(L_0\right)_{rR}
	\left(R_K\right)_{aA}
	\left(R_K\right)_{bB}.
\label{E_gP_FR_KA_KB}
\end{eqnarray}

\begin{eqnarray}
	\gamma{_{_{_{F_{_R} \eta_{_S} \eta_{_T} }}}} & = & 
	\left\langle
	{ {\partial^3 V}
		\over 
		{
			\partial  {\boldsymbol F}_{_R} 
			\partial  {\boldsymbol \eta}_{_S}  
			\partial  {\boldsymbol \eta}_{_T} }} 
	\right\rangle_0
	\nonumber \\
	&=& \sum_{r,a,b} 
	\left\langle
	{ {\partial^3 V}
		\over 
		{
			\partial \left( {\boldsymbol F_0}\right)_r 
			\partial  \left( {\boldsymbol \eta_0}\right)_s  
			\partial \left( {\boldsymbol \eta_0}\right)_t }} 
	\right\rangle_0 
	\left(L_0\right)_{rR}
	\left(R_0\right)_{sS}
	\left(R_0\right)_{tT}.
\label{E_gP_FR_ES_ET}
\end{eqnarray}

\subsection{Bare couplings}

The bare coupling constants that are needed  in calculation of the  physical coupling constats that appear in decay formulas (\ref{E_gP_FR_PA_PB}),(\ref{E_gP_FR_KA_KB}) and (\ref{E_gP_FR_ES_ET}), are  directly calculated from the Lagrangian (\ref{E_L_Def_MMp2g}) and are given below.

\begin{eqnarray}
    \left\langle
{ {\partial^3 V}
	\over 
	{
		\partial \left( {\boldsymbol F_0}\right)_1
		\partial  \left( {\boldsymbol \Pi_0^+}\right)_1  
		\partial \left( {\boldsymbol \Pi_0^-}\right)_1 }} 
\right\rangle_0 
	&=& 4 \sqrt{2} \alpha_1 u_2
\end{eqnarray}

\begin{eqnarray}
	\left\langle
	{ {\partial^3 V}
		\over 
		{
			\partial \left( {\boldsymbol F_0}\right)_2
			\partial  \left( {\boldsymbol \Pi_0^+}\right)_1  
			\partial \left( {\boldsymbol \Pi_0^-}\right)_2 }} 
	\right\rangle_0 
	&=&
4 h_0 u_4 
\end{eqnarray}

\begin{eqnarray}
	\left\langle
	{ {\partial^3 V}
		\over 
		{
			\partial \left( {\boldsymbol F_0}\right)_4
			\partial  \left( {\boldsymbol \Pi_0^+}\right)_1  
			\partial \left( {\boldsymbol \Pi_0^-}\right)_1 }} 
	\right\rangle_0 
	&=&
4 h_0 u_4 
\end{eqnarray}

\begin{eqnarray}
	\left\langle
	{ {\partial^3 V}
		\over 
		{
			\partial \left( {\boldsymbol F_0}\right)_5
			\partial  \left( {\boldsymbol \Pi_0^+}\right)_1  
			\partial \left( {\boldsymbol \Pi_0^-}\right)_1 }} 
	\right\rangle_0 
	&=&
	4 h_0 u_1 + 4 \beta_3 u_4 
\end{eqnarray}

\begin{eqnarray}
	\left\langle
	{ {\partial^3 V}
		\over 
		{
			\partial \left( {\boldsymbol F_0}\right)_5
			\partial  \left( {\boldsymbol \Pi_0^+}\right)_1  
			\partial \left( {\boldsymbol \Pi_0^-}\right)_2 }} 
	\right\rangle_0 
	&=&
	4 \alpha_3 u_4 
\end{eqnarray}	

\begin{eqnarray}
	\left\langle
	{ {\partial^3 V}
		\over 
		{
			\partial \left( {\boldsymbol F_0}\right)_5
			\partial  \left( {\boldsymbol \Pi_0^+}\right)_2  
			\partial \left( {\boldsymbol \Pi_0^-}\right)_2 }} 
	\right\rangle_0 
	&=&
	4 h_0 u_3 
\end{eqnarray}	

%%%%%%%%%%%%%%%%%%%%%%%%%%%%%%%%%%%%%%%%%%%%%%%%%%%%%%%%%%%

\begin{eqnarray}
	\left\langle
{ {\partial^3 V}
	\over 
	{
		\partial \left( {\boldsymbol F_0}\right)_1
		\partial  \left( {\boldsymbol K_0^+}\right)_1  
		\partial \left( {\boldsymbol K_0^-}\right)_1 }} 
\right\rangle_0 
&=&
2 \sqrt{2} \left(2 \alpha_1 - \alpha_3 \right) u_2 
\end{eqnarray}

\begin{eqnarray}
	\left\langle
	{ {\partial^3 V}
		\over 
		{
			\partial \left( {\boldsymbol F_0}\right)_1
			\partial  \left( {\boldsymbol K_0^+}\right)_1  
			\partial \left( {\boldsymbol K_0^-}\right)_2}
    }		 
	\right\rangle_0 
	&=&
	2 \sqrt{2} h_0  u_4 
\end{eqnarray}

\begin{eqnarray}
	\left\langle
	{ {\partial^3 V}
		\over 
		{
			\partial \left( {\boldsymbol F_0}\right)_2
			\partial  \left( {\boldsymbol K_0^+}\right)_1  
			\partial \left( {\boldsymbol K_0^-}\right)_1}
	}		 
	\right\rangle_0 
	&=&
	4 \left(2 \alpha_3 - \alpha_1  \right) u_2 
	\end{eqnarray}

\begin{eqnarray}
	\left\langle
	{ {\partial^3 V}
		\over 
		{
			\partial \left( {\boldsymbol F_0}\right)_3
			\partial  \left( {\boldsymbol K_0^+}\right)_1  
			\partial \left( {\boldsymbol K_0^-}\right)_1}
	}		 
	\right\rangle_0 
	&=&
	2 \sqrt{2} h_0  u_4 
\end{eqnarray}

\begin{eqnarray}
	\left\langle
	{ {\partial^3 V}
		\over 
		{
			\partial \left( {\boldsymbol F_0}\right)_5
			\partial  \left( {\boldsymbol K_0^+}\right)_1  
			\partial \left( {\boldsymbol K_0^-}\right)_1}
	}		 
	\right\rangle_0 
	&=&
	4 \beta_1 u_4  + 4 h_0 u_1  
\end{eqnarray}

\begin{eqnarray}
	\left\langle
	{ {\partial^3 V}
		\over 
		{
			\partial \left( {\boldsymbol F_0}\right)_5
			\partial  \left( {\boldsymbol K_0^+}\right)_1  
			\partial \left( {\boldsymbol K_0^-}\right)_2}
	}		 
	\right\rangle_0 
	&=&
	4 \alpha_1 u_4 
\end{eqnarray}

\begin{eqnarray}
	\left\langle
	{ {\partial^3 V}
		\over 
		{
			\partial \left( {\boldsymbol F_0}\right)_5
			\partial  \left( {\boldsymbol K_0^+}\right)_2  
			\partial \left( {\boldsymbol K_0^-}\right)_2}
	}		 
	\right\rangle_0 
	&=&
	4  h_0  u_3 
\end{eqnarray}

%%%%%%%%%%%%%%%%%%%%%%%%%%%%%%%%%%%%%%%%%%%%%%%%%%%%%%%%%%%%
\begin{eqnarray}
		\left\langle
	{ {\partial^3 V}
		\over 
		{
			\partial \left( {\boldsymbol F_0}\right)_1 
			\partial  \left( {\boldsymbol \eta_0}\right)_1  
			\partial \left( {\boldsymbol \eta_0}\right)_1 }} 
	\right\rangle_0 
	&=&
\frac{32 \sqrt{2}}
	{
		\alpha_1^3 \left(2 \alpha_1  \beta_1  + \alpha_3  \beta_3  \right)^3
	} 
	\Bigg[
	\alpha_1^7 \beta_1^3 u_2 
	+\frac{3}{2} \alpha_1^6 \alpha_3  \,\beta_1^2 \beta_3 u_2	+\frac{3}{4} \alpha_1^5 \alpha_3^2 \beta_1  \,\beta_3^2 u_2
	\nonumber \\
	&&
	+\frac{1}{8}\alpha_1^4 \alpha_3^3 \beta_3^3 u_2
	+ c_3 \,\xi^{2} \beta_1^3 \left(\gamma_1  +1 \right)^2 \alpha_1^3
	+3 c_3 \,\xi^2 \beta_1^2 \beta_3  \alpha_3  \gamma_1  
	\left(\gamma_1  +1 \right) \alpha_1^2
	\nonumber \\
&&
	+\frac{5}{2} \beta_1  \left(\gamma_1  +\frac{1}{5}\right) \alpha_3^2 \beta_3^2 \xi^{2} \gamma_1  c_3 \alpha_1 
	+\frac{1}{2}\alpha_3^3 \beta_3^3 c_3 \,\gamma_1^2 \xi^2
	\Bigg]
\end{eqnarray}

\begin{eqnarray}
	\left\langle
	{ {\partial^3 V}
		\over 
		{
			\partial \left( {\boldsymbol F_0}\right)_1 
			\partial  \left( {\boldsymbol \eta_0}\right)_1  
			\partial \left( {\boldsymbol \eta_0}\right)_2 }} 
	\right\rangle_0 
	&=&
\frac{32 \xi^2 c_3}
{\alpha_3 \alpha_1^2 \left(2 \alpha_1  \beta_1  +\alpha_3 \beta_3  \right)^3 }
	\Bigg[\alpha_1^3\beta_1^3 \gamma_1  \left(1+ \gamma_1\right) 
	 + \frac{3}{2} \beta_3  \alpha_3  \,\alpha_1^2\beta_1^2 
		\left(\gamma_1^2 + \frac{1}{3} \gamma_1  +\frac{2}{3}\right) 
		\nonumber\\
		&&+\frac{3}{2} \alpha_1  \,\alpha_3^2 \beta_1  \,\beta_3^2 \gamma_1 +
		\frac{1}{4}\alpha_3^3 \beta_3^3 \gamma_1 \Bigg]
\end{eqnarray}

\begin{eqnarray}
	\left\langle
	{ {\partial^3 V}
		\over 
		{
			\partial \left( {\boldsymbol F_0}\right)_1 
			\partial  \left( {\boldsymbol \eta_0}\right)_1  
			\partial \left( {\boldsymbol \eta_0}\right)_3 }} 
	\right\rangle_0 
	&=&
\frac{16 \sqrt{2}\xi^{2} \beta_1 c_3 \left(\gamma_1 - 1\right)}
{\left(2 \alpha \mathit{1}  \beta \mathit{1}  +\alpha \mathit{3}  \beta \mathit{3}  \right)^{3}}
  \left[\alpha_1\beta_1  \left(\gamma_1   + 1 \right)  +\frac{3}{2} \alpha_3\beta_3 \left(\gamma_1   -  \frac{1}{3}\right)  \right] 
\end{eqnarray}

\begin{eqnarray}
	\left\langle
	{ {\partial^3 V}
		\over 
		{
			\partial \left( {\boldsymbol F_0}\right)_1 
			\partial  \left( {\boldsymbol \eta_0}\right)_1  
			\partial \left( {\boldsymbol \eta_0}\right)_4 }} 
	\right\rangle_0 
	&=&
- \frac{8}
     {
     	\left(2 \alpha_1 \beta_1  +\alpha_3  \beta_3  \right)^{3} \alpha_1^2
     } 
\Bigg[
4 \alpha_1^5 \beta_1^3 u_4 h_0 + 6 \alpha_1^4 \alpha_3  \,\beta_1^2 \beta_3  u_4 h_0 
+ 3 \alpha_1^3 \alpha_3^2 \beta_1  \beta_3^{2} u_4 h_0 
\nonumber \\
&&
- 4 \alpha_3\alpha_1^2  \left(-\frac{1}{8}u_4 h_0 \beta_3^3 \alpha_3^2 + c_3 \xi^2 \beta_1^2 \left(\gamma_1 - 1 \right) \left(\gamma_1 + 1\right)\right) 
\nonumber \\
&&
-6 c_3 \xi^2 \beta_1 \beta_3 \alpha_3^2  \gamma_1  \left(\gamma_1 - 1   \right) \alpha_1  -  c_3 \xi^2 \beta_3^2 \alpha_3^3 \gamma_1  \left(\gamma_1 -1  \right)
\Bigg]
\end{eqnarray}

\begin{eqnarray}
	\left\langle
	{ {\partial^3 V}
		\over 
		{
			\partial \left( {\boldsymbol F_0}\right)_1 
			\partial  \left( {\boldsymbol \eta_0}\right)_1  
			\partial \left( {\boldsymbol \eta_0}\right)_5 }} 
	\right\rangle_0 
	&=&
\frac{1}{6} \left(\xi -1 \right) h_0^3 
\left(\frac{\gamma_1}{\alpha_1^2} + \frac{2 \left(1-\gamma_1 \right) \beta_1^2}{\left(2 \alpha_1 \beta_1  + \alpha_3 \beta_3  \right)^2}\right)
\end{eqnarray}

\begin{eqnarray}
	\left\langle
{ {\partial^3 V}
	\over 
	{
		\partial \left( {\boldsymbol F_0}\right)_1 
		\partial  \left( {\boldsymbol \eta_0}\right)_2  
		\partial \left( {\boldsymbol \eta_0}\right)_2 }} 
\right\rangle_0 
&=&
\frac{16 \sqrt{2} c_3 \xi^2 \beta_1 \beta_3 \left(1 - \gamma_1 \right) \left(2 \alpha_1  \beta_1  \gamma_1 +\alpha_3  
\beta_3 \right) }
{\alpha_3 \left(2 \alpha_1 \beta_1  +\alpha_3 \beta_3 \right)^3 }
\end{eqnarray}

\begin{eqnarray}
	\left\langle
	{ {\partial^3 V}
		\over 
		{
			\partial \left( {\boldsymbol F_0}\right)_1 
			\partial  \left( {\boldsymbol \eta_0}\right)_2  
			\partial \left( {\boldsymbol \eta_0}\right)_3 }} 
	\right\rangle_0 
	&=&
- \frac{4}{{\left(2 \alpha_1 \beta_1 +\alpha_3 \beta_3 \right)^3}}
\Bigg[
\alpha_3^3  \beta_3^3 u_4 h_0 
+ 2 \alpha_3   \beta_3^2  \left(3 \alpha_1 \alpha_3 \beta_1  u_4 h_0 + c_3 \xi^2 \left(\gamma_1 -1   \right)\right) 
\nonumber\\
&&
+ 8  \alpha_1\beta_1 \beta_3 
\left(
\frac{3}{2} \alpha_1 \alpha_3 \beta_1 u_4 h_0
+ \xi^2 c_3 \left(\gamma_1   -\frac{1}{2}\right)  \left(\gamma_1 - 1  \right)\right)   
+ 8 \alpha_1^3 \beta_1^3 u_4 h_0
\Bigg]
\end{eqnarray}

\begin{eqnarray}
	\left\langle
	{ {\partial^3 V}
		\over 
		{
			\partial \left( {\boldsymbol F_0}\right)_1 
			\partial  \left( {\boldsymbol \eta_0}\right)_2  
			\partial \left( {\boldsymbol \eta_0}\right)_4 }} 
	\right\rangle_0 
	&=&
\frac{
	   8 \sqrt{2}\xi^{2} \beta_1 c_3 \left( \gamma_1 -1  \right)
       \left[
              2 \alpha_1 \beta_1 \gamma_1 - \alpha_3 \beta_3 \left(\gamma_1  -2\right) 
       \right]       
      }
      {\left(2 \alpha_1  \beta_1  + \alpha_3 \beta_3 \right)^3
      }
\end{eqnarray}

\begin{eqnarray}
	\left\langle
	{ {\partial^3 V}
		\over 
		{
			\partial \left( {\boldsymbol F_0}\right)_1 
			\partial  \left( {\boldsymbol \eta_0}\right)_2  
			\partial \left( {\boldsymbol \eta_0}\right)_5 }} 
	\right\rangle_0 
	&=&
	\frac{
		\sqrt{2} \beta_1\beta_3 h_0^3 \left(1 - \xi \right)  \left(\gamma_1 -1  \right)  
	     }
	     {
	      6 \left(2 \alpha_1  \beta_1  +\alpha_3 \beta_3  \right)^2
	     }
	\end{eqnarray}

% ----------------------------------------------------------

\begin{eqnarray}
	\left\langle
	{ {\partial^3 V}
		\over 
		{
			\partial \left( {\boldsymbol F_0}\right)_1 
			\partial  \left( {\boldsymbol \eta_0}\right)_3  
			\partial \left( {\boldsymbol \eta_0}\right)_3 }} 
	\right\rangle_0 
	&=&
	-\frac{
		16 \sqrt{2}\alpha_1 \alpha_3 \beta_3 c_3 \xi^2 \left(\gamma_1 - 1  \right)^2 
	}
		  {		 
		   \left( 2 \alpha_1 \beta_1 + \alpha_3 \beta_3 \right)^3 
	    }
\end{eqnarray}

\begin{eqnarray}
	\left\langle
	{ {\partial^3 V}
		\over 
		{
			\partial \left( {\boldsymbol F_0}\right)_1 
			\partial  \left( {\boldsymbol \eta_0}\right)_3  
			\partial \left( {\boldsymbol \eta_0}\right)_4 }} 
	\right\rangle_0 
	&=&
\frac{
	8 \alpha_3 c_3 \xi^{2} \left(\gamma_1 - 1  \right)^2  
	\left(2 \alpha_1 \beta_1  -\alpha_3 \beta_3  \right)
	}
	{\left(2 \alpha_1 \beta_1  +\alpha_3 \beta_3 \right)^3
	}
\end{eqnarray}

\begin{eqnarray}
	\left\langle
	{ {\partial^3 V}
		\over 
		{
			\partial \left( {\boldsymbol F_0}\right)_1 
			\partial  \left( {\boldsymbol \eta_0}\right)_3  
			\partial \left( {\boldsymbol \eta_0}\right)_5 }} 
	\right\rangle_0 
	&=&
	\frac{\alpha_3 \beta_3 h_0^3 \left(1 - \xi \right) 
		\left(\gamma_1  - 1  \right) 
		}
		{
		  6 \left(2 \alpha_1 \beta_1  +\alpha_3 \beta_3\right)^2
		}
\end{eqnarray}

%---------------------------------------------------------------

\begin{eqnarray}
	\left\langle
	{ {\partial^3 V}
		\over 
		{
			\partial \left( {\boldsymbol F_0}\right)_1 
			\partial  \left( {\boldsymbol \eta_0}\right)_4  
			\partial \left( {\boldsymbol \eta_0}\right)_4 }} 
	\right\rangle_0 
	&=&
\frac{
	16 \sqrt{2} \beta_1 \alpha_3^2 c_3 \xi^2 \left(\gamma_1 - 1  \right)^{2} 
     }
     {\left(2 \alpha_1 \beta_1  +\alpha_3  \beta_3  \right)^3
     }
\end{eqnarray}

\begin{eqnarray}
	\left\langle
	{ {\partial^3 V}
		\over 
		{
			\partial \left( {\boldsymbol F_0}\right)_1 
			\partial  \left( {\boldsymbol \eta_0}\right)_4  
			\partial \left( {\boldsymbol \eta_0}\right)_5 }} 
	\right\rangle_0 
	&=&
\frac{
	   \sqrt{2} \beta_1 \alpha_3 h_0^3 \left(\xi -1 \right) \left(\gamma_1 - 1 \right) 
	 }
	 {
	     6 \left(2 \alpha_1 \beta_1   + \alpha_3 \beta_3  \right)^2
	  }
\end{eqnarray}

%----------------------------------------------------------------

\begin{eqnarray}
	\left\langle
	{ {\partial^3 V}
		\over 
		{
			\partial \left( {\boldsymbol F_0}\right)_1 
			\partial  \left( {\boldsymbol \eta_0}\right)_5  
			\partial \left( {\boldsymbol \eta_0}\right)_5 }} 
	\right\rangle_0 
	&=& 0
\end{eqnarray}

% -------------------------------------------------------------

%----------------------------------------------------------------

\begin{eqnarray}
	\left\langle
	{ {\partial^3 V}
		\over 
		{
			\partial \left( {\boldsymbol F_0}\right)_2 
			\partial  \left( {\boldsymbol \eta_0}\right)_1  
			\partial \left( {\boldsymbol \eta_0}\right)_1 }} 
	\right\rangle_0 
	&=&
\frac{
	32 \beta_1 \beta_3  c_3 \left(1 - \gamma_1 \right) \xi^2 
	\left[
	  \alpha_1 \beta_1 \left( 1 + \gamma_1 \right)  
	  + \alpha_3 \beta_3  \gamma_1  
	\right] 
     }
	{
	\alpha_1 \left(2 \alpha_1 \beta_1 + \alpha_3\beta_3 \right)^3
	}
\end{eqnarray}

% -------------------------------------------------------------
\begin{eqnarray}
	\left\langle
	{ {\partial^3 V}
		\over 
		{
			\partial \left( {\boldsymbol F_0}\right)_2 
			\partial  \left( {\boldsymbol \eta_0}\right)_1  
			\partial \left( {\boldsymbol \eta_0}\right)_2 }} 
	\right\rangle_0 
	&=&
\frac{  8  \sqrt{2} c_3 \xi^2 }
      {
     	\alpha_1\alpha_3^2  
     	\left(2 \alpha_1 \beta_1 +  \alpha_3 \beta_3 \right)^3
     }
     \Bigg[ 
	         4 \alpha_1^3 \beta_1^3 \gamma_1 \left(1 + \gamma_1 \right) 
	         + 6  \alpha_1^2 \beta_1^2 \beta_3 \alpha_3 \gamma_1 
	           \left(1 + \gamma_1 \right)
\nonumber\\
&&
	         + 2 \alpha_1 \beta_1 \beta_3^2 \alpha_3^2   
	           \left(2 \gamma_1^2 + 1 \right) 
	         + \alpha_3^3 \beta_3^3 \gamma_1
	   \Bigg]
	 \end{eqnarray}

% -------------------------------------------------------------
\begin{eqnarray}
	\left\langle
	{ {\partial^3 V}
		\over 
		{
			\partial \left( {\boldsymbol F_0}\right)_2 
			\partial  \left( {\boldsymbol \eta_0}\right)_1  
			\partial \left( {\boldsymbol \eta_0}\right)_3 }} 
	\right\rangle_0 
	&=&
-\frac{4}
     {\left(2 \alpha_1  \beta_1  +\alpha_3  \beta_3 \right)^3}
\Bigg[
       \alpha_3^3 \beta_3^3  u_4 h_0 
       - 2 \alpha_3 \beta_3^2 \left(
                  -3 \alpha_1 \alpha_3 \beta_1 u_4 h_0
           \right.
\nonumber\\
&&
            \left.
                   + 2   c_3\xi^2 \gamma_1  
                     \left(\gamma_1 -1  \right)
           \right)
           -4  \alpha_1 \beta_1   \beta_3 \left(- 3 \alpha_1 \alpha_3 \beta_1 u_4 h_0
               \right.
\nonumber\\
&&      
                 \left.
                       +2   c_3 \xi^2 
                       \left(\gamma_1 - 1  \right)
                \right) 
          + 8 \alpha_1^3 \beta_1^3 u_4 h_0
\Bigg]
\end{eqnarray}

% -------------------------------------------------------------
\begin{eqnarray}
	\left\langle
	{ {\partial^3 V}
		\over 
		{
			\partial \left( {\boldsymbol F_0}\right)_2 
			\partial  \left( {\boldsymbol \eta_0}\right)_1  
			\partial \left( {\boldsymbol \eta_0}\right)_4 }} 
	\right\rangle_0 
	&=&
     \frac{8 \sqrt{2} \xi^2 \beta_1 c_3 
         \left(1-\gamma_1 \right)       
     	 \left[2 \alpha_1 \beta_1  
     	       \left(1 + \gamma_1  \right)  +  
     	        \alpha_3 \beta_3 \left(3 \gamma_1 -  1 \right)       
     	  \right]  
     	 }
     	 {
     	  \left(2 \alpha_1  \beta_1  +\alpha_3 \beta_3  \right)^3
     	 }
	\end{eqnarray}

% -------------------------------------------------------------
\begin{eqnarray}
	\left\langle
	{ {\partial^3 V}
		\over 
		{
			\partial \left( {\boldsymbol F_0}\right)_2 
			\partial  \left( {\boldsymbol \eta_0}\right)_1  
			\partial \left( {\boldsymbol \eta_0}\right)_5 }} 
	\right\rangle_0 
	&=&
	\frac{
		\sqrt{2}  \beta_1 \beta_3 h_0^3 \left(1 - \xi \right) \left(\gamma_1 -1 \right) 
	}
	{
		6 \left(2 \alpha_1 \beta_1  +\alpha_3 \beta_3  \right)^2
	}
	\end{eqnarray}

% -------------------------------------------------------------
\begin{eqnarray}
	\left\langle
	{ {\partial^3 V}
		\over 
		{
			\partial \left( {\boldsymbol F_0}\right)_2 
			\partial  \left( {\boldsymbol \eta_0}\right)_2  
			\partial \left( {\boldsymbol \eta_0}\right)_2 }} 
	\right\rangle_0 
	&=&
8 u_2 \alpha_3   +
\frac{16 c_3 \xi^2}
     {
      \alpha_3^2 \left(
                       2 \alpha_1 \beta_1 + \alpha_3 \beta_3 
                 \right)^3
     }
	  \left(
	          2 \alpha_1 \beta_1 \gamma_1  + \alpha_3  \beta_3
	   \right) 
	   \left(
	         4 \alpha_1^2 \beta_1^2 \gamma_1  
        \right.
\nonumber\\
&&
        \left.
	         + 4 \alpha_1 \alpha_3 \beta_1 \beta_3 \gamma_1   
	         + \alpha_3^2 \beta_3^2 
	   \right)
\end{eqnarray}

% -------------------------------------------------------------
\begin{eqnarray}
	\left\langle
	{ {\partial^3 V}
		\over 
		{
			\partial \left( {\boldsymbol F_0}\right)_2 
			\partial  \left( {\boldsymbol \eta_0}\right)_2  
			\partial \left( {\boldsymbol \eta_0}\right)_3 }} 
	\right\rangle_0 
	&=&
\frac{
	   16 \sqrt{2} \alpha_1  c_3 \xi^{2} 
	   \left(\gamma_1 - 1 \right)   
	   \left(
	         2 \alpha_1^2 \beta_1^2 \gamma_1  
	         +3 \alpha_1 \alpha_3 \beta_1 \beta_3  \gamma_1 
	         +\alpha_3^2 \beta_3^2
	  \right)
	 }
	 {
	   \alpha_3^2 
	   \left(
	         2 \alpha_1 \beta_1  +\alpha_3  \beta_3  
	   \right)^3 
	 }
\end{eqnarray}

% -------------------------------------------------------------
\begin{eqnarray}
	\left\langle
	{ {\partial^3 V}
		\over 
		{
			\partial \left( {\boldsymbol F_0}\right)_2 
			\partial  \left( {\boldsymbol \eta_0}\right)_2  
			\partial \left( {\boldsymbol \eta_0}\right)_4 }} 
	\right\rangle_0 
	&=&
	\frac{
		   8 \beta_3 c_3 \xi^2 \left(\gamma_1 - 1   \right) 
		   \left(
		          2 \alpha_1 \beta_1 
		          \left(
		                  2 \gamma_1  - 1
		          \right)  
		          +\alpha_3 \beta_3 
		   \right) 
		 }
		 {
		   \left(2 \alpha_1 \beta_1  +\alpha_3 \beta_3 
		   \right)^3
		 }
\end{eqnarray}

% -------------------------------------------------------------
\begin{eqnarray}
	\left\langle
	{ {\partial^3 V}
		\over 
		{
			\partial \left( {\boldsymbol F_0}\right)_2 
			\partial  \left( {\boldsymbol \eta_0}\right)_2  
			\partial \left( {\boldsymbol \eta_0}\right)_5 }} 
	\right\rangle_0 
	&=&
\frac{\mathrm{1}}{6} 
\left(\xi -1 \right) h_0^3 
\left(
   	  \frac{\gamma_1}{\alpha_3^2}
	  +\frac{
			   \beta_3^2 \left(1-\gamma_1 \right) 		    	   
			}
		    {
		      	\left(2 \alpha_1\beta_1 +\alpha_3 \beta_3 
		      	\right)^2
		    }
\right)
	\end{eqnarray}

% -------------------------------------------------------------
\begin{eqnarray}
	\left\langle
	{ {\partial^3 V}
		\over 
		{
			\partial \left( {\boldsymbol F_0}\right)_2 
			\partial  \left( {\boldsymbol \eta_0}\right)_3  
			\partial \left( {\boldsymbol \eta_0}\right)_3 }} 
	\right\rangle_0 
	&=&
\frac{
	  32 \alpha_1^2 \beta_3 c_3 \xi^2 \left(\gamma_1 -1 \right)^2 
	  }
	  {
	  	\left(2 \alpha_1 \beta_1  +\alpha_3 \beta_3 \right)^3
	  }
\end{eqnarray}
% -------------------------------------------------------------

\begin{eqnarray}
	\left\langle
	{ {\partial^3 V}
		\over 
		{
			\partial \left( {\boldsymbol F_0}\right)_2 
			\partial  \left( {\boldsymbol \eta_0}\right)_3  
			\partial \left( {\boldsymbol \eta_0}\right)_4}} 
	\right\rangle_0 
	&=&
\frac{
	   8\sqrt{2}  \alpha_1 \xi^{2} c_3 
	   \left(\gamma_1 - 1   \right)^2  
       \left(\alpha_3 \beta_3 - 2 \alpha_1 \beta_1\right)
      }
      {
        \left(
              2 \alpha_1  \beta_1  + \alpha_3 \beta_3  
       \right)^3
      }
\end{eqnarray}

% -------------------------------------------------------------

\begin{eqnarray}
	\left\langle
	{ {\partial^3 V}
		\over 
		{
			\partial \left( {\boldsymbol F_0}\right)_2 
			\partial  \left( {\boldsymbol \eta_0}\right)_3  
			\partial \left( {\boldsymbol \eta_0}\right)_5 }} 
	\right\rangle_0 
	&=&
\frac{
	   \sqrt{2} \alpha_1 \beta_3 h_0^3 \left(1 - \xi \right) 
	   \left(1 - \gamma_1 \right) 
	 }
	 {
	  6 \left(2 \alpha_1  \beta_1  +\alpha_3  \beta_3  \right)^2
	 }
\end{eqnarray}

% -------------------------------------------------------------
\begin{eqnarray}
	\left\langle
	{ {\partial^3 V}
		\over 
		{
			\partial \left( {\boldsymbol F_0}\right)_2 
			\partial  \left( {\boldsymbol \eta_0}\right)_4  
			\partial \left( {\boldsymbol \eta_0}\right)_4}} 
	\right\rangle_0 
	&=&
-\frac{
	    32\alpha_1 \alpha_3 \beta_1 c_3 \xi^2 
	    \left(\gamma_1 - 1 \right)^2 
	  }
	  {
	  	\left(2 \alpha_1  \beta_1  + \alpha_3  \beta_3  \right)^3
	  }
\end{eqnarray}

% -------------------------------------------------------------
\begin{eqnarray}
	\left\langle
	{ {\partial^3 V}
		\over 
		{
			\partial \left( {\boldsymbol F_0}\right)_2 
			\partial  \left( {\boldsymbol \eta_0}\right)_4  
			\partial \left( {\boldsymbol \eta_0}\right)_5 }} 
	\right\rangle_0 
	&=&
\frac{
	   \alpha_1 \beta_1 h_0^3 
	   \left(1 - \xi \right) 
	   \left(\gamma_1 - 1 \right) 
	 }
	 {3 \left(2 \alpha_1 \beta_1  + \alpha_3 \beta_3 
	 	\right)^2
	 }
\end{eqnarray}

% -------------------------------------------------------------

\begin{eqnarray}
	\left\langle
	{ {\partial^3 V}
		\over 
		{
			\partial \left( {\boldsymbol F_0}\right)_3
			\partial  \left( {\boldsymbol \eta_0}\right)_1  
			\partial \left( {\boldsymbol \eta_0}\right)_1 }} 
	\right\rangle_0 
	&=&
\frac{
	   16 \sqrt{2} \alpha_3 \beta_3 c_3 \xi^{2} 
	   \left[
	         \alpha_1 \beta_1 \left(1 + \gamma_1 \right)  
	         +\alpha_3 \beta_3 \gamma_1  
	   \right]
	   \left(\gamma_1 - 1 \right) 
	  }
	  {
	  	\alpha_1 
	  	\left(
	  	       2 \alpha_1 \beta_1 + \alpha_3 \beta_3  
	  	\right)^3
	  }
\end{eqnarray}

% -------------------------------------------------------------

\begin{eqnarray}
	\left\langle
	{ {\partial^3 V}
		\over 
		{
			\partial \left( {\boldsymbol F_0}\right)_3
			\partial  \left( {\boldsymbol \eta_0}\right)_1  
			\partial \left( {\boldsymbol \eta_0}\right)_2 }} 
	\right\rangle_0 
	&=&
-\frac{4}
     {\left(2 \alpha_1\beta_1  +\alpha_3 \beta_3  \right)^3}
\Bigg[
	 \alpha_3^3 \beta_3^3 u_4 h_0 
	+ 2 \alpha_3  \beta_3^2 
	      \left[
	            3 \alpha_1 \alpha_3 \beta_1 u_4 h_0
	       \right.
\nonumber \\ 
&&	           
           \left.       
	            +  \xi^2 c_3  \left(2 \gamma_1  - 1 \right) 
	              \left(\gamma_1 -1   \right)
	      \right] 
	       +4  \alpha_1 \beta_1  \beta_3 
	       \left[
	                3 \alpha_1 \alpha_3  \beta_1  u_4 h_0 
	         \right.
\nonumber\\
&&
             \left.
	                + c_3\xi^2 
	                \left(\gamma_1 - 1   \right)
	          \right]   
	       + 8 \alpha_1^3 \beta_1^3 u_4 h_0
\Bigg]
\end{eqnarray}

% -------------------------------------------------------------

\begin{eqnarray}
	\left\langle
	{ {\partial^3 V}
		\over 
		{
			\partial \left( {\boldsymbol F_0}\right)_3
			\partial  \left( {\boldsymbol \eta_0}\right)_1  
			\partial \left( {\boldsymbol \eta_0}\right)_3 }} 
	\right\rangle_0 
	&=&
\frac{
	   8\sqrt{2} \alpha_1 c_3  \xi^{2} 
	   \left(\gamma_1 - 1  \right)  
	   \left[
	         2 \alpha_1 \beta_1  \left(1 + \gamma_1 \right)
	         + \alpha_3 \beta_3 \left( 3\gamma_1 - 1 \right) 
	  \right]  
	  }
    {
	 \left(2 \alpha_1 \beta_1  + \alpha_3  \beta_3  \right)^{3}
	}
\end{eqnarray}

% -------------------------------------------------------------

\begin{eqnarray}
	\left\langle
	{ {\partial^3 V}
		\over 
		{
			\partial \left( {\boldsymbol F_0}\right)_3
			\partial  \left( {\boldsymbol \eta_0}\right)_1  
			\partial \left( {\boldsymbol \eta_0}\right)_4 }} 
	\right\rangle_0 
	&=&
\frac{
	  8 \alpha_3 c_3 \xi^{2} \left(\gamma_1 -1 \right)   
	  \left[
	        2 \alpha_1\beta_1  \left(1 + \gamma_1\right)   
	        + \alpha_3 \beta_3 \left(3 \gamma_1  -   1 \right)   
	  \right] 
	 }
	 {
	   \left(2 \alpha_1  \beta_1  + \alpha_3  \beta_3  
	   \right)^3
	 }
\end{eqnarray}

% -------------------------------------------------------------

\begin{eqnarray}
	\left\langle
	{ {\partial^3 V}
		\over 
		{
			\partial \left( {\boldsymbol F_0}\right)_3
			\partial  \left( {\boldsymbol \eta_0}\right)_1  
			\partial \left( {\boldsymbol \eta_0}\right)_5 }} 
	\right\rangle_0 
	&=&
\frac{
	   \alpha_3 \beta_3 h_0^3 \left(\xi  -1 \right)  
	   \left(\gamma_1 - 1  \right)  
	  }
	  {
	   6 \left(2 \alpha_1 \beta_1   + \alpha_3  \beta_3   \right)^2
	  }
\end{eqnarray}

% -------------------------------------------------------------

\begin{eqnarray}
	\left\langle
	{ {\partial^3 V}
		\over 
		{
			\partial \left( {\boldsymbol F_0}\right)_3
			\partial  \left( {\boldsymbol \eta_0}\right)_2  
			\partial \left( {\boldsymbol \eta_0}\right)_2 }} 
	\right\rangle_0 
	&=&
\frac{
	  16 \sqrt{2} \alpha_1 \beta_3 c_3 \xi^2 
	  \left(1 - \gamma_1 \right) 
	  \left(
	        2 \alpha_1  \beta_1  \gamma_1   + \alpha_3 \beta_3 
	  \right) 
	 }
	 {
	  \alpha_3 \left(2 \alpha_1 \beta_2  + \alpha_3 \beta_3 \right)^3  
	 }
\end{eqnarray}

% -------------------------------------------------------------

\begin{eqnarray}
	\left\langle
	{ {\partial^3 V}
		\over 
		{
			\partial \left( {\boldsymbol F_0}\right)_3
			\partial  \left( {\boldsymbol \eta_0}\right)_2  
			\partial \left( {\boldsymbol \eta_0}\right)_3 }} 
	\right\rangle_0 
	&=&
\frac{
	  16 \alpha_1^2 c_3 \xi^{2} 
	  \left(1 - \gamma_1 \right)
	  \left[
	        \alpha_3 \beta_3 \left(\gamma_1  -2\right)  
	        - 2 \alpha_1 \beta_1 \gamma_1 
	 \right]  
	 }
	 {
	   \alpha_3 \left(2 \alpha_1 \beta_1 +\alpha_3 \beta_3  \right)^3  
	 }
\end{eqnarray}

% -------------------------------------------------------------

\begin{eqnarray}
	\left\langle
	{ {\partial^3 V}
		\over 
		{
			\partial \left( {\boldsymbol F_0}\right)_3
			\partial  \left( {\boldsymbol \eta_0}\right)_2  
			\partial \left( {\boldsymbol \eta_0}\right)_4 }} 
	\right\rangle_0 
	&=&
\frac{
	  8\sqrt{2} \alpha_1 c_3 \xi^2  
	  \left(1  - \gamma_1 \right) 	
	  \left[ \alpha_3 \beta_3 \left(\gamma_1 -2 \right) 
	         - 2 \alpha_1 \beta_1 \gamma_1  
	  \right]   
	 }
	 {
	   \left(2 \alpha_1 \beta_1  + \alpha_3 \beta_3  \right)^3
	 }
\end{eqnarray}

% -------------------------------------------------------------

\begin{eqnarray}
	\left\langle
	{ {\partial^3 V}
		\over 
		{
			\partial \left( {\boldsymbol F_0}\right)_3
			\partial  \left( {\boldsymbol \eta_0}\right)_2  
			\partial \left( {\boldsymbol \eta_0}\right)_5 }} 
	\right\rangle_0 
	&=&
\frac{
	   \sqrt{2} \alpha_1 \beta_3 h_0^3 \left(1 - \xi \right) \left(\gamma_1 -1 \right) 
	  }
	  {
	  	6 \left(2 \alpha_1  \beta_1  + \alpha_3 \beta_3 \right)^2
	  }
\end{eqnarray}

% -------------------------------------------------------------

\begin{eqnarray}
	\left\langle
	{ {\partial^3 V}
		\over 
		{
			\partial \left( {\boldsymbol F_0}\right)_3
			\partial  \left( {\boldsymbol \eta_0}\right)_3 
			\partial \left( {\boldsymbol \eta_0}\right)_3}} 
	\right\rangle_0 
	&=&
\frac{
	   32 \sqrt{2} \alpha_1^3 c_3 \xi^{2} \left(\gamma_1 - 1\right)^2 
	 }
	 {
	   \left(2 \alpha_1 \beta_1   + \alpha_3  \beta_3 \right)^3
	 }
\end{eqnarray}

% -------------------------------------------------------------

\begin{eqnarray}
	\left\langle
	{ {\partial^3 V}
		\over 
		{
			\partial \left( {\boldsymbol F_0}\right)_3
			\partial  \left( {\boldsymbol \eta_0}\right)_3
			\partial \left( {\boldsymbol \eta_0}\right)_4 }} 
	\right\rangle_0 
	&=&
\frac{
	   32 \alpha_1^2 \alpha_3 c_3 \xi^2 
	   \left( \gamma_1 - 1 \right)^2  
	  }
	  {
	  	\left(2 \alpha_1 \beta_1  + \alpha_3  \beta_3  \right)^3
	  }
\end{eqnarray}

% -------------------------------------------------------------

\begin{eqnarray}
	\left\langle
	{ {\partial^3 V}
		\over 
		{
			\partial \left( {\boldsymbol F_0}\right)_3
			\partial  \left( {\boldsymbol \eta_0}\right)_3 
			\partial \left( {\boldsymbol \eta_0}\right)_5 }} 
	\right\rangle_0 
	&=&
\frac{
	   \alpha_1^2 h_0^3 \left(1 - \xi \right) 
	   \left(1 - \gamma_1 \right) 
	  }
	  {
	  	3 \left(2 \alpha_! \beta_1  + \alpha_3  \beta_3  \right)^2
	  }
\end{eqnarray}

% -------------------------------------------------------------

\begin{eqnarray}
	\left\langle
	{ {\partial^3 V}
		\over 
		{
			\partial \left( {\boldsymbol F_0}\right)_3
			\partial  \left( {\boldsymbol \eta_0}\right)_4 
			\partial \left( {\boldsymbol \eta_0}\right)_4 }} 
	\right\rangle_0 
	&=&
\frac{
	   16 \sqrt{2} \alpha_1 \alpha_3^2 c_3 \xi^2 
	   \left(\gamma_1 - 1 \right)^2 
	  }
	 {
	  	\left(2 \alpha_1 \beta_1 + \alpha_3 \beta_3 \right)^3
	 }
\end{eqnarray}

% -------------------------------------------------------------

\begin{eqnarray}
	\left\langle
	{ {\partial^3 V}
		\over 
		{
			\partial \left( {\boldsymbol F_0}\right)_3
			\partial  \left( {\boldsymbol \eta_0}\right)_4  
			\partial \left( {\boldsymbol \eta_0}\right)_5 }} 
	\right\rangle_0 
	&=&
\frac{
	  \sqrt{2}\alpha_1 \alpha_3 h_0^3 \left( 1  - \xi \right)  
	  \left(1 - \gamma_1 \right)
	 }
	 {
	 	6 \left(2 \alpha_1 \beta_1   + \alpha_3 \beta_3 \right)^2
	 }
\end{eqnarray}

% -------------------------------------------------------------

\begin{eqnarray}
	\left\langle
	{ {\partial^3 V}
		\over 
		{
			\partial \left( {\boldsymbol F_0}\right)_3
			\partial  \left( {\boldsymbol \eta_0}\right)_5  
			\partial \left( {\boldsymbol \eta_0}\right)_5 }} 
	\right\rangle_0 
	&=& 0
\end{eqnarray}

% -------------------------------------------------------------

% -------------------------------------------------------------

\begin{eqnarray}
	\left\langle
	{ {\partial^3 V}
		\over 
		{
			\partial \left( {\boldsymbol F_0}\right)_4
			\partial  \left( {\boldsymbol \eta_0}\right)_1  
			\partial \left( {\boldsymbol \eta_0}\right)_1 }} 
	\right\rangle_0 
	&=&
-\frac{4}
     {\alpha_1  \left(2 \alpha_1 \beta_1 + \alpha_3 \beta_3  \right)^3}
\Bigg[ 
      8 \alpha_1^4 \beta_1^3 u_4 h_0 
      + 12 \alpha_1^3 \alpha_3 \beta_1^2 \beta_3 u_4 h_0 
\nonumber\\
&&
      +6 \alpha_1^2 \alpha_3^2 \beta_1\beta_3^2 u_4 h_0 
      + \alpha_1 \alpha_3 
        \left[
               u_4 h_0 \beta_3^3 \alpha_3^2
               + 8 c_3 \xi^2 \beta_1^2 
                 \left(\gamma_1 - 1 \right) 
                 \left(\gamma_1 + 1 \right)
         \right]  
\nonumber\\
&&
      + 8 c_3 \xi^2 \beta_1  \beta_3 \alpha_3^2 \gamma_1  
      \left(\gamma_1 -1 \right)
\Bigg] 
\end{eqnarray}

% -------------------------------------------------------------

\begin{eqnarray}
	\left\langle
	{ {\partial^3 V}
		\over 
		{
			\partial \left( {\boldsymbol F_0}\right)_4
			\partial  \left( {\boldsymbol \eta_0}\right)_1  
			\partial \left( {\boldsymbol \eta_0}\right)_2 }} 
	\right\rangle_0 
	&=&
\frac{   
	  8 \sqrt{2} \beta_1 c_3 \xi^2 (\gamma_1  - 1)    
	  \left[ 
	        \alpha_3 \beta_3 
	        \left(2 \gamma_1   -   1   \right) 
	        +2 \alpha_1 \beta_1  
	  \right]   
	  } 
	  {
	  	\left(2 \alpha_1  \beta_1  + \alpha_3  \beta_3 \right)^3
	  }
\end{eqnarray}

% -------------------------------------------------------------

\begin{eqnarray}
	\left\langle
	{ {\partial^3 V}
		\over 
		{
			\partial \left( {\boldsymbol F_0}\right)_4
			\partial  \left( {\boldsymbol \eta_0}\right)_1  
			\partial \left( {\boldsymbol \eta_0}\right)_3 }} 
	\right\rangle_0 
	&=&
\frac{
	  16 \alpha_3 c_3 \xi^2 \left(\gamma_1 - 1 \right) 
	  \left(\alpha_3 \beta_3 \gamma_1  + 2 \alpha_1 \beta_1 \right)
	  }
	  {
	  	\left(2 \alpha_1 \beta_1  + \alpha_3  \beta_3 \right)^3
	  }
\end{eqnarray}

% -------------------------------------------------------------

\begin{eqnarray}
	\left\langle
	{ {\partial^3 V}
		\over 
		{
			\partial \left( {\boldsymbol F_0}\right)_4
			\partial  \left( {\boldsymbol \eta_0}\right)_1  
			\partial \left( {\boldsymbol \eta_0}\right)_4 }} 
	\right\rangle_0 
	&=&
\frac{
	   8 \sqrt{2} \alpha_3^2 c_3 \xi^2  \left(\gamma_1 - 1 \right) \left(\alpha_3  \beta_3  \gamma_1  + 2 \alpha_1 \beta_1 \right)
	 }
	 {
	   \alpha_1 \left(2 \alpha_1 \beta_1 + \alpha_3 \beta_3 \right)^3
	 }
\end{eqnarray}

% -------------------------------------------------------------

\begin{eqnarray}
	\left\langle
	{ {\partial^3 V}
		\over 
		{
			\partial \left( {\boldsymbol F_0}\right)_4
			\partial  \left( {\boldsymbol \eta_0}\right)_1  
			\partial \left( {\boldsymbol \eta_0}\right)_5 }} 
	\right\rangle_0 
	&=&
\frac{
	   \sqrt{2} \alpha_3 \beta_1 h_0^3 
	   \left(1 - \xi \right) \left(\gamma_1 - 1   \right)  
	  }
	  {
	  	6 \left(2 \alpha_1  \beta_1  +\alpha_3 \beta_3 \right)^2
	  }
\end{eqnarray}

% -------------------------------------------------------------

\begin{eqnarray}
	\left\langle
	{ {\partial^3 V}
		\over 
		{
			\partial \left( {\boldsymbol F_0}\right)_4
			\partial  \left( {\boldsymbol \eta_0}\right)_2  
			\partial \left( {\boldsymbol \eta_0}\right)_2 }} 
	\right\rangle_0 
	&=&
\frac{
	  32 \alpha_1 \beta_1 c_3 \xi^2 
	  \left(\gamma_1 - 1 \right)  
	  \left(2 \alpha_1 \beta_1 \gamma_1 +\alpha_3 \beta_3 \right) 
	  }
	  {
	    \alpha_3 \left(2 \alpha_1 \beta_1  +\alpha_3  \beta_3\right)^3  
	  }
\end{eqnarray}

% -------------------------------------------------------------

\begin{eqnarray}
	\left\langle
	{ {\partial^3 V}
		\over 
		{
			\partial \left( {\boldsymbol F_0}\right)_4
			\partial  \left( {\boldsymbol \eta_0}\right)_2  
			\partial \left( {\boldsymbol \eta_0}\right)_3}} 
	\right\rangle_0 
	&=&
\frac{
	  8 \sqrt{2} \alpha_1 c_3 \xi^{2} \left(\gamma_1 -1   \right) \left[2 \alpha_1 \beta_1 
	        \left(2 \gamma_1  - 1  \right) 
	        + \alpha_3 \beta_3
	  \right] 
	 }
	 {
	   \left(2 \alpha_1 \beta_1 + \alpha_3 \beta_3  \right)^3
	 }
\end{eqnarray}

% -------------------------------------------------------------

\begin{eqnarray}
	\left\langle
	{ {\partial^3 V}
		\over 
		{
			\partial \left( {\boldsymbol F_0}\right)_4
			\partial  \left( {\boldsymbol \eta_0}\right)_2  
			\partial \left( {\boldsymbol \eta_0}\right)_4 }} 
	\right\rangle_0 
	&=&
\frac{
	8 \alpha_3 c_3 \xi^{2} \left(\gamma_1 -1   \right) \left[2 \alpha_1 \beta_1 
	\left(2 \gamma_1  - 1  \right) 
	+ \alpha_3 \beta_3
	\right] 
}
{
	\left(2 \alpha_1 \beta_1 + \alpha_3 \beta_3  \right)^3
}
\end{eqnarray}

% -------------------------------------------------------------

\begin{eqnarray}
	\left\langle
	{ {\partial^3 V}
		\over 
		{
			\partial \left( {\boldsymbol F_0}\right)_4
			\partial  \left( {\boldsymbol \eta_0}\right)_2  
			\partial \left( {\boldsymbol \eta_0}\right)_5 }} 
	\right\rangle_0 
	&=&
\frac{
	   \alpha_1 \beta_1 h_0^3 \left(\xi - 1 \right) 
	    \left(\gamma_1 - 1 \right)  
	 }
	 {
	   3 \left(2 \alpha_1 \beta_!   + \alpha_3 \beta_3  \right)^2
	 }
\end{eqnarray}

% -------------------------------------------------------------

\begin{eqnarray}
	\left\langle
	{ {\partial^3 V}
		\over 
		{
			\partial \left( {\boldsymbol F_0}\right)_4
			\partial  \left( {\boldsymbol \eta_0}\right)_3  
			\partial \left( {\boldsymbol \eta_0}\right)_3 }} 
	\right\rangle_0 
	&=&
\frac{
	  32\alpha_1^2 \alpha_3 c_3 \xi^2 
	  \left(\gamma_1 - 1 \right)^2  
	 }
	 {
	   \left(2 \alpha_1 \beta_1  +  \alpha_3  \beta_3 \right)^3
	 }	  
\end{eqnarray}

% -------------------------------------------------------------

\begin{eqnarray}
	\left\langle
	{ {\partial^3 V}
		\over 
		{
			\partial \left( {\boldsymbol F_0}\right)_4
			\partial  \left( {\boldsymbol \eta_0}\right)_3  
			\partial \left( {\boldsymbol \eta_0}\right)_4 }} 
	\right\rangle_0 
	&=&
\frac{
	  16 \sqrt{2} \alpha_1 \alpha_3^2 c_3 \xi^2 
	  \left(\gamma_1 - 1   \right)^2 
	  }
	  {
	  	\left(2 \alpha_1  \beta_1  + \alpha_3  \beta_3  \right)^3
	 }
\end{eqnarray}

% -------------------------------------------------------------

\begin{eqnarray}
	\left\langle
	{ {\partial^3 V}
		\over 
		{
			\partial \left( {\boldsymbol F_0}\right)_4
			\partial  \left( {\boldsymbol \eta_0}\right)_3  
			\partial \left( {\boldsymbol \eta_0}\right)_5 }} 
	\right\rangle_0 
	&=&
\frac{\sqrt{2} \alpha_1 \alpha_3 h_0^3
	  \left(\xi - 1 \right) \left(\gamma_1 - 1  \right) 
	 }
	 {
	   6 \left(2 \alpha_1  \beta_1  + \alpha_3  \beta_3 \right)^2
	 }
\end{eqnarray}

% -------------------------------------------------------------

\begin{eqnarray}
	\left\langle
	{ {\partial^3 V}
		\over 
		{
			\partial \left( {\boldsymbol F_0}\right)_4
			\partial  \left( {\boldsymbol \eta_0}\right)_4  
			\partial \left( {\boldsymbol \eta_0}\right)_4 }} 
	\right\rangle_0 
	&=&
\frac{
	  16\alpha_3^3 c_3 \xi^{2} 
	  \left(\gamma_1 - 1  \right)^2 
	 }
	 {
	   \left(2 \alpha_1 \beta_1   +\alpha_3 \beta_3 \right)^3
	 }
\end{eqnarray}

% -------------------------------------------------------------

\begin{eqnarray}
	\left\langle
	{ {\partial^3 V}
		\over 
		{
			\partial \left( {\boldsymbol F_0}\right)_4
			\partial  \left( {\boldsymbol \eta_0}\right)_4  
			\partial \left( {\boldsymbol \eta_0}\right)_5 }} 
	\right\rangle_0 
	&=&
\frac{
	\alpha_3^2 h_0^3   
	\left(\xi  - 1  \right) 
	\left(\gamma_1 - 1  \right) 
}
{
	6 \left(2 \alpha_1 \beta_1   +\alpha_3 \beta_3 \right)^2
}
\end{eqnarray}

% -------------------------------------------------------------

\begin{eqnarray}
	\left\langle
	{ {\partial^3 V}
		\over 
		{
			\partial \left( {\boldsymbol F_0}\right)_4
			\partial  \left( {\boldsymbol \eta_0}\right)_5  
			\partial \left( {\boldsymbol \eta_0}\right)_5 }} 
	\right\rangle_0 
	&=& 0
\end{eqnarray}

% -------------------------------------------------------------
% -------------------------------------------------------------

\begin{eqnarray}
	\left\langle
	{ {\partial^3 V}
		\over 
		{
			\partial \left( {\boldsymbol F_0}\right)_5
			\partial  \left( {\boldsymbol \eta_0}\right)_1  
			\partial \left( {\boldsymbol \eta_0}\right)_1 }} 
	\right\rangle_0 
	&=& 
4 u_1 h_0  - 4 \beta_3  u_4 
\end{eqnarray}

\begin{eqnarray}
	\left\langle
	{ {\partial^3 V}
		\over 
		{
			\partial \left( {\boldsymbol F_0}\right)_5
			\partial  \left( {\boldsymbol \eta_0}\right)_1  
			\partial \left( {\boldsymbol \eta_0}\right)_2 }} 
	\right\rangle_0 
	&=& 
-4 \sqrt{2} \beta_1 u_4 
\end{eqnarray}

\begin{eqnarray}
	\left\langle
	{ {\partial^3 V}
		\over 
		{
			\partial \left( {\boldsymbol F_0}\right)_5
			\partial  \left( {\boldsymbol \eta_0}\right)_1  
			\partial \left( {\boldsymbol \eta_0}\right)_3 }} 
	\right\rangle_0 
	&=& 
- 4 \alpha_3 u_4
\end{eqnarray}

\begin{eqnarray}
\left\langle
{ {\partial^3 V}
	\over 
	{
		\partial \left( {\boldsymbol F_0}\right)_5
		\partial  \left( {\boldsymbol \eta_0}\right)_1  
		\partial \left( {\boldsymbol \eta_0}\right)_4 }} 
\right\rangle_0 
&=& 
- 4 \sqrt{2} \alpha_1 u_4 
\end{eqnarray}

\begin{eqnarray}
\left\langle
{ {\partial^3 V}
	\over 
	{
		\partial \left( {\boldsymbol F_0}\right)_5
		\partial  \left( {\boldsymbol \eta_0}\right)_1  
		\partial \left( {\boldsymbol \eta_0}\right)_5 }} 
\right\rangle_0 
&=& 
\frac{  
	   \sqrt{2} h_0^2  \left(1 - \xi \right)
	   \left[
	         \alpha_1 \beta_1  
	         \left(1 + \gamma_1 \right) 
	         +\alpha_3 \beta_3 \gamma_1   
	   \right] 
	  }
	  {  
	  	2 \alpha_1 
	  	\left(2 \alpha_1  \beta_1  + \alpha_3  \beta_3 \right)
	  	  }
\end{eqnarray}

\begin{eqnarray}
	\left\langle
	{ {\partial^3 V}
		\over 
		{
			\partial \left( {\boldsymbol F_0}\right)_5
			\partial  \left( {\boldsymbol \eta_0}\right)_2  
			\partial \left( {\boldsymbol \eta_0}\right)_2 }} 
	\right\rangle_0 
	&=& 
4 u_1 h_0 
\end{eqnarray}

\begin{eqnarray}
	\left\langle
	{ {\partial^3 V}
		\over 
		{
			\partial \left( {\boldsymbol F_0}\right)_5
			\partial  \left( {\boldsymbol \eta_0}\right)_2  
			\partial \left( {\boldsymbol \eta_0}\right)_3 }} 
	\right\rangle_0 
	&=& 
- 4 \sqrt{2} \alpha_1 u_4  
\end{eqnarray}

\begin{eqnarray}
\left\langle
{ {\partial^3 V}
	\over 
	{
		\partial \left( {\boldsymbol F_0}\right)_5
		\partial  \left( {\boldsymbol \eta_0}\right)_2  
		\partial \left( {\boldsymbol \eta_0}\right)_4 }} 
\right\rangle_0 
&=& 
0
\end{eqnarray}

\begin{eqnarray}
\left\langle
{ {\partial^3 V}
	\over 
	{
		\partial \left( {\boldsymbol F_0}\right)_5
		\partial  \left( {\boldsymbol \eta_0}\right)_2  
		\partial \left( {\boldsymbol \eta_0}\right)_5 }} 
\right\rangle_0 
&=& 
\frac{  
	 h_0^2  \left(1 - \xi \right)
	\left[
        	2 \alpha_1 \beta_1  \gamma_1  + \alpha_3 \beta_3    
	\right] 
}
{  
	2 \alpha_3 
	\left(2 \alpha_1  \beta_1  + \alpha_3  \beta_3 \right)
}
\end{eqnarray}

\begin{eqnarray}
	\left\langle
	{ {\partial^3 V}
		\over 
		{
			\partial \left( {\boldsymbol F_0}\right)_5
			\partial  \left( {\boldsymbol \eta_0}\right)_3  
			\partial \left( {\boldsymbol \eta_0}\right)_3 }} 
	\right\rangle_0 
	&=& 
4 h_0 u_3 
\end{eqnarray}

\begin{eqnarray}
	\left\langle
	{ {\partial^3 V}
		\over 
		{
			\partial \left( {\boldsymbol F_0}\right)_5
			\partial  \left( {\boldsymbol \eta_0}\right)_3  
			\partial \left( {\boldsymbol \eta_0}\right)_4 }} 
	\right\rangle_0 
	&=& 
0
\end{eqnarray}

\begin{eqnarray}
	\left\langle
	{ {\partial^3 V}
		\over 
		{
			\partial \left( {\boldsymbol F_0}\right)_5
			\partial  \left( {\boldsymbol \eta_0}\right)_3  
			\partial \left( {\boldsymbol \eta_0}\right)_5 }} 
	\right\rangle_0 
	&=& 
\frac{  
	\sqrt{2} \alpha_1  h_0^2  \left(1 - \xi \right) 
	\left(\gamma_1 - 1 \right)
}
{  
	2  
	\left(2 \alpha_1  \beta_1  + \alpha_3  \beta_3 \right)
}
\end{eqnarray}

\begin{eqnarray}
	\left\langle
	{ {\partial^3 V}
		\over 
		{
			\partial \left( {\boldsymbol F_0}\right)_5
			\partial  \left( {\boldsymbol \eta_0}\right)_4
			\partial \left( {\boldsymbol \eta_0}\right)_4 }} 
	\right\rangle_0 
	&=& 
	4 h_0 u_3 
\end{eqnarray}

\begin{eqnarray}
	\left\langle
	{ {\partial^3 V}
		\over 
		{
			\partial \left( {\boldsymbol F_0}\right)_5
			\partial  \left( {\boldsymbol \eta_0}\right)_4
			\partial \left( {\boldsymbol \eta_0}\right)_5 }} 
	\right\rangle_0 
	&=& 
\frac{  
	\alpha_3  h_0^2  \left(1 - \xi \right) 
	\left(\gamma_1 - 1 \right)
}
{  
	2 
	\left(2 \alpha_1  \beta_1  + \alpha_3  \beta_3 \right)
}
\end{eqnarray}

\begin{eqnarray}
	\left\langle
	{ {\partial^3 V}
		\over 
		{
			\partial \left( {\boldsymbol F_0}\right)_5
			\partial  \left( {\boldsymbol \eta_0}\right)_5
			\partial \left( {\boldsymbol \eta_0}\right)_5 }} 
	\right\rangle_0 
	&=& 
4 h_0 u_6.
\end{eqnarray}

\end{document}